\newif\ifPDFLaTeX
\expandafter\ifx\csname pdfoutput\endcsname\relax\else\PDFLaTeXtrue\fi

\ifPDFLaTeX
  \documentclass[10pt,a4paper,notitlepage,chapters,openany]{book}
  \usepackage{type1cm}
 \usepackage[pdftex,colorlinks]{hyperref}
  \usepackage[pdftex]{graphicx,feynmp}
   \usepackage{hypernat}
\else
\documentclass[a4paper,10pt,openany,oneside]{book}
\usepackage{graphicx,feynmp}
\fi

\newlength\graphwidth
\label{math}

\usepackage{amsmath,amssymb,amsthm,mathrsfs}
\usepackage{url,thophys,thohacks}
\usepackage[numbers,square,sort&compress]{natbib}

\usepackage[plain]{fancyref}
\allowdisplaybreaks

\DeclareMathOperator{\tr}{tr}

\newcommand{\ii}{\mathrm{i}}

\renewcommand{\Re}{\text{Re}}
\renewcommand{\Im}{\text{Im}}

\newcommand{\Tr}{ {\rm Tr} }
\newcommand{\dmu}{\partial_\mu}

\newcommand{\admu}{\overleftrightarrow\dmu}
\newcommand{\deltabrs}{\delta_{\text{BRS}}}
\newcommand{\gz} {\left(\scriptstyle{\frac{1}{2}-
    \sin^2\theta_w}\right)}
\newcommand{\Out}{\text{out}}
\newcommand{\In}{\text{in}}
\newcommand{\onepi}[1]{\Braket{#1}^{1\text{PI}}}
\newcommand{\me}{\mathcal{M}}
\newcommand{\fmfbrs}[1]{\fmfv{decor.shape=square,decor.filled=empty,decor.size=5}{#1}}

\newcommand\OMEGA{O'Mega}

\newcommand{\textbooks}{Itzykson:1980,Peskin:1995,Weinberg:1995,Kugo:1997,Boehm:2001,Cheng:1984}
\newcommand{\omegacite}{\OMEGA \cite{Moretti:2001,OMega}}
\newcommand*{\fancyrefpartlabelprefix}{part}
\newcommand*{\frefpartname}{part}
\frefformat{plain}{\fancyrefpartlabelprefix}{%
        \frefpartname \fancyrefdefaultspacing#1%
        }%
\newcommand*{\Frefpartname}{Part}
\Frefformat{plain}{\fancyrefpartlabelprefix}{%
        \Frefpartname \fancyrefdefaultspacing#1%
        }%
\newcommand*{\fancyrefsubeqlabelprefix}{subeq}
\newcommand*{\frefsubeqname}{equations}
\frefformat{plain}{\fancyrefsubeqlabelprefix}{%
        \frefsubeqname \fancyrefdefaultspacing(#1)%
        }%
\newcommand*{\fancyrefapplabelprefix}{app}
\newcommand*{\frefappname}{appendix}
\frefformat{plain}{\fancyrefapplabelprefix}{%
        \frefappname \fancyrefdefaultspacing#1%
        }%
\newcommand*{\Frefappname}{Appendix}
\Frefformat{plain}{\fancyrefapplabelprefix}{%
        \Frefappname \fancyrefdefaultspacing#1%
        }%

\frefformat{vario}{\fancyrefapplabelprefix}{%
        \frefappname \fancyrefdefaultspacing#1\fancyrefdefaultspacing
        \frefonname\fancyrefdefaultspacing\frefpgname\fancyrefdefaultspacing
        #2%
        }%
\newcommand*{\fancyreftheolabelprefix}{theo}
\newcommand*{\freftheoname}{theorem}
\frefformat{plain}{\fancyreftheolabelprefix}{%
        {\it\freftheoname \fancyrefdefaultspacing#1}%
        }%
\frefformat{vario}{\fancyreftheolabelprefix}{%
        {\bf\freftheoname \fancyrefdefaultspacing#1}\fancyrefdefaultspacing
        \frefonname\fancyrefdefaultspacing\frefpgname\fancyrefdefaultspacing
        #2%
        }%
\newcommand*{\fancyrefdeflabelprefix}{def}
\newcommand*{\frefdefname}{definition}
\frefformat{plain}{\fancyrefdeflabelprefix}{%
        {\frefdefname \fancyrefdefaultspacing{\it #1}}%
        }%
\makeatletter
\renewcommand\thesubsubsection{(\@roman\c@subsubsection)}

\makeatother

\newcommand{\hepcite}[1]
{ \ifPDFLaTeX
\href{http://xxx.lanl.gov/abs/#1}{[#1]},
\else [#1]
\fi}

\allowdisplaybreaks
\numberwithin{equation}{section}

\ifPDFLaTeX
\theoremstyle{definition}
\fi

\newtheorem{theorem}{Theorem}[chapter]
\newtheorem{definition}{Definition}[chapter]

\renewcommand{\thanks}[1]{
\def\thefootnote{\fnsymbol{footnote}}\footnote{#1}
\def\thefootnote{\fnsymbol{footnote}}}
\makeatletter
  \@ifundefined{frontmatter}%
    {%
     \def\mainmatter{\cleardoublepage\pagenumbering{arabic}}}
    {}
  \@ifundefined{abstract}%
    {%
      {\newpage\section*{Abstract}\begin{quote}}%
      {\end{quote}\newpage}}
    {}
\makeatother

\begin{document}
\setcounter{tocdepth}{1}
\setcounter{secnumdepth}{3}

\begin{fmffile}{diss_pics}
\fmfset{arrow_len}{2mm}
\let\today\hfil

\thispagestyle{empty}
\begin{center}
{\bf\Large Gauge checks, consistency of approximation schemes 
and numerical evaluation of realistic scattering amplitudes\thanks{Based on PhD
  thesis, Darmstadt University of Technology, June 2003. Advisor: P.Manakos}}\\
\vspace{1cm}
{\large Christian Schwinn\thanks{schwinn@physik.uni-wuerzburg.de}}\\
\vspace{1cm}
{ Institut f\"ur Theoretische~Physik und Astrophysik}\\
{    Universit\"at~W\"urzburg}\\{ Am~Hubland, D-97074~W\"urzburg,
  Germany}\hfil\\\hfil\\
and\\\hfil\\
{ Institut f\"ur Kernphysik}\\ { Darmstadt University of Technology}\\
{Schlo\ss{}gartenstr. 9 D-64289 Darmstadt, Germany}

\vspace{1cm}

hep-ph/0307057
\end{center}

\section*{Abstract}
{
We discuss both theoretical tools to verify gauge invariance in numerical
calculations of cross sections and the consistency of approximation schemes used in realistic calculations. 

A finite set of Ward Identities for 4 point scattering amplitudes is
determined, that is sufficient to verify the correct
implementation of Feynman rules of a spontaneously broken gauge theory in a model independent way. These identities have been implemented in the matrix element generator O'Mega and have been used to verify the implementation of the complete Standard Model in $R_\xi$ gauge. 

The consistency of approximation schemes in tree level
calculations is discussed in the last part of this work.
 We determine the gauge invariance classes of spontaneously broken gauge theories, providing a new proof for the formalism of gauge and flavor flips.

The schemes for finite width effects that have been implemented in O'Mega are
reviewed. As a comparison with existing calculations, we study the consistency of these schemes in the process $e^-e^+\to e^- \bar \nu_e u\bar d$. The violations of gauge invariance caused by the introduction of running coupling constants are  analyzed.}

\tableofcontents
\mainmatter
\chapter{Introduction}
\begin{motto}{Galileo Galiliei}{zitiert nach: \\Albrecht F\"olsing: Galileo Galiliei\\
Proze\ss ohne Ende, Eine Biographie}
Die Philosophie steht in diesem gro\ss{}en Buch geschrieben, dem Universum, das unserem Blick st\"andig offenliegt. Aber das Buch ist nicht zu verstehen, wenn man nicht zuvor die Sprache erlernt und sich mit den Buchstaben vertraut gemacht hat, in denen es geschrieben ist. Es ist in der Sprache der Mathematik geschrieben, (\dots), ohne die es dem Menschen unm\"oglich ist, ein einziges Wort davon zu verstehen; ohne diese irrt man in einem dunklen Labyrinth umher.
\end{motto}

 The agreement between the theoretical
predictions of the  Standard Model (SM) of the electroweak
interactions and experiment is established to an impressive
degree \cite{PDG}. The only missing ingredient is the
Higgs boson 
that yet has to be discovered. However, there are  compelling theoretical reasons for believing that the
electroweak Standard Model is merely a low energy approximation to a more
fundamental theory that should become visible at TeV scale energies. Thus indications on the underlying theory should be found by experiments at the LHC or at  future linear  colliders (see e.g. \cite{Abe:2001}). These future experiments pose the challenge to theorists to make predictions for processes
with many particles in the final state.  This is a general signature for
processes with heavy, unstable particles in intermediate states that have to
be considered to identify the nature of the new physics phenomena. Because the number of  (tree-level) Feynman diagrams
contributing to the scattering amplitude is
growing rapidly with the number of external particles (it can be shown, that
this growth is \emph{factorial} in an unflavored $\phi^3$ theory \cite{Moretti:2001}\nocite{OMega}), it is necessary to perform completely
automatized numerical calculations \cite{Ohl:2002sr}. 

An important example is given by the study of the nature of electroweak
symmetry breaking. Assuming a Higgs boson is found in future experiments,  determining its
properties like the form of the self interaction  and the
Yukawa couplings will require the study of processes with many fermions in the
final state. In \fref{tab:top-higgs} we show the number of diagrams contributing to  associated top-Higgs production that
can be used to measure the top-quark Yukawa coupling.
\begin{table}[htbp]
  \begin{center}
    \begin{tabular}{c|c}
        Process &Diagrams \\
      $e^+e^-\to t \bar t H$ & 5 \\
      $e^+e^-\to t \bar t b \bar b$ & 45 \\  
      $e^+e^-\to  b W^+ \bar b W^- b \bar b $ &  8314\\  
      $e^+e^-\to  b \mu^+\nu_\mu \bar b d \bar u b \bar b $ & 38232  
       \end{tabular}\caption{Associated top-Higgs production}\label{tab:top-higgs}
     \end{center}
\end{table}
While the number of `signal' diagrams is very small, almost forty-thousand  diagrams contribute to the physical final state.

If no light Higgs boson is found, the scattering of longitudinal gauge bosons
can be used to detect signals of  electroweak symmetry breaking by strong
interactions at the TeV scale. The study of
quartic gauge boson scattering involves processes with 6 fermions in the final
states, so again a large number of Feynman diagrams contributes.

As a final example we mention theories with Supersymmetry (SUSY) where 
(assuming
$R$-parity conservation) supersymmetric partner particles can only be produced
in pairs and decay through cascade decays. Numerical examples for the drastic
growth of the number of Feynman diagrams with the number of external particles in the
minimal supersymmetric Standard Model have been
given in \cite{Reuter:2002}. \nocite{ReuterOhl:2002} 

To calculate cross sections for scattering processes with many final state particles in an efficient way, it is mandatory to use a matrix
  element generator that generates compact code without redundancy. The algorithm of the program \OMEGA~(An Optimizing Matrix Element Generator) \cite{Moretti:2001,OMega} solves this problem and suppresses the factorial growth of complexity to an  exponential one. As the second step in the calculation of the cross section  one needs to perform the integration over the phase space of the final state particles, using an automatized phase space generator like WHIZARD \cite{Kilian:WHIZARD}. The calculation of cross sections with more than 4 fermions in the final state is currently limited to tree level precision, the calculation of loop corrections using \OMEGA~and WHIZARD is currently being studied \cite{Kilian:loops}.  

The use of automatized calculation systems also implies the need for automatized
consistency checks of the numerical calculations that ensure both the validity
of the  Feynman rules of the particle physics model  and the numerical stability of the
algorithm. A natural choice for these consistency checks is the use of the Ward Identities associated with the gauge invariance of the particle physics model used in the calculation. Since gauge invariance is intimately connected to 
(tree-level) unitarity \cite{LlewellynSmith:1973,Cornwall:1973},
\nocite{Lee:1977,Horejsi:1994} large numerical errors can be caused by
violations of gauge invariance and it is important to maintain gauge invariance in the numerical
calculations. This is also a challenge to approximation schemes that include higher order effects like finite widths of unstable particles or running coupling constants in effective tree level calculations. 

In this work we discuss both the tools to verify gauge invariance in numerical
calculations of cross sections and the consistency of approximation schemes. 
 In \fref{chap:intro} we will give a more detailed discussion of
the importance of gauge invariance and introduce the Ward Identities that 
express the gauge invariance of physical scattering amplitudes. 
 
In \fref{part:reconstruction} we  determine a \emph{finite}
set of Ward Identities for scattering amplitudes that is sufficient to verify the correct
implementation of Feynman rules of a spontaneously broken gauge theory in a model independent way. As described in \fref{chap:implementation}, these identities have been implemented in
\OMEGA~and have been used to verify the implementation of the complete
Standard Model in $R_\xi$ gauge.  

The problem of the consistency of approximation schemes in tree level
calculations is discussed in
\fref{part:resummation}. The factorial growth of the  number of Feynman diagrams with the number of external particles motivates the search for gauge invariant  subsets of Feynman diagrams. 
 The decomposition of the amplitude into several separately gauge
invariant sets of diagrams has been initiated in 
\cite{Bardin:1994} and a complete classification has been achieved in
\cite{Boos:1999}.  \nocite{Ohl:1999,Ondreka:2000,Ondreka:2003} In
\fref{chap:ssb_groves} we extend the formalism of \cite{Boos:1999} to spontaneously broken gauge theories. 
These results have already appeared in \cite{OS:FGH}.

Approximation schemes also have to be used to include 
finite gauge boson widths and running coupling constants in tree level
calculations. The notorious violations of gauge invariance from inconsistent
prescriptions for finite width effects can lead to errors
of orders of magnitudes \cite{Argyres:1995}. In \fref{chap:width} we review schemes that have been suggested to treat unstable particles in tree level calculations  \cite{LopezCastro:1991,Baur:1991,Baur:1995,Denner:1999,Beenakker:1999} \nocite{Heide:2000,Dittmaier:2002,Beenakker:2003,Beenakker:1997,Aeppli:1994rs,Denner:1996,Papavassiliou:1998,Accomando:1999,Passarino:2000} and that we have implemented in \OMEGA. As a comparison to the literature, we study the consistency of these schemes in the single $W$ production process $e^-e^+\to e^- \bar \nu_e u\bar d$. The violations of gauge invariance caused by the introduction of running coupling constants are analyzed in \fref{chap:running}. 

In \fref{part:tools} we collect necessary tools for the remainder of the work,
including a new diagrammatic proof of the formalism of
\cite{Boos:1999} for gauge invariance classes \cite{OS:FGH} and a identity for
vertex functions with several momentum contractions that to my knowledge
previously hasn't been derived in the literature.

\chapter{Gauge invariance in numerical calculations: tool and challenge }\label{chap:intro}\chaptermark{Gauge invariance: tool and challenge}
In this chapter we review the importance of gauge invariance in
theoretical models in particle physics, stressing both theoretical and
numerical aspects that will become important in later parts of this work. A
main motivation for the formulation of elementary particle physics in terms of
gauge theories is the well known  connection
between gauge invariance and (tree-level-) unitarity \cite{LlewellynSmith:1973,Cornwall:1973} that we briefly review in 
\fref{sec:unitarity}. 

The importance of gauge invariance for unitarity and therefore good high
energy behavior of scattering amplitudes leads naturally to the question how to \emph{check} gauge invariance in numerical calculations of scattering amplitudes. The gauge invariance of a theory manifests itself in the Ward Identities that  are reviewed in \fref{sec:intro-wi}. In \fref{sec:rec-motivate} we address the question how the consistent implementation of a particle physics model can be verified using the Ward Identities. This discussion will be the subject of \fref{part:reconstruction}.

Having discussed the reasons for insisting on gauge invariant calculations, we
turn to the challenges this poses in actual computations of cross sections. In general it is important, to consider \emph{all} Feynman diagrams
contributing to an amplitude to obtain gauge invariant results. Because of the factorial growth of the number of diagrams with the number of external particles, the computation of scattering
amplitudes with many particles in the final state involves thousands of Feynman diagrams. In some cases,
however, the set of diagrams can be decomposed into several, separately gauge
invariant, subsets. In \fref{sec:groves} we sketch the formalism of
\emph{gauge and flavour flips} \cite{Boos:1999}, \nocite{Ohl:bocages,Ondreka:2000,Ondreka:2003} to obtain these so called \emph{groves}. A
derivation of this formalism from the Slavnov-Taylor Identities (STIs)
 is given in
\fref{chap:diagrammatica}. In \fref{chap:ssb_groves} we will discuss the
peculiarities of groves in spontaneously broken gauge theories.

We conclude our survey of challenges posed by gauge invariance in \fref{sec:intro-widths},  sketching
the severe problems caused by finite width effects in realistic calculations involving unstable particles. The subject of maintaining gauge invariance while including higher order effects in  realistic calculations are discussed more throughly in \fref{part:resummation} of this work.

\section{Tree-level unitarity and gauge invariance}\label{sec:unitarity}
In \cite{LlewellynSmith:1973,Cornwall:1973} it was shown that a spontaneously broken gauge theory is the only theory of massive vector bosons satisfying tree-level unitarity. For a didactic derivation of the Standard Model Lagrangian using tree level unitarity see \cite{Horejsi:1994}. 
Tree-level unitarity is defined by the requirement\footnote{
One can go further and demand not only the correct scaling properties of the matrix elements but also that the partial waves are bounded by $1$. Using this approach, an upper bound of the standard-model Higgs mass of $\sim 1$ TeV can be derived \cite{Lee:1977}. We don't consider this approach to tree level unitarity in the following discussion.} that tree level matrix elements for $N$-particle scattering amplitudes scale for high energies at most as $E^{4-N}$:
\begin{equation}\label{eq:unitarity-bound}
     \me(\Phi_{1}\Phi_{2}\to
     \Phi_{3}\dots\Phi_{N})\stackrel{E\to\infty}{\lesssim} E^{4-N}
\end{equation}
To show that this requirements uniquely singles out spontaneously broken gauge theories, one has to consider 4 particle Green's functions and also the 5 particle amplitudes $WW\to HHH$ and $WW\to WWH$ to determine the scalar self interaction. 

The problems concerning unitarity originate from the high energy behavior of the longitudinal polarization vectors of massive gauge bosons that become proportional to the momentum:
\begin{equation}\label{eq:longi_app}
  \epsilon_L= \frac{p}{m_W}+O\left(\frac{m_W}{E}\right)
\end{equation}
Since the triple gauge boson vertices contain one momentum, one expects the matrix element for 4 gauge boson scattering to grow like $E^4$ (including the terms $\propto p^\mu p^\nu$ in the propagator even like $E^6$), violating the bound \eqref{eq:unitarity-bound}. Similarly the matrix element for 2 fermion 2 gauge boson scattering and the matrix element for 2 scalar - 2 gauge boson scattering can expected to diverge quadratically. For fermions the fact that polarization spinors scale like  $\sqrt E$ has to be used to arrive at that result. 

As a simple example we consider the process $\bar f_i f_j \to W_a W_b$ in a toy model  of massive fermions and gauge bosons
\begin{multline}\label{eq:ym-toy}
 \mathscr{L}=-\frac{1}{4}(\dmu W_{\nu a}-\partial_\nu W_{\mu a})(\partial^\mu W^\nu_a-\partial^\nu W^\mu_a)+m_a^2 W_a^2 +\bar\psi_i(\ii\fmslash\partial-m_i)\psi_i\\
+\bar\psi_i\fmslash W_a\tau^{a}_{ij}\psi_j-f^{abc}W_{b\mu} W_{c\nu}\partial^\mu
  W_a^\nu-\frac{1}{4}g_{W^4}^{abcd}(W_a\cdot W_b)(W_b\cdot W_c)
\end{multline}
where we don't suppose that the couplings are of Yang-Mills form.

We will show, that  the bad high energy behavior associated with longitudinal gauge bosons can only be avoided if the couplings are of the Yang Mills type. 

At energies much larger that the masses of all particles, we can drop the fermion masses and use the approximation
\begin{equation}
  \epsilon_L\approx \frac{p}{m_W}
\end{equation}
for the longitudinal polarization vector of the massive gauge bosons. 

A calculation similar to the evaluation of the $\bar f_i f_j \to W_a W_b$ Ward Identity in \fref{app:ffww-wi} gives the result
\begin{multline}
  \epsilon_L^{\mu_a}\epsilon_L^{\mu_b}\me_{\mu_a\mu_b}(\bar f_i f_j
  W_aW_b)
\approx \frac{1}{m_{W_a}m_{W_b}}\left(\bar v_i \fmslash p_a
  (\lbrack\tau^a,\tau^b\rbrack _{ij}u_j\right)\\
-\frac{(p_a+p_b)^2}{m_{W_a}m_{W_b}}f^{abc}\frac{1}{(p_a+p_b)^2-m_{W_c^2}}\left(\bar v_i \fmslash p_a \tau_{ij}^c u_j\right)
\end{multline}
As expected, this diverges quadratically with $E$ for $E\to\infty$.

If the $f^{abc}$ are the structure constants of the Lie algebra of the $\tau$, this can be simplified to 
\begin{equation}
  \epsilon_L^{\mu_a}\epsilon_L^{\mu_b}\me_{\mu_a\mu_b}(\bar f_i f_j
  W_aW_b)
\approx-\frac{f^{abc} }{m_{W_a}m_{W_b}}\left(\bar v_i\fmslash p_a \tau_{ij}^c u_j\right) \frac{m_{W_c}^2}{(p_a+p_b)^2-m_{W_c}^2}
\end{equation}
so the dangerous divergence is removed. 

If we don't neglect the masses, however, a linear divergence remains that has to be canceled by including Higgs bosons with the appropriate couplings (see e.g. \cite{Horejsi:1994}).  In  \cite{Gunion:1991} the constraints on the Higgs-gauge boson couplings arising from tree level unitarity were used as 'Higgs sum rules' and phenomenological applications of these relations in constructing non-minimal  Higgs models are reviewed in \cite{Gunion:1989}.
\sectionmark{Ward Identities}
\section{Consequences of gauge invariance: Ward Identities}\sectionmark{Ward Identities}\label{sec:intro-wi}
According to  the Noether
theorem, the symmetry of a theory implies the existence of a conserved current. The consequences of current conservation for the Green's functions in quantum field theory are the so called \emph{Ward Identities.} We will discuss their form first for the case of global symmetries in \fref{sec:global:wi} before turning to QED in \fref{sec:sti-intro}. We then briefly review Ward Identities in unbroken and spontaneously broken nonabelian theories, leaving a more detailed discussion to \fref{chap:sti}.
\subsection{Global Symmetries}\label{sec:global:wi}
We begin our discussion of Ward Identities by considering a global infinitesimal transformation of the fields
\begin{equation}
\Phi\to\Delta\Phi
\end{equation}
that leaves  the  action  $S[\Phi]$  invariant. This invariance manifests itself in relations among the observables of the
theory. As a consequence of the symmetry, the scattering
matrix elements of
different fields are related by the identity 
\begin{equation}\label{eq:global-me-wi}
  0=\Delta\left(\me(\Phi_1(p_1)\dots\dots \Phi_n(p_n))\right)=\sum_{\Phi_i} \me(\Phi_1(p_1)\dots\Delta\Phi_j(p_j)\dots \Phi_n(p_n))
\end{equation}
This relation is in fact a special
case of a more general identity that involves  the conserved Noether current $j^\mu$.  Either from canonical
commutation relations \cite{Weinberg:1995,Itzykson:1980} or using path
integral methods  \cite{Peskin:1995,Boehm:2001} \nocite{\textbooks} one can derive  the so called \emph{Ward Identity} for
insertions of the current into Green's functions:
\begin{multline}
  \label{eq:global-wi}
  \dmu^x\Greensfunc{j^\mu(x)\Phi_1(y_1)\dots\Phi_n(y_n)}\\
=\sum_i\Greensfunc{\Phi_i(y_1)\dots\Delta\Phi_i(y_i)\dots \Phi_n(y_n)}\delta^4(x-y_i)
\end{multline}

Since the left hand side of \fref{eq:global-wi} is a total derivative, it drops out if this equation
is integrated  over the whole of spacetime. Applying the LSZ formula to get
scattering matrix elements we then obtain \fref{eq:global-me-wi}\footnote{For two reasons, this argument goes only through if
the symmetry is unbroken: The integral over the current cannot be performed,
since the charge $Q=\int d^3 x j_0(x)$ doesn't exist for spontaneously broken
symmetries. Also, the masses of the particles in a broken multiplet need not
be equal. Therefore the Green's functions on the right hand side of
\fref{eq:global-wi} may have a different pole structure and we can't amputate
all Green's functions with the LSZ formula.}.

A symmetry of the theory imposes also conditions on the irreducible vertices
of the theory. To obtain these relations, we note that the invariance of the
action under symmetry transformations leads to the condition
\begin{equation}
\int d^4x\sum_{\phi} \Delta\Phi \,\frac{\delta S}{\delta\Phi}=0
\end{equation}
It can be shown that in theories without anomalies the effective action
satisfies the same identity \cite{Piguet:1981}, called the Ward Identity of
the effective action:
\begin{equation}\label{eq:wi-irr}
\int d^4x \sum_{\Phi}\Delta\Phi_{cl} \frac{\delta \Gamma}{\delta\Phi_{cl}}=0
\end{equation}
Since the effective action is the generating functional for the irreducible
vertices (some properties of the effective action are reviewed in \fref{sec:sti_irr}) one can derive the Ward Identities for the irreducible vertices by taking
derivatives of this equation with respect to a suitable set of classical
fields.

\subsection{Quantum electrodynamics}\label{sec:sti-intro}
Since a massless vector particle like the photon has 2 degrees of freedom, the
description in terms of a vector field with 4 components is redundant. The
only consistent treatment of this redundancy is to insist on a gauge invariant
coupling of photons to matter \cite{Weinberg:1995}. On the quantum level, the
decoupling of the unphysical degrees of freedom is guaranteed by the Ward Identities of the theory.

The Ward Identity for the insertion of currents into Green's functions has the same form as in the
case of global symmetries \cite{Itzykson:1980,Peskin:1995,Boehm:2001}. From \fref{eq:global-wi} we obtain, specializing to
$U(1)$ transformations:
\begin{multline}\label{eq:qed-wti}
\partial^x_\mu\Greensfunc{j^\mu(x)\psi(x_1)\bar\psi(y_1)\dots A_{\mu_n}(z_n) }\\
= e\Greensfunc{\psi(x_1)\bar\psi(y_1)\dots A_{\mu_n}(z_n)
  }\sum_i\left(\delta^4(x-y_i) -\delta^4(x-x_i)\right)
\end{multline}
The most famous special case of \fref{eq:qed-wti} is the original Ward Identity that provides a connection between the electron self-energy $\Sigma$ and the electron-photon vertex function $\Lambda$ : 
\begin{equation}\label{eq:ward}
  \Gamma^{e \bar e A}_\mu(p,p,0)=\frac{\partial}{\partial p^\mu}\Sigma(p)
\end{equation}
Like global symmetries the gauge symmetry implies a Ward Identity of the effective action. Inserting the gauge transformations of the fields into \fref{eq:wi-irr} and taking the derivative with respect to the gauge fixing parameter, we find that the effective action of QED satisfies the
Ward Identity (neglecting the effects of gauge fixing) 
\begin{equation}\label{eq:qed-sti}
 \dmu\frac{\delta \Gamma}{\delta A_\mu (x)}+\ii e
\frac{\Gamma\overleftarrow\delta}{\delta\psi(x)}\psi(x)-\ii e\bar\psi(x)\frac{\delta\Gamma}{\delta\bar\psi}=0
\end{equation}
Differentiating with respect to  $\psi$ and $\bar \psi$ gives an identity for
the electron photon vertex that can be used to give another derivation of \fref{eq:ward}.
\subsection{Yang-Mills-Theory}\label{sec:ym-wi}
The above discussion glossed over an important point that can be circumvented
in QED but makes the derivation of the STIs, the
identities corresponding to the Ward Identity  in nonabelian gauge theories, much more
complicated. Because of  gauge fixing, the Lagrangian used in
perturbative calculations is in fact \emph{not} gauge invariant and the naive
current conservation breaks down. In QED the gauge transformation of the gauge
fixing term is  noninteracting so this can be ignored in most cases. This is not possible in nonabelian gauge theories and new methods have to be used, including the introduction of ghost fields \cite{Feynman:1963,Faddeev:1967}. 

The original treatment of the STIs involved either complicated diagrammatic arguments \cite{'tHooft:1972} or nonlocal gauge transformations \cite{Taylor:1971}. 
 For a systematic discussion, the discovery of a  new unbroken global symmetry of the gauge fixed Lagrangian, the so called BRS (Becchi-Rouet-Stora) 
symmetry  \cite{Becchi:1975} was essential.

The BRS symmetry is a \emph{global} symmetry with a \emph{fermionic}, nilpotent generator $Q$. The BRS transformation of physical fields is obtained by replacing the gauge transformation parameter $\omega_a(x)$ by  a constant Grassmann number $\epsilon$ times a Faddeev Popov ghost field $c_a$, For fields with a linear transformation law under gauge transformations
\begin{equation}
\delta\Phi_i = \omega_a T^a_{ij}\Phi_j
\end{equation}
the BRS Transformation therefore reads
\begin{equation}
\Delta_{\text{BRS}}\Phi_i\equiv\epsilon\deltabrs\Phi_i\equiv\lbrack\ii\epsilon Q,\Phi_i\rbrack=\epsilon c_a T^a_{ij}\Phi_j
\end{equation}
The transformation laws of the gauge fields and ghosts are given in \fref{app:swi}. For the derivation of the Ward Identities, the transformation law of the antighost\footnote{We have used an equation of motion and we work in a general linear t'Hooft gauge.} 
\begin{equation}
\deltabrs \bar c_a =-\frac{1}{\xi}\dmu G_a^\mu
\end{equation}
is important. Here $G_a^\mu$ is the gluon field. Since physical states in the BRS formalism satisfy the so called \emph{Kugo-Ojima relation}
\begin{equation}\label{eq:kugo}
Q\ket{\In}=0\qquad \bra{\Out}Q=0
\end{equation}
we obtain the identity:
\begin{equation}\label{eq:qcd-wi}
\braket{\Out|\dmu G_a^\mu|\In}=0
\end{equation}
called Ward Identity, like in QED.

The STI for off-shell Green's functions, corresponding to \fref{eq:qed-wti}, and the generalization of the Ward Identity \eqref{eq:qed-sti}, the so called Zinn-Justin (ZJ)-equation \cite{Zinn-Justin:1974}, are more complicated than in QED. These identities are reviewed in \fref{chap:sti}, in this introduction we discuss only the simple Ward Identity for on-shell scattering amplitudes. 
\subsection{Massive vector bosons}\label{sec:sbgt-wi}
Since explicit mass terms for vector bosons break gauge invariance, Yang Mills
theories cannot be straightforwardly generalized to massive vector bosons. Naive
theories for massive vector bosons suffer from nonrenormalizability and
unitarity-violation. 

The only consistent way to describe massive vector bosons is through a spontaneously broken gauge theory. We will discuss only spontanteous symmetry breaking  by fundamental scalars in the main part of this work. Here the symmetry group $G$ of the Lagrangian is broken to a subgroup $H$ that leaves the vacuum expectation value of some scalar fields invariant. The gauge bosons associated to the broken generators of $G$ acquire a mass. For every massive gauge boson, a so called Goldstone boson $\phi$ gets `eaten', i.e. it becomes unphysical and provides the longitudinal degree of freedom for the gauge boson. 

The Goldstone bosons are connected
to the massive  gauge bosons by the identity\footnote{This simple form is only valid on tree level, loop corrections have been considered in \cite{Yao:1988}} \cite{Cornwall:1973,Lee:1977,Chanowitz:1985}
\begin{equation}\label{eq:gf-wi}
-\ii k_\mu\me^\mu(\In+W\to\Out)
=m_w\me(\In+\phi\to\Out) 
\end{equation}
called (like in QED) `Ward Identity' in the following. We sketch the
derivation of \fref{eq:gf-wi} from BRS-invariance briefly in \fref{sec:sti-gf}.

The Ward Identity \eqref{eq:gf-wi} can be generalized to several gauge bosons contracted with a momentum. For two contracted gauge bosons the Ward Identity reads
\begin{multline}\label{eq:chanowitz2}
(-\ii)^2 p_a^\mu p_b^\nu\me_{\mu\nu}(\In +W_aW_b \to \Out)-m_{W_a}m_{W_b}\me(\In +
\phi_a\phi_b\to \Out)\\
+\ii m_{W_a}p_b^\mu\me_\mu(
\In +\phi_a W_b\to \Out )+\ii m_{W_b}p_a^\mu\me_\mu(\In +W_a\phi_b \to \Out)=0
\end{multline}
Similar identities can be obtained for more contractions. 

Since for high energies the longitudinal gauge boson vector approaches the momentum  (see \fref{eq:longi_app}), the \emph{exact} Ward Identity\eqref{eq:chanowitz}  can be used to derive the \emph{approximate} formula\footnote{There are some modifications of this simple identity on loop level \cite{Yao:1988}, in effective field theories \cite{Veltman:1990} and for off shell gauge bosons \cite{Schwinn:2000}.}
\begin{equation}\label{eq:et}
\epsilon_{L}^\mu\me_{\mu}(\In+ W \to
  \Out)
= \ii\me(\In+ \phi \to
  \Out)+O\left(\frac{m_W}{E}\right)
\end{equation}
called the Goldstone boson equivalence theorem
\cite{Chanowitz:1985}
(see also \cite{Peskin:1995,Boehm:2001}). A similar  relation holds for more
than one gauge boson. This theorem can be used to express \emph{physical}
scattering amplitudes for longitudinal gauge bosons approximately by simpler
matrix elements for Goldstone bosons.

\section{Ward Identities as tool in numerical calculations}\label{sec:rec-motivate}\sectionmark{Ward Identities in numerical calculations}
Because violations of gauge invariance can result in numerical instabilities associated with violation of unitarity, it is very important to verify gauge invariance in large scale numerical calculations of cross sections. The Ward Identities resulting from gauge invariance are a natural tool for such consistency checks. 

In numerical calculations, it is important to separate violations of the Ward Identities caused by numerical instabilities from those originating from  wrong implementation of the Feynman rules. One would like to restrict the gauge checks to on-shell scattering amplitudes since those are the natural objects appearing in tree level calculations. Furthermore, this allows to use the simple Ward Identities \eqref{eq:gf-wi} that  provide universal, model independent checks, while the identities for off -shell Green's functions (like the QED Ward Identity \eqref{eq:qed-wti} and the STIs discussed in \fref{chap:sti}) depend on the gauge
group and the representations. Also the complete ghost Lagrangian is needed to check the STIs and one prefers to postpone the introduction of  ghosts as long as possible. This suggests to determine a \emph{finite} set of on-shell Ward Identities that are sufficient to verify the Feynman rules of the theory.

\subsection{Reconstruction of the Feynman rules?}
It is a well known textbook  example \cite{Peskin:1995,Kugo:1997,Cheng:1984} that the Ward Identity for the 2 quark 2 gluon amplitude in Yang Mills theory requires the existence of a 3 gluon vertex with the coupling constants given by the structure constants of the gauge group. 

Evaluating the diagrams contributing to the Ward Identity \eqref{eq:qcd-wi} for the $\bar q q \to g g $ scattering amplitude, using the Feynman rules from \fref{eq:ym-toy} and setting the gauge boson masses to zero,  gives (the calculation is done in a more general case in \fref{app:ffww-wi}) for the two Compton diagrams:
\begin{equation}\label{eq:compton}
\parbox{26mm}{\fmfframe(2,2)(2,4){
\begin{fmfchar*}(20,15)
  \fmfleft{f1,f2} \fmfright{A,H} \fmf{fermion,label=$q_j$}{f1,a}
  \fmf{fermion,label=$\bar q_i$, label.side=left}{b,f2}
  \fmf{fermion,label=$q_l$}{a,b} \fmf{photon}{A,a} \fmf{photon}{H,b}
 \fmfdot{a}
  \fmfdot{b}  
  \fmfv{label=$p_a^\mu$,la.di=0.1}{A}
  \fmfv{label=$\epsilon_b$,la.di=0.1}{H}
 \end{fmfchar*}}}+
\parbox{26mm}{\fmfframe(2,2)(2,4){
\begin{fmfchar*}(20,15)
  \fmfleft{f1,f2} \fmfright{A,H} \fmf{fermion,label=$q_j$}{f1,a}
  \fmf{fermion,label=$\bar q_i$, label.side=left}{b,f2}
  \fmf{fermion,label=$q_l$}{a,b} 
\fmf{phantom}{A,a} \fmf{phantom}{H,b}\fmffreeze 
  \fmf{photon}{A,b} \fmf{photon}{H,a}
 \fmfdot{a}
  \fmfdot{b}  
 \fmfv{label=$p_a^\mu$,la.di=0.1}{A}
  \fmfv{label=$\epsilon_b$,la.di=0.1}{H}
\end{fmfchar*}}}
=\bar v_i(p_i)\fmslash\epsilon_b[\tau^a, \tau^b]_{ij}u_j(p_j)
\end{equation}
In contrast to the case of QED, the sum of the Compton diagrams doesn't satisfy the Ward Identity by itself since the commutator $[\tau^a, \tau^b]$ is nonzero in a nonabelian theory. From the Lagrangian \eqref{eq:ym-toy} there is another diagram involving the triple gauge boson vertex:
\begin{equation}
\parbox{26mm}{\fmfframe(2,2)(2,4){
\begin{fmfchar*}(20,15)
  \fmfleft{f1,f2} \fmfright{A,H} \fmf{fermion,label=$q_j$,label.side=left}{f1,a}
  \fmf{fermion,label=$\bar q_i$, label.side=left}{a,f2}
  \fmf{photon,label=$G_c$}{a,b} \fmf{photon}{A,b} \fmf{photon}{H,b}
 \fmfdot{a}
 \fmfdot{b}  
 \fmfv{label=$p_a^\mu$,la.di=0.1}{A}
  \fmfv{label=$\epsilon_b$,la.di=0.1}{H}
\end{fmfchar*}}}
=-if^{abc}\bar v_i(p_i)\fmslash\epsilon_b\tau^c_{ij}u_j(p_j)
\end{equation}
Therefore \emph{imposing} the Ward Identity \eqref{eq:qcd-wi} on the $qq\to g g $ scattering amplitude implies that the quark-gluon coupling matrices $\tau$ and the triple gauge boson coupling $f^{abc}$ must be connected via the Lie algebra
\begin{equation}
[\tau^a, \tau^b]=\ii f^{abc}\tau^c
\end{equation}
Similarly, demanding that the 4 gluon scattering amplitude satisfies the Ward Identity \eqref{eq:qcd-wi}
implies that the $f^{abc}$ satisfy the Jacobi Identity
\begin{equation}
f^{abe}f^{cde}+f^{cae}f^{bde}+f^{ade}f^{bce}=0
\end{equation}
and that the quartic gauge boson coupling is given by
\begin{equation}
g_{W^4}^{abcd}=f^{abe}f^{cde}
\end{equation}
This calculation is presented in \fref{app:wwww-wi} for the case of spontaneously broken gauge theories.

Therefore we have `reconstructed' the Yang-Mills structure of the Lagrangian \eqref{eq:ym-toy} by imposing the Ward Identity \eqref{eq:qcd-wi} on the 4 point amplitudes of the theory.
  
Unfortunately this result can not be easily extended to a spontaneously broken gauge theory. The difficulties result from the unphysical nature of the Goldstone bosons. To be more specific, the origin of the complications is the  inhomogeneous term in the BRS transformation law of the Goldstone bosons:
\begin{equation}
\deltabrs\phi_a=-m_{W_a}c_a+\dots 
\end{equation}
 If we consider the STI for an physical amplitude with insertions of an unphysical gauge boson and an additional Goldstone boson:
\begin{multline}\label{eq:gold-sti}
0=\deltabrs\Braket{\Out|\bar c_a \phi_b|\In}\\
=\Braket{\Out|(\dmu W_a-
 \xi m_w\phi_a)\phi_b|\In}+m_w\braket{\Out|\bar c_a c_b|\In}+ \cdots
\end{multline}
we see that both terms have a pole at the unphysical Goldstone boson mass that survives amputation with the LSZ formula. To evaluate the second term, we must use the ghost Lagrangian. Since ghosts don't appear in tree-level calculations of physical amplitudes, one would like to avoid the use of \fref{eq:gold-sti}.  

As we will show in \fref{part:reconstruction} of this work, gauge checks of 4 point functions without external Goldstone bosons are sufficient to verify the Feynman rules (apart from quartic Higgs selfcouplings) \emph{as long as the Ward Identities  \eqref{eq:chanowitz2} with up to 4  contractions are used}. 

As sketched in \fref{sec:unitarity}, the Feynman rules spontaneously broken gauge theories can also be reconstructed from tree level unitarity of 4 and 5 point functions.
 Unfortunately, the use of \fref{eq:unitarity-bound} as check in numerical calculations requires to determine the energy dependence of scattering amplitudes at very high energies, while the Ward Identities are valid \emph{at every point} in phase space. This allows to perform gauge checks at the matrix element level, independent of phase space integration.
\subsection{Numerical checks of Ward identities}\label{sec:numerical-wi}
We will now  sketch some aspects of numerical checks of Ward Identities and our implementation in the matrix element generator \omegacite. \OMEGA~is especially suited for the implementation of gauge checks since it uses gauge invariant subamplitudes to construct the scattering amplitude. Therefore the simple Ward Identity \eqref{eq:gf-wi} can be used for gauge checks in \OMEGA~in a very natural way. Gauge invariance can be checked for every \emph{internal} gauge boson contributing to the amplitude. The architecture of \OMEGA~and the implementation of the Ward Identities  is discussed in more detail in \fref{chap:implementation}.

Other matrix element generators that use Feynman diagram as building blocks like \texttt{CompHEP} \cite{Pukhov:1999} allow checks of Ward Identities only for external gauge bosons. Another approach is to verify the BRS invariance of the Feynman rules directly \cite{Semenov:2002}, but this makes it necessary to use the explicit form of the BRS transformation so no model independent checks are possible. The purely numerical algorithm of \texttt{Alpha}  \cite{Caravaglios:1995} works also with gauge invariant subamplitudes but it is the symbolic nature of \OMEGA~that makes a transparent implementation of gauge checks possible.

According to the results discussed in \fref{sec:rec-motivate}, the Ward Identities for 4 point amplitudes with on-shell particles are sufficient to reconstruct the Feynman rules
of a spontaneously broken gauge theory, as long as the  identities with several contractions \eqref{eq:chanowitz2} are considered. The implementation described in \fref{sec:omega-wi} allows to verify \fref{eq:chanowitz2} for external gauge bosons, but checks of internal gauge bosons are not possible in this case. 

The numerical checks of the Ward Identities \eqref{eq:gf-wi} and \eqref{eq:chanowitz2} have proved very useful in debugging the implementation of the Standard Model Feynman rules in \OMEGA. The sensitivity of these gauge checks is so large that relative errors in coupling constants as small as $\mathcal{O}(10^{-8})$ can be detected.

\section{Gauge invariance classes}\label{sec:groves}
 Since the number of diagrams in processes with a large number of external
 particles will be rather huge, one would like to select some \emph{signal}
 diagrams that dominate over the remaining \emph{background} diagrams  because
 of their pole structure. However, in general \emph{all} Feynman diagrams contributing to
 a Green's function must be considered to obtain a gauge invariant expression, satisfying
 the Ward Identity \eqref{eq:qcd-wi}. Nevertheless, in some cases it is possible to find
 separately gauge invariant subsets of Feynman diagrams, the so called `gauge invariance
 classes' or `groves'. After a classifications of gauge invariance classes in 4
 fermion production processes  \cite{Bardin:1994}, a systematic procedure to construct this subsets
 using  the formalism of `gauge and flavor flips' has been found in \cite{Boos:1999}. Before we introduce this formalism, we will first describe some simple examples in QED and QCD. 
\subsection{QED}\label{sec:qed-groves}
As a first, very simple example, consider Bhabbha-scattering in QED. Here a $s$- and a $t$-channel  diagram contribute:
\begin{equation}\label{eq:bhabbha}
\parbox{20mm}{
\begin{fmfchar*}(20,15)
\fmftopn{t}{2}
\fmfbottomn{b}{2}
\fmfright{r}
\fmf{fermion,label=$e^-$,la.si=left}{t1,t}
\fmf{fermion,label=$e^+$,la.si=left}{t,b1}
\fmf{photon}{v,t}
\fmf{fermion,label=$e^-$,la.si=right}{t2,v}
\fmf{fermion,label=$e^+$,la.si=right}{v,b2}
\fmfdot{v,t}
\end{fmfchar*}}
\qquad
\parbox{20mm}{
\begin{fmfchar*}(20,15)
\fmftopn{t}{2}
\fmfbottomn{b}{2}
\fmfright{r}
\fmf{fermion,label=$e^-$}{t1,t}
\fmf{fermion,label=$e^-$}{t,t2}
\fmf{photon}{v,t}
\fmf{fermion,label=$e^+$}{b1,v}
\fmf{fermion,label=$e^+$,la.si=left}{v,b2}
\fmfdot{v,t}
\end{fmfchar*}}
\end{equation}
Both diagrams are separately gauge invariant (i.e. independent of the gauge parameter of the photon propagator) as can be seen by the following observation, providing a simple example for a so called \emph{flavor selection rule}:

In the similar process $e^+ e^-\to \mu^+ \mu^-$ where both fermion pairs belong to different families, only the $s$-channel diagram appears, while in the case of  $e^+ \mu^-\to e^+ \mu^-$ only the $t$-channel diagram appears. Since the scattering amplitudes for both processes are gauge invariant, both diagrams in \fref{eq:bhabbha} must also be gauge invariant by themselves. 

In this simple example, this can be of course verified easily by explicit calculation, but the argument carries over to more complicated situations and to other theories than QED. 

Turning to larger diagrams, it is well known in QED that the expression obtained from a given Feynman diagram by summing over all possible insertions of  a photon  \emph{along a charge carrying fermion line} going through the diagram, satisfies the Ward Identity by itself (see e.g. the graphical proof of the Ward Identity in \cite{Peskin:1995}).   
Thus the amplitude for $e^+ e^- \to \mu^+ \mu^-\gamma$ can be separated in two gauge invariant subsets:
\begin{equation}\label{eq:qed-groves}
\begin{aligned}
G_1&= \left\{\parbox{20mm}{
\begin{fmfchar*}(20,15)
\fmftopn{t}{2}
\fmfbottomn{b}{2}
\fmfright{r}
\fmf{fermion,label=$e^-$,la.si=left}{t1,t}
\fmf{fermion,label=$e^+$,la.si=left}{t,b1}
\fmf{photon}{v2,t}
\fmf{fermion,tension=2}{t2,v1}
\fmf{plain,tension=2,label=$\mu^-$,la.si=right}{v1,v2}
\fmf{fermion,label=$\mu^+$,la.si=right}{v2,b2}
\fmffreeze
\fmf{photon}{r,v1}
\fmfdot{v1,v2,t}
\end{fmfchar*}}\, ,\, 
\parbox{20mm}{
\begin{fmfchar*}(20,15)
\fmftopn{t}{2}
\fmfbottomn{b}{2}
\fmfright{r}
\fmf{fermion}{t1,t,b1}
\fmf{photon}{v1,t}
\fmf{fermion}{t2,v1}
\fmf{plain,tension=2}{v1,v2}
\fmf{fermion,tension=2}{v2,b2}
\fmffreeze
\fmf{photon}{r,v2}
\fmfdot{v1,v2,t}
\end{fmfchar*}}\right\}\\
G_2&= \left\{\parbox{25mm}{
\begin{fmfchar*}(20,15)
\fmfbottomn{t}{3}
\fmftopn{b}{2}
\fmf{fermion}{t3,t,b2}
\fmf{photon}{v2,t}
\fmf{fermion,tension=2}{t1,v1}
\fmf{plain,tension=2}{v1,v2}
\fmf{fermion}{v2,b1}
\fmffreeze
\fmf{photon}{t2,v1}
\fmfdot{v1,v2,t}
\end{fmfchar*}}, 
\parbox{20mm}{
\begin{fmfchar*}(20,15)
\fmftopn{b}{3}
\fmfbottomn{t}{2}
\fmfright{r}
\fmf{fermion}{t2,t,b3}
\fmf{photon}{v1,t}
\fmf{fermion}{t1,v1}
\fmf{plain,tension=2}{v1,v2}
\fmf{fermion,tension=2}{v2,b1}
\fmffreeze
\fmf{photon}{b2,v2}
\fmfdot{v1,v2,t}
\end{fmfchar*}}\right\}
\end{aligned}
\end{equation}
This allows the separate treatment of initial-state and final-state Bremsstrahlung. If we consider instead the process $e^+ e^- \to e^+ e^- \gamma$, i.e. Bhabbha scattering with an additional Bremsstrahlungs-photon, we can appeal to the flavor selection rules discussed above and see that we get altogether 4 gauge invariance classes.
\subsection{Nonabelian gauge theories}\label{sec:qcd-groves}
The situation in nonabelian gauge theories is considerably more complicated then in QED, since the gauge bosons carry charges of the gauge group themselves. Considering the process $\bar qq \to \bar q q g$, as an analog of  the QED example $e^+ e^- \to \mu^+ \mu^-\gamma $ in \fref{eq:qed-groves}, we get an additional diagram since the gluon can be inserted into a gluon propagator:
\begin{equation*}
\parbox{20mm}{
\begin{fmfchar*}(20,15)
\fmftopn{b}{3}
\fmfbottomn{t}{2}
\fmfright{r}
\fmf{fermion}{t2,t,b3}
\fmf{photon,tension=2}{v1,v2}
\fmf{photon,tension=2}{v2,t}
\fmf{fermion}{t1,v1,b1}
\fmffreeze
\fmf{photon}{b2,v2}
\fmfdot{v1,v2,t}
\end{fmfchar*}}
\end{equation*}
This diagram has to be considered together with \emph{all four} diagrams in \fref{eq:qed-groves} to obtain a gauge invariant expression. Therefore the decomposition into gauge invariance classes in \fref{eq:qed-groves} breaks down. However, the argument leading to the flavor selection rules remains valid, since one can always introduce fictitious additional generations of quarks, so one gets a conserved quantum number that has to be conserved along quark-lines going through the diagrams. Therefore the amplitude for the process $\bar qq \to \bar q q g$ contains  2 gauge invariance classes, resulting from the insertion of the gluon into the $s$ and $t$- channel diagrams for the process $\bar qq \to \bar q q$. The gauge invariance class $G_s$ obtained by inserting the gluon in the $s$ channel diagram looks like 
\begin{equation}\label{eq:qcd-grove}
G_s= \left\{
\begin{aligned}&\parbox{20mm}{
 \begin{fmfchar*}(20,15)
\fmftopn{t}{2}
\fmfbottomn{b}{2}
\fmfright{r}
\fmf{fermion}{t1,t}
\fmf{fermion}{t,b1}
\fmf{photon}{v2,t}
\fmf{fermion,tension=2}{t2,v1}
\fmf{plain,tension=2}{v1,v2}
\fmf{fermion}{v2,b2}
\fmffreeze
\fmf{photon}{r,v1}
\fmfdot{v1,v2,t}
\end{fmfchar*}}\, ,\, 
\parbox{20mm}{
\begin{fmfchar*}(20,15)
\fmftopn{t}{2}
\fmfbottomn{b}{2}
\fmfright{r}
\fmf{fermion}{t1,t,b1}
\fmf{photon}{v1,t}
\fmf{fermion}{t2,v1}
\fmf{plain,tension=2}{v1,v2}
\fmf{fermion,tension=2}{v2,b2}
\fmffreeze
\fmf{photon}{r,v2}
\fmfdot{v1,v2,t}
\end{fmfchar*}}\,,\, 
\parbox{20mm}{
\begin{fmfchar*}(20,15)
\fmfbottomn{t}{3}
\fmftopn{b}{2}
\fmf{fermion}{t3,t,b2}
\fmf{photon}{v2,t}
\fmf{fermion,tension=2}{t1,v1}
\fmf{plain,tension=2}{v1,v2}
\fmf{fermion}{v2,b1}
\fmffreeze
\fmf{photon}{t2,v1}
\fmfdot{v1,v2,t}
\end{fmfchar*}}\,,\\ 
&\parbox{20mm}{
\begin{fmfchar*}(20,15)
\fmftopn{b}{3}
\fmfbottomn{t}{2}
\fmfright{r}
\fmf{fermion}{t2,t,b3}
\fmf{photon}{v1,t}
\fmf{fermion}{t1,v1}
\fmf{plain,tension=2}{v1,v2}
\fmf{fermion,tension=2}{v2,b1}
\fmffreeze
\fmf{photon}{b2,v2}
\fmfdot{v1,v2,t}
\end{fmfchar*}}\, , \,
\parbox{20mm}{
\begin{fmfchar*}(20,15)
\fmftopn{b}{3}
\fmfbottomn{t}{2}
\fmfright{r}
\fmf{fermion}{t2,t,b3}
\fmf{photon,tension=2}{v1,v2}
\fmf{photon,tension=2}{v2,t}
\fmf{fermion}{t1,v1,b1}
\fmffreeze
\fmf{photon}{b2,v2}
\fmfdot{v1,v2,t}
\end{fmfchar*}}
\end{aligned}\right\}
\end{equation}
Similar considerations apply to scalar particles in nonabelian gauge theories. 

In spontaneously broken gauge theories the role of Higgs bosons is not clear a priori. Because of the presence of a $WWH$-vertex
\begin{equation*}
\parbox{15mm}{
\begin{fmfchar*}(15,15)
\fmfleft{a,b}
\fmfright{h}
\fmf{photon}{a,ghg,b}
\fmf{dashes}{ghg,h}
\fmfdot{ghg}
\end{fmfchar*}}
\end{equation*}
neutral Higgs bosons cannot be assigned a (fictitious) conserved charge as we have done above to derive the flavor selection rules. We discuss groves in spontaneously broken gauge theories in \fref{chap:ssb_groves} in detail. 
\subsection{Forests, Groves and flips}
In order to derive gauge invariance classes in nonabelian gauge theories systematically, in \cite{Boos:1999} the formalism of `gauge and flavor flips' was introduced. We will sketch the formalism for the case of a unbroken Yang-Mills theory, the application to  spontaneously broken gauge theories is discussed in \fref{chap:ssb_groves}. 
 
An elementary \emph{flavor flip} is defined as a exchange of two diagrams in the set $T_F$, defined as follows: 
\begin{equation}\label{eq:flavor_flips}
T^F_4=\{ t_{4}^{F,1}, t_{4}^{F,2}, t_{4}^{F,3} \}=
\left\{
\parbox{15mm}{
\begin{fmfchar*}(15,15)
\fmfleft{a,b}
\fmfright{f1,f2}
\fmf{fermion}{a,fwf1}
\fmf{photon}{fwf1,fwf2}
\fmf{fermion}{b,fwf2}
\fmf{fermion}{fwf1,f1}
\fmf{fermion}{fwf2,f2}
\fmfdot{fwf1}
\fmfdot{fwf2}
\end{fmfchar*}}\,,\,
\parbox{15mm}{
\begin{fmfchar*}(15,15)
\fmfleft{a,b}
\fmfright{f1,f2}
\fmf{fermion}{a,fwf1}
\fmf{photon}{fwf1,fwf2}
\fmf{fermion}{b,fwf2}
\fmf{phantom}{fwf2,f2}
\fmf{phantom}{fwf1,f1}
\fmffreeze
\fmf{fermion}{fwf2,f1}
\fmf{fermion}{fwf1,f2}
\fmfdot{fwf1}
\fmfdot{fwf2}
\end{fmfchar*}}\, ,\, 
\parbox{15mm}{
\begin{fmfchar*}(15,15)
\fmfleft{a,b}
\fmfright{f1,f2}
\fmf{fermion}{a,fwf}
\fmf{fermion}{fwf,b}
\fmf{photon}{fwf,www}
\fmf{fermion}{f1,www}
\fmf{fermion}{www,f2}
\fmfdot{fwf}
\fmfdot{www}
\end{fmfchar*}}
\right \}
\end{equation}
Flips between  larger diagrams can be obtained by  applying elementary flips  to 4 particle subdiagrams. 

Returning to the example of Bhabbha scattering, we see that the two diagrams in \fref{eq:bhabbha} are connected by a flavor flip. 

Elementary \emph{gauge flips} are defined as exchanges among the diagrams in the sets
\begin{subequations}\label{subeq:gauge_flips}
\begin{equation}
\{ t_{4}^{G,1}, t_{4}^{G,2}, t_{4}^{G,3},t_{4}^{G,4} \}=
\left\{
\parbox{15mm}{
\begin{fmfchar*}(15,15)
\fmfleft{a,b}
\fmfright{f1,f2}
\fmf{photon}{a,fwf1}
\fmf{photon}{fwf1,fwf2}
\fmf{photon}{fwf2,b}
\fmf{photon}{fwf1,f1}
\fmf{photon}{fwf2,f2}
\fmfdot{fwf1}
\fmfdot{fwf2}
\end{fmfchar*}}\,,\,
\parbox{15mm}{
\begin{fmfchar*}(15,15)
\fmfleft{a,b}
\fmfright{f1,f2}
\fmf{photon}{a,fwf1}
\fmf{photon}{fwf1,fwf2}
\fmf{photon}{fwf2,b}
\fmf{phantom}{fwf2,f2}
\fmf{phantom}{fwf1,f1}
\fmffreeze
\fmf{photon}{fwf2,f1}
\fmf{photon}{fwf1,f2}
\fmfdot{fwf1}
\fmfdot{fwf2}
\end{fmfchar*}}\, ,\, 
\parbox{15mm}{
\begin{fmfchar*}(15,15)
\fmfleft{a,b}
\fmfright{f1,f2}
\fmf{photon}{a,fwf}
\fmf{photon}{fwf,b}
\fmf{photon}{fwf,www}
\fmf{photon}{www,f1}
\fmf{photon}{www,f2}
\fmfdot{fwf}
\fmfdot{www}
\end{fmfchar*}}\, , \, 
\parbox{15mm}{
\begin{fmfchar*}(15,15)
\fmfleft{a,b}
\fmfright{f1,f2}
\fmf{photon}{a,c}
\fmf{photon}{c,b}
\fmf{photon}{c,f1}
\fmf{photon}{c,f2}
\fmfdot{c}
\end{fmfchar*}}\right \}
\end{equation}
and
\begin{equation}\label{eq:gauge_flips2}
\{ t_{4}^{G,5}, t_{4}^{G,6}, t_{4}^{G,7} \}=
\left\{
\parbox{15mm}{
\begin{fmfchar*}(15,15)
\fmfleft{a,b}
\fmfright{f1,f2}
\fmf{fermion}{a,fwf1}
\fmf{fermion}{fwf1,fwf2}
\fmf{fermion}{fwf2,b}
\fmf{photon}{fwf1,f1}
\fmf{photon}{fwf2,f2}
\fmfdot{fwf1,fwf2}
\end{fmfchar*}}\quad,\quad
\parbox{15mm}{
\begin{fmfchar*}(15,15)
\fmfleft{a,b}
\fmfright{f1,f2}
\fmf{fermion}{a,fwf1}
\fmf{fermion}{fwf1,fwf2}
\fmf{fermion}{fwf2,b}
\fmf{phantom}{fwf2,f2}
\fmf{phantom}{fwf1,f1}
\fmffreeze
\fmf{photon}{fwf2,f1}
\fmf{photon}{fwf1,f2}
\fmfdot{fwf1,fwf2}
\end{fmfchar*}}\quad,\quad 
\parbox{15mm}{
\begin{fmfchar*}(15,15)
\fmfleft{a,b}
\fmfright{f1,f2}
\fmf{fermion}{a,fwf}
\fmf{fermion}{fwf,b}
\fmf{photon}{fwf,www}
\fmf{photon}{www,f1}
\fmf{photon}{www,f2}
\fmfdot{www,fwf}
\end{fmfchar*}}\right \}
\end{equation}
\end{subequations}
The diagrams in the gauge invariance class \eqref{eq:qcd-grove} are connected by gauge flips of $\bar q q \to g g$ subdiagrams. 

The set of diagrams connected by flavor \emph{and} gauge flips is called \emph{forest}, a set of diagrams connected by gauge flips only is called \emph{grove}. 

The example of the  $\bar q q \to \bar q q  g$ amplitude suggests, that the groves can be identified with gauge invariance classes, while the flavor flips are connected to  flavor selection rules. Indeed it has been shown in \cite{Boos:1999}, that the groves are the \emph{minimal gauge invariance classes} of Feynman diagrams. In \fref{chap:diagrammatica} we will provide a more explicit proof, directly based on the STIs, that justifies our treatment of flips in spontaneously broken gauge theories in \fref{chap:ssb_groves}.

Further results and examples for the structure of groves can be found in
\cite{Boos:1999,Ohl:1999,Ondreka:2000}. The formalism is applied to loop diagrams in \cite{Ondreka:2003}.
 
\section{Gauge invariance and finite widths}\label{sec:intro-widths}
The calculation of realistic scattering amplitudes makes it necessary to include the finite decay widths of unstable particles. The simplest method to include finite width effects is the use of propagators in Breit-Wigner form. Field theoretically the propagators of unstable particles can be obtained by resumming  the self-energy insertions into the propagators. However, this procedure violates gauge invariance. Even tiny violations of gauge invariance can have disastrous effects, if the subamplitudes that violate the Ward Identity are contracted with propagators of almost resonant gauge bosons and  it is well known that the inconsistent treatment of finite width effects can result in result that are wrong by orders of magnitude \cite{Argyres:1995}.

The fact that the naive introduction of finite gauge boson widths in tree
level calculation violates electromagnetic gauge invariance can be seen from  the observation that the introduction of a finite gauge boson width using a naive Breit-Wigner propagator
in unitarity gauge:
\begin{equation}\label{eq:uni-width}
G^{\mu\nu}=  \frac{(-\ii)}{q^2-M_W^2+\ii M_W\Gamma_W}\left(g^{\mu\nu}-\frac{q^\mu
    q^\nu}{M_W^2}\right)
\end{equation} 
results in a violation of the Ward Identity for the 
$WW\gamma$ vertex:
\begin{multline}\label{eq:aww-wi}
\partial^x_\mu\Greensfunc{A^\mu(x)W^+_\nu(y_1)W^-_\rho(y_2)}\neq\\
\sum \pm e\Greensfunc{W^+_\nu(y_1)W^-_\rho(y_2)}\delta^4(x-y_i)
\end{multline}
Several schemes have been proposed to overcome this problem. Some schemes introduce simple prescriptions to introduce finite widths in tree level calculations\cite{LopezCastro:1991,Baur:1991,Baur:1995,Denner:1999,Beenakker:1999}. These simple schemes have all been implemented in \OMEGA~and will be discussed in \fref{chap:width} where also  numerical results for 4 fermion production processes are presented. 

Given the present capabilities of \OMEGA, we have not implemented schemes that involve loop calculations \cite{Beenakker:1997,Aeppli:1994rs,Denner:1996,Papavassiliou:1998}. 

Similar problems with gauge invariance arise from the use of running coupling
constants as will be discussed in \fref{chap:running}.

\part{Gauge invariance of tree level amplitudes}\label{part:tools}
\chapter{Slavnov Taylor Identities}\label{chap:sti}
In \fref{chap:intro} we have sketched the importance of maintaining gauge invariance in numerical
calculations in gauge theories. We now turn to a more formal review of the Ward Identities and STIs that express gauge invariance on the Green's functions and irreducible vertices.

In numerical applications, we are concerned with gauge checks for amputated Green's functions and we will discuss the STIs satisfied by these functions in  \fref{sec:sti-gf}. 

 To make the  discussion of the reconstruction of the Feynman rules from the Ward Identities in \fref{part:reconstruction} more transparent and for the discussion of gauge invariance classes in spontaneously broken gauge theories, it will be useful to  get insight in the mechanisms that ensure the validity of the Ward Identities from a diagrammatic point of view. To prepare for this graphical analysis in \fref{chap:diagrammatica}, we review the STIs of the irreducible vertices that direct the gauge cancellations among different Feynman diagrams. In \fref{sec:sti_irr} we will sketch the Zinn-Justin equation  in spontaneously broken gauge theories that is used in \fref{sec:sti} to obtain STIs for physical vertices. 

In \fref{sec:2cov-sti} we derive the STIs for irreducible vertices where two gauge bosons are contracted with their momenta. These STIs are used in \fref{chap:reconstruction} to simplify the evaluation of the Ward Identities with several contracted gauge bosons \eqref{eq:chanowitz2}. To my knowledge the STI for irreducible vertices with two momentum contractions has not appeared in the literature previously. 

We will postpone the detailed discussion of the Lagrangian of spontaneously broken gauge theories until \fref{chap:ssb-lag}, using some general facts about this theories in this chapter that can be found in textbooks like \cite{\textbooks}. 
 
\section{STI for Green's functions}\label{sec:sti-gf}
As we will show in \fref{part:reconstruction}, the  Ward Identities \eqref{eq:chanowitz2}---in which only the Gauge bosons are off their mass shell---are sufficient to verify all Feynman rules of a spontaneously broken gauge theory. 

Nevertheless it is important to discuss the generalizations of these identities to off-shell particles, the so called Slavnov Taylor Identities. Concerning the discussion in \fref{part:reconstruction}, the STIs  are important tools to simplify and structure the evaluations of Ward Identities for Green's functions. 

While the off-shell identities are not necessary for numerical gauge checks in tree level calculations of ordinary gauge theories,  the STIs are relevant for consistency checks in loop calculations. Thus their discussion is also useful as a preparation for future applications. Also the STIs following from SUSY-BRS transformations \cite{Maggiore:1996} are needed to verify the Feynman rules of supersymmetric gauge theories with broken supersymmetry \cite{ReuterOhl:2002,Reuter:2002}.
\subsection{The general STI of Green's functions}
The simple Ward Identities sketched in \fref{sec:intro-wi} are only valid if all external particles (except the gauge bosons) are on shell. We will now briefly review the systematic derivation of the STIs  for Green's functions, that are the generalization of the Ward Identities to off shell particles. 

From the general STI for Green's functions \eqref{eq:gf-sti-app} 
\begin{equation}\label{eq:gf-sti-a}
0=\sum_i(-)^{s(i)}\braket{\Out|\tprod{\Psi_1\dots\deltabrs\Psi_i\dots \Psi_n}|\In}
\end{equation}
we can obtain the identity corresponding to the QED Ward Identity \eqref{eq:qed-wti} if we choose one field to be an antighost and let the rest of the fields be physical fields (from now on denoted generically by $\Phi_i$). We will first consider only fermions and physical scalar fields,  gauge and Goldstone bosons are more involved and we will return to them later. The STI then reads
\begin{multline}\label{eq:local-wi1}
0=\braket{\Out|\tprod{\lbrack \ii Q,\bar
    c(x)\Phi_1(y_1)\dots\Phi_n(y_n)\rbrack}\In}\\
=\braket{\Out|\tprod{B(x)\Phi_1(y_1)\dots\Phi_n(y_n)}\In}\\
+\sum_i\braket{\Out|\tprod{c(y_i)\bar c(x)
    \Phi_i(y_1)\dots\Delta\Phi_i(y_i)\dots \Phi_n(y_n)}|\In}
\end{multline}
We have written the BRS transformation \eqref{eq:ssb-brs} of the  fermions and physical scalars schematically as 
\begin{equation}
\deltabrs\Phi=c\Delta\Phi
\end{equation}
 where $\Delta\Phi$ is the gauge transformation of the field $\Phi$.
After using the equation of motion for the Nakanishi-Lautrup field \eqref{eq:b-eom-app} in a spontaneously broken gauge theory in $R_\xi$ gauge
\begin{equation}\label{eq:b-eom}
B=-\frac{1}{\xi}(\dmu W^\mu-\xi m_{W}\phi)
\end{equation}
 this turns into
\begin{multline}\label{eq:local-wi}
\braket{\Out|\tprod{ \frac{1}{\xi}(\dmu W^\mu(x)-m_w\phi(x))\Phi_1(y_1)\dots\Phi_n(y_n)}\In}\\
=\sum_i\braket{\Out|\tprod{c(y_i)\bar c(x) \Phi_i(y_1)\dots\Delta\Phi_i(y_i)\dots\Phi_n(y_n)}|\In}
\end{multline}
We don't consider nonlinear gauge fixing functions, that result in similar but more complicated relations.

Comparison with the QED Ward Identity \eqref{eq:qed-wti} shows that we arrived at a similar equation where the insertion of the divergence of the current $\dmu J^\mu$  is replaced by $\frac{1}{\xi}(\dmu W^\mu(x)-\xi m_W \phi)$  and a  ghost-antighost  pair appears instead of the delta function. 

 Using the graphical notation of  \fref{app:brs-graph} we can represent the identities \eqref{eq:local-wi} as 
\begin{equation}\label{eq:sti-diag}
-\parbox{20mm}{
\begin{fmfchar*}(20,20)
\fmfbottomn{v}{6}
\fmftop{t}
\fmf{double,tension=6}{t,v}
\begin{fmffor}{i}{1}{1}{6}
\fmf{plain}{v[i],v}
\end{fmffor}
\fmfv{d.sh=circle,d.f=empty,d.size=30pt,l=$N$,l.d=0}{v}
\end{fmfchar*}}\quad
=\quad\sum_{\Phi_i}\quad
\parbox{30mm}{\fmfframe(5,5)(5,5){%
\begin{fmfchar*}(20,25)
\fmfbottomn{v}{6}
\fmftop{t}
\fmf{ghost,tension=8}{t,v}
\begin{fmffor}{i}{1}{1}{4}
\fmf{plain}{v[i],v}
\end{fmffor}
\fmf{phantom}{v,v5}
\fmf{plain}{v6,v}
\fmfv{d.sh=diamond,d.f=empty,d.size=30pt}{v}
\fmf{ghost,left}{v,v5}
\fmf{plain}{v,v5}
\fmfv{label=$\Phi_i$}{v5}
\fmfv{decor.shape=square,decor.filled=empty,decor.size=5}{v5}
\end{fmfchar*}}}
\end{equation}
The `contact terms' on the right hand side of \fref{eq:local-wi} give rise to disconnected terms so they are represented by a diamond shaped blob. The form  of these terms will be discussed in more detail below (see \fref{eq:contact-graph}).

Since the BRS transformation of the gauge bosons \eqref{eq:gauge-brs}
\begin{equation}
 \deltabrs W_{\mu a}=(D_\mu c)_a=\dmu c_a+f^{abc}W_{\mu b}c_c
\end{equation}
is inhomogeneous, gauge bosons appearing as off-shell particles in the STI \eqref{eq:local-wi} have to be treated separately. The graphical representation of the STI with one external gauge boson is
\begin{equation}\label{eq:sti-gauge}
-\parbox{20mm}{ 
\begin{fmfchar*}(20,25)
\fmfbottomn{v}{6}
\fmftop{t}
\fmf{double,tension=8}{t,v}
\begin{fmffor}{i}{1}{1}{4}
\fmf{plain}{v[i],v}
\end{fmffor}
\fmf{plain}{v6,v}
\fmfv{d.sh=circle,d.f=empty,d.size=30pt}{v}
\fmf{photon}{v,v5}
\end{fmfchar*}}\quad = \quad
\parbox{20mm}{ 
\begin{fmfchar*}(20,25)
\fmfbottomn{v}{6}
\fmftop{t}
\fmf{ghost,tension=8}{t,v}
\begin{fmffor}{i}{1}{1}{4}
\fmf{plain}{v[i],v}
\end{fmffor}
\fmf{ghost}{v,v5}
\fmf{plain}{v6,v}
\fmfv{d.sh=circle,d.f=empty,d.size=30pt}{v}
\fmfv{decor.shape=square,decor.filled=full,decor.size=5}{v5}
\end{fmfchar*}}\quad +\quad\sum_{\Phi_i}\qquad 
\parbox{20mm}{ 
\begin{fmfchar*}(20,25)
\fmfbottomn{v}{6}
\fmftop{t}
\fmf{ghost,tension=8}{t,v}
\begin{fmffor}{i}{1}{1}{4}
\fmf{plain}{v[i],v}
\end{fmffor}
\fmf{plain}{v6,v}
\fmfv{d.sh=diamond,d.f=empty,d.size=30pt}{v}
\fmf{ghost,left}{v,v5}
\fmf{photon}{v,v5}
\fmfv{decor.shape=square,decor.filled=empty,decor.size=5}{v5}
\end{fmfchar*}}
\end{equation}
The generalization to more external gauge bosons should be obvious. 

We can also consider Green's functions with several unphysical gauge bosons and off-shell particles, i.e. the generalization of the Ward Identity \eqref{eq:chanowitz2}. The STI for this case reads
\begin{multline}\label{eq:ncov-sti-gf}
0=\Greensfunc{\lbrack \ii Q,\bar
    c(x_1)B(x_2)\dots B(x_m)\Phi_1(y_1)\dots\Phi_n(y_n)}\\
=\Greensfunc{B(x_1)B(x_2)\dots B(x_m)\Phi_1(y_1)\dots\Phi_n(y_n)}\\
+\sum_i\Greensfunc{c(y_i)\bar c(x)B(x_2)\dots B(x_m)
    \Phi_i(y_1)\dots\Delta\Phi_i(y_i)\dots \Phi_n(y_n)}
\end{multline}
and can be represented graphically as
\begin{equation}\label{eq:sti-diag-n}
-\parbox{20mm}{
\begin{fmfchar*}(20,20)
\fmfbottomn{v}{6}
\fmftopn{t}{4}
\begin{fmffor}{i}{1}{1}{4}
\fmf{double,tension=1.5}{t[i],v}
\end{fmffor}
\begin{fmffor}{i}{1}{1}{6}
\fmf{plain}{v[i],v}
\end{fmffor}
\fmfv{d.sh=circle,d.f=empty,d.size=30pt}{v}
\end{fmfchar*}}\quad
=\quad\sum_{\Phi_i}\quad
\parbox{30mm}{
\fmfframe(5,5)(5,5){%
\begin{fmfchar*}(20,25)
\fmfbottomn{v}{6}
\fmftopn{t}{4}
\fmf{ghost,tension=1.5}{t1,v}
\begin{fmffor}{i}{2}{1}{4}
\fmf{double,tension=1.5}{t[i],v}
\end{fmffor}
\begin{fmffor}{i}{1}{1}{6}
\fmf{plain}{v[i],v}
\end{fmffor}
\fmfv{d.sh=diamond,d.f=empty,d.size=30pt}{v}
\fmf{ghost,left}{v,v5}
\fmf{plain}{v,v5}
\fmfv{label=$\Phi_i$}{v5}
\fmfv{decor.shape=square,decor.filled=empty,decor.size=5}{v5}
\end{fmfchar*}}}
\end{equation}

\subsection{STIs for amputated Green's functions}\label{sec:sti-amp}
To turn the STI  \eqref{eq:local-wi} into a form that applies to scattering matrix elements, it is necessary to amputate the propagators of the external particles. The resulting identities are those that are relevant for numerical gauge checks. 

If all particles with the possible exception of the gauge boson are on-shell, the contact terms on the right hand side of \fref{eq:local-wi} vanish and one arrives at the simple identity \eqref{eq:gf-wi}
 \begin{equation}
-\ii k_\mu\me^\mu(\In+W\to\Out)
=m_w\me(\In+\phi\to\Out) 
\end{equation}
that we have already given in \fref{chap:intro}. In the derivation one has to  use the (tree level) relation \cite{Chanowitz:1985,Yao:1988} 
\begin{equation}\label{eq:prop-rel}
k_\mu D^{\mu\nu}_W=-\xi D_c k^\nu
\end{equation}
that is a consequence of the STIs for the propagators as we review in \fref{app:amputate}. 

To discuss the  STI with several insertions of the auxiliary field $B$ given in \fref{eq:ncov-sti-gf}, we will introduce the shorthand
\begin{equation}\label{eq:def-contraction}
\me(\mathcal{D}_a(p_a)\dots)\equiv
-\ii p_{a\mu} \me^\mu(W_a^\mu(p_a)\dots)-m_{W_a}\me(\phi_a(p_a)\dots)
\end{equation}
For outgoing gauge bosons, the sign of the momentum has to be changed. We will take this as understood in the definition of $\mathcal{D}$.

Using this notation, one can obtain from \eqref{eq:ncov-sti-gf} the identity
\begin{equation}\label{eq:chanowitz}
\me(\mathcal{D}\dots\mathcal{D}\Phi\dots\Phi)=0
\end{equation}
where only the contracted gauge bosons  are off-shell. 

If other external particles are off shell, the contact terms have to be taken into account. We have already discussed the STIs for amputated Green's functions with off-shell
particles in \cite{Schwinn:2000} in the context of corrections to the equivalence theorem for off-shell gauge
bosons, so we just sketch the results. Our  conventions in the definitions of the amputated Green's functions are given in \fref{app:amp}, the implementation in \OMEGA~will be discussed in \fref{sec:omega-sti}. 

 At tree level, the contact terms can be written as a sum over factorized \emph{connected} diagrams, interconnected only by the BRS-vertices:
\begin{equation}\label{eq:contact-graph} 
\parbox{30mm}{\fmfframe(5,5)(5,5){%
\begin{fmfchar*}(20,25)
\fmfbottomn{v}{6}
\fmftop{t}
\fmf{ghost,tension=8}{t,v}
\begin{fmffor}{i}{1}{1}{6}
\fmf{plain}{v[i],v}
\end{fmffor}
\fmffreeze
\fmfv{d.sh=diamond,d.f=empty,d.size=30pt,l=$N$,l.di=0}{v}
\fmf{ghost,left}{v,v4}
\fmfv{decor.shape=square,decor.filled=empty,decor.size=5,label=$\Phi_i$}{v4}
\end{fmfchar*}}}\quad \overset{\text{tree level}}{=} \quad\sum_{k+l=N+1}
\parbox{30mm}{\fmfframe(5,5)(5,5){%
\begin{fmfchar*}(20,20)
\fmfbottomn{v}{6}
\fmftopn{t}{3}
\fmf{phantom,tension=4}{t1,v}
\fmf{phantom,tension=4}{t3,u}
\fmf{phantom}{u,v}
\begin{fmffor}{i}{1}{1}{3}
\fmf{plain}{v[i],v}
\fmf{plain}{v[i+3],u}
\end{fmffor}
\fmfv{d.sh=circle,d.f=empty,d.size=15pt,l=$k$,l.d=0}{v}
\fmfv{d.sh=circle,d.f=empty,d.size=15pt,l=$l$,l.d=0}{u}
\fmffreeze
\fmf{phantom}{t2,u}
\fmf{ghost}{t2,v,v4}
\fmfv{decor.shape=square,decor.filled=empty,decor.size=5,label=$\Phi_i$}{v4}
\end{fmfchar*}}} 
\end{equation}
 As an example, consider a 4 point contact term with two external fermions and one gauge boson:
\begin{multline}\label{eq:4-contact} 
\Greensfunc{c(x_1)\bar c(x)\Delta\psi(x_1)\bar \psi(x_2) W(x_3)}=
\parbox{20mm}{
\begin{fmfchar*}(20,20)
\fmfleft{l1,l2}
\fmfright{r1,r2}
\fmf{fermion}{l1,m,l2}
\fmf{photon}{r1,m}
\fmf{ghost}{r2,m}
\fmffreeze
\fmf{ghost,right=0.5}{m,l1}
\fmfv{decor.shape=square,decor.filled=empty,decor.size=5,label=$p_1$}{l1}
\fmfv{decor.shape=diamond,decor.filled=empty,d.si=15}{m}
\fmfv{label=$p_2$,l.di=0.6}{l2}
\fmfv{label=$p_3$,l.di=0.6}{r1}
\fmfv{label=$k$,l.di=0.6}{r2}
\end{fmfchar*}}\\[5mm]
 \overset{\text{Tree level}}{=} \quad
\parbox{20mm}{
\begin{fmfchar*}(20,20)
\fmfleft{l1,l2}
\fmfright{r1,r2}
\fmf{fermion,label=${\scriptstyle p_1+k}$,l.si=left,l.di=0.6}{l1,m}
\fmf{fermion}{m,l2}
\fmf{photon}{r1,m}
\fmf{phantom}{r2,m}
\fmffreeze
\fmf{ghost,left=0.5}{r2,l1}
\fmfv{decor.shape=square,decor.filled=empty,decor.size=5}{l1}
\fmfdot{m}
\end{fmfchar*}}\qquad+ \qquad
\parbox{20mm}{
\begin{fmfchar*}(20,20)
\fmfleft{l1,l2}
\fmfright{r1,r2}
\fmf{ghost}{r2,m}
\fmf{ghost,label=${\scriptstyle p_3+k}$,l.si=left,l.di=0.6}{m,l1}
\fmf{photon}{r1,m}
\fmf{phantom}{l2,m}
\fmffreeze
\fmf{fermion,label=$p_2$}{l1,l2}
\fmfv{decor.shape=square,decor.filled=empty,decor.size=5}{l1}
\fmfdot{m}
\end{fmfchar*}}
\end{multline} 
The first term corresponds to $k=2,l=3$, the second one to $k=3,l=2$ in \fref{eq:contact-graph}.

We see from the example and the general expression, that in the contact terms there is no propagator with the momentum of the BRS transformed external particle. Therefore the operator $(c\Delta\Phi)(x)$ has to be treated as  an insertion that is \emph{not} amputated. When we go over to amputated Green's functions in the STI \eqref{eq:global-wi}, because of this missing propagator the contact terms appear with the inverse propagator of the BRS transformed particle. The amputation of the contracted gauge boson line is again performed using \fref{eq:prop-rel} and  we obtain the \emph{STI for amputated Green's functions}:
\begin{equation}\label{eq:lsz-wi-boson}
\me(\mathcal{D}(k)\dots \Phi\dots)=\sum_j
D_{\Phi_j}^{-1}(p_j) \me(\bar c(k)\dots (c\Delta\Phi_j)(p_j)\dots) 
\end{equation}
In the case of external fermions, one has to distinguish carefully between particles and antiparticles, the correct expressions are given in \fref{eq:lsz-ferm}. 
If the external particles (except the unphysical gauge boson) are on shell, the
inverse propagators vanish and we obtain again \eqref{eq:local-wi}. Note that already at tree level we need the complete ghost Lagrangian to evaluate the contact terms \eqref{eq:contact-graph}. 

As we have seen in \fref{eq:sti-gauge}, in the case of external gauge bosons the STI gets additional contributions from the term $\propto \dmu c_a$ in the BRS transformation of the gauge bosons. The amputation of these terms has  been done already in \cite{Schwinn:2000} and is reviewed briefly in \fref{app:ghost-amp}. 

Compared to the STI without off-shell gauge bosons \eqref{eq:lsz-wi-boson} we get an additional contribution for every gauge boson\footnote{Going from Green's functions to amputated matrix elements,  (incoming) ghosts and antighosts have to be exchanged since ghost field operators create incoming antighosts and vice versa.}: 
\begin{multline}\label{eq:lsz-wi-gauge}
\me(\mathcal{D}_a(k) \dots W_b(p_b)\dots  )
=-\frac{\ii}{\xi}p_b^\nu\me(c_a(p) \bar c_b(p_b)\dots)\\
+f^{bcd}D_{W_b}^{-1}(p_b) \me^\nu(\bar c_a(k)\dots (c_d W_c)(k)\dots)+\text{remaining contact terms}
\end{multline}
Here we have to use the inverse propagator of the gauge boson in $R_\xi$ gauge.If  the matrix elements are contracted with a physical polarization vector of the gauge boson $W_a$, satisfying 
 \begin{equation}
   \epsilon_a\cdot p_a =0 
 \end{equation}
the additional term in the STI proportional to $p_a^\nu$ drops out. Also the gauge dependent terms in the inverse gauge boson propagator vanish upon contraction with $\epsilon_a$ and we can use the simpler STI \eqref{eq:lsz-wi-boson} taking  
\begin{equation}
  \left(D_{W_j}^{-1}(p_j)\right)_{\mu\nu}=\ii (p_j^2-m_{W}^2)g_{\mu\nu}
\end{equation}
as the inverse $W$ propagator. 

In \fref{app:uni-sti} we clarify the meaning of the  STI
\eqref{eq:lsz-wi-boson} in unitarity gauge. Taking the limit $\xi\to\infty$ \emph{after} the amputation of the external
legs, we find that Goldstone bosons have to be treated as external, nonpropagating
particles. The ghosts decouple from the gauge bosons but a coupling to the scalars remains. The ghost Lagrangian is given by
\begin{equation}\label{eq:uni-ghosts-main}
\mathscr{L}_{FP}^U=-m_{W_a}\bar c_a c_a+g_{\bar c H c}^{aib} \bar c_a H_i c_b+g_{\bar c \phi c}^{abc}\bar c_a \phi_b c_c
\end{equation}
that results in a constant ghost propagator. The ghost vertices in unitarity
gauge can be obtained from those in $R_\xi$ gauge by 
\begin{equation}
g_{\bar c \Phi c}^U=-\frac{1}{\xi m_{W}^2}g_{\bar c \Phi c}^{R_\xi}
\end{equation}
It is well known  that the
Higgs-ghost coupling in the above form has to be used in loop diagrams in
unitarity gauge  (see  \cite{Grosse-Knetter:1993} for a review), but the
correct use of the STIs and the form of the Goldstone boson coupling has not been discussed in the literature to my knowledge. 
\section{Zinn-Justin equation}\label{sec:sti_irr} 
The STIs for (amputated) Green's functions studied in the previous section are the relevant objects for numerical checks of gauge invariance of scattering amplitudes. From the point of view of (algebraic) renormalization theory \cite{Piguet:1981}, the natural object to study is the 
\emph{effective action} $\Gamma$, the generating
functional of the irreducible vertices. It is  connected to the generating functional of connected Green's functions by a Legendre transformation, as is described in standard 
quantum field theory text books (see e.g. \cite{\textbooks}).

The symmetry of the effective action leads to the so called \emph{Zinn-Justin equation} \cite{Zinn-Justin:1974} that implies the STIs for the irreducible vertices. As we will discuss in \fref{chap:diagrammatica}, the STIs of the Green's functions are a consequence of the STIs for the irreducible vertices so one might regard the ZJ-equation as  more fundamental than the remaining identities. 

 In this section we will  give a brief review of the formalism of the effective action and the ZJ-equation before we turn to the STIs for effective vertices in \fref{sec:sti}. 

To establish the notation we will first recall some properties of the effective action. We will use the definition 
\begin{multline}
\ii \Gamma(\Phi_{cl})=\Greensfunc{e^{ \int d^4 x \sum_{\Phi}\Phi(x) \Phi_{cl}(x)}}^{1\text{PI}}\\
=\sum_{n=0}^{\infty}\frac{1}{n!}\int \left(\prod_{j=0}^n d^4 x_j\right)\Phi_{cl}(x_1)\dots\Phi_{cl}(x_n)\onepi{0|\Tprod{\Phi(x_1)\dots \Phi(x_n)}|0}
\end{multline}
 for a set of quantum fields $\Phi$ and the corresponding \emph{classical
   fields} $\Phi_{cl}$ that act as sources for the quantum fields. It can be
   shown that on tree level the effective action agrees with the classical
   action, i.e. 
   \begin{equation}
     \Gamma=S+\text{ loop corrections}
   \end{equation}
The one particle irreducible vertices can be obtained by taking functional
derivatives with respect to the classical fields:
\begin{align}
  \frac{\ii \delta^n}{\delta\Phi_{cl}(x_1)\dots\delta\Phi_{cl}(x_n)}\Gamma(\Phi_{cl})&\equiv
\Gamma_{\Phi_1\cdots\Phi_n}(x_1\dots x_n)\\
&=\onepi{0|\Tprod{\Phi(x_1)\dots\Phi(x_n)}|0}\nonumber
\end{align} 
One can show that the irreducible two point function is the inverse of the propagator:
\begin{equation}\label{eq:inv-prop}
\ii \frac{\delta^2\Gamma}{\delta\Phi_{cl}(x)\delta\Phi_{cl}(y)}=\Gamma_{\Phi\Phi}(x,y)=-D_{\Phi\Phi}^{-1}(x,y)
\end{equation}
This relation will be essential in relating the STIs for Green's functions and the STIs for irreducible vertices. 

Since the BRS transformations are nonlinear, the form
\eqref{eq:wi-irr} of the STI for the effective action cannot be used on the quantum level because operators
nonlinear in the fields require additional renormalization. The correct way to
handle these nonlinear operators is to add sources  $\Psi^\star$ for the BRS
transformations of the fields $\Psi$ to the action $S_0$ :
\begin{multline}\label{eq:antifields}
 S=S_0+\int d^4x\Tr\lbrack {W^\star}^\mu (\deltabrs W_\mu)\rbrack +
\bar\psi^\star(\deltabrs \psi)+(\deltabrs\bar\psi) \psi^\star\\
+\phi^\star( \deltabrs\phi)
+H^\star(\deltabrs H+)\Tr\lbrack c^\star (\deltabrs c)\rbrack
\end{multline}
The invariance of the action under BRS transformations is now expressed by the equation\footnote{The compact notation has to be used with care in the case of fermions, because here the order of the two factors is reversed:
\begin{equation*}\label{eq:fermion-sti}
\frac{\delta_L S}{\delta \bar\psi^\star}\frac{\delta_R
  S}{\delta \bar\psi}+\frac{\delta_L S}{\delta\psi}\frac{\delta_R
  S}{\delta \psi^\star}
\end{equation*}
In the following this will always be understood.} 
\begin{equation}\label{eq:sti-action}
\sum_\Psi\int d^4 x  \frac{\delta_L S}{\delta \Psi^\star}\frac{\delta_R
  S}{\delta \Psi}+\Tr B\frac{\delta_R S}{\delta\bar c}=0
\end{equation}
Here the sum runs over physical and unphysical fields.

As in the case of QED, in anomaly free theories the effective action satisfies the same equation,  called the \emph{Zinn-Justin equation}:
\begin{equation}\label{eq:sti}
\sum_\Psi\int d^4 x \frac{\delta_L \Gamma}{\delta \Psi^\star}\frac{\delta_R
  \Gamma}{\delta \Psi}+B\frac{\delta_R \Gamma}{\delta\bar c}=0
\end{equation}
The gauge fixing condition is implemented by choosing a gauge fixing function $G_a$ and demanding 
\begin{equation}\label{eq:g-fix}
\frac{\delta\Gamma}{\delta B_a}=\xi B_a+ G_a
\end{equation}
This equation can be implemented to all orders of perturbation theory \cite{Boehm:2001,Piguet:1981}. We will always use a linear $R_\xi$ gauge  with the gauge fixing function as given in \fref{eq:rxi-gf}. This leads to the equation of motion \eqref{eq:b-eom} for the Nakanishi-Lautrup field.

The ZJ-equation \eqref{eq:sti} can be also written a little bit more intuitively as
\begin{equation}\label{eq:sti-irr-gf}
\sum_\Psi\int d^4 x \onepi{\Psi^\star(x)}_{\Psi_{cl}, \Psi^\star}\onepi{ \Psi(x)}_{\Psi_{cl},\Psi^\star}+B_a(x)\onepi{\bar c_a(x)} =0
\end{equation}
where we have introduced the notation 
\begin{equation}
\onepi{\mathcal{O}(\Psi)}_{\Psi_{cl},
  \Psi^\star}=\Greensfunc{\mathcal{O}(\Psi)e^{ \int d^4 x \sum_{\Phi}\Phi(x) \Phi_{cl}(x)}}^{1\text{PI}}
\end{equation}
 to denote the generating functional of irreducible vertices with the insertion of a composite operator $\mathcal{O}(\Psi)$. The notation
\begin{equation*}
\mathcal{O}(\Psi)\cdot \Gamma
\end{equation*}
for the same object is more common in the literature \cite{Piguet:1981} but less intuitive. A simplification of \eqref{eq:sti} can be obtained if one defines an effective action without the  gauge fixing term (see e.g. \cite{Boehm:2001,Piguet:1981}). This  is very convenient in the analysis of the renormalizability of the effective action, however, calculation of Feynman diagrams becomes awkward because of the missing gauge fixing so we won't use this simplified ZJ-equation in this work.

\section{STI for physical vertices}\label{sec:sti}
We now use the ZJ-equation to derive the STIs of the physical vertices of the theory. 
This is usually done in the context of loop calculations, where one can formulate an algorithm \cite{Grassi:2001} that consists of an iterative application of the  STIs  until a closed set is found that constrains all
occurring vertex functions. 
Since we are mainly interested in tree level applications, we will only sketch the first step of the procedure of \cite{Grassi:2001}. To obtain STIs for physical vertex functions, one has to differentiate the ZJ-equation \eqref{eq:sti} with respect to classical fields. 
   The BRS transformation of the physical fields increases the ghost number by one, so we have to differentiate \fref{eq:sti} with respect to a ghost field to obtain relations among non-vanishing vertex functions. 
To derive the STI for the three point vertex, we take the derivative of \fref{eq:sti} with respect to two  fields $\Phi$ and one ghost field and set the classical fields and sources to zero:
\begin{multline}\label{eq:sti-irr-3point}
0=\sum_\Psi\int d^4 x \Biggl\{ \onepi{ c_a(y) \Psi^\star(x)}\onepi{ \Psi(x)\Phi_1(x_1)\Phi_2(x_2)}\\
+\Bigl[\onepi{c_a(y)\Psi^\star(x)\Phi_1(x_1)}\onepi{\Psi(x)\Phi_2(x_2)}\\
+\frac{\delta
  B(x)}{\delta\Phi_1(x_1)}\onepi{c_a(y)\bar c(x)\Phi_2(x_2)} +\quad (1\leftrightarrow 2)\Bigr]\Biggr\}
\end{multline}

Repeating the procedure with an additional derivative, we can derive the relation for the irreducible 4 point function:
\begin{multline}\label{eq:sti-irr-4point}
0=\sum_\Psi\int d^4 x \Biggl\{ \onepi{c_a\Psi^\star(x)}
\onepi{ \Psi(x)\Phi_1 \Phi_2\Phi_3}\\
+\onepi{c_a\Psi^\star(x)\Phi_1}\onepi{ \Psi(x)\Phi_2\Phi_3}\\
+\onepi{c_a\Psi^\star(x)\Phi_1\Phi_2}\onepi{\Psi(x)\Phi_3}+\frac{\delta B(x)}{\delta\Phi_1}\onepi{c_a\bar c(x)\Phi_2\Phi_3} \Biggr\}+\text{permutations}
\end{multline}
The generalization to higher order vertex functions is obvious but increasingly tedious to write down. 

These STIs are the analogs of their counterparts for Green's functions with the
insertions of the Nakanishi-Lautrup field $B$ \eqref{eq:local-wi}. To see
this, we note that the vertex $\onepi{c_a(y)\Psi^\star(x)}$ is only present for gauge bosons and Goldstone bosons since only those fields have inhomogeneous terms in the transformation laws:
\begin{equation}
\deltabrs W_a =\dmu c_a+\dots \quad \deltabrs\phi_a=-m_{W_a}c_a+\dots
\end{equation}
Since these terms are linear in the fields, they don't receive additional radiative
corrections and we get to all orders
\begin{multline}\label{eq:fourier-linear-brs}
\text{F.T.}\sum_\Psi\int d^4 x \onepi{c_a(y)\Psi^\star(x)}\onepi{
  \Psi(x)\dots}\\
=-\ii p_\mu \onepi{ W_a^\mu(p)\dots}-m_{W_a}\onepi{ \phi_a(p)\dots}=\onepi{ \mathcal{D}_a^\mu(p)\dots}
\end{multline}
Here we have again used the abbreviation $\mathcal{D}$ from  \eqref{eq:def-contraction}. These terms just correspond to an (amputated) insertion of the
Nakanishi-Lautrup  field $B$, although they arise from the BRS transformations of $W$ and $\phi$ and not from the transformation of an antighost as in the case of the STIs for Green's functions. 
\subsection{Graphical notation}
To illustrate the  STIs \fref{eq:sti-irr-3point} and \fref{eq:sti-irr-4point}, we introduce the following graphical notation:
\begin{equation}
\begin{aligned}
\Gamma_{\Phi\Phi}(x,y)=-D_\Phi^{-1}(x-y)&=\qquad 
\parbox{10mm}{
\begin{fmfchar*}(10,10)
\fmfleft{l}
\fmfright{r}
\fmf{zigzag}{l,r}
\fmfv{label=x}{l}
\fmfv{label=y}{r}
\end{fmfchar*}
}\\
\onepi{c_a(y)\Psi^\star(x)\Phi_1(x_1)\dots\Phi_n(x_n)}&=\quad 
\parbox{15mm}{
\begin{fmfchar*}(15,15)
\fmfleftn{g}{3}
\fmfrightn{r}{2}
\fmf{ghost}{m,r1}
\fmf{plain}{g1,m,g3}
\fmf{phantom}{m,r2}
\fmffreeze
\fmf{plain,right=0.5}{m,r2}
\fmf{ghost,right=0.5}{r2,m}
\fmfv{decor.shape=square,decor.filled=empty,decor.size=5}{r2}
\fmfv{label=$\vdots$,la.di=0}{g2}
\fmfblob{10}{m}
\end{fmfchar*}}
\label{eq:antifield-graph}
\end{aligned}
\end{equation}
In this notation the STI for the 3 point function \fref{eq:sti-irr-3point} becomes
\begin{equation}
\parbox{20mm}{
\begin{fmfchar*}(15,15)
\fmfleft{l}
\fmfright{r1,r2}
\fmfpolyn{smooth,filled=shaded}{g}{3}
\fmf{double}{l,g1}
\fmf{plain}{r1,g2}
\fmf{plain}{r2,g3}
\end{fmfchar*}}
=\qquad \parbox{20mm}{\begin{fmfchar*}(20,20)
\fmfleft{l}
\fmfright{r1,r2}
\fmf{plain}{r1,brs}
\fmf{ghost}{l,brs}
\fmf{plain,right=0.75}{brs,i}
\fmf{ghost,right=0.75}{i,brs}
\fmf{zigzag,tension=2}{i,r2}
\fmfv{decor.shape=circle,decor.filled=shaded,decor.size=10}{brs}
\fmfbrs{i}
\end{fmfchar*}}
\qquad +\qquad \parbox{20mm}{\begin{fmfchar*}(20,20)
\fmfleft{l}
\fmfright{r1,r2}
\fmf{ghost}{l,brs}
\fmf{zigzag,tension=2}{r1,i}
\fmf{plain,right=0.5}{brs,i}
\fmf{ghost,right=0.5}{i,brs}
\fmf{plain}{brs,r2}
\fmfbrs{i}
\fmfv{decor.shape=circle,decor.filled=shaded,decor.size=10}{brs}
\end{fmfchar*}}
\end{equation}
We have not displayed the term $\propto \frac{\delta B}{\delta\Phi}$ whose significance will be discussed in \fref{sec:sti-tree}. The STI for the 4 point function \fref{eq:sti-irr-4point} is written as
\begin{equation*} 
\parbox{20mm}{
\begin{fmfchar*}(20,20)
\fmfleft{g,r3}
\fmfrightn{r}{2}
\fmf{double}{g,b}
\begin{fmffor}{i}{1}{1}{3}
\fmf{plain}{r[i],b}
\end{fmffor}
\fmfblob{10}{b}
\end{fmfchar*}}
=\sum_{\Phi_i}
\parbox{35mm}{
\fmfframe(0,0)(5,0){
\begin{fmfchar*}(30,20)
\fmfleftn{g}{2}
\fmfrightn{r}{2}
\fmf{ghost}{m,g1}
\fmf{plain}{m,g2}
\fmf{phantom}{m,brs}
\fmf{plain,tension=1.5}{brs,b}
\begin{fmffor}{i}{1}{1}{2}
\fmf{plain}{r[i],b}
\end{fmffor}
\fmffreeze
\fmf{plain,right=0.75}{m,brs}
\fmf{ghost,right=0.75}{brs,m}
\fmfv{decor.shape=square,decor.filled=empty,decor.size=5}{brs}
\fmfblob{10}{m}
\fmfblob{10}{b}
\fmfv{l=$\Phi_i$,l.si=right,l.d=0.3}{g2}
\end{fmfchar*}}}
+
\sum_{\Phi_i}
\parbox{25mm}{
\fmfframe(0,5)(0,0){
\begin{fmfchar*}(20,20)
\fmfleftn{g}{2}
\fmfrightn{r}{2}
\fmf{ghost}{m,g1}
\fmf{plain}{r2,m,r1}
\fmf{phantom,tension=2}{m,brs}
\fmf{zigzag,tension=2}{brs,g2}
\fmffreeze
\fmf{plain,right}{m,brs}
\fmf{ghost,right}{brs,m}
\fmfv{decor.shape=square,decor.filled=empty,decor.size=5}{brs}
\fmfblob{10}{m}
\fmfv{l=$\Phi_i$,l.si=right,l.d=0.1}{r2}
\end{fmfchar*}}}
\end{equation*}
Using this graphical representation one must keep in mind that these are \emph{not} Feynman diagrams: an internal line represents a summation over all particles, not a propagator connecting the irreducible vertices. 

For later use, we  also display the  STIs for amputated Green's functions \fref{eq:lsz-wi-boson} in this graphical notation
\begin{equation}\label{eq:sti-graph}
\parbox{20mm}{
\begin{fmfchar*}(20,20)
\fmfbottomn{v}{6}
\fmftop{t}
\fmf{double,tension=6}{t,v}
\begin{fmffor}{i}{1}{1}{6}
\fmf{plain}{v[i],v}
\end{fmffor}
\fmfv{d.sh=circle,d.f=empty,d.size=30pt}{v}
\end{fmfchar*}}\quad
=\quad\sum_{\Phi_i}\quad
\parbox{20mm}{ 
\begin{fmfchar*}(20,25)
\fmfbottomn{v}{6}
\fmftop{t}
\fmf{ghost,tension=8}{t,v}
\begin{fmffor}{i}{1}{1}{4}
\fmf{plain}{v[i],v}
\end{fmffor}
\fmf{phantom}{v,v5}
\fmf{plain}{v6,v}
\fmfv{d.sh=diamond,d.f=empty,d.size=25pt}{v}
\fmf{plain}{v,i}
\fmf{zigzag,tension=2}{i,v5}
\fmffreeze
\fmf{ghost,left}{v,i}
\fmfv{label=$\phi_i$,l.d=0.6}{v5}
\fmfv{decor.shape=square,decor.filled=empty,decor.size=5}{i}
\end{fmfchar*}}
\end{equation}
\subsection{Tree level}\label{sec:sti-tree}
On tree level the content of the STIs of the irreducible vertices is more intuitive, since the vertex functions with the insertions of sources of the BRS transformed fields can be read from the BRS transformation laws given in \fref{eq:ssb-brs}. This follows from the identity
\begin{equation}
\frac{i\delta\Gamma}{\delta \Phi_i^\star(x)}\overset{\text{Tree level}}{=}\delta_{BRS}\Phi_i(x)
\end{equation}
If we write the BRS transformation of the fields (apart from the inhomogeneous terms) schematically as
\begin{equation}
\deltabrs \Phi_i=c_a T_{ij}^a\Phi_j
\end{equation}
we can simplify the terms appearing in the STIs:
\begin{multline}
\text{F.T}\int d^4 x\, \onepi{c_a(y)\Phi_i^\star(x)\Phi_j(x_j)}\onepi{ \Phi_i(x)\dots}\\
=\text{F.T} \,\left( T^a_{ij}\onepi{
    \Phi_i(x_j)\dots)}\delta(y-x_j)\right)=T^a_{ij}\onepi{ \Phi_i(p+k_j)\dots}
\end{multline}
Therefore the STI for the three point function \eqref{eq:sti-irr-3point} becomes
\begin{multline}\label{eq:sti-irr-3tree}
-\onepi{ \mathcal{D}_a(p)\Phi_i(k_i)\Phi_j(k_j)}\\
= T^a_{ki}\onepi{\Phi_k(p+k_i)\Phi_j(k_j)}+T^a_{kj}\onepi{\Phi_i(k_i)\Phi_k(p+k_j)}
\end{multline}
In the graphical notation of \fref{sec:sti_irr} this can be written as
\begin{equation}\label{eq:sti-3graph}
\parbox{25mm}{\fmfframe(0,5)(0,5){
\begin{fmfchar*}(15,15)
\fmfleft{l}
\fmfright{r1,r2}
\fmfpolyn{smooth,filled=shaded}{g}{3}
\fmf{double}{l,g1}
\fmf{plain}{r1,g2}
\fmf{plain}{r2,g3}
\fmfv{l=$a$}{l}
\fmfv{l=$i$}{r1}
\fmfv{l=$j$}{r2}
\end{fmfchar*}}}
=\qquad \parbox{25mm}{\fmfframe(0,5)(0,5){\begin{fmfchar*}(15,15)
\fmfleft{l}
\fmfright{r1,r2}
\fmf{plain}{r1,i}
\fmf{ghost}{l,i}
\fmf{zigzag}{i,r2}
\fmfv{l=$a$}{l}
\fmfv{decor.shape=square,decor.filled=empty,decor.size=5,label=$k$}{i}
\fmfv{l=$i$}{r1}
\fmfv{l=$j$}{r2}
\end{fmfchar*}}}
\qquad +\qquad \parbox{25mm}{\fmfframe(0,5)(0,5){\begin{fmfchar*}(15,15)
\fmfleft{l}
\fmfright{r1,r2}
\fmf{ghost}{l,i}
\fmf{zigzag}{r1,i}
\fmf{plain}{i,r2}
\fmfv{l=$a$}{l}
\fmfv{decor.shape=square,decor.filled=empty,decor.size=5,label=$k$}{i}
\fmfv{l=$i$}{r1}
\fmfv{l=$j$}{r2}
\end{fmfchar*}}}
\end{equation}

For vertices involving gauge or Goldstone bosons we get additional contributions from the
term $B\frac{\partial \Gamma}{\partial\phi}$ in the ZJ-equation. Using the equation of motion of the  Nakanishi-Lautrup field \eqref{eq:b-eom} we find that the STI for 3 point functions with two gauge bosons and one arbitrary physical field is modified to 
\begin{multline}\label{eq:2gb-sti}
-\onepi{ \mathcal{D}_a(p_a) W_b^\nu(p_b)\Phi_i(k_i)}= f^{acb}\onepi{W_c(p_a+p_b)\Phi_i(k_i)}\\
+T^a_{ki}\onepi{W_b^\nu(p_b)\Phi_k(p_a+k_i)} +\ii \frac{1}{\xi}p_b^\nu \onepi { c_a(p_a)\bar c_b(p_b)\Phi_i(k_i) }
\end{multline}
The graphical representation of this identity is
\begin{equation}\label{eq:2gb-sti-graph}
\parbox{20mm}{
\begin{fmfchar*}(15,15)
\fmfleft{l}
\fmfright{r1,r2}
\fmfpolyn{smooth,filled=shaded}{g}{3}
\fmf{double}{l,g1}
\fmf{photon}{r1,g2}
\fmf{plain}{r2,g3}
\end{fmfchar*}}
=\qquad \parbox{15mm}{\begin{fmfchar*}(15,15)
\fmfleft{l}
\fmfright{r1,r2}
\fmf{plain}{r1,i}
\fmf{ghost}{l,i}
\fmf{zigzag}{i,r2}
\fmfv{decor.shape=square,decor.filled=empty,decor.size=5}{i}
\end{fmfchar*}}
\qquad +\qquad \parbox{15mm}{\begin{fmfchar*}(15,15)
\fmfleft{l}
\fmfright{r1,r2}
\fmf{ghost}{l,i}
\fmf{zigzag}{r1,i}
\fmf{plain}{i,r2}
\fmfv{decor.shape=square,decor.filled=empty,decor.size=5}{i}
\end{fmfchar*}}
\qquad +\qquad \frac{1}{\xi}\quad
\parbox{25mm}{
\fmfframe(0,5)(0,5){
\begin{fmfchar*}(15,15)
\fmfleft{l}
\fmfright{r1,r2}
\fmfpolyn{smooth,filled=shaded}{g}{3}
\fmf{ghost}{l,g1}
\fmf{ghost}{g2,r1}
\fmf{plain}{r2,g3}
\fmfv{decor.shape=square,decor.filled=full,decor.size=5}{r1}
\end{fmfchar*}}}
\end{equation}
Similarly, the vertex function of one gauge boson, one Goldstone boson and one physical field satisfies
\begin{multline}\label{eq:phi-sti}
- \onepi{ \mathcal{D}_a(p_a) \phi_b(p_b)\Phi_i(k_i)}= T^a_{kb}\onepi{\Phi_k(p_a+p_b)\Phi_i(k_i)}\\+T^a_{ki}\onepi{\phi_b(p_b)\Phi_k(p_a+k_i)}
 +m_{W_b}\onepi {c_a(p_a) \bar c_b(p_b)\Phi_i(k_i) }
\end{multline}
If we replace the physical field $\Phi_i$ by a gauge- or Goldstone boson, we have to add another ghost term of the same form as above. 

In the STI for the 4 point function  \fref{eq:sti-irr-4point} the terms
involving the BRS vertices  $\onepi{c_a(y)\deltabrs
  \Phi(x)\Phi_1(x_1)\Phi_2(x_2)}$ vanish on tree level and no ghost terms appear for a linear gauge fixing. Thus this
equation simplifies to
 \begin{multline}\label{eq:sti-irr-4tree}
- \onepi{ \mathcal{D}_a(p)\Phi_i(k_i)\Phi_j(k_j)\Phi_k(k_k)}= T^a_{li}\onepi{\Phi_l(p+k_i)\Phi_j(k_j)\Phi_k(k_k)}\\
+ T^a_{lj}\onepi{\Phi_i(k_i)\Phi_l(p+k_j)\Phi_k(k_k)}+ T^a_{lk}\onepi{\Phi_i(k_i)\Phi_j(k_j)\Phi_l(p+k_k)}
\end{multline}
We see that these STIs indeed are very similar  to the STIs for amputated Green's functions \eqref{eq:lsz-wi-boson}. In the graphical notation of \fref{app:brs-graph} the terms on the right hand side of \fref{eq:sti-irr-4tree} can be written as
\begin{equation}
T^a_{ij} \onepi{\Phi_j\Phi_k\Phi_l}=\quad 
\parbox{20mm}{
\begin{fmfchar*}(15,15)
\fmfleft{l}
\fmfright{r1,r2}
\fmfpolyn{smooth,filled=shaded}{g}{3}
\fmf{plain,tension=2}{l,i,g1}
\fmf{plain}{r1,g2}
\fmf{plain}{r2,g3}
\fmfv{label=j,label.angle=135,l.d=0.8}{g1}
\fmfv{label=k,l.d=0.6}{r1}
\fmfv{label=l,l.d=0.6}{r2}
\fmfv{label=i,l.d=0.6}{l}
\fmfv{decor.shape=square,decor.filled=empty,decor.size=5}{i}
\end{fmfchar*}}
\end{equation}
so the graphical representation of the STI for the 4 point function is given by
\begin{equation}\label{eq:sti4-graph}
\parbox{15mm}{
\begin{fmfchar*}(15,15)
\fmfleft{l1,l2}
\fmfright{r1,r2}
\fmfpolyn{smooth,filled=shaded}{g}{4}
\fmf{double}{l1,g2}
\fmf{plain}{l2,g1}
\fmf{plain}{r1,g3}
\fmf{plain}{r2,g4}
\end{fmfchar*}}
=\qquad \parbox{15mm}{\begin{fmfchar*}(15,15)
\fmfleft{l1,l2}
\fmfright{r1,r2}
\fmf{phantom}{l1,g}
\fmf{plain,tension=2}{i,l2}
\fmf{plain,tension=2}{i,g}
\fmf{plain}{r1,g}
\fmf{plain}{r2,g}
\fmfv{decor.shape=square,decor.filled=empty,decor.size=5}{i}
\fmfv{decor.shape=circle,decor.filled=shaded,decor.size=9}{g}
\fmffreeze
\fmf{ghost}{l1,i}
 \end{fmfchar*}}
+
\qquad \parbox{15mm}{\begin{fmfchar*}(15,15)
\fmfleft{l1,l2}
\fmfright{r1,r2}
\fmf{phantom}{l1,g}
\fmf{plain}{l2,g}
\fmf{plain,tension=2}{r1,i}
\fmf{plain,tension=2}{i,g}
\fmf{plain}{r2,g}
\fmfv{decor.shape=square,decor.filled=empty,decor.size=5}{i}
\fmfv{decor.shape=circle,decor.filled=shaded,decor.size=9}{g}
\fmffreeze
\fmf{ghost}{l1,i}
 \end{fmfchar*}}
+
\qquad \parbox{15mm}{\begin{fmfchar*}(15,15)
\fmfleft{l1,l2}
\fmfright{r1,r2}
\fmf{phantom}{l1,g}
\fmf{plain}{l2,g}
\fmf{plain}{r1,g}
\fmf{plain,tension=2}{r2,i}
\fmf{plain,tension=2}{i,g}
\fmfv{decor.shape=square,decor.filled=empty,decor.size=5}{i}
\fmfv{decor.shape=circle,decor.filled=shaded,decor.size=9}{g}
\fmffreeze
\fmf{ghost,right=.3}{l1,i}
 \end{fmfchar*}}
\end{equation}
In a renormalizable theory there is no $5$ point vertex on tree level, so taking 4 derivatives of \fref{eq:sti} with respect to physical fields we get
\begin{equation}\label{eq:sti5-graph}
0=\parbox{20mm}{
\begin{fmfchar*}(20,20)
\fmfleftn{l}{2}
\fmfrightn{r}{3}
\fmf{plain}{l1,g}
\fmf{plain}{l2,g}
\fmf{plain}{r1,g}
\fmf{plain,tension=2}{r3,i}
\fmf{plain,tension=2}{i,g}
\fmffreeze
\fmf{ghost}{r2,i}
\fmfv{decor.shape=circle,decor.filled=shaded,decor.size=15}{g}
\fmfv{decor.shape=square,decor.filled=empty,decor.size=5}{i}
\end{fmfchar*}}\quad+\quad
\parbox{20mm}{
\begin{fmfchar*}(20,20)
\fmfleft{l1,l2}
\fmfright{r1,r2,r3}
\fmf{plain}{l1,g}
\fmf{plain}{l2,g}
\fmf{plain}{r3,g}
\fmf{plain,tension=2}{r1,i}
\fmf{plain,tension=2}{i,g}
\fmffreeze
\fmf{ghost}{r2,i}
\fmfv{decor.shape=square,decor.filled=empty,decor.size=5}{i}
\fmfv{decor.shape=circle,decor.filled=shaded,decor.size=15}{g}
\end{fmfchar*}}
\quad+\quad\parbox{20mm}{
\begin{fmfchar*}(20,20)
\fmfleft{l1,l2}
\fmfright{r1,r2,r3}
\fmf{plain}{l2,g}
\fmf{plain}{r1,g}
\fmf{plain,tension=2}{l1,i}
\fmf{plain,tension=2}{i,g}
\fmf{plain}{g,r3}
\fmffreeze
\fmf{ghost,left=0.3}{r2,i}
\fmfv{decor.shape=circle,decor.filled=shaded,decor.size=15}{g}
\fmfv{decor.shape=square,decor.filled=empty,decor.size=5}{i}
\end{fmfchar*}}\quad+\quad
\parbox{20mm}{
\begin{fmfchar*}(20,20)
\fmfleft{l1,l2}
\fmfright{r1,r2,r3}
\fmf{plain}{l1,g}
\fmf{plain}{r1,g}
\fmf{plain}{r3,g}
\fmf{plain,tension=2}{l2,i}
\fmf{plain,tension=2}{i,g}
\fmffreeze
\fmf{ghost}{r2,i}
\fmfv{decor.shape=square,decor.filled=empty,decor.size=5}{i}
\fmfv{decor.shape=circle,decor.filled=shaded,decor.size=15}{g}
\end{fmfchar*}}
\end{equation}

\section{STI for vertices with several contractions}\label{sec:2cov-sti}
For the reconstruction of the Feynman rules of a spontaneously broken gauge theory in \fref{part:reconstruction} we will need the Ward Identity \eqref{eq:chanowitz} with up to 4 contractions. The derivation of this identities from the STI  \fref{eq:ncov-sti-gf}  it is rather straightforward. One would expect that a similar identity for irreducible vertices can be derived from the ZJ-equation but to my knowledge this hasn't been done in the literature yet. 

Such an identity cannot be derived in the same way as the STIs with one contraction, since the dependence of the vertices on  $B$ is at most linear. Formally this can be seen from the gauge-fixing condition \eqref{eq:g-fix} that implies, taking two derivatives with respect to $B$:
\begin{equation}
\frac{\delta^2\Gamma}{\delta B_a\delta B_b}=\xi \delta_{ab}
\end{equation}
Therefore we cannot derive STIs with more than one contracted gauge boson by taking derivatives with respect to $B$. Also derivations of the the ZJ-equation \eqref{eq:sti} with respect to BRS-sources cannot be used in this context since this leads to no relations involving physical vertices. 

To arrive at the desired  identity for vertices with two contractions, we make use of a trick similar to that  used in the derivation of ordinary STIs in the formalism of BRS transformations without the Nakanishi-Lautrup field $B$ \cite{Boehm:2001,Boehm:1986} \nocite{Hollik:1993} and act on  \fref{eq:sti} with the operator
\begin{equation}
\mathcal{D}_a(x)=-\left(\dmu \frac{\delta}{\delta W_a(x)}+m_{W_a}\frac{\delta}{\delta \phi_a(x)}\right)
\end{equation}
This gives
\begin{multline}\label{eq:2cov-sti}
0=\sum_\Phi \onepi{ \mathcal{D}_a(p_a)\Phi^\star(p)}_{\Psi_{cl},  \Psi^\star}\onepi{\Phi(p)}_{\Psi_{cl},
  \Psi^\star}  +\onepi{ \Phi^\star(p)}_{\Psi_{cl},
  \Psi^\star}\onepi{ \mathcal{D}_a(p_a)\Phi(p)}_{\Psi_{cl},
  \Psi^\star}\\
+\left(\mathcal{D}_a(x)B_b(x)\right)\onepi{\bar c_b}_{\Psi_{cl},
  \Psi^\star}+B(x)\onepi{ \mathcal{D}_a(p_a)\bar c}
\end{multline}
The first term of \fref{eq:2cov-sti} where the operator $\mathcal{D}$ is inserted into a BRS-vertex appears because of the nonlinearity of the ZJ-equation \eqref{eq:sti}. In a similar way, we can obtain relations for vertices with 3 or 4 contractions by hitting \eqref{eq:2cov-sti} again with the operator $\mathcal{D}$. 
\subsection{3 point function}
We now derive the STI for 3 point vertices with 2 unphysical gauge bosons. Taking the derivative of  \eqref{eq:2cov-sti} with respect to a ghost field $c$ and an  arbitrary field $\Phi$ and setting classical fields and BRS-sources to zero yields 
\begin{multline}\label{eq:2cov-3}
0=\sum_\Phi\Biggl[\onepi{c_b (p_b)\mathcal{D}_a(p_a)\Phi^\star(p)}\onepi{\Phi(p)\Phi(k)}\\
+\onepi{ c_b (p_b)\Phi^\star(p)}\onepi{ \mathcal{D}_a(p_a)\Phi(p)\Phi(k)}
+\frac{\delta B}{\delta\Phi}\onepi{c(p_b) \mathcal{D}_a(p_a)\bar c(p)}\Bigr]
\end{multline}
Here we have omitted some steps that are given in detail in \fref{app:2cov-ghosts}.

Proceeding in the same way as in \fref{sec:sti-tree} we can now obtain tree level identities for the irreducible vertices. For the three point function with no additional gauge or vector boson we obtain
\begin{multline}\label{eq:3point-2cov}
\onepi{ \mathcal{D}_a(p_a)\mathcal{D}_b(p_b)\Phi(k))}\\ 
=\ii f^{cab}p_{a\mu}\onepi{W_c^\mu(p_a+p_b)\Phi(k)}+m_{W_a}T^b_{ia}\onepi{\Phi_i(p_a+p_b)\Phi(k)}
\end{multline}
Note that the last line arises from the insertion of the operator $\mathcal{D}$ into a BRS vertex and therefore this identity cannot be obtained from STIs with one contraction. At tree level the first term of the right hand side will only be present for $\Phi=W$  while the second term will be present for particles that are connected to the Goldstone bosons by gauge transformations (i.e. Higgs or Goldstone bosons). 

We introduce the following graphical representation for \fref{eq:3point-2cov}:
\begin{equation}\label{eq:3-2cov-graph}
\parbox{20mm}{
\begin{fmfchar*}(15,15)
\fmfleft{l}
\fmfright{r1,r2}
\fmf{double}{l,g}
\fmf{double}{r2,g}
\fmf{plain}{r1,g}
\fmfblob{10}{g}
\end{fmfchar*}}
=\qquad \parbox{15mm}{\begin{fmfchar*}(15,15)
\fmfleft{l}
\fmfright{r1,r2}
\fmf{ghost}{l,g}
\fmf{double}{r2,g}
\fmf{zigzag,tension=2}{r1,i}
\fmf{plain,tension=2}{i,g}
\fmfv{decor.shape=square,decor.filled=empty,decor.size=5}{g}
\end{fmfchar*}}
\end{equation}
For the three gauge boson vertex the ghost term in \fref{eq:3point-2cov} gives an additional contribution to the STI
\begin{multline}
\onepi{ \mathcal{D}_a(p_a)\mathcal{D}_b(p_b)W_{c\nu}(p_c))}\\ 
=\ii f^{dab}p_{a\mu}\onepi{W_c^\mu(p_a+p_b)W_{c\nu}(p_c)}
-\ii\frac{1}{\xi} p_{c\nu}\onepi{\mathcal{D}_a(p_a)\bar c_c(p_c)c_b(p_b)} 
 \end{multline}
that can be written diagrammatically as 
\begin{equation}\label{eq:3-2cov-ghost}
\parbox{20mm}{
\begin{fmfchar*}(15,15)
\fmfleft{l}
\fmfright{r1,r2}
\fmfpolyn{smooth,filled=shaded}{g}{3}
\fmf{double}{l,g1}
\fmf{photon}{r1,g2}
\fmf{double}{r2,g3}
\end{fmfchar*}}\quad =\quad
\parbox{15mm}{\begin{fmfchar*}(15,15)
\fmfleft{l}
\fmfright{r1,r2}
\fmf{ghost}{l,g}
\fmf{double}{r2,g}
\fmf{zigzag,tension=2}{r1,i}
\fmf{photon,tension=2}{i,g}
\fmfv{decor.shape=square,decor.filled=empty,decor.size=5}{g}
\end{fmfchar*}}
\qquad +\qquad \frac{1}{\xi}\quad
\parbox{25mm}{
\fmfframe(0,5)(0,5){
\begin{fmfchar*}(15,15)
\fmfleft{l}
\fmfright{r1,r2}
\fmfpolyn{smooth,filled=shaded}{g}{3}
\fmf{ghost}{l,g1}
\fmf{ghost}{g2,r1}
\fmf{double}{r2,g3}
\fmfv{decor.shape=square,decor.filled=full,decor.size=5}{r1}
\end{fmfchar*}}}
\end{equation}
The identity for the 2 gauge boson-Goldstone boson vertex becomes
\begin{multline}
\onepi{ \mathcal{D}_a(p_a)\mathcal{D}_b(p_b)\phi_c(p_c))}\\ 
=m_{W_a}T^b_{da}\onepi{\phi_d(p_a+p_b)\phi_c(p_c)}-m_{W_c}\onepi{\mathcal{D}_a(p_a)\bar c_c(p_c)c_b(p_b)} 
\end{multline}
\subsection{4 point function}
To derive the STI for 4 point vertices with 2 unphysical gauge bosons, we  take the derivative of  \eqref{eq:2cov-sti} with respect to a ghost field $c$ and two arbitrary fields $\Phi$. Setting classical fields and BRS-sources to zero we get 
\begin{equation}
\begin{aligned}
0=\sum_\Phi\Biggl[&\onepi{c_b (p_b)\mathcal{D}_a(p_a)\Phi^\star(p)}\onepi{\Phi(p)\Phi(k_i)\Phi(k_j)}\\
&+\onepi{ c_b (p_b)\Phi^\star(p)}\onepi{ \mathcal{D}_a(p_a)\Phi(p)\Phi(k_i)\Phi(k_j)}\\
&+\onepi{ c_b (p_b)\Phi^\star(p)\Phi(k_j)}\onepi{ \mathcal{D}_a(p_a)\Phi(p)\Phi(k_i)} + (i\leftrightarrow j)
\Biggr]
\end{aligned}
\end{equation}
On tree level this gives
\begin{multline}\label{eq:4point-2cov}
\onepi{ \mathcal{D}_a(p_a)\mathcal{D}_b(p_b)\Phi_i(k_i)\Phi_j(k_j)}=\ii f^{cab}p_{a\mu}\onepi{W_c^\mu(p_a+p_b)\Phi_i(k_i)\Phi_j(k_j)}\\
+m_{W_a}T^b_{ka}\onepi{\Phi_k(p_a+p_b)\Phi_i(k_i)\Phi_j(k_j)}
- T^b_{ki}\onepi{ \mathcal{D}_a(p_a)\Phi_k(p_b+k_i)\Phi_j(k_j)}\\
-T^b_{kj}\onepi{ \mathcal{D}_a(p_a)\Phi_i(k_i)\Phi_k(p_b+k_j)}
\end{multline}
Here the last two terms are similar to the structure of the STI with one momentum contraction \fref{eq:sti-irr-4tree} while the first two terms of the rhs arises from the insertion of the operator $\mathcal{D}$ into a BRS vertex. 

The graphical representation of \fref{eq:4point-2cov} is
\begin{equation}\label{eq:4-2cov-graph}
\parbox{15mm}{
\begin{fmfchar*}(15,15)
\fmfleft{l1,l2}
\fmfright{r1,r2}
\fmfpolyn{smooth,filled=shaded}{g}{4}
\fmf{double}{l1,g2}
\fmf{double}{l2,g1}
\fmf{plain}{r1,g3}
\fmf{plain}{r2,g4}
\end{fmfchar*}}
=\qquad \parbox{15mm}{\begin{fmfchar*}(15,15)
\fmfleft{l1,l2}
\fmfright{r1,r2}
\fmf{phantom}{l1,g}
\fmf{double,tension=2}{i,l2}
\fmf{plain,tension=2}{i,g}
\fmf{plain}{r1,g}
\fmf{plain}{r2,g}
\fmfv{decor.shape=square,decor.filled=empty,decor.size=5}{i}
\fmfv{decor.shape=circle,decor.filled=shaded,decor.size=9}{g}
\fmffreeze
\fmf{ghost}{l1,i}
 \end{fmfchar*}}
+
\qquad \parbox{15mm}{\begin{fmfchar*}(15,15)
\fmfleft{l1,l2}
\fmfright{r1,r2}
\fmf{phantom}{l1,g}
\fmf{double}{l2,g}
\fmf{plain,tension=2}{r1,i}
\fmf{plain,tension=2}{i,g}
\fmf{plain}{r2,g}
\fmfv{decor.shape=square,decor.filled=empty,decor.size=5}{i}
\fmfv{decor.shape=circle,decor.filled=shaded,decor.size=9}{g}
\fmffreeze
\fmf{ghost}{l1,i}
 \end{fmfchar*}}
+
\qquad \parbox{15mm}{\begin{fmfchar*}(15,15)
\fmfleft{l1,l2}
\fmfright{r1,r2}
\fmf{phantom}{l1,g}
\fmf{double}{l2,g}
\fmf{plain}{r1,g}
\fmf{plain,tension=2}{r2,i}
\fmf{plain,tension=2}{i,g}
\fmfv{decor.shape=square,decor.filled=empty,decor.size=5}{i}
\fmfv{decor.shape=circle,decor.filled=shaded,decor.size=9}{g}
\fmffreeze
\fmf{ghost,right=.3}{l1,i}
 \end{fmfchar*}}
\end{equation}

\chapter{Diagrammatical analysis of STIs}\label{chap:diagrammatica}
In \fref{chap:sti} we have reviewed the STIs for Green's functions and for irreducible vertices that are both consequences of BRS symmetry but were derived using different formalisms. On a formal level, the equivalence of  the STIs for those different objects follows from the fact that the generating functional of connected Green's functions and the effective action are Legendre transforms of one another. Both for the reconstruction of the Feynman rules from the Ward Identities and for the question of the correct definition of the gauge flips, a more explicit understanding of this connection seems desirable. We will provide a diagrammatic analysis of these subjects in this chapter. After the groundbreaking work of 't Hooft and Veltman  \cite{'tHooft:1972}, diagrammatic  methods have been somewhat superseded by algebraic and operator  methods  \cite{Becchi:1975,Piguet:1981} but they are---almost by definition---indispensable for the discussion of gauge invariance classes of Feynman diagrams. Furthermore,  since practical calculations in perturbation theory are done using Feynman diagrams, a diagrammatic understanding of the STIs is useful to see `how things really work'. 

For later use in the reconstruction of the Feynman rules, in \fref{sec:skeleton} we analyze the connection between  the on-shell Ward Identities of the 4 point functions and the STIs of irreducible vertices, both for the STIs with one contraction and the identities with several contractions derived in \fref{sec:2cov-sti}. 
  
 We will then turn to the gauge parameter independence of physical matrix elements in \fref{sec:gpi}. We show on tree level that the Ward Identities of subamplitudes imply the gauge parameter independence of physical scattering amplitudes. 

In \fref{sec:gi-classes} we discuss the STIs of general Green's functions and the
appearance of gauge invariant subsets of Feynman diagrams. We will establish a connection
to the formalism of gauge flips \cite{Boos:1999}  reviewed in
\fref{sec:groves} and provide a precise definition of gauge flips that is
applied to spontaneously broken gauge theories in \fref{sec:ssb-flips}.  This proof has 
already appeared in less detailed form in \cite{OS:FGH} for the case of Green's functions where the external legs
are not amputated.
\section{4 point functions}\label{sec:skeleton}
As discussed in \fref{sec:rec-motivate}, our aim is to check the validity of the Feynman rules from a finite set of Ward Identities for on shell amplitudes. It is known from algebraic renormalization theory\cite{Piguet:1981} that the STIs of the irreducible vertices determine the form of the action. The generating functionals of connected Green's functions and of the irreducible vertices are connected by a Legendre transformation. This suggests that at tree level in a renormalizable theory the STIs for 4 point Green's functions are sufficient to check the Feynman rules of the theory. However, this argument doesn't show that the Ward Identities for on shell Green's functions are sufficient for the reconstruction of the Feynman rules. To see if this can be achieved nevertheless, we analyze the connection between the STIs of the irreducible vertices and the Ward Identities in this section. 

\subsection{Ward Identities}\label{sec:wi-4} 
We use the \emph{skeleton series} to express
Green's functions in terms of irreducible vertices. The 3 point vertex function is just the amputated 3 point Green's function:
\begin{equation}
  \onepi{\Phi_i(p_i)\Phi_j(p_j)\Phi_k(p_k)}=\me(\Phi_i(p_i)\Phi_j(p_j)\Phi_k(p_k))
\end{equation}

\begin{figure}[htbp]
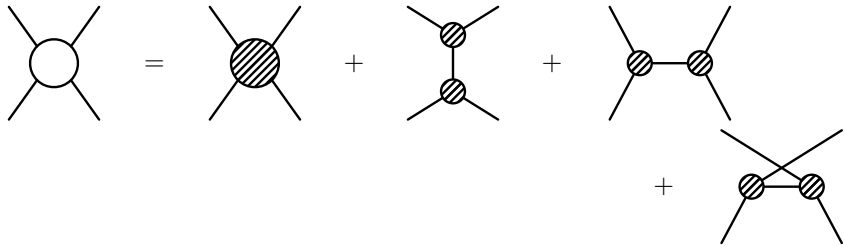

\begin{center}
\begin{multline*}
\parbox{15mm}{
\begin{fmfchar*}(15,15)
\fmfleft{A1,A2}
\fmfright{A3,A4}
\fmf{plain}{A1,a2}
\fmf{plain}{A2,a1}
\fmf{plain}{A3,a3}
\fmf{plain}{A4,a4}
\fmfpolyn{smooth}{a}{4}
\end{fmfchar*}}
\quad=\quad
\parbox{15mm}{
\begin{fmfchar*}(15,15)
\fmfleft{A1,A2}
\fmfright{A3,A4}
\fmf{plain}{A1,a2}
\fmf{plain}{A2,a1}
\fmf{plain}{A3,a3}
\fmf{plain}{A4,a4}
\fmfpolyn{shade,smooth}{a}{4}
\end{fmfchar*}}
\quad+\quad
\parbox{15mm}{
\begin{fmfchar*}(15,15)
\fmfleft{A1,A2}
\fmfright{A3,A4}
\fmf{plain}{A1,a}
\fmf{plain}{A2,b}
\fmf{plain}{b,a}
\fmf{plain}{a,A3}
\fmf{plain}{b,A4}
\fmfv{decor.shape=circle,decor.filled=shaded,decor.size=9}{a}
\fmfv{decor.shape=circle,decor.filled=shaded,decor.size=9}{b}
\end{fmfchar*}}
\quad+\quad
\parbox{20mm}{
\begin{fmfchar*}(20,15)
\fmfleft{A1,A2}
\fmfright{A3,A4}
\fmf{plain}{A1,a}
\fmf{plain}{A2,a}
\fmf{plain}{a,b}
\fmf{plain}{A3,b}
\fmf{plain}{A4,b}
\fmfv{decor.shape=circle,decor.filled=shaded,decor.size=9}{a}
\fmfv{decor.shape=circle,decor.filled=shaded,decor.size=9}{b}
\end{fmfchar*}}
\\
+\quad
\parbox{20mm}{
\begin{fmfchar*}(20,15)
\fmfleft{A1,A2}
\fmfright{A3,A4}
\fmf{plain}{A1,a}
\fmf{phantom}{A2,a}
\fmf{plain}{a,b}
\fmf{plain}{A3,b}
\fmf{phantom}{A4,b}
\fmffreeze
\fmf{plain}{A2,b}
\fmf{plain}{A4,a}
\fmfv{decor.shape=circle,decor.filled=shaded,decor.size=9}{a}
\fmfv{decor.shape=circle,decor.filled=shaded,decor.size=9}{b}
\end{fmfchar*}}
\end{multline*}
\caption{Skeleton series}\label{fig:skeleton-graph}
\end{center}
\end{figure}

The skeleton series expresses the amputated 4 point Green's function in terms
of vertex functions and propagators  (see \fref{fig:skeleton-graph}):
 \begin{equation}\label{eq:4-skeleton}
\begin{aligned}
 \me(\Phi_i\Phi_j\Phi_k\Phi_l)=& \onepi{\Phi_i\Phi_j\Phi_k\Phi_l}\\
 &+\onepi{\Phi_i\Phi_j\Phi_m}D_{\Phi_m}(p_i+p_j)
\onepi{\Phi_m\Phi_k\Phi_l}\\
&+\onepi{\Phi_i\Phi_k\Phi_m}D_{\Phi_m}(p_i+p_k)
\onepi{\Phi_m\Phi_j\Phi_l}\\
&+\onepi{\Phi_i\Phi_l\Phi_m}D_{\Phi_m}(p_i+p_l)
\onepi{\Phi_m\Phi_j\Phi_k}
\end{aligned}
 \end{equation}
 
Using the skeleton series for the 4 point function and assuming that the STI for
the 3 point vertex \eqref{eq:sti-irr-3tree} is satisfied we can now show that
the STI for the 4 point vertex implies the Ward Identity for the 4 point Green's function
and vice versa. 

If (at least) one of the fields in the Green's function is a gauge boson, we
can contract it with the momentum vector and add the corresponding Goldstone boson diagram so we obtain 
 \begin{equation}\label{eq:4-skeleton-wi}
\me(\mathcal{D}_a\Phi_i\Phi_j\Phi_k)
= \onepi{\mathcal{D}_a\Phi_i\Phi_j\Phi_k}+\onepi{\mathcal{D}_a\Phi_i\Phi_l}D_{\Phi_l}(p_a+p_i) \onepi{\Phi_l\Phi_j\Phi_k}+\dots
 \end{equation}
In the next step we simplify this expression using the STI for
the 3 point vertex derived in \fref{eq:sti-irr-3tree}. Apart from the ghost
terms for internal gauge and Goldstone bosons, the STI \eqref{eq:sti-irr-3tree} has the
same form for all particles. We will show below, that the ghost terms drop out
of physical amplitudes and concentrate on the generic features
first. Neglecting the ghost terms, we  can use \eqref{eq:sti-irr-3tree} to simplify the terms with particle exchange in \fref{eq:4-skeleton-wi} to:
\begin{multline}\label{eq:4-skeleton-sti}
\onepi{\mathcal{D}_a\Phi_i\Phi_l}D_{\Phi_l}(p_a+p_i) \onepi{\Phi_l\Phi_j\Phi_k}\\
= \Bigl(T^a_{li} D^{-1}_{\Phi_l}(p_i+p_a)+T^a_{il}D^{-1}_{\Phi_i}(p_i)\Bigr)D_{\Phi_l}(p_i+p_a) \onepi{\Phi_l\Phi_j\Phi_k}\\
=T^a_{li} \onepi{\Phi_l\Phi_j\Phi_k} + D^{-1}_{\Phi_i}(p_i)D_{\Phi_l}(p_i+p_a) T^a_{il}\onepi{\Phi_l\Phi_j\Phi_k}
 \end{multline}
If all external particles (with the possible exception of $W_a$) are  on-shell, we
have $D^{-1}_{\Phi_j}(p_j)=0$ and therefore we get for the amplitude
 \begin{multline}
 0=\me(\mathcal{D}_a\Phi_i\Phi_j\Phi_k)= \onepi{\mathcal{D}_a\Phi_i\Phi_j\Phi_k} \\
+T^a_{li}\onepi{\Phi_l(p_i+p_a)\Phi_j(p_j)\Phi_k(p_k)}+T^a_{lj}\onepi{\Phi_i(p_i)\Phi_l(p_j+p_a)\Phi_k(p_k)}\\
+T^a_{lk}\onepi{\Phi_i(p_i)\Phi_j(p_j)\Phi_k(p_k+p_a)}
 \end{multline}
This is just the STI for the 4 point function, as given in \fref{eq:sti-irr-4tree}. 

For internal gauge bosons we have to take the ghost terms from \fref{eq:2gb-sti} into account. This gives additional terms compared to \fref{eq:4-skeleton-sti} that can be shown to  vanish if the external particles are  on-shell, using the Ward Identities for the 3 point functions: 
\begin{multline}\label{eq:wi-internal}
\ii \frac{1}{\xi} \onepi { c_a \Phi_l\bar c_b}p_b^\mu {D_{W_b}}_\mu(p_a+p_i) \onepi{W_b\Phi_j\Phi_k}\\
 +m_{W_b}\onepi {c_a(p_a) \Phi_l\bar c_b}D_{\phi_b}(p_a+p_i) \onepi{\phi_b\Phi_j\Phi_k}\\
=-\onepi { c_a \Phi_l\bar c_b}D_{c_b}(p_a+p_i) \onepi{\mathcal{D}_b\Phi_j\Phi_k}=0
\end{multline}
Here we have used the relation \eqref{eq:prop-rel} to express the internal propagators in terms of the ghost propagator. Note that the internal gauge boson has to be regarded as outgoing at one and as incoming at the other vertex. 

Therefore we have proven the following theorem:
\begin{theorem}\label{theo:sti-wi-equiv}
The  STIs for the three point and four vertices \eqref{eq:sti-irr-3tree} and  \eqref{eq:sti-irr-4tree}  imply the Ward Identity for the 4 point function \eqref{eq:gf-wi}:
\begin{subequations}
\begin{equation}
\begin{aligned}
\left(\begin{aligned}
&-\onepi{ \mathcal{D}_a(p)\Phi_i\Phi_j}\\
&=T^a_{ki}D^{-1}(k_j)+T^a_{kj}D^{-1}(k_i)
\end{aligned}\right)&\text{and}
\left(
\begin{aligned}
&-\onepi{ \mathcal{D}_a(p)\Phi_i\Phi_j\Phi_k}\\
&= T^a_{li}\onepi{\Phi_l(p+k_i)\Phi_j\Phi_k}+\dots
\end{aligned}\right)\\
&\Downarrow\\
\me( \mathcal{D}_a(p)\Phi_i\Phi_j\Phi_k)&=0
\end{aligned}
\end{equation}
Conversely, the three particle STI  \eqref{eq:sti-irr-3tree} and the Ward Identity
\eqref{eq:gf-wi} for the 4 point Green's function imply the STI
\eqref{eq:sti-irr-4tree} for the 4 point vertex:
\begin{equation}
\begin{aligned}
 \me( \mathcal{D}_a(p)\Phi_i\Phi_j\Phi_k)=0\quad 
&\text{and}\quad
\left(\begin{aligned}
&\onepi{ \mathcal{D}_a(p)\Phi_i\Phi_j}\\
&=- T^a_{ki}D^{-1}(k_j)- T^a_{kj}D^{-1}(k_i)
\end{aligned}\right)\\
&\Downarrow\\
 \onepi{ \mathcal{D}_a(p)\Phi_i\Phi_j\Phi_k}&= T^a_{li}\onepi{\Phi_l(p+k_i)\Phi_j\Phi_k}+\dots
\end{aligned}
\end{equation}
\qed
\end{subequations}
\end{theorem} 
\subsection{Slavnov Taylor Identity}\label{sec:sti-4}
Theorem \ref{theo:sti-wi-equiv} is the main result that is relevant for the discussion of the reconstruction of the Feynman rules in \fref{part:reconstruction}. But---as a preparation of the discussion of gauge invariance classes in \fref{sec:gi-classes}---we can also show that a more general result is true and the STIs of the 3 and 4 point vertices imply also the \emph{STIs} of the 4 point Green's functions. To show this, it is more convenient to use  the graphical notation introduced in 
\fref{sec:sti-tree}. Our treatment is a generalization of the  graphical discussion of the $\bar q q\to gg$ amplitude in \cite{Kugo:1997} to general four point functions with off shell particles in spontaneously broken gauge theories.

An important diagrammatic identity is the graphical expression of the fact that the
irreducible 2 point function is the inverse propagator \eqref{eq:inv-prop}:
\begin{equation}
\parbox{15mm}{
\begin{fmfchar*}(15,15)
\fmfleft{l}
\fmfright{r1,r2}
\fmf{plain,tension=1.5}{l,v}
\fmf{phantom}{v,r1}
\fmf{phantom}{v,r2}
\fmffreeze
\fmf{zigzag}{v,r1}
\fmf{ghost}{r2,r1}
\end{fmfchar*}}
\quad =\quad-\quad
\parbox{15mm}{
\begin{fmfchar*}(15,15)
\fmfleft{l}
\fmfright{r1,r2}
\fmf{phantom}{l,r1}
\fmf{phantom}{r2,r1}
\fmffreeze
\fmf{ghost}{r2,l}
\end{fmfchar*}}
\end{equation}

If we use this together with the STI for the 3 point vertex
\eqref{eq:sti-3graph} we get for an $s$-channel exchange diagram (the ghost terms in the STIs \fref{eq:2gb-sti} and \eqref{eq:phi-sti} lead to contact terms only and don't involve  cancellations with the 4 point vertex. They will be discussed in the sequel.):
\begin{multline}
\parbox{15mm}{
\begin{fmfchar*}(15,15)
\fmfleft{A1,A2}
\fmfright{A3,A4}
\fmf{double}{A1,a}
\fmf{plain}{A2,b}
\fmf{plain}{b,a}
\fmf{plain}{a,A3}
\fmf{plain}{b,A4}
\fmfv{decor.shape=circle,decor.filled=shaded,decor.size=9}{a}
\fmfv{decor.shape=circle,decor.filled=shaded,decor.size=9}{b}
\end{fmfchar*}}\qquad =
\qquad
\parbox{15mm}{
\begin{fmfchar*}(15,15)
\fmfleft{A1,A2}
\fmfright{A3,A4}
\fmf{ghost}{A1,a}
\fmf{plain}{A2,b}
\fmf{zigzag}{a,i}
\fmf{plain}{b,i}
\fmf{plain}{a,A3}
\fmf{plain}{b,A4}
\fmfv{decor.shape=square,decor.filled=empty,decor.size=5}{a}
\fmfv{decor.shape=circle,decor.filled=shaded,decor.size=9}{b}
\end{fmfchar*}}
\qquad +\qquad 
\parbox{15mm}{
\begin{fmfchar*}(15,15)
\fmfleft{A1,A2}
\fmfright{A3,A4}
\fmf{ghost}{A1,a}
\fmf{plain}{A2,b}
\fmf{plain}{b,a}
\fmf{zigzag}{a,A3}
\fmf{plain}{b,A4}
\fmfv{decor.shape=square,decor.filled=empty,decor.size=5}{a}
\fmfv{decor.shape=circle,decor.filled=shaded,decor.size=9}{b}
\end{fmfchar*}}\\
=-\qquad 
\parbox{15mm}{
\begin{fmfchar*}(15,15)
\fmfleft{l1,l2}
\fmfright{r1,r2}
\fmfv{decor.shape=circle,decor.filled=shaded,decor.size=9}{g}
\fmf{phantom}{l1,g}
\fmf{plain}{l2,g}
\fmf{plain,tension=2}{r1,i,g}
\fmf{plain}{r2,g}
\fmffreeze
\fmfv{decor.shape=square,decor.filled=empty,decor.size=5}{i}
\fmf{ghost}{l1,i}
\end{fmfchar*}}
\qquad +\qquad 
\parbox{15mm}{
\begin{fmfchar*}(15,15)
\fmfleft{A1,A2}
\fmfright{A3,A4}
\fmf{ghost}{A1,a}
\fmf{plain}{A2,b}
\fmf{plain}{b,a}
\fmf{zigzag}{a,A3}
\fmf{plain}{b,A4}
\fmfv{decor.shape=square,decor.filled=empty,decor.size=5}{a}
\fmfv{decor.shape=circle,decor.filled=shaded,decor.size=9}{b}
\end{fmfchar*}}
\end{multline}
The same manipulations can be done in the $s$ and $u$ channel diagrams.
Using the skeleton expansion from \fref{fig:skeleton-graph} and the STI \eqref{eq:sti4-graph} we see that the first term cancels against a term from the STI for the 4-particle vertex. The same happens for the remaining diagrams and therefore we obtain an STI for the amplitude:
\begin{multline}\label{eq:sti4}
\parbox{15mm}{
\begin{fmfchar*}(15,15)
\fmfleft{A1,A2}
\fmfright{A3,A4}
\fmf{double}{A1,a2}
\fmf{plain}{A2,a1}
\fmf{plain}{A3,a3}
\fmf{plain}{A4,a4}
\fmfpolyn{smooth}{a}{4}
\end{fmfchar*}}
=\parbox{15mm}{
\begin{fmfchar*}(15,15)
\fmfleft{A1,A2}
\fmfright{A3,A4}
\fmf{ghost}{A1,a}
\fmf{plain}{A2,b}
\fmf{plain}{b,a}
\fmf{zigzag}{a,A3}
\fmf{plain}{b,A4}
\fmfv{decor.shape=square,decor.filled=empty,decor.size=5}{a}
\fmfv{decor.shape=circle,decor.filled=shaded,decor.size=9}{b}
\end{fmfchar*}}
+\parbox{20mm}{
\begin{fmfchar*}(20,15)
\fmfleft{A1,A2}
\fmfright{A3,A4}
\fmf{ghost}{A1,a}
\fmf{plain}{A3,b}
\fmf{plain}{b,a}
\fmf{zigzag}{a,A2}
\fmf{plain}{b,A4}
\fmfv{decor.shape=square,decor.filled=empty,decor.size=5}{a}
\fmfv{decor.shape=circle,decor.filled=shaded,decor.size=9}{b}
\end{fmfchar*}}
+\parbox{20mm}{
\begin{fmfchar*}(20,15)
\fmfleft{A1,A2}
\fmfright{A3,A4}
\fmf{ghost}{A1,a}
\fmf{plain}{A3,b}
\fmf{plain}{b,a}
\fmf{phantom}{a,A2}
\fmf{phantom}{b,A4}
\fmffreeze
\fmf{zigzag}{a,A4}
\fmf{plain}{b,A2}
\fmfv{decor.shape=square,decor.filled=empty,decor.size=5}{a}
\fmfv{decor.shape=circle,decor.filled=shaded,decor.size=9}{b}
\end{fmfchar*}}
\end{multline}
Comparing with the structure of the contact terms of the STI sketched in \fref{eq:4-contact} we see that the terms on the right hand side are the contact terms of the STI \eqref{eq:sti-graph} that arise if the ghosts are connected to the rest of the diagram only by a BRS-vertex. The remaining contact terms will arise if we consider internal gauge bosons. 

Next, we turn to the case of external gauge bosons. For every external gauge boson, we get an additional term from the ghost contribution to the STI of the three point vertex \eqref{eq:2gb-sti-graph}. The $s$-channel exchange diagram considered above now becomes:
\begin{equation}
\parbox{15mm}{
\begin{fmfchar*}(15,15)
\fmfleft{A1,A2}
\fmfright{A3,A4}
\fmf{double}{A1,a}
\fmf{plain}{A2,b}
\fmf{plain}{b,a}
\fmf{photon}{a,A3}
\fmf{plain}{b,A4}
\fmfv{decor.shape=circle,decor.filled=shaded,decor.size=9}{a}
\fmfv{decor.shape=circle,decor.filled=shaded,decor.size=9}{b}
\end{fmfchar*}}\quad =
-\quad 
\parbox{15mm}{
\begin{fmfchar*}(15,15)
\fmfleft{l1,l2}
\fmfright{r1,r2}
\fmfv{decor.shape=circle,decor.filled=shaded,decor.size=9}{g}
\fmf{phantom}{l1,g}
\fmf{plain}{l2,g}
\fmf{photon}{r1,i}
\fmf{plain}{i,g}
\fmf{plain}{r2,g}
\fmffreeze
\fmfv{decor.shape=square,decor.filled=empty,decor.size=5}{i}
\fmf{ghost}{l1,i}
\end{fmfchar*}}
\quad +\quad 
\parbox{15mm}{
\begin{fmfchar*}(15,15)
\fmfleft{A1,A2}
\fmfright{A3,A4}
\fmf{ghost}{A1,a}
\fmf{plain}{A2,b}
\fmf{plain}{b,a}
\fmf{zigzag}{a,A3}
\fmf{plain}{b,A4}
\fmfv{decor.shape=square,decor.filled=empty,decor.size=5}{a}
\fmfv{decor.shape=circle,decor.filled=shaded,decor.size=9}{b}
\end{fmfchar*}}
\quad+\frac{1}{\xi}\quad
\parbox{15mm}{
\begin{fmfchar*}(15,15)
\fmfleft{A1,A2}
\fmfright{A3,A4}
\fmf{ghost}{A1,a}
\fmf{plain}{A2,b}
\fmf{plain}{b,a}
\fmf{ghost}{a,A3}
\fmf{plain}{b,A4}
\fmfv{decor.shape=circle,decor.filled=shaded,decor.size=9}{a}
\fmfv{decor.shape=circle,decor.filled=shaded,decor.size=9}{b}
\fmfv{decor.shape=square,decor.filled=full,decor.size=5}{A3}
\end{fmfchar*}}
\end{equation}
The additional terms are the same ones that appear in the STI for amputated Green's functions \eqref{eq:lsz-wi-gauge}. 
 
In the case of internal gauge bosons we also have to consider the corresponding diagrams with internal Goldstone bosons to cancel the ghost diagrams of the STI \eqref{eq:2gb-sti-graph}. The diagram with an internal gauge boson gives
\begin{equation}\label{eq:t-channel-gauge}
\parbox{15mm}{
\begin{fmfchar*}(15,15)
\fmfleft{A1,A2}
\fmfright{A3,A4}
\fmf{double}{A1,a}
\fmf{plain}{A2,b}
\fmf{photon}{b,a}
\fmf{plain}{a,A3}
\fmf{plain}{b,A4}
\fmfv{decor.shape=circle,decor.filled=shaded,decor.size=9}{a}
\fmfv{decor.shape=circle,decor.filled=shaded,decor.size=9}{b}
\end{fmfchar*}}\quad =
-\quad 
\parbox{15mm}{
\begin{fmfchar*}(15,15)
\fmfleft{l1,l2}
\fmfright{r1,r2}
\fmfv{decor.shape=circle,decor.filled=shaded,decor.size=9}{g}
\fmf{phantom}{l1,g}
\fmf{plain}{l2,g}
\fmf{plain}{r1,i}
\fmf{photon}{i,g}
\fmf{plain}{r2,g}
\fmffreeze
\fmfv{decor.shape=square,decor.filled=empty,decor.size=5}{i}
\fmf{ghost}{l1,i}
\end{fmfchar*}}
\quad +\quad 
\parbox{15mm}{
\begin{fmfchar*}(15,15)
\fmfleft{A1,A2}
\fmfright{A3,A4}
\fmf{ghost}{A1,a}
\fmf{plain}{A2,b}
\fmf{photon}{b,a}
\fmf{zigzag}{a,A3}
\fmf{plain}{b,A4}
\fmfv{decor.shape=square,decor.filled=empty,decor.size=5}{a}
\fmfv{decor.shape=circle,decor.filled=shaded,decor.size=9}{b}
\end{fmfchar*}}
\quad+\frac{1}{\xi}\quad
\parbox{20mm}{
\begin{fmfchar*}(15,20)
\fmfleft{A1,A2}
\fmfright{A3,A4}
\fmf{ghost}{A1,a}
\fmf{plain}{A2,b}
\fmf{ghost,tension=1}{a,i}
\fmf{photon}{i,b}
\fmf{plain}{a,A3}
\fmf{plain}{b,A4}
\fmfv{decor.shape=circle,decor.filled=shaded,decor.size=9}{a}
\fmfv{decor.shape=circle,decor.filled=shaded,decor.size=9}{b}
\fmfv{decor.shape=square,decor.filled=full,decor.size=5}{i}
\end{fmfchar*}}
\end{equation}
Here  the internal ghost line in the last diagram has to be understood as amputated, no ghost propagator appears. 

Similarly, the corresponding Goldstone boson diagram is 
\begin{equation}
\parbox{20mm}{
\begin{fmfchar*}(20,20)
\fmfleft{A1,A2}
\fmfright{A3,A4}
\fmf{double}{A1,a}
\fmf{plain}{A2,b}
\fmf{dashes}{b,a}
\fmf{plain}{a,A3}
\fmf{plain}{b,A4}
\fmfv{decor.shape=circle,decor.filled=shaded,decor.size=9}{a}
\fmfv{decor.shape=circle,decor.filled=shaded,decor.size=9}{b}
\end{fmfchar*}}\quad =
-\quad 
\parbox{15mm}{
\begin{fmfchar*}(15,15)
\fmfleft{l1,l2}
\fmfright{r1,r2}
\fmfv{decor.shape=circle,decor.filled=shaded,decor.size=9}{g}
\fmf{phantom}{l1,g}
\fmf{plain}{l2,g}
\fmf{plain}{r1,i}
\fmf{dashes}{i,g}
\fmf{plain}{r2,g}
\fmffreeze
\fmfv{decor.shape=square,decor.filled=empty,decor.size=5}{i}
\fmf{ghost}{l1,i}
\end{fmfchar*}}
\quad +\quad 
\parbox{15mm}{
\begin{fmfchar*}(15,15)
\fmfleft{A1,A2}
\fmfright{A3,A4}
\fmf{ghost}{A1,a}
\fmf{plain}{A2,b}
\fmf{dashes}{b,a}
\fmf{zigzag}{a,A3}
\fmf{plain}{b,A4}
\fmfv{decor.shape=square,decor.filled=empty,decor.size=5}{a}
\fmfv{decor.shape=circle,decor.filled=shaded,decor.size=9}{b}
\end{fmfchar*}}
\quad -m_W\quad
\parbox{20mm}{
\begin{fmfchar*}(15,20)
\fmfleft{A1,A2}
\fmfright{A3,A4}
\fmf{ghost}{A1,a}
\fmf{plain}{A2,b}
\fmf{ghost,tension=1}{a,i}
\fmf{dashes}{i,b}
\fmf{plain}{a,A3}
\fmf{plain}{b,A4}
\fmfv{decor.shape=circle,decor.filled=shaded,decor.size=9}{a}
\fmfv{decor.shape=circle,decor.filled=shaded,decor.size=9}{b}
\fmfv{decor.shape=cross,decor.size=5}{i}
\end{fmfchar*}}
\end{equation}
The additional contributions from both diagrams can be simplified using the STI \eqref{eq:local-wi}:
\begin{equation}\label{eq:sub3-sti}
\frac{1}{\xi}\quad
\parbox{20mm}{
\begin{fmfchar*}(15,20)
\fmfleft{A1,A2}
\fmfright{A3,A4}
\fmf{ghost}{A1,a}
\fmf{plain}{A2,b}
\fmf{ghost,tension=1}{a,i}
\fmf{photon}{i,b}
\fmf{plain}{a,A3}
\fmf{plain}{b,A4}
\fmfv{decor.shape=circle,decor.filled=shaded,decor.size=9}{a}
\fmfv{decor.shape=circle,decor.filled=shaded,decor.size=9}{b}
\fmfv{decor.shape=square,decor.filled=full,decor.size=5}{i}
\end{fmfchar*}}
\quad-m_W\quad
\parbox{20mm}{
\begin{fmfchar*}(15,20)
\fmfleft{A1,A2}
\fmfright{A3,A4}
\fmf{ghost}{A1,a}
\fmf{plain}{A2,b}
\fmf{ghost,tension=1.0}{a,i}
\fmf{dashes}{i,b}
\fmf{plain}{a,A3}
\fmf{plain}{b,A4}
\fmfv{decor.shape=circle,decor.filled=shaded,decor.size=9}{a}
\fmfv{decor.shape=circle,decor.filled=shaded,decor.size=9}{b}
\fmfv{decor.shape=cross,decor.size=5}{i}
\end{fmfchar*}}
\quad =\quad
\parbox{15mm}{
\begin{fmfchar*}(15,15)
\fmfleft{A1,A2}
\fmfright{A3,A4}
\fmf{ghost}{A1,a}
\fmf{plain}{A3,a}
\fmf{ghost}{a,i}
\fmf{plain}{A2,i}
\fmf{zigzag}{i,A4}
\fmfv{decor.shape=circle,decor.filled=shaded,decor.size=9}{a}
\fmfv{decor.shape=square,decor.size=5,d.fi=empty}{i}
\end{fmfchar*}}\quad+\quad
\parbox{15mm}{
\begin{fmfchar*}(15,15)
\fmfleft{A1,A2}
\fmfright{A3,A4}
\fmf{ghost}{A1,a}
\fmf{plain}{A3,a}
\fmf{ghost}{a,i}
\fmf{zigzag}{A2,i}
\fmf{plain}{i,A4}
\fmfv{decor.shape=circle,decor.filled=shaded,decor.size=9}{a}
\fmfv{decor.shape=square,decor.size=5,d.fi=empty}{i}
\end{fmfchar*}}
\end{equation}
This manipulation is the graphical representation of \eqref{eq:wi-internal}
for the case of external off-shell particles. We see that the diagrams that arise because of the ghost terms in the STIs for the gauge bosons, are exactly the contributions of the interactions from the ghost-Lagrangian to the contact terms (compare with \fref{eq:4-contact}). The first diagram on the right hand side together with a diagram from the $s$-channel and from the $u$ channel gives a contact term of the STI:
\begin{equation} 
\parbox{20mm}{
\begin{fmfchar*}(20,15)
\fmfbottom{A1,A2}
\fmftop{A3,A4}
\fmf{ghost}{A1,a}
\fmf{plain}{A3,a}
\fmf{ghost}{a,i}
\fmf{plain}{A2,i}
\fmf{zigzag}{i,A4}
\fmfv{decor.shape=circle,decor.filled=shaded,decor.size=9}{a}
\fmfv{decor.shape=square,decor.size=5,d.fi=empty}{i}
\end{fmfchar*}}\quad + \quad 
\parbox{15mm}{
\begin{fmfchar*}(15,15)
\fmfleft{A1,A2}
\fmfright{A3,A4}
\fmf{ghost}{A1,a}
\fmf{plain}{A3,a}
\fmf{ghost}{a,i}
\fmf{plain}{A2,i}
\fmf{zigzag}{i,A4}
\fmfv{decor.shape=circle,decor.filled=shaded,decor.size=9}{a}
\fmfv{decor.shape=square,decor.size=5,d.fi=empty}{i}
\end{fmfchar*}}\quad+\quad
\parbox{20mm}{
\begin{fmfchar*}(20,15)
\fmfleft{A1,A2}
\fmfright{A3,A4}
\fmf{ghost}{A1,a}
\fmf{plain}{A3,b}
\fmf{plain}{b,a}
\fmf{phantom}{a,A2}
\fmf{phantom}{b,A4}
\fmffreeze
\fmf{zigzag}{a,A4}
\fmf{plain}{b,A2}
\fmfv{decor.shape=square,decor.filled=empty,decor.size=5}{a}
\fmfv{decor.shape=circle,decor.filled=shaded,decor.size=9}{b}
\end{fmfchar*}}
=\parbox{20mm}{
\begin{fmfchar*}(20,20)
\fmfbottom{A1,A2}
\fmftop{A3,A4}
\fmf{ghost}{A1,a}
\fmf{plain}{A3,a}
\fmf{plain}{a,A2}
\fmf{phantom}{a,A4}
\fmffreeze
\fmf{plain}{a,i}
\fmf{zigzag}{i,A4}
\fmffreeze
\fmf{ghost,left}{a,i}
\fmfv{decor.shape=diamond,decor.filled=empty,decor.size=15}{a}
\fmfv{decor.shape=square,decor.size=5,d.fi=empty}{i}
\end{fmfchar*}}
\end{equation}

Similarly, the remaining terms of \fref{eq:sti4} and \fref{eq:sub3-sti} and the corresponding $s$- and $u$-channel diagrams add up to contact terms so altogether we have derived the STI:
\begin{equation}
\parbox{20mm}{
\begin{fmfchar*}(20,20)
\fmfleft{A1,A2}
\fmfright{A3,A4}
\fmf{double}{A1,a2}
\fmf{plain}{A2,a1}
\fmf{plain}{A3,a3}
\fmf{plain}{A4,a4}
\fmfpolyn{smooth}{a}{4}
\end{fmfchar*}}
\quad=\quad\parbox{20mm}{
\begin{fmfchar*}(20,20)
\fmfbottom{A1,A2}
\fmftop{A3,A4}
\fmf{ghost}{A1,a}
\fmf{plain}{A3,a}
\fmf{plain}{a,A2}
\fmf{phantom}{a,A4}
\fmffreeze
\fmf{plain}{a,i}
\fmf{zigzag}{i,A4}
\fmffreeze
\fmf{ghost,left}{a,i}
\fmfv{decor.shape=diamond,decor.filled=empty,decor.size=15}{a}
\fmfv{decor.shape=square,decor.size=5,d.fi=empty}{i}
\end{fmfchar*}}\quad +\quad
\parbox{20mm}{
\begin{fmfchar*}(20,20)
\fmfbottom{A1,A2}
\fmftop{A3,A4}
\fmf{ghost}{A1,a}
\fmf{plain}{A2,a}
\fmf{plain}{a,A4}
\fmf{phantom}{a,A3}
\fmffreeze
\fmf{plain}{a,i}
\fmf{zigzag}{i,A3}
\fmffreeze
\fmf{ghost,left}{a,i}
\fmfv{decor.shape=diamond,decor.filled=empty,decor.size=15}{a}
\fmfv{decor.shape=square,decor.size=5,d.fi=empty}{i}
\end{fmfchar*}}\quad +\quad
\parbox{20mm}{
\begin{fmfchar*}(20,20)
\fmfbottom{A1,A2}
\fmftop{A3,A4}
\fmf{ghost}{A1,a}
\fmf{plain}{A3,a}
\fmf{plain}{a,A4}
\fmf{phantom}{a,A2}
\fmffreeze
\fmf{plain}{a,i}
\fmf{zigzag}{i,A2}
\fmffreeze
\fmf{ghost,left}{a,i}
\fmfv{decor.shape=diamond,decor.filled=empty,decor.size=15}{a}
\fmfv{decor.shape=square,decor.size=5,d.fi=empty}{i}
\end{fmfchar*}}
\end{equation}

\subsection{Green's function with 2 unphysical gauge bosons}
A similar analysis can be performed for the STI with 2 unphysical gauge bosons. We can show that the STIs for the three and four point vertices with one and two  contractions imply the STI for the Green's function with two contractions.  

The corresponding STI for the Green's function with 2 contractions is (see \fref{eq:ncov-sti-gf})
\begin{equation}\label{eq:2cov-sti4-diag}
\parbox{15mm}{
\begin{fmfchar*}(15,15)
\fmfleft{A1,A2}
\fmfright{A3,A4}
\fmf{double}{A1,a2}
\fmf{double}{A2,a1}
\fmf{plain}{A3,a3}
\fmf{plain}{A4,a4}
\fmfpolyn{smooth}{a}{4}
\end{fmfchar*}}
\quad=\quad\parbox{20mm}{
\begin{fmfchar*}(20,20)
\fmfbottom{A1,A2}
\fmftop{A3,A4}
\fmf{ghost}{A1,a}
\fmf{double}{A3,a}
\fmf{plain}{a,A2}
\fmf{phantom}{a,A4}
\fmffreeze
\fmf{plain}{a,i}
\fmf{zigzag}{i,A4}
\fmffreeze
\fmf{ghost,left}{a,i}
\fmfv{decor.shape=diamond,decor.filled=empty,decor.size=15}{a}
\fmfbrs{i}
\end{fmfchar*}}\quad +\quad
\parbox{20mm}{
\begin{fmfchar*}(20,20)
\fmfbottom{A1,A2}
\fmftop{A3,A4}
\fmf{ghost}{A1,a}
\fmf{double}{A3,a}
\fmf{plain}{a,A4}
\fmf{phantom}{a,A2}
\fmffreeze
\fmf{plain}{a,i}
\fmf{zigzag}{i,A2}
\fmffreeze
\fmf{ghost,left}{a,i}
\fmfv{decor.shape=diamond,decor.filled=empty,decor.size=15}{a}
\fmfbrs{i}
\end{fmfchar*}}
\end{equation}
The $t$ channel diagram in the skeleton expansion of the 4 point function (\fref{fig:skeleton-graph}) can be treated similarly to the case with one unphysical gauge boson:
\begin{equation}
\parbox{15mm}{
\begin{fmfchar*}(15,15)
\fmfleft{A1,A2}
\fmfright{A3,A4}
\fmf{double}{A1,a}
\fmf{double}{A2,b}
\fmf{plain}{b,a}
\fmf{plain}{a,A3}
\fmf{plain}{b,A4}
\fmfv{decor.shape=circle,decor.filled=shaded,decor.size=9}{a}
\fmfv{decor.shape=circle,decor.filled=shaded,decor.size=9}{b}
\end{fmfchar*}}\quad =
-\quad 
\parbox{15mm}{
\begin{fmfchar*}(15,15)
\fmfleft{l1,l2}
\fmfright{r1,r2}
\fmfv{decor.shape=circle,decor.filled=shaded,decor.size=9}{g}
\fmf{phantom}{l1,g}
\fmf{double}{l2,g}
\fmf{plain,tension=2}{r1,i,g}
\fmf{plain}{r2,g}
\fmffreeze
\fmfv{decor.shape=square,decor.filled=empty,decor.size=5}{i}
\fmf{ghost}{l1,i}
\end{fmfchar*}}
\quad +\quad 
\parbox{15mm}{
\begin{fmfchar*}(15,15)
\fmfleft{A1,A2}
\fmfright{A3,A4}
\fmf{ghost}{A1,a}
\fmf{double}{A2,b}
\fmf{plain}{b,a}
\fmf{zigzag}{a,A3}
\fmf{plain}{b,A4}
\fmfv{decor.shape=square,decor.filled=empty,decor.size=5}{a}
\fmfv{decor.shape=circle,decor.filled=shaded,decor.size=9}{b}
\end{fmfchar*}}\quad +\quad
\parbox{15mm}{
\begin{fmfchar*}(15,15)
\fmfleft{A1,A2}
\fmfright{A3,A4}
\fmf{ghost}{A1,a}
\fmf{plain}{A3,a}
\fmf{ghost}{a,i}
\fmf{double}{A2,i}
\fmf{zigzag}{i,A4}
\fmfv{decor.shape=circle,decor.filled=shaded,decor.size=9}{a}
\fmfv{decor.shape=square,decor.size=5,d.fi=empty}{i}
\end{fmfchar*}}
\end{equation}
The first term cancels a term from the STI for the 4 point vertex \eqref{eq:4-2cov-graph} while the other terms contribute to the contact terms on the right hand side of \fref{eq:2cov-sti4-diag}. The last term arises from the ghost terms in the STIs for the gauge and Goldstone boson vertices that can be treated like in \fref{eq:sub3-sti}, this time using the STI with 2 contractions \eqref{eq:3-2cov-graph}:
\begin{equation}
\parbox{20mm}{
\begin{fmfchar*}(15,20)
\fmfleft{A1,A2}
\fmfright{A3,A4}
\fmf{ghost}{A1,a}
\fmf{double}{A2,b}
\fmf{ghost,tension=1}{a,i}
\fmf{double}{i,b}
\fmf{plain}{a,A3}
\fmf{plain}{b,A4}
\fmfv{decor.shape=circle,decor.filled=shaded,decor.size=9}{a,b}
\fmfdot{i}
\end{fmfchar*}}
\quad =\quad
\parbox{20mm}{
\begin{fmfchar*}(15,20)
\fmfleft{A1,A2}
\fmfright{A3,A4}
\fmf{ghost}{A1,a}
\fmf{plain}{A3,a}
\fmf{ghost}{a,i}
\fmf{double}{A2,i}
\fmf{zigzag}{i,A4}
\fmfv{decor.shape=circle,decor.filled=shaded,decor.size=9}{a}
\fmfv{decor.shape=square,decor.size=5,d.fi=empty}{i}
\end{fmfchar*}}
\end{equation}
The result for the $u$-channel diagram is similar. 

The $s$-channel diagram has a different structure. Using the STI for the 3 point vertex with two contractions \eqref{eq:3-2cov-graph} we find
\begin{equation}
\parbox{20mm}{
\begin{fmfchar*}(20,15)
\fmfleft{A1,A2}
\fmfright{A3,A4}
\fmf{double}{A1,a}
\fmf{double}{A2,a}
\fmf{plain}{b,a}
\fmf{plain}{b,A3}
\fmf{plain}{b,A4}
\fmfv{decor.shape=circle,decor.filled=shaded,decor.size=9}{a}
\fmfv{decor.shape=circle,decor.filled=shaded,decor.size=9}{b}
\end{fmfchar*}}\quad =\quad 
-\quad 
\parbox{20mm}{
\begin{fmfchar*}(20,15)
\fmfleft{l1,l2}
\fmfright{r1,r2}
\fmfv{decor.shape=circle,decor.filled=shaded,decor.size=9}{g}
\fmf{double,tension=2}{l2,i}
\fmf{plain,tension=2}{i,g}
\fmf{phantom}{l1,g}
\fmf{plain}{r1,g}
\fmf{plain}{r2,g}
\fmffreeze
\fmfv{decor.shape=square,decor.filled=empty,decor.size=5}{i}
\fmf{ghost}{l1,i}
\end{fmfchar*}}+\parbox{20mm}{
\begin{fmfchar*}(20,15)
\fmfleft{A1,A2}
\fmfright{A3,A4}
\fmf{ghost}{A1,a}
\fmf{double}{A2,a}
\fmf{ghost}{a,b}
\fmf{plain}{b,A3}
\fmf{zigzag}{b,A4}
\fmfv{decor.shape=circle,decor.filled=shaded,decor.size=9}{a}
\fmfv{decor.shape=square,decor.filled=empty,decor.size=5}{b}
\end{fmfchar*}}\quad +\parbox{20mm}{
\begin{fmfchar*}(20,15)
\fmfleft{A1,A2}
\fmfright{A3,A4}
\fmf{ghost}{A1,a}
\fmf{double}{A2,a}
\fmf{ghost}{a,b}
\fmf{plain}{b,A4}
\fmf{zigzag}{b,A3}
\fmfv{decor.shape=circle,decor.filled=shaded,decor.size=9}{a}
\fmfv{decor.shape=square,decor.filled=empty,decor.size=5}{b}
\end{fmfchar*}}
\end{equation}
The first diagram cancels the remaining term from the STI of the 4 point vertex \eqref{eq:4-2cov-graph}. Like in \fref{eq:sub3-sti}, the other diagrams arise from the ghost terms in the STIs for gauge and Goldstone bosons \eqref{eq:3-2cov-ghost} and  contribute to the contact terms . Thus we have derived the STI \eqref{eq:2cov-sti4-diag} from the STIs for the 3 and  4 point vertices with two contractions.

Like in \fref{sec:wi-4}, for external on-shell particles we obtain the following result:

\begin{theorem}\label{theo:sti-wi-equiv2}
The Ward Identities for the 4 point Green's functions with two unphysical gauge bosons and the STIs for three point vertices with one and two unphysical gauge bosons imply the STIs for four point vertices with two unphysical gauge bosons
\qed
\end{theorem} 
\section{Gauge parameter independence}\label{sec:gpi}
We will now show on tree level, that the gauge parameter independence of physical amplitudes is a
consequence of the Ward Identities of the theory. 

To obtain gauge parameter independent amplitudes, the $\xi$ dependence of the propagators must cancel among the gauge boson
and the Goldstone boson exchange diagrams. To see how this works, we note that the gauge boson propagator in $R_\xi$ gauge can be written as the propagator in unitarity gauge plus a term proportional to the Goldstone boson propagator (see \fref{eq:rxi-prop}):
\begin{align}\label{eq:rxi-uni-prop}
  \ii D_W^{\mu\nu}(q)&=\frac{1}{q^2-m_W^2}\left(g^{\mu\nu}-\frac{q^\mu
  q^\nu}{m_W^2}\right) +\frac{q^\mu q^\nu}{m_W^2}\frac{1}{q^2-\xi m_W^2}\nonumber\\
&=\ii D^{\mu\nu}_{W,U}-\frac{q^\mu q^\nu}{m_W^2}(\ii D_{\phi}) 
\end{align} 

If we consider a gauge boson that is exchanged between two subamplitudes together with the corresponding Goldstone boson, we see that the gauge parameter dependence cancels between the unphysical part of the gauge boson propagator and the Goldstone boson propagator if the subamplitudes satisfy the Ward Identity \eqref{eq:gf-wi}:
\begin{equation}\label{eq:xi-cancel}
\parbox{25mm}{
\begin{fmfchar*}(25,20)
\fmfstraight
\fmfleftn{a}{4}
\fmfrightn{b}{4}
\begin{fmffor}{i}{1}{1}{4}
\fmf{plain}{a[i],x}
\fmf{plain}{b[i],y}
\end{fmffor}
\fmf{boson,tension=2,label=$W/ R_\xi$}{x,y}
\fmfv{d.sh=circle,d.f=empty,d.size=15pt}{x}
\fmfv{d.sh=circle,d.f=empty,d.size=15pt}{y}
\end{fmfchar*}}+
\parbox{25mm}{
\begin{fmfchar*}(25,20)
\fmfstraight
\fmfleftn{a}{4}
\fmfrightn{b}{4}
\begin{fmffor}{i}{1}{1}{4}
\fmf{plain}{a[i],x}
\fmf{plain}{b[i],y}
\end{fmffor}
\fmf{scalar,tension=2,label=$\phi/ R_\xi$}{x,y}
\fmfv{d.sh=circle,d.f=empty,d.size=15pt}{x}
\fmfv{d.sh=circle,d.f=empty,d.size=15pt}{y}
\end{fmfchar*}}\quad =\quad
\parbox{25mm}{
\begin{fmfchar*}(25,20)
\fmfstraight
\fmfleftn{a}{4}
\fmfrightn{b}{4}
\begin{fmffor}{i}{1}{1}{4}
\fmf{plain}{a[i],x}
\fmf{plain}{b[i],y}
\end{fmffor}
\fmf{boson,tension=2,label=$W/ U$}{x,y}
\fmfv{d.sh=circle,d.f=empty,d.size=15pt}{x}
\fmfv{d.sh=circle,d.f=empty,d.size=15pt}{y}
\end{fmfchar*}}
\end{equation}
However, in general we \emph{cannot} decompose an scattering matrix element into a sum over subamplitudes, connected by \emph{one} propagator:
\begin{equation}
\quad\parbox{20mm}{
\begin{fmfchar*}(20,20)
\fmfsurroundn{v}{6}
\begin{fmffor}{i}{1}{1}{6}
\fmf{plain}{v[i],v}
\end{fmffor}
\fmfv{d.sh=circle,d.f=empty,d.size=30pt,l=$N$,l.d=0}{v}
\end{fmfchar*}}\neq \sum_{i+j=N+2}
\parbox{25mm}{\begin{fmfchar*}(25,20)
\fmfleftn{l}{3}
\fmfrightn{r}{3}
\begin{fmffor}{i}{1}{1}{3}
\fmf{plain}{l[i],i1}
\fmf{plain}{r[i],i2}
\end{fmffor}
\fmf{plain}{i1,i2}
\fmfv{d.sh=circle,d.f=empty,d.size=20pt,l=$i$,l.d=0}{i1}
\fmfv{d.sh=circle,d.f=empty,d.size=20pt,l=$j$,l.d=0}{i2}
\end{fmfchar*}}
\end{equation}
The trouble causing the inequality comes from the double counting of diagrams. Consider a 5
point amplitude $\bar f f\to \bar f f W$ as an example. The grove $G_s$ from
\fref{eq:qcd-grove} cannot be factorized into subamplitudes:
 \begin{equation}
G_s\neq\quad   
\parbox{20mm}{
\begin{fmfchar*}(20,20)
\fmfleft{a,b}
\fmfright{f1,f2}
\fmftop{z}
\fmf{fermion,tension=1.0}{eae,a}
\fmf{photon,tension=1.0}{uwd,eae}
\fmf{fermion}{uwd,f1}
\fmf{fermion}{b,eae}
\fmf{fermion}{f2,uwd}
\fmfv{d.sh=circle,d.f=empty,d.size=20pt}{uwd}
\fmfdot{eae}
\fmffreeze
\fmf{photon}{z,uwd}
\end{fmfchar*}}\qquad +\qquad
\parbox{20mm}{
\begin{fmfchar*}(20,20)
\fmfright{a,b}
\fmfleft{f1,f2}
\fmftop{z}
\fmf{fermion,tension=1.0}{eae,a}
\fmf{photon,tension=1.0}{uwd,eae}
\fmf{fermion}{uwd,f1}
\fmf{fermion}{b,eae}
\fmf{fermion}{f2,uwd}
\fmfv{d.sh=circle,d.f=empty,d.size=20pt}{uwd}
\fmfdot{eae}
\fmffreeze
\fmf{photon}{z,uwd}
\end{fmfchar*}}
\end{equation}
because the diagram
\begin{equation}\label{eq:double_diag}
\parbox{20mm}{
\begin{fmfchar*}(20,15)
\fmftopn{b}{3}
\fmfbottomn{t}{2}
\fmfright{r}
\fmf{fermion}{t2,t,b3}
\fmf{photon,tension=2}{v1,v2}
\fmf{photon,tension=2}{v2,t}
\fmf{fermion}{t1,v1,b1}
\fmffreeze
\fmf{photon}{b2,v2}
\fmfdot{v1,v2,t}
\end{fmfchar*}}
\end{equation}
contributes to both subamplitudes and would thus be double-counted.

Therefore we cannot just use the result from \fref{sec:sbgt-wi} directly. This problem can be avoided if we consider an  infinitesimal change in the gauge parameter
\begin{equation*}
\xi = \xi_0+\delta\xi
\end{equation*}
and work to first order in $\delta\xi$. This is sufficient, since finite changes of $\xi$ can be generated by successive infinitesimal transformations. 

Under an infinitesimal variation of $\xi$ , the gauge boson propagator changes as  (see \fref{eq:rxi-uni-prop})
\begin{equation}\label{eq:dxi-gauge}
D_{W,\xi}^{\mu\nu}(q)= D_{W,\xi_0}^{\mu\nu}(q)- q^\mu q^\nu \frac{\ii}{(q^2-\xi_0 m_W^2)^2} (\delta\xi)+\mathcal{O}(\delta^2)
\end{equation}
We will  represent  this decomposition  graphically as
\begin{equation}\label{eq:uni-xi-graph}
\begin{array}{rcl}
\parbox{15mm}{
\begin{fmfchar*}(15,15)
\fmfleft{l}
\fmfright{r}
\fmf{photon}{l,r}  
\end{fmfchar*}}
&=
\parbox{15mm}{
\begin{fmfchar*}(15,15)
\fmfleft{l}
\fmfright{r}
\fmf{gluon}{l,r}  
\end{fmfchar*}}
&+\parbox{15mm}{
\begin{fmfchar*}(15,15)
\fmfleft{l}
\fmfright{r}
\fmf{dbl_dots}{l,r}  
\end{fmfchar*}}\\
 D_{W,\xi}^{\mu\nu}(q)&= D^{\mu\nu}_{W,\xi_0}&-q^\mu q^\nu D_{c,\xi_0}^2(q)\delta\xi 
\end{array}
\end{equation}
Similarly the Goldstone boson propagator becomes
\begin{equation}\label{eq:dxi-gold}
D_{\phi \xi}= D_{\phi \xi_0}+m_W^2 \frac{\ii}{(q^2-\xi_0 m_W^2)^2} +\mathcal{O}(\delta^2)
\end{equation}
Inserting the decomposition \eqref{eq:uni-xi-graph} into the diagram
\eqref{eq:double_diag} we get \emph{two} contributions linear in $\delta\xi$:

\begin{equation}  
\parbox{20mm}{
\begin{fmfchar*}(20,15)
\fmftopn{b}{3}
\fmfbottomn{t}{2}
\fmfright{r}
\fmf{fermion}{t2,t,b3}
\fmf{photon,tension=2}{v1,v2}
\fmf{photon,tension=2}{v2,t}
\fmf{fermion}{t1,v1,b1}
\fmffreeze
\fmf{photon}{b2,v2}
\fmfdot{v1,v2,t}
\end{fmfchar*}}\,=\,
\parbox{20mm}{
\begin{fmfchar*}(20,15)
\fmftopn{b}{3}
\fmfbottomn{t}{2}
\fmfright{r}
\fmf{fermion}{t2,t,b3}
\fmf{gluon,tension=2}{v1,v2}
\fmf{gluon,tension=2}{v2,t}
\fmf{fermion}{t1,v1,b1}
\fmffreeze
\fmf{photon}{b2,v2}
\fmfdot{v1,v2,t}
\end{fmfchar*}}\, +\,  
\parbox{20mm}{
\begin{fmfchar*}(20,15)
\fmftopn{b}{3}
\fmfbottomn{t}{2}
\fmfright{r}
\fmf{fermion}{t2,t,b3}
\fmf{gluon,tension=2}{v1,v2}
\fmf{dbl_dots,tension=2}{v2,t}
\fmf{fermion}{t1,v1,b1}
\fmffreeze
\fmf{photon}{b2,v2}
\fmfdot{v1,v2,t}
\end{fmfchar*}}\,+\,
\parbox{20mm}{
\begin{fmfchar*}(20,15)
\fmftopn{b}{3}
\fmfbottomn{t}{2}
\fmfright{r}
\fmf{fermion}{t2,t,b3}
\fmf{dbl_dots,tension=2}{v1,v2}
\fmf{gluon,tension=2}{v2,t}
\fmf{fermion}{t1,v1,b1}
\fmffreeze
\fmf{photon}{b2,v2}
\fmfdot{v1,v2,t}
\end{fmfchar*}}+\mathcal{O}(\delta\xi^2)
\end{equation}
Therefore we \emph{can} factorize the contributions linear in $\delta \xi$:
 \begin{equation}
\partial_\xi G_s=\quad   
\parbox{20mm}{
\begin{fmfchar*}(20,20)
\fmfleft{a,b}
\fmfright{f1,f2}
\fmftop{z}
\fmf{fermion,tension=1.0}{a,eae}
\fmf{dbl_dots,tension=1.0}{uwd,eae}
\fmf{fermion}{f1,uwd}
\fmf{fermion}{eae,b}
\fmf{fermion}{uwd,f2}
\fmfv{d.sh=circle,d.f=empty,d.size=20pt}{uwd}
\fmfdot{eae}
\fmffreeze
\fmf{photon}{z,uwd}
\end{fmfchar*}}\qquad +\qquad
\parbox{20mm}{
\begin{fmfchar*}(20,20)
\fmfright{a,b}
\fmfleft{f1,f2}
\fmftop{z}
\fmf{fermion,tension=1.0}{a,eae}
\fmf{dbl_dots,tension=1.0}{uwd,eae}
\fmf{fermion}{f1,uwd}
\fmf{fermion}{eae,b}
\fmf{fermion}{uwd,f2}
\fmfv{d.sh=circle,d.f=empty,d.size=20pt}{uwd}
\fmfdot{eae}
\fmffreeze
\fmf{photon}{z,uwd}
\end{fmfchar*}}
\end{equation}
Using the Ward Identities for the subamplitudes we now see that the gauge parameter
dependence of the grove $G_s$ vanishes. 

The terms linear in $\delta\xi$ of general amplitudes can be factorized in a similar way: 
\begin{equation}\label{eq:xi-factorize}
\partial_\xi \quad \parbox{20mm}{
\begin{fmfchar*}(20,20)
\fmfsurroundn{v}{6}
\begin{fmffor}{i}{1}{1}{6}
\fmf{plain}{v[i],v}
\end{fmffor}
\fmfv{d.sh=circle,d.f=empty,d.size=30pt,l=$\xi$,l.d=0}{v}
\end{fmfchar*}}
\quad = \sum_{i+j=N+2}
\parbox{25mm}{\begin{fmfchar*}(25,20)
\fmfleftn{l}{3}
\fmfrightn{r}{3}
\begin{fmffor}{i}{1}{1}{3}
\fmf{plain}{l[i],i1}
\fmf{plain}{r[i],i2}
\end{fmffor}
\fmf{dbl_dots}{i1,i2}
\fmfv{d.sh=circle,d.f=empty,d.size=20pt,l=$i$,l.d=0}{i1}
\fmfv{d.sh=circle,d.f=empty,d.size=20pt,l=$j$,l.d=0}{i2}
\end{fmfchar*}}
\end{equation}
Double counting doesn't occur for the diagrams linear in $\delta \xi$. To
see this, we regard the unphysical propagators as new `particles' with the appropriate Feynman rules. The $\mathcal{O}(\delta)$ contribution to the Green's function consists of Feynman diagrams where the new particle appears \emph{exactly} once. Therefore no double counting can occur and the decomposition given in \fref{eq:xi-factorize} is  unique. 

The same reasoning can be applied to the variation of the Goldstone boson propagator. 

We have just shown that the linear variation of the gauge parameter leads us to
the  situation already discussed in \fref{eq:xi-cancel}: in the terms linear in
$\delta \xi$, the propagators connect \emph{complete} subamplitudes that satisfy the Ward Identities. As in \fref{eq:xi-cancel}, the contributions from the gauge bosons and the Goldstone bosons cancel and we find that the derivative of a physical amplitude with respect to $\xi$ vanishes:
 \begin{equation}\label{eq:dxi-amp}
\partial_\xi\quad \parbox{20mm}{
\begin{fmfchar*}(20,20)
\fmfsurroundn{v}{6}
\begin{fmffor}{i}{1}{1}{6}
\fmf{plain}{v[i],v}
\end{fmffor}
\fmfv{d.sh=circle,d.f=empty,d.size=30pt,l=$\xi$,l.d=0}{v}
\end{fmfchar*}}=0
 \end{equation}
Note that only the Ward Identities with one unphysical gauge boson \eqref{eq:gf-wi} were needed in this argument. 

Nevertheless, the Ward Identities with more unphysical gauge bosons are needed for the consistency of the theory since they assure that the Ward Identities themselves are independent of the gauge parameter:

\begin{equation}
\partial_\xi \quad \parbox{20mm}{
\begin{fmfchar*}(20,20)
\fmfsurroundn{v}{6}
\begin{fmffor}{i}{1}{1}{5}
\fmf{plain}{v[i],v}
\end{fmffor}
\fmf{double}{v6,v}
\fmfv{d.sh=circle,d.f=empty,d.size=30pt,l=$\xi$,l.d=0}{v}
\end{fmfchar*}}
\quad = \sum_{i+j=N+2}
\parbox{25mm}{\begin{fmfchar*}(25,20)
\fmfleftn{l}{3}
\fmfrightn{r}{3}
\fmf{double}{r1,i2}
\fmf{plain}{l1,i1}
\begin{fmffor}{i}{2}{1}{3}
\fmf{plain}{l[i],i1}
\fmf{plain}{r[i],i2}
\end{fmffor}
\fmf{dbl_dots}{i1,i2}
\fmfv{d.sh=circle,d.f=empty,d.size=20pt,l=$i$,l.d=0}{i1}
\fmfv{d.sh=circle,d.f=empty,d.size=20pt,l=$j$,l.d=0}{i2}
\end{fmfchar*}}=0
\end{equation}
If the external particles are off-shell, we have to use the STIs instead of the Ward Identities and the Green's functions become $\xi$ dependent. The form of the additional off-shell terms can be obtained by considering changes in the gauge fixing functional and use of the STIs \cite{Boehm:2001,Kugo:1997}. Similarly, the effective action is gauge dependent and the explicit form is determined by the so called \emph{Nielsen Identity} \cite{Piguet:1985}. The extension of our graphical  approach to these issues would be interesting but is beyond the scope of the present work.
\section{Gauge invariance classes}\label{sec:gi-classes}
In \fref{sec:skeleton} we have derived the STI for the 4 point function from the STI of the irreducible vertices, using a diagrammatic approach. In this section, we will extend this discussion to Green's functions with more external particles and  to establish a connection with the formalism of gauge flips \cite{Boos:1999} reviewed in \fref{sec:groves}. 

The aim of this discussion is obtain a definition of the gauge flips that
covers also the case of spontaneously broken gauge theories. Our approach might also be useful for the
extensions of groves to loop diagrams \cite{Ondreka:2003} and to
supersymmetric theories, using the results of
\cite{ReuterOhl:2002,Reuter:2002}.

\subsection{Definition of gauge invariance classes}\label{sec:groves-def}
It appears natural to define  a gauge invariance class of Feynman diagrams in the following way: 
\begin{definition}\label{def:gic} A gauge invariance class is a subset of Feynman diagrams that is independent of the gauge parameter and satisfies the STIs. 
\end{definition}

To use this definition, we have to introduce a notion of a subset of diagrams satisfying a STI. The meaning of this is not clear a priori, since we don't know what contact terms have to appear on the right hand side of the STI \eqref{eq:sti-graph} if we select a subset of diagrams on the left hand side.

To motivate our definition for the STIs for subsets of Feynman diagrams, we first consider the 5 point functions where we have discussed the groves already in \fref{sec:qcd-groves}. We begin with the amplitude for $q \bar q\to  q\bar q g$. 
The STI for this 5 point function is
\begin{equation}\label{eq:4fw-sti}
\parbox{20mm}{
\begin{fmfchar*}(20,20)
\fmftopn{l}{3}
\fmfbottomn{r}{3}
\fmf{double}{l2,v}
\fmf{fermion}{l1,v}
\fmf{fermion}{v,l3}
\fmf{fermion}{v,r1}
\fmf{phantom}{r2,v}
\fmf{fermion}{v,r3}
\fmfv{d.sh=circle,d.f=empty,d.size=25pt}{v}
\end{fmfchar*}}\quad
=\quad\sum_{f_i}\quad
\parbox{25mm}{ 
\fmfframe(0,5)(0,0){
\begin{fmfchar*}(20,20)
\fmftopn{l}{3}
\fmfbottomn{r}{3}
\fmf{ghost}{l2,v}
\fmf{phantom}{r2,v}
\fmf{fermion}{l1,v}
\fmf{fermion}{v,l3}
\fmf{fermion}{v,r1}
\fmfv{d.sh=diamond,d.f=empty,d.size=25pt}{v}
\fmf{fermion}{i,v}
\fmf{zigzag,tension=2}{i,r3}
\fmffreeze
\fmf{ghost,left}{v,i}
\fmfv{label=$f_i$,l.d=0.6}{r3}
\fmfv{decor.shape=square,decor.filled=empty,decor.size=5}{i}
\end{fmfchar*}}}
\end{equation}
Since the fermions don't couple to ghosts, the contact terms on the right hand side can only consist of diagrams where the ghost is connected to the BRS-transformation only or interacts with the internal scalar or gauge boson:
\begin{equation}\label{eq:5-point-contact}
 \parbox{17mm}{
\begin{fmfchar*}(20,15)
\fmfright{a,b}
\fmfleft{f1,f2}
\fmftop{z}
\fmf{fermion}{ewn,a}
\fmf{fermion}{uwd,f1}
\fmf{plain}{uwd,ewn}
\fmf{zigzag,tension=2.0}{b,nan}
\fmf{fermion}{f2,uwd}
\fmf{fermion,tension=2.0}{nan,ewn}
\fmffreeze
\fmfdot{uwd}
\fmfdot{ewn}
\fmf{ghost}{z,nan}
\fmfv{decor.shape=square,decor.filled=empty,decor.size=5}{nan}
\end{fmfchar*}}
\qquad ,\qquad
 \parbox{15mm}{
\begin{fmfchar*}(20,15)
\fmfright{a,b}
\fmfleft{f1,f2}
\fmftop{z}
\fmf{fermion}{ewn,b}
\fmf{fermion}{uwd,f1}
\fmf{plain,tension=1.5}{uwd,i}
\fmf{ghost,tension=1.5}{i,ewn}
\fmf{zigzag}{a,ewn}
\fmf{fermion}{f2,uwd}
\fmffreeze
\fmfdot{uwd,i}
\fmf{ghost}{z,i}
\fmfv{decor.shape=square,decor.filled=empty,decor.size=5}{ewn}
\end{fmfchar*}}
\end{equation}

 Comparing the contact terms \eqref{eq:5-point-contact} to the diagrams contributing to the grove \eqref{eq:qcd-grove},  we can define the contact terms corresponding to a grove as those diagrams obtained by replacing the external gauge boson by a ghost and one external particle by an inverse propagator connected to a BRS-transformation. 

We will introduce a mapping $\mathscr{F}$ that maps every Feynman diagram to the corresponding contact term. Note that this is a purely \emph{formal} mapping and in general it is not true that a contraction of a gauge boson in the original diagram results in the contact terms generated by this mapping. 
\begin{definition}
\begin{subequations}\label{subeq:generate-contact}
The action of  $\mathscr{F}$ on a diagram with the insertion of a gauge boson into a external line is given by
\begin{equation}
\parbox{20mm}{
\begin{fmfchar*}(20,15)
\fmfright{a,b}
\fmfleft{f1,f2}
\fmftop{z}
\fmf{fermion}{ewn,a}
\fmf{fermion}{uwd,f1}
\fmf{plain}{uwd,ewn}
\fmf{fermion,tension=2.0}{b,nan}
\fmf{fermion}{f2,uwd}
\fmf{fermion,tension=2.0}{nan,ewn}
\fmffreeze
\fmfdot{uwd,ewn,nan}
\fmf{photon}{z,nan}
\end{fmfchar*}}
\xrightarrow{\mathscr{F}}
 \parbox{20mm}{
\begin{fmfchar*}(20,15)
\fmfright{a,b}
\fmfleft{f1,f2}
\fmftop{z}
\fmf{fermion}{ewn,a}
\fmf{fermion}{uwd,f1}
\fmf{plain}{uwd,ewn}
\fmf{zigzag,tension=2.0}{b,nan}
\fmf{fermion}{f2,uwd}
\fmf{fermion,tension=2.0}{nan,ewn}
\fmffreeze
\fmfdot{uwd}
\fmfdot{ewn}
\fmf{ghost}{z,nan}
\fmfv{decor.shape=square,decor.filled=empty,decor.size=5}{nan}
\end{fmfchar*}}
\end{equation}
The action of $\mathscr{F}$ on diagrams with a insertion of a gauge boson  into an internal gauge boson line is defined as follows: also the internal gauge bosons have to be replaced by a ghost until the external particles are reached. In this case, one original diagram can correspond to more than one contact term:
\begin{multline}
\parbox{20mm}{
\begin{fmfchar*}(20,15)
\fmfright{a,b}
\fmfleft{f1,f2}
\fmftop{z}
\fmf{fermion}{ewn,b}
\fmf{fermion}{f1,uwd}
\fmf{photon,tension=1.5}{uwd,i}
\fmf{photon,tension=1.5}{i,ewn}
\fmf{fermion}{ewn,a}
\fmf{fermion}{uwd,f2}
\fmffreeze
\fmfdot{uwd,i,ewn}
\fmf{photon}{z,i}
\end{fmfchar*}}\quad \xrightarrow{\mathscr{F}}\quad
\parbox{20mm}{
\begin{fmfchar*}(20,15)
\fmfright{a,b}
\fmfleft{f1,f2}
\fmftop{z}
\fmf{fermion}{ewn,b}
\fmf{fermion}{f1,uwd}
\fmf{photon,tension=1.5}{uwd,i}
\fmf{ghost,tension=1.5}{i,ewn}
\fmf{zigzag}{a,ewn}
\fmf{fermion}{uwd,f2}
\fmffreeze
\fmfdot{uwd,i}
\fmf{ghost}{z,i}
\fmfv{decor.shape=square,decor.filled=empty,decor.size=5}{ewn}
\end{fmfchar*}}
\quad +\quad
\parbox{20mm}{
\begin{fmfchar*}(20,15)
\fmfright{a,b}
\fmfleft{f1,f2}
\fmftop{z}
\fmf{fermion}{ewn,a}
\fmf{fermion}{f1,uwd}
\fmf{photon,tension=1.5}{uwd,i}
\fmf{ghost,tension=1.5}{i,ewn}
\fmf{zigzag}{b,ewn}
\fmf{fermion}{uwd,f2}
\fmffreeze
\fmfdot{uwd,i}
\fmf{ghost}{z,i}
\fmfv{decor.shape=square,decor.filled=empty,decor.size=5}{ewn}
\end{fmfchar*}}\\
+\quad \parbox{20mm}{
\begin{fmfchar*}(20,15)
\fmfleft{a,b}
\fmfright{f1,f2}
\fmftop{z}
\fmf{fermion}{ewn,b}
\fmf{fermion}{f1,uwd}
\fmf{photon,tension=1.5}{uwd,i}
\fmf{ghost,tension=1.5}{i,ewn}
\fmf{zigzag}{a,ewn}
\fmf{fermion}{uwd,f2}
\fmffreeze
\fmfdot{uwd,i}
\fmf{ghost}{z,i}
\fmfv{decor.shape=square,decor.filled=empty,decor.size=5}{ewn}
\end{fmfchar*}}
\quad +\quad
\parbox{20mm}{
\begin{fmfchar*}(20,15)
\fmfleft{a,b}
\fmfright{f1,f2}
\fmftop{z}
\fmf{fermion}{ewn,a}
\fmf{fermion}{f1,uwd}
\fmf{photon,tension=1.5}{uwd,i}
\fmf{ghost,tension=1.5}{i,ewn}
\fmf{zigzag}{b,ewn}
\fmf{fermion}{uwd,f2}
\fmffreeze
\fmfdot{uwd,i}
\fmf{ghost}{z,i}
\fmfv{decor.shape=square,decor.filled=empty,decor.size=5}{ewn}
\end{fmfchar*}}
\end{multline}
\end{subequations}
\end{definition}

By continuation, the same rule applies to larger diagrams where gauge bosons lines,  connecting the external gauge boson to the BRS transformed particle, have to be replaced by  ghost lines. In this process the Feynman rules of the ghosts must of course be taken into account, i.e. diagrams not allowed by the Feynman rules have to be omitted. 

In this way we can associate to every Feynman diagram some contact diagrams. Conversely, replacing a ghost line by a gauge boson line and the inverse propagator by an external particle, we can associate a Feynman diagram to every contact term. 

 Since (in a linear $R_\xi$ gauge) to every Feynman rule of the ghosts corresponds a Feynman rule of the gauge bosons, the contact terms generated in that way from the complete set of Feynman diagrams must indeed be all the contact terms required by the STI. 

Therefore it is sensible to define:
\begin{definition}\label{def:sti} A subset of diagrams satisfies a STI if the contact terms obtained by the  mapping $\mathscr{F}$ agree with the result of contracting an external gauge boson.
\end{definition}
This makes it possible to use \fref{def:gic} for the gauge invariance classes.
\subsection{Definition of gauge flips}\label{sec:flip-def}
We will see below, that we have to define the elementary gauge flips  as the \emph{minimal set of 4 point diagrams} with a given set of external particles and at least one external gauge boson,  \emph{satisfying the STI}. In a generic graphical notation, the gauge flips are denoted as:
\begin{equation}\label{eq:generic-flips}
T_4^G=\left\{\parbox{15mm}{
\begin{fmfchar*}(15,15)
\fmfleft{a,b}
\fmfright{f1,f2}
\fmf{photon}{a,fwf1}
\fmf{plain}{fwf1,fwf2}
\fmf{plain}{fwf2,b}
\fmf{plain}{fwf1,f1}
\fmf{plain}{fwf2,f2}
\fmfdot{fwf1}
\fmfdot{fwf2}
\end{fmfchar*}}\,,\,
\parbox{15mm}{
\begin{fmfchar*}(15,15)
\fmfleft{a,b}
\fmfright{f1,f2}
\fmf{photon}{a,fwf1}
\fmf{plain}{fwf1,fwf2}
\fmf{plain}{fwf2,b}
\fmf{phantom}{fwf2,f2}
\fmf{phantom}{fwf1,f1}
\fmffreeze
\fmf{plain}{fwf2,f1}
\fmf{plain}{fwf1,f2}
\fmfdot{fwf1}
\fmfdot{fwf2}
\end{fmfchar*}}\, ,\, 
\parbox{15mm}{
\begin{fmfchar*}(15,15)
\fmfleft{a,b}
\fmfright{f1,f2}
\fmf{photon}{a,fwf}
\fmf{plain}{fwf,b}
\fmf{plain}{fwf,www}
\fmf{plain}{www,f1}
\fmf{plain}{www,f2}
\fmfdot{fwf}
\fmfdot{www}
\end{fmfchar*}}\, , \, 
\parbox{15mm}{
\begin{fmfchar*}(15,15)
\fmfleft{a,b}
\fmfright{f1,f2}
\fmf{photon}{a,c}
\fmf{plain}{c,b}
\fmf{plain}{c,f1}
\fmf{plain}{c,f2}
\fmfdot{c}
\end{fmfchar*}}\right\}
\end{equation}
The internal particles appearing in these diagrams are determined by the requirement that the flips are the \emph{minimal} set of diagrams satisfying the STIs. For example, the question if the Higgs bosons in spontaneously broken gauge theories have to be included in the gauge flips is discussed in detail in \fref{chap:ssb_groves}.

There is a subtlety that will become important in
the discussion of gauge flips in spontaneously broken gauge theories in \fref{chap:ssb_groves}. In $R_\xi$ gauge also the corresponding 4 point functions with some or all external gauge bosons replaced by Goldstone bosons appear as subamplitudes in larger diagrams. It may happen, that the minimal gauge invariance class for the gauge boson subamplitude does not coincide with the minimal gauge invariance class for the Goldstone boson subamplitude. In this case the gauge flips have to be defined in such a way that not only the gauge boson amplitudes but also \emph{all corresponding Goldstone boson amplitudes} satisfy the STI. An example will be discussed in \fref{sec:2f2w-flips}.

In the presence of quartic Higgs vertices,
we will also need elementary flips among 5 point functions:
\begin{equation}\label{eq:generic-flips5}
T_5^G=\left\{\,
\parbox{15mm}{
\begin{fmfchar*}(15,15)
\fmfleftn{l}{2}
\fmfrightn{r}{3}
\fmf{dashes}{l1,g}
\fmf{dashes}{l2,g}
\fmf{dashes}{r1,g}
\fmf{plain,tension=2}{r3,i}
\fmf{dashes,tension=2}{i,g}
\fmffreeze
\fmf{photon}{r2,i}
\fmfdot{g,i}
\end{fmfchar*}}\quad,\quad
\parbox{15mm}{
\begin{fmfchar*}(15,15)
\fmfleft{l1,l2}
\fmfright{r1,r2,r3}
\fmf{dashes}{l1,g}
\fmf{dashes}{l2,g}
\fmf{dashes}{r3,g}
\fmf{plain,tension=2}{r1,i}
\fmf{dashes,tension=2}{i,g}
\fmffreeze
\fmf{photon}{r2,i}
\fmfdot{g,i}
\end{fmfchar*}}
\quad,\quad\parbox{15mm}{
\begin{fmfchar*}(15,15)
\fmfleft{l1,l2,l3}
\fmfright{r1,r3}
\fmf{dashes}{l3,g}
\fmf{dashes}{r1,g}
\fmf{plain,tension=2}{l1,i}
\fmf{dashes,tension=2}{i,g}
\fmf{dashes}{g,r3}
\fmffreeze
\fmf{photon}{l2,i}
\fmfdot{g,i}
\end{fmfchar*}}\quad,\quad
\parbox{15mm}{
\begin{fmfchar*}(15,15)
\fmfleft{l1,l2,l3}
\fmfright{r1,r2}
\fmf{dashes}{l1,g}
\fmf{dashes}{r1,g}
\fmf{dashes}{r2,g}
\fmf{plain,tension=2}{l3,i}
\fmf{dashes,tension=2}{i,g}
\fmffreeze
\fmf{photon}{l2,i}
\fmfdot{g,i}
\end{fmfchar*}}\,\right\}
 \end{equation}
\subsection{$\bar f f \to \bar f f W$}
Before we turn to the general 5 point function, we will consider  the 5 point Green's function with one external
gauge boson and 4 fermions.  In the example of QCD, the groves have been discussed already in \fref{sec:qcd-groves}. 

We will show, using the graphical approach, that the grove $G_s$ from \fref{eq:qcd-grove} satisfies the STI. The independence of the gauge parameter was already established in \fref{sec:gpi}. 
 An insertion of the gauge boson into an external fermion line becomes, using the STI for the 3 particle vertex \eqref{eq:sti-3graph} :
\begin{equation}\label{eq:5-point-external}
\parbox{20mm}{
\begin{fmfchar*}(20,15)
\fmfright{a,b}
\fmfleft{f1,f2}
\fmftop{z}
\fmf{fermion,tension=2.0}{eae,b}
\fmf{fermion,tension=2.0}{ewn,eae}
\fmf{fermion}{uwd,f1}
\fmf{plain}{ewn,uwd}
\fmf{fermion}{a,ewn}
\fmf{fermion}{f2,uwd}
\fmfdot{uwd}
\fmfdot{ewn}
\fmffreeze
\fmf{double}{z,eae}
\fmfdot{eae}
\end{fmfchar*}}
=-
\parbox{20mm}{
\begin{fmfchar*}(20,15)
\fmfright{a,b}
\fmfleft{f1,f2}
\fmftop{z}
\fmf{fermion,tension=2.0}{eae,b}
\fmf{fermion,tension=2.0}{ewn,eae}
\fmf{fermion}{uwd,f1}
\fmf{plain}{ewn,uwd}
\fmf{fermion}{a,ewn}
\fmf{fermion}{f2,uwd}
\fmfdot{uwd}
\fmfdot{ewn}
\fmffreeze
\fmf{ghost}{z,eae}
\fmfv{decor.shape=square,decor.filled=empty,decor.size=5}{eae}
\end{fmfchar*}}
\quad +\quad
\parbox{20mm}{
\begin{fmfchar*}(20,15)
\fmfright{a,b}
\fmfleft{f1,f2}
\fmftop{z}
\fmf{zigzag,tension=2.0}{b,eae}
\fmf{fermion,tension=2.0}{ewn,eae}
\fmf{fermion}{uwd,f1}
\fmf{plain}{ewn,uwd}
\fmf{fermion}{ewn,a}
\fmf{fermion}{f2,uwd}
\fmfdot{uwd}
\fmfdot{ewn}
\fmffreeze
\fmf{ghost}{z,eae}
\fmfv{decor.shape=square,decor.filled=empty,decor.size=5}{eae}
\end{fmfchar*}}
\end{equation}
The second term is the contact term corresponding to the original diagram according to the mapping \eqref{subeq:generate-contact}. To satisfy the STI, the first term must cancel against contributions from other diagrams. 

Similarly we get for the diagram where the gauge boson is inserted into the
internal Higgs  line
\begin{equation}\label{eq:5-point-internal}
\parbox{20mm}{
\begin{fmfchar*}(20,15)
\fmfright{a,b}
\fmfleft{f1,f2}
\fmftop{z}
\fmf{fermion}{ewn,a}
\fmf{fermion}{uwd,f1}
\fmf{dashes}{ewn,waw}
\fmf{dashes}{waw,uwd}
\fmf{fermion}{b,ewn}
\fmf{fermion}{f2,uwd}
\fmfdot{uwd}
\fmfdot{ewn}
\fmffreeze
\fmf{double}{z,waw}
\fmfdot{waw}
\end{fmfchar*}}
=\quad-\quad
\parbox{20mm}{
\begin{fmfchar*}(20,15)
\fmfright{a,b}
\fmfleft{f1,f2}
\fmftop{z}
\fmf{fermion}{ewn,a}
\fmf{fermion}{uwd,f1}
\fmf{dashes,tension=2.0}{ewn,waw}
\fmf{dashes}{waw,uwd}
\fmf{fermion}{b,ewn}
\fmf{fermion}{f2,uwd}
\fmfdot{uwd}
\fmfdot{ewn}
\fmffreeze
\fmf{ghost}{z,waw}
\fmfv{decor.shape=square,decor.filled=empty,decor.size=5}{waw}
\end{fmfchar*}}
\quad-\quad
\parbox{20mm}{
\begin{fmfchar*}(20,15)
\fmfright{a,b}
\fmfleft{f1,f2}
\fmftop{z}
\fmf{fermion}{ewn,a}
\fmf{fermion}{uwd,f1}
\fmf{dashes}{ewn,waw}
\fmf{dashes,tension=2}{waw,uwd}
\fmf{fermion}{b,ewn}
\fmf{fermion}{f2,uwd}
\fmfdot{uwd}
\fmfdot{ewn}
\fmffreeze
\fmf{ghost}{z,waw}
\fmfv{decor.shape=square,decor.filled=empty,decor.size=5}{waw}
\end{fmfchar*}}
\end{equation}
For any internal gauge boson, we get an additional contribution compared to  \fref{eq:5-point-internal} from the ghost term in the STI \eqref{eq:2gb-sti-graph}:
\begin{equation}\label{eq:5-internal-gauge}
\parbox{20mm}{
\begin{fmfchar*}(20,15)
\fmfright{a,b}
\fmfleft{f1,f2}
\fmftop{z}
\fmf{fermion}{ewn,a}
\fmf{fermion}{uwd,f1}
\fmf{plain}{ewn,waw}
\fmf{photon}{waw,uwd}
\fmf{fermion}{b,ewn}
\fmf{fermion}{f2,uwd}
\fmfdot{uwd}
\fmfdot{ewn}
\fmffreeze
\fmf{double}{z,waw}
\fmfdot{waw}
\end{fmfchar*}}
=\quad-\quad
\parbox{20mm}{
\begin{fmfchar*}(20,15)
\fmfright{a,b}
\fmfleft{f1,f2}
\fmftop{z}
\fmf{fermion}{ewn,a}
\fmf{fermion}{uwd,f1}
\fmf{plain,tension=2.0}{ewn,waw}
\fmf{photon}{waw,uwd}
\fmf{fermion}{b,ewn}
\fmf{fermion}{f2,uwd}
\fmfdot{uwd}
\fmfdot{ewn}
\fmffreeze
\fmf{ghost}{z,waw}
\fmfv{decor.shape=square,decor.filled=empty,decor.size=5}{waw}
\end{fmfchar*}}
\quad-\quad
\parbox{20mm}{
\begin{fmfchar*}(20,15)
\fmfright{a,b}
\fmfleft{f1,f2}
\fmftop{z}
\fmf{fermion}{ewn,a}
\fmf{fermion}{uwd,f1}
\fmf{plain}{ewn,waw}
\fmf{photon,tension=2}{waw,uwd}
\fmf{fermion}{b,ewn}
\fmf{fermion}{f2,uwd}
\fmfdot{uwd}
\fmfdot{ewn}
\fmffreeze
\fmf{ghost}{z,waw}
\fmfv{decor.shape=square,decor.filled=empty,decor.size=5}{waw}
\end{fmfchar*}}\quad +\quad
 \parbox{20mm}{
\begin{fmfchar*}(20,15)
\fmfright{a,b}
\fmfleft{f1,f2}
\fmftop{z}
\fmf{fermion,tension=1.5}{ewn,a}
\fmf{fermion,tension=1.5}{uwd,f1}
\fmf{photon,tension=2}{uwd,i}
\fmf{ghost,tension=1.5}{j,i}
\fmf{plain,tension=2}{j,ewn}
\fmf{fermion,tension=1.5}{b,ewn}
\fmf{fermion,tension=1.5}{f2,uwd}
\fmffreeze
\fmfdot{uwd,j,ewn}
\fmf{ghost}{z,j}
\fmfv{decor.shape=square,decor.filled=full,decor.size=5}{i}
\end{fmfchar*}}
\end{equation}
In this diagrams, the straight internal line can be a gauge or a Higgs boson.
The additional term gives rise to contact terms of the form of the second term in \fref{eq:5-point-contact} as we will discuss shortly. Let us first turn to the cancellation of the remaining terms. 

Since there is no 4 point vertex involving fermions,  the STI for the 4 point vertex \eqref{eq:sti4-graph} implies the relation
\begin{equation}\label{eq:3sti-int}
0 \quad= \
\parbox{20mm}{
\begin{fmfchar*}(20,15)
\fmfright{a,b}
\fmfleft{f1,f2}
\fmftop{z}
\fmf{fermion,tension=2.0}{eae,a}
\fmf{fermion,tension=2.0}{ewn,eae}
\fmf{fermion}{uwd,f1}
\fmf{plain}{ewn,uwd}
\fmf{fermion}{b,ewn}
\fmf{fermion}{f2,uwd}
\fmfdot{uwd}
\fmfdot{ewn}
\fmffreeze
\fmf{ghost,right=0.5}{z,eae}
\fmfv{decor.shape=square,decor.filled=empty,decor.size=5}{eae}
\end{fmfchar*}}
\quad+\quad
\parbox{20mm}{
\begin{fmfchar*}(20,15)
\fmfright{a,b}
\fmfleft{f1,f2}
\fmftop{z}
\fmf{fermion}{ewn,a}
\fmf{fermion}{uwd,f1}
\fmf{plain,tension=2.0}{ewn,waw}
\fmf{plain}{waw,uwd}
\fmf{fermion}{b,ewn}
\fmf{fermion}{f2,uwd}
\fmfdot{uwd}
\fmfdot{ewn}
\fmffreeze
\fmf{ghost}{z,waw}
\fmfv{decor.shape=square,decor.filled=empty,decor.size=5}{waw}
\end{fmfchar*}}
\quad+\quad
 \parbox{20mm}{
\begin{fmfchar*}(20,15)
\fmfright{a,b}
\fmfleft{f1,f2}
\fmftop{z}
\fmf{fermion}{ewn,a}
\fmf{fermion}{uwd,f1}
\fmf{plain}{ewn,uwd}
\fmf{fermion,tension=2.0}{b,nan}
\fmf{fermion}{f2,uwd}
\fmf{fermion,tension=2.0}{nan,ewn}
\fmffreeze
\fmfdot{uwd}
\fmfdot{ewn}
\fmf{ghost}{z,nan}
\fmfv{decor.shape=square,decor.filled=empty,decor.size=5}{nan}
\end{fmfchar*}}
\end{equation}
and a similar relation for the other vertex. Adding up all diagrams of the grove \eqref{eq:qcd-grove}, all terms except the contact terms cancel because of this identity. We see that we need contributions from three diagrams connected by gauge flips to get a cancellation because of the STI. 

We still have to discuss  the additional terms for internal gauge bosons from \fref{eq:5-internal-gauge}:
Like in \fref{eq:sub3-sti}, we can add the corresponding Goldstone boson diagram and use the STI for the three point function:
\begin{equation}\label{eq:int-sti}
 \parbox{20mm}{
\begin{fmfchar*}(20,15)
\fmfright{a,b}
\fmfleft{f1,f2}
\fmftop{z}
\fmf{fermion,tension=1.5}{b,ewn,a}
\fmf{fermion,tension=1.5}{f2,uwd,f1}
\fmf{double,tension=2}{uwd,i}
\fmf{ghost,tension=1.5}{j,i}
\fmf{plain,tension=2}{j,ewn}
\fmffreeze
\fmfdot{uwd,j,i,ewn}
\fmf{ghost}{z,j}
\end{fmfchar*}}\quad =\quad
 \parbox{20mm}{
\begin{fmfchar*}(20,15)
\fmfleft{a,b}
\fmfright{f1,f2}
\fmftop{z}
\fmf{fermion}{ewn,a}
\fmf{fermion}{uwd,f1}
\fmf{plain,tension=1.5}{uwd,i}
\fmf{ghost,tension=1.5}{i,ewn}
\fmf{zigzag}{b,ewn}
\fmf{fermion}{f2,uwd}
\fmf{plain,tension=2.0}{nan,ewn}
\fmffreeze
\fmfdot{uwd,i,ewn}
\fmf{ghost}{z,i}
\fmfv{decor.shape=square,decor.filled=empty,decor.size=5}{nan}
\end{fmfchar*}}\quad+ \quad
\parbox{20mm}{
\begin{fmfchar*}(20,15)
\fmfleft{a,b}
\fmfright{f1,f2}
\fmftop{z}
\fmf{fermion}{b,ewn}
\fmf{fermion}{uwd,f1}
\fmf{plain,tension=1.5}{uwd,i}
\fmf{ghost,tension=1.5}{i,ewn}
\fmf{zigzag}{a,ewn}
\fmf{fermion}{f2,uwd}
\fmffreeze
\fmfdot{uwd,i,ewn}
\fmf{ghost}{z,i}
\fmfv{decor.shape=square,decor.filled=empty,decor.size=5}{ewn}
\end{fmfchar*}}
\end{equation}
This shows that the additional terms are the contact terms of  the form of the second term in \fref{eq:5-point-contact} that are generated from the original diagram by the mapping $\mathscr{F}$. In contrast to the simpler case discussed above, this can be regarded as a `second order' cancellation because it involves not only the STI for the vertex with the gauge boson insertion but also a STI for a larger subamplitude. However, ultimately also the second order cancellations boil down to `first order' ones since the STI of the subamplitude is satisfied because of the STIs for the vertices. 

We have thus demonstrated, using the STIs,  that the groves in the amplitude $\bar f f \to \bar f f W$, obtained in \fref{sec:qcd-groves} from flavor selection rules, satisfy the STI for the 5 point function by themselves. 
\subsection{General 5 point amplitudes}\label{sec:5-point-groves}
The decomposition of the amplitude into groves gets destroyed for 5 point functions with external Higgs bosons or several external gauge bosons. This can be seen either from the absence of flavor selection rules or the formalism of flips sketched in \fref{sec:groves}. We will now discuss  the reasons for this directly from the STIs. 

There are some differences compared to the case of external fermions. First, since the STI must be satisfied for all external gauge bosons, we have to apply the gauge flips to every external gauge boson. 

Furthermore, there is the possibility of quartic vertices, that result in additional diagrams:
\begin{equation}
\parbox{15mm}{
\begin{fmfchar*}(15,15)
\fmfright{a,b}
\fmfleft{f1,f2}
\fmftop{z}
\fmf{plain}{ewn,a}
\fmf{plain}{uwd,f1}
\fmf{plain}{ewn,uwd}
\fmf{plain}{b,ewn}
\fmf{plain}{f2,uwd}
\fmfdot{uwd}
\fmfdot{ewn}
\fmffreeze
\fmf{photon}{z,ewn}
\end{fmfchar*}}
\quad\parbox{15mm}{
\begin{fmfchar*}(15,15)
\fmfright{a,b}
\fmfleft{f1,f2}
\fmftop{z}
\fmf{plain}{ewn,a}
\fmf{plain}{uwd,f1}
\fmf{plain}{ewn,uwd}
\fmf{plain}{b,ewn}
\fmf{plain}{f2,uwd}
\fmfdot{uwd}
\fmfdot{ewn}
\fmffreeze
\fmf{photon}{z,uwd}
\end{fmfchar*}}
\end{equation}
This doesn't change our argument of the fermion case, as can be seen using the STI for the 4 point vertex \eqref{eq:sti4-graph}:
\begin{multline}\label{eq:sti4-int}
\parbox{15mm}{
\begin{fmfchar*}(15,15)
\fmfright{a,b}
\fmfleft{f1,f2}
\fmftop{z}
\fmf{plain}{ewn,a}
\fmf{plain}{uwd,f1}
\fmf{plain}{ewn,uwd}
\fmf{plain}{b,ewn}
\fmf{plain}{f2,uwd}
\fmfdot{uwd}
\fmfdot{ewn}
\fmffreeze
\fmf{double}{z,ewn}
\end{fmfchar*}}
 \quad= \
\parbox{15mm}{
\begin{fmfchar*}(15,15)
\fmfright{a,b}
\fmfleft{f1,f2}
\fmftop{z}
\fmf{plain,tension=2.0}{eae,a}
\fmf{plain,tension=2.0}{ewn,eae}
\fmf{plain}{uwd,f1}
\fmf{plain}{ewn,uwd}
\fmf{plain}{b,ewn}
\fmf{plain}{f2,uwd}
\fmfdot{uwd}
\fmfdot{ewn}
\fmffreeze
\fmf{ghost,right=0.5}{z,eae}
\fmfv{decor.shape=square,decor.filled=empty,decor.size=5}{eae}
\end{fmfchar*}}
\quad+\quad
\parbox{15mm}{
\begin{fmfchar*}(15,15)
\fmfright{a,b}
\fmfleft{f1,f2}
\fmftop{z}
\fmf{plain}{ewn,a}
\fmf{plain}{uwd,f1}
\fmf{plain,tension=2.0}{ewn,waw}
\fmf{plain}{waw,uwd}
\fmf{plain}{b,ewn}
\fmf{plain}{f2,uwd}
\fmfdot{uwd}
\fmfdot{ewn}
\fmffreeze
\fmf{ghost}{z,waw}
\fmfv{decor.shape=square,decor.filled=empty,decor.size=5}{waw}
\end{fmfchar*}}
\quad+\quad
 \parbox{15mm}{
\begin{fmfchar*}(15,15)
\fmfright{a,b}
\fmfleft{f1,f2}
\fmftop{z}
\fmf{plain}{ewn,a}
\fmf{plain}{uwd,f1}
\fmf{plain}{ewn,uwd}
\fmf{plain,tension=2.0}{b,nan}
\fmf{plain}{f2,uwd}
\fmf{plain,tension=2.0}{nan,ewn}
\fmffreeze
\fmfdot{uwd}
\fmfdot{ewn}
\fmf{ghost}{z,nan}
\fmfv{decor.shape=square,decor.filled=empty,decor.size=5}{nan}
\end{fmfchar*}}
\end{multline}
These terms cancel the contributions from the three point vertices \eqref{eq:5-point-external} and \eqref{eq:5-point-internal} since they appear with the opposite sign. 

Similarly, the diagrams connected by the 5 point flips
\eqref{eq:generic-flips5} lead only to contact terms because of the STI \eqref{eq:sti5-graph}
\begin{multline}\label{eq:5-contact}
\parbox{15mm}{
\begin{fmfchar*}(15,15)
\fmfleftn{l}{2}
\fmfrightn{r}{3}
\fmf{dashes}{l1,g}
\fmf{dashes}{l2,g}
\fmf{dashes}{r1,g}
\fmf{plain,tension=2}{r3,i}
\fmf{dashes,tension=2}{i,g}
\fmffreeze
\fmf{double}{r2,i}
\fmfdot{g,i}
\end{fmfchar*}}\quad+\quad
\parbox{15mm}{
\begin{fmfchar*}(15,15)
\fmfleft{l1,l2}
\fmfright{r1,r2,r3}
\fmf{dashes}{l1,g}
\fmf{dashes}{l2,g}
\fmf{dashes}{r3,g}
\fmf{plain,tension=2}{r1,i}
\fmf{dashes,tension=2}{i,g}
\fmffreeze
\fmf{double}{r2,i}
\fmfdot{g,i}
\end{fmfchar*}}
\quad+\quad\parbox{15mm}{
\begin{fmfchar*}(15,15)
\fmfleft{l1,l2,l3}
\fmfright{r1,r2}
\fmf{dashes}{l3,g}
\fmf{dashes}{r1,g}
\fmf{plain,tension=2}{l1,i}
\fmf{dashes,tension=2}{i,g}
\fmf{plain}{g,r2}
\fmffreeze
\fmf{double}{l2,i}
\fmfdot{g,i}
\end{fmfchar*}}\quad+\quad
\parbox{15mm}{
\begin{fmfchar*}(15,15)
\fmfleft{l1,l2,l3}
\fmfright{r1,r2}
\fmf{dashes}{l1,g}
\fmf{dashes}{r1,g}
\fmf{dashes}{r2,g}
\fmf{plain,tension=2}{l3,i}
\fmf{dashes,tension=2}{i,g}
\fmffreeze
\fmf{double}{l2,i}
\fmfdot{g,i}
\end{fmfchar*}}\\
=\quad\parbox{15mm}{
\begin{fmfchar*}(15,15)
\fmfleftn{l}{2}
\fmfrightn{r}{3}
\fmf{dashes}{l1,g}
\fmf{dashes}{l2,g}
\fmf{dashes}{r1,g}
\fmf{zigzag,tension=2}{r3,i}
\fmf{dashes,tension=2}{i,g}
\fmffreeze
\fmf{ghost}{r2,i}
\fmfdot{g}
\fmfv{decor.shape=square,decor.filled=empty,decor.size=5}{i}
\end{fmfchar*}}\quad+\quad
\parbox{15mm}{
\begin{fmfchar*}(15,15)
\fmfleft{l1,l2}
\fmfright{r1,r2,r3}
\fmf{dashes}{l1,g}
\fmf{dashes}{l2,g}
\fmf{dashes}{r3,g}
\fmf{zigzag,tension=2}{r1,i}
\fmf{dashes,tension=2}{i,g}
\fmffreeze
\fmf{ghost}{r2,i}
\fmfdot{g}
\fmfv{decor.shape=square,decor.filled=empty,decor.size=5}{i}
\end{fmfchar*}}
\quad+\quad\parbox{15mm}{
\begin{fmfchar*}(15,15)
\fmfleft{l1,l2,l3}
\fmfright{r1,r2}
\fmf{dashes}{l3,g}
\fmf{dashes}{r1,g}
\fmf{zigzag,tension=2}{l1,i}
\fmf{dashes,tension=2}{i,g}
\fmf{dashes}{g,r2}
\fmffreeze
\fmf{ghost}{l2,i}
\fmfdot{g}
\fmfv{decor.shape=square,decor.filled=empty,decor.size=5}{i}
\end{fmfchar*}}\quad+\quad
\parbox{15mm}{
\begin{fmfchar*}(15,15)
\fmfleft{l1,l2,l3}
\fmfright{r1,r2}
\fmf{dashes}{l1,g}
\fmf{dashes}{r1,g}
\fmf{dashes}{r2,g}
\fmf{zigzag,tension=2}{l3,i}
\fmf{dashes,tension=2}{i,g}
\fmffreeze
\fmf{ghost}{l2,i}
\fmfdot{g}
\fmfv{decor.shape=square,decor.filled=empty,decor.size=5}{i}
\end{fmfchar*}}
 \end{multline}
If the external particle is a gauge boson, we get additional terms of the form 
\begin{equation}
\parbox{15mm}{
\begin{fmfchar*}(15,15)
\fmfleft{l1,l2,l3}
\fmfright{r1,r2}
\fmf{dashes}{l1,g}
\fmf{dashes}{r1,g}
\fmf{dashes}{r2,g}
\fmf{ghost,tension=2}{i,l3}
\fmf{dashes,tension=2}{i,g}
\fmffreeze
\fmf{ghost}{l2,i}
\fmfdot{g,i}
\fmfv{decor.shape=square,decor.filled=full,decor.size=5}{l3}
\end{fmfchar*}}
\end{equation}
as required by the STIs. 

The remaining differences are connected and more serious. We get a contribution form another topology to the contact terms if the external particles couple to ghosts: 
\begin{equation}\label{eq:5-point-contact-b}
\parbox{20mm}{
\begin{fmfchar*}(20,15)
\fmfleft{f1,w,a}
\fmfright{f2,b}
\fmf{plain,tension=2}{a,i}
\fmf{ghost,tension=2}{i,fwf}
\fmf{zigzag}{fwf,b}
\fmf{plain}{fwf,www}
\fmf{plain}{www,f1}
\fmf{plain}{www,f2}
\fmffreeze
\fmf{ghost}{w,i}
\fmfdot{www,i}
\fmfv{decor.shape=square,decor.filled=empty,decor.size=5}{fwf}
\end{fmfchar*}}
\end{equation}
Since---in contrast to fermion number---the number of gauge bosons and Higgs bosons is not conserved, we get
 additional contributions to the terms of the form \eqref{eq:5-point-external} for internal gauge bosons:
\begin{equation}
\parbox{20mm}{
\begin{fmfchar*}(20,15)
\fmfleft{a,z,b}
\fmfright{f1,f2}
\fmf{plain,tension=2.0}{eae,a}
\fmf{photon,tension=2.0}{ewn,eae}
\fmf{plain}{uwd,f1}
\fmf{plain}{ewn,uwd}
\fmf{plain}{b,ewn}
\fmf{plain}{f2,uwd}
\fmfdot{uwd}
\fmfdot{ewn}
\fmffreeze
\fmf{double}{z,eae}
\fmfdot{eae}
\end{fmfchar*}}\quad =\quad
\parbox{20mm}{
\begin{fmfchar*}(20,15)
\fmfleft{a,z,b}
\fmfright{f1,f2}
\fmf{plain,tension=2.0}{eae,a}
\fmf{photon,tension=2.0}{ewn,eae}
\fmf{plain}{uwd,f1}
\fmf{plain}{ewn,uwd}
\fmf{plain}{b,ewn}
\fmf{plain}{f2,uwd}
\fmfdot{uwd}
\fmfdot{ewn}
\fmffreeze
\fmf{ghost}{z,eae}
\fmfv{decor.shape=square,decor.filled=empty,decor.size=5}{eae}
\end{fmfchar*}}
\quad +\quad
\parbox{20mm}{
\begin{fmfchar*}(20,15)
\fmfleft{a,z,b}
\fmfright{f1,f2}
\fmf{zigzag,tension=2.0}{eae,a}
\fmf{photon,tension=2.0}{ewn,eae}
\fmf{plain}{uwd,f1}
\fmf{plain}{ewn,uwd}
\fmf{plain}{b,ewn}
\fmf{plain}{f2,uwd}
\fmfdot{uwd}
\fmfdot{ewn}
\fmffreeze
\fmf{ghost}{z,eae}
\fmfv{decor.shape=square,decor.filled=empty,decor.size=5}{eae}
\end{fmfchar*}}\quad +\quad
\parbox{20mm}{
\begin{fmfchar*}(20,15)
\fmfleft{a,z,b}
\fmfright{f1,f2}
\fmf{plain,tension=2.0}{eae,a}
\fmf{dots,tension=2}{eae,p}
\fmf{photon,tension=2.0}{p,ewn}
\fmf{plain,tension=2}{uwd,f1}
\fmf{plain,tension=2}{ewn,uwd}
\fmf{plain,tension=0.66}{b,ewn}
\fmf{plain,tension=2}{f2,uwd}
\fmfdot{uwd}
\fmfdot{ewn}
\fmffreeze
\fmf{ghost}{z,eae}
\fmfdot{eae}
\fmfv{decor.shape=square,decor.filled=full,decor.size=5}{p}
\end{fmfchar*}}
\end{equation}
Note that these diagrams only appear for particles that couple both to two gauge bosons and to ghosts. In a spontaneously broken gauge theory in $R_\xi$ gauge this applies to gauge bosons and Higgs bosons (or more precisely scalar bosons that mix with Goldstone bosons under BRS-transformations). 

To proceed further, we need to use a STI for a 4 particle subamplitude. This is only possible provided we  add additional diagrams that---by definition--- can be found by applying the elementary gauge flips to the \emph{internal} gauge boson:
\begin{multline}\label{eq:5-internal-flips}
\parbox{20mm}{
\begin{fmfchar*}(20,15)
\fmfleft{a,z,b}
\fmfright{f1,f2}
\fmf{plain,tension=2.0}{eae,a}
\fmf{photon,tension=2.0}{ewn,eae}
\fmf{plain}{uwd,f1}
\fmf{plain}{ewn,uwd}
\fmf{plain}{b,ewn}
\fmf{plain}{f2,uwd}
\fmfdot{uwd}
\fmfdot{ewn}
\fmffreeze
\fmf{photon}{z,eae}
\fmfdot{eae}
\end{fmfchar*}}
\xrightarrow{T_4^G}
\left\{\,
\parbox{20mm}{
\begin{fmfchar*}(20,15)
\fmfleft{a,w,b}
\fmfright{f1,f2}
\fmf{plain,tension=2}{a,i}
\fmf{photon,tension=2}{i,fwf1}
\fmf{plain}{fwf1,fwf2}
\fmf{plain}{fwf2,b}
\fmf{phantom}{fwf2,f2}
\fmf{phantom}{fwf1,f1}
\fmffreeze
\fmf{photon}{w,i}
\fmf{plain}{fwf2,f1}
\fmf{plain}{fwf1,f2}
\fmfdot{fwf1,fwf2,i}
\end{fmfchar*}}\quad ,\quad 
\parbox{20mm}{
\begin{fmfchar*}(20,15)
\fmfbottom{a,b}
\fmftop{f1,f2}
\fmfleft{w}
\fmf{plain,tension=2}{a,i}
\fmf{photon,tension=2}{i,fwf}
\fmf{plain}{fwf,b}
\fmf{plain}{fwf,www}
\fmf{plain}{www,f1}
\fmf{plain}{www,f2}
\fmffreeze
\fmf{photon}{w,i}
\fmfdot{fwf,www,i}
\end{fmfchar*}}\quad ,\quad 
\parbox{15mm}{
\begin{fmfchar*}(15,15)
\fmfleft{a,w,b}
\fmfright{f1,f2}
\fmf{plain,tension=2}{a,i}
\fmf{photon,tension=2}{i,c}
\fmf{plain}{c,b}
\fmf{plain}{c,f1}
\fmf{plain}{c,f2}
\fmffreeze
\fmf{photon}{w,i}
\fmfdot{c,i}
\end{fmfchar*}}
\right\}
\end{multline}
This brings in $t$ and $u$ channel diagrams and the quartic vertices, so the simple structure of the groves is destroyed. Adding these diagrams and the corresponding Goldstone boson diagrams, we can use the STI for the 4 point function:
\begin{multline}
\parbox{20mm}{
\begin{fmfchar*}(20,20)
\fmfbottom{a,b}
\fmftop{f1,f2}
\fmfleft{z}
\fmf{plain,tension=2.0}{eae,a}
\fmf{dots,tension=2}{eae,p}
\fmf{double,tension=2.0}{p,x}
\fmf{plain,tension=1}{x,f1}
\fmf{plain,tension=0.66}{b,x}
\fmf{plain,tension=1}{f2,x}
\fmffreeze
\fmf{ghost}{z,eae}
\fmfdot{eae,p}
\fmfv{d.sh=circle,d.f=empty,d.size=15pt}{x}
\end{fmfchar*}}\quad =\quad
\parbox{20mm}{
\begin{fmfchar*}(20,15)
\fmfleft{a,z,b}
\fmfright{f1,f2}
\fmf{plain,tension=2.0}{eae,a}
\fmf{ghost,tension=2.0}{eae,ewn}
\fmf{plain}{uwd,f1}
\fmf{plain}{ewn,uwd}
\fmf{zigzag}{b,ewn}
\fmf{plain}{f2,uwd}
\fmfdot{uwd,eae}
\fmffreeze
\fmf{ghost}{z,eae}
\fmfv{decor.shape=square,decor.filled=empty,decor.size=5}{ewn}
\end{fmfchar*}}\quad + \quad
\parbox{20mm}{
\begin{fmfchar*}(20,15)
\fmfleft{a,w,b}
\fmfright{f1,f2}
\fmf{plain,tension=2}{a,i}
\fmf{ghost,tension=2}{i,fwf1}
\fmf{plain}{fwf1,fwf2}
\fmf{plain}{fwf2,b}
\fmf{phantom}{fwf2,f2}
\fmf{phantom}{fwf1,f1}
\fmffreeze
\fmf{ghost}{w,i}
\fmf{plain}{fwf2,f1}
\fmf{zigzag}{fwf1,f2}
\fmfdot{fwf2,i}
\fmfv{decor.shape=square,decor.filled=empty,decor.size=5}{fwf1}
\end{fmfchar*}}\quad +\quad 
\parbox{20mm}{
\begin{fmfchar*}(20,15)
\fmfbottom{a,b}
\fmftop{f1,f2}
\fmfleft{w}
\fmf{plain,tension=2}{a,i}
\fmf{ghost,tension=2}{i,fwf}
\fmf{zigzag}{fwf,b}
\fmf{plain}{fwf,www}
\fmf{plain}{www,f1}
\fmf{plain}{www,f2}
\fmffreeze
\fmf{ghost}{w,i}
\fmfdot{www,i}
\fmfv{decor.shape=square,decor.filled=empty,decor.size=5}{fwf}
\end{fmfchar*}}
\end{multline}
These are some of the contact terms obtained by applying the mapping \eqref{subeq:generate-contact} to the diagrams of \fref{eq:5-internal-flips} and therefore ---according to our definition---together with those of the form \eqref{eq:5-point-external} and \eqref{eq:5-contact} the contact terms needed to satisfy the STI. Like in \fref{eq:int-sti}, this is an example for a `second order' cancellation. Of course now we have to flip the external gauge boson through all 4 diagrams in \fref{eq:5-internal-flips} and this will in general result in \emph{all} diagrams of the amplitude. 
  
\subsection{$N$ point diagrams}
We are now ready to show that the groves obtained by the gauge flips defined as in \fref{sec:flip-def} are indeed the minimal gauge invariance classes according to \fref{def:gic}. The necessary steps are essentially the same as in the case of the general 5 point function in \fref{sec:5-point-groves}. 

Let us first turn to the subject of gauge parameter independence. We have seen
 for the 5 point function, that the prescription to apply gauge flips to all
 external and internal gauge bosons of a diagram results in an expression
 satisfying the STI. We now show, that the flips of the internal gauge bosons
 lead to the gauge parameter independence of the groves with
all external particles on their mass shell: From
\fref{eq:xi-factorize} we know that the parts of the propagators linear in
$\delta\xi$ must connect subamplitudes satisfying the Ward Identities to obtain gauge
parameter independent quantities. We assume it has been shown that the $N-1$
particle diagrams connected by gauge flips satisfy the STI and are gauge
parameter independent. Therefore applying gauge flips to all internal gauge bosons
of a $N$-point function ensures its gauge parameter independence. This covers also the case of amplitudes without external gauge bosons that is thus reduced to the discussion of amplitudes with fewer external particles and external gauge bosons. 

In the following we consider the insertion of a gauge boson into a Feynman diagram with $N-1$ external particles. 

We pick out an arbitrary vertex. For simplicity, we discuss the case of a
cubic vertex, for quartic vertices no new features appear. According to the gauge flips, we have to insert the gauge boson into all three legs of the vertex and include a quartic vertex (if allowed by the Feynman rules):
\begin{multline}\label{eq:n-ampl-ternary}
\parbox{35\unitlength}{
  \fmfframe(2,3)(2,1){
      \begin{fmfgraph*}(35,20)
       \fmftop{x}
       \fmfbottomn{n}{6}
       \fmfleft{li}
       \fmfright{ri}
       \fmf{phantom,tension=8}{li,i}
       \fmf{phantom,tension=8}{ri,l}  
       \fmf{phantom,tension=6}{x,i}
       \fmf{phantom,tension=6}{x,l}        
       \fmf{phantom}{i,l}       
        \begin{fmffor}{i}{1}{1}{2}
         \fmf{plain,tension=3}{i,n[i]}
       \end{fmffor}        
       \begin{fmffor}{i}{5}{1}{6}
         \fmf{plain,tension=3}{l,n[i]}
       \end{fmffor}
       \fmf{plain,tension=1}{i,n}
       \fmf{plain,tension=1}{n,l}        
        \fmffreeze
        \begin{fmffor}{i}{3}{1}{4}
         \fmf{plain}{j,n[i]}
       \end{fmffor} 
        \fmf{plain}{n,j} 
      \fmfdot{n}
      \fmffreeze
      \fmf{photon}{x,y}
      \fmf{phantom,label=$\otimes$,la.di=0}{y,n}
      \fmfv{d.sh=circle,d.f=30,d.si=12}{i}
      \fmfv{d.sh=circle,d.f=30,d.si=12}{j}
      \fmfv{d.sh=circle,d.f=30,d.si=12}{l}
      \end{fmfgraph*}}}\, \equiv\,
\parbox{35\unitlength}{
  \fmfframe(2,3)(2,1){
      \begin{fmfgraph*}(35,20)
       \fmftop{x}
       \fmfbottomn{n}{6}
       \fmfleft{li}
       \fmfright{ri}
       \fmf{phantom,tension=8}{li,i}
       \fmf{phantom,tension=8}{ri,l}  
       \fmf{phantom,tension=6}{x,i}
       \fmf{phantom,tension=6}{x,l}        
       \fmf{phantom}{i,l}
        \begin{fmffor}{i}{1}{1}{2}
         \fmf{plain,tension=3}{i,n[i]}
       \end{fmffor}        
       \begin{fmffor}{i}{5}{1}{6}
         \fmf{plain,tension=3}{l,n[i]}
       \end{fmffor}
       \fmf{plain,tension=2}{i,y,n}
       \fmf{plain,tension=1}{n,l}        
        \fmffreeze
       \fmf{photon,tension=6}{x,y}
        \begin{fmffor}{i}{3}{1}{4}
         \fmf{plain}{j,n[i]}
       \end{fmffor} 
        \fmf{plain}{n,j} 
      \fmfdot{n,y}
      \fmfv{d.sh=circle,d.f=30,d.si=12pt}{i}
      \fmfv{d.sh=circle,d.f=30,d.si=12pt}{j}
      \fmfv{d.sh=circle,d.f=30,d.si=12pt}{l}
      \end{fmfgraph*}}}
+
\parbox{35\unitlength}{
  \fmfframe(2,3)(2,1){
      \begin{fmfgraph*}(35,20)
       \fmftop{x}
       \fmfbottomn{n}{6}
        \fmfleft{li}
       \fmfright{ri}
       \fmf{phantom,tension=8}{li,i}
       \fmf{phantom,tension=8}{ri,l}   
       \fmf{phantom,tension=6}{x,i}
       \fmf{phantom,tension=6}{x,l}        
       \fmf{phantom}{i,l}       
        \begin{fmffor}{i}{1}{1}{2}
         \fmf{plain,tension=3}{i,n[i]}
       \end{fmffor}        
       \begin{fmffor}{i}{5}{1}{6}
         \fmf{plain,tension=3}{l,n[i]}
       \end{fmffor}
       \fmf{plain,tension=1}{i,n}
       \fmf{plain,tension=2}{n,y,l}        
        \fmffreeze
        \fmf{photon}{x,y}
        \begin{fmffor}{i}{3}{1}{4}
         \fmf{plain}{j,n[i]}
       \end{fmffor} 
        \fmf{plain}{n,j} 
      \fmfdot{n,y}
      \fmfv{d.sh=circle,d.f=30,d.si=12}{i}
      \fmfv{d.sh=circle,d.f=30,d.si=12}{j}
      \fmfv{d.sh=circle,d.f=30,d.si=12}{l}
      \end{fmfgraph*}}}\\
+\parbox{35\unitlength}{
  \fmfframe(2,3)(2,1){
      \begin{fmfgraph*}(35,20)
       \fmftop{x}
       \fmfbottomn{n}{6}
        \fmfleft{li}
       \fmfright{ri}
       \fmfleft{li}
       \fmfright{ri}
       \fmf{phantom,tension=8}{li,i}
       \fmf{phantom,tension=8}{ri,l}  
       \fmf{phantom,tension=8}{li,i}
       \fmf{phantom,tension=8}{ri,l}  
       \fmf{phantom,tension=6}{x,i}
       \fmf{phantom,tension=6}{x,l}        
       \fmf{phantom}{i,l}       
        \begin{fmffor}{i}{1}{1}{2}
         \fmf{plain,tension=3}{i,n[i]}
       \end{fmffor}        
       \begin{fmffor}{i}{5}{1}{6}
         \fmf{plain,tension=3}{l,n[i]}
       \end{fmffor}
       \fmf{plain,tension=1}{i,n}
       \fmf{plain,tension=1}{n,l}        
        \fmffreeze
        \begin{fmffor}{i}{3}{1}{4}
         \fmf{plain}{j,n[i]}
       \end{fmffor} 
        \fmf{plain}{n,y,j} 
      \fmfdot{n,y}
      \fmffreeze
      \fmf{photon,left=0.3}{x,y}
      \fmfv{d.sh=circle,d.f=30,d.si=12}{i}
      \fmfv{d.sh=circle,d.f=30,d.si=12}{j}
      \fmfv{d.sh=circle,d.f=30,d.si=12}{l}
      \end{fmfgraph*}}}
+
\parbox{35\unitlength}{
  \fmfframe(2,3)(2,1){
      \begin{fmfgraph*}(35,20)
       \fmftop{x}
       \fmfbottomn{n}{6}
       \fmfleft{li}
       \fmfright{ri}
       \fmf{phantom,tension=8}{li,i}
       \fmf{phantom,tension=8}{ri,l}  
       \fmf{phantom,tension=6}{x,i}
       \fmf{phantom,tension=6}{x,l}        
       \fmf{phantom}{i,l}       
        \begin{fmffor}{i}{1}{1}{2}
         \fmf{plain,tension=3}{i,n[i]}
       \end{fmffor}        
       \begin{fmffor}{i}{5}{1}{6}
         \fmf{plain,tension=3}{l,n[i]}
       \end{fmffor}
       \fmf{plain,tension=1}{i,n}
       \fmf{plain,tension=1}{n,l}        
        \fmffreeze
        \begin{fmffor}{i}{3}{1}{4}
         \fmf{plain}{j,n[i]}
       \end{fmffor} 
        \fmf{plain}{n,j} 
      \fmfdot{n}
      \fmffreeze
      \fmf{photon}{x,n}
      \fmfv{d.sh=circle,d.f=30,d.si=12}{i}
      \fmfv{d.sh=circle,d.f=30,d.si=12}{j}
      \fmfv{d.sh=circle,d.f=30,d.si=12}{l}
      \end{fmfgraph*}}}
\end{multline}
Here the gray blobs denote a subdiagram, while white blobs continue to mean sub~\emph{amplitudes} or subgroves, i.e. sets of diagrams. 

The case of the `first order' cancellations involving only the STI for three and 4 point vertices is straightforward. Neglecting the ghost terms in the STI for internal gauge and Goldstone bosons  for the moment and repeating the steps leading from \fref{eq:5-point-external} to \fref{eq:3sti-int} or \eqref{eq:sti4-int},  we find that everything cancels apart from the terms
\begin{multline}\label{eq:sti-int-n}
\parbox{35\unitlength}{
  \fmfframe(2,3)(2,1){
      \begin{fmfgraph*}(35,20)
       \fmftop{x}
       \fmfbottomn{n}{6}
       \fmfleft{li}
       \fmfright{ri}
       \fmf{phantom,tension=8}{li,i}
       \fmf{phantom,tension=8}{ri,l}  
       \fmf{phantom,tension=6}{x,i}
       \fmf{phantom,tension=6}{x,l}        
       \fmf{phantom}{i,l}       
        \begin{fmffor}{i}{1}{1}{2}
         \fmf{plain,tension=3}{i,n[i]}
       \end{fmffor}        
       \begin{fmffor}{i}{5}{1}{6}
         \fmf{plain,tension=3}{l,n[i]}
       \end{fmffor}
       \fmf{plain,tension=1}{i,n}
       \fmf{plain,tension=1}{n,l}        
        \fmffreeze
        \begin{fmffor}{i}{3}{1}{4}
         \fmf{plain}{j,n[i]}
       \end{fmffor} 
        \fmf{plain}{n,j} 
      \fmfdot{n}
      \fmffreeze
      \fmf{double}{x,y}
      \fmf{phantom,label=$\otimes$,la.di=0}{y,n}
      \fmfv{d.sh=circle,d.f=30,d.si=12}{i}
      \fmfv{d.sh=circle,d.f=30,d.si=12}{j}
      \fmfv{d.sh=circle,d.f=30,d.si=12}{l}
      \end{fmfgraph*}}}\,=\,
-\parbox{35\unitlength}{
  \fmfframe(2,3)(2,1){
      \begin{fmfgraph*}(35,20)
       \fmftop{x}
       \fmfbottomn{n}{6}
       \fmfleft{li}
       \fmfright{ri}
       \fmf{phantom,tension=8}{li,i}
       \fmf{phantom,tension=8}{ri,l}  
       \fmf{phantom,tension=6}{x,i}
       \fmf{phantom,tension=6}{x,l}        
       \fmf{phantom}{i,l}
        \begin{fmffor}{i}{1}{1}{2}
         \fmf{plain,tension=3}{i,n[i]}
       \end{fmffor}        
       \begin{fmffor}{i}{5}{1}{6}
         \fmf{plain,tension=3}{l,n[i]}
       \end{fmffor}
       \fmf{plain,tension=2}{i,y,n}
       \fmf{plain,tension=1}{n,l}        
        \fmffreeze
       \fmf{ghost,tension=6}{x,y}
        \begin{fmffor}{i}{3}{1}{4}
         \fmf{plain}{j,n[i]}
       \end{fmffor} 
        \fmf{plain}{n,j} 
      \fmfdot{n}
       \fmfv{decor.shape=square,decor.filled=empty,decor.size=5}{y}
      \fmfv{d.sh=circle,d.f=30,d.si=12pt}{i}
      \fmfv{d.sh=circle,d.f=30,d.si=12pt}{j}
      \fmfv{d.sh=circle,d.f=30,d.si=12pt}{l}
      \end{fmfgraph*}}}
-
\parbox{35\unitlength}{
  \fmfframe(2,3)(2,1){
      \begin{fmfgraph*}(35,20)
       \fmftop{x}
       \fmfbottomn{n}{6}
        \fmfleft{li}
       \fmfright{ri}
       \fmf{phantom,tension=8}{li,i}
       \fmf{phantom,tension=8}{ri,l}   
       \fmf{phantom,tension=6}{x,i}
       \fmf{phantom,tension=6}{x,l}        
       \fmf{phantom}{i,l}       
        \begin{fmffor}{i}{1}{1}{2}
         \fmf{plain,tension=3}{i,n[i]}
       \end{fmffor}        
       \begin{fmffor}{i}{5}{1}{6}
         \fmf{plain,tension=3}{l,n[i]}
       \end{fmffor}
       \fmf{plain,tension=1}{i,n}
       \fmf{plain,tension=2}{n,y,l}        
        \fmffreeze
        \fmf{ghost}{x,y}
        \begin{fmffor}{i}{3}{1}{4}
         \fmf{plain}{j,n[i]}
       \end{fmffor} 
        \fmf{plain}{n,j} 
      \fmfdot{n}
       \fmfv{decor.shape=square,decor.filled=empty,decor.size=5}{y}
      \fmfv{d.sh=circle,d.f=30,d.si=12}{i}
      \fmfv{d.sh=circle,d.f=30,d.si=12}{j}
      \fmfv{d.sh=circle,d.f=30,d.si=12}{l}
      \end{fmfgraph*}}}\\
-\parbox{35\unitlength}{
  \fmfframe(2,3)(2,1){
      \begin{fmfgraph*}(35,20)
       \fmftop{x}
       \fmfbottomn{n}{6}
        \fmfleft{li}
       \fmfright{ri}
       \fmfleft{li}
       \fmfright{ri}
       \fmf{phantom,tension=8}{li,i}
       \fmf{phantom,tension=8}{ri,l}  
       \fmf{phantom,tension=6}{x,i}
       \fmf{phantom,tension=6}{x,l}        
       \fmf{phantom}{i,l}       
        \begin{fmffor}{i}{1}{1}{2}
         \fmf{plain,tension=3}{i,n[i]}
       \end{fmffor}        
       \begin{fmffor}{i}{5}{1}{6}
         \fmf{plain,tension=3}{l,n[i]}
       \end{fmffor}
       \fmf{plain,tension=1}{i,n}
       \fmf{plain,tension=1}{n,l}        
        \fmffreeze
        \begin{fmffor}{i}{3}{1}{4}
         \fmf{plain}{j,n[i]}
       \end{fmffor} 
        \fmf{plain}{n,y,j} 
      \fmfdot{n}
       \fmfv{decor.shape=square,decor.filled=empty,decor.size=5}{y}
      \fmffreeze
      \fmf{ghost,left=0.3}{x,y}
      \fmfv{d.sh=circle,d.f=30,d.si=12}{i}
      \fmfv{d.sh=circle,d.f=30,d.si=12}{j}
      \fmfv{d.sh=circle,d.f=30,d.si=12}{l}
      \end{fmfgraph*}}}
\end{multline}
To cancel these remaining diagrams, we have to `zoom' into the blobs, insert the external gauge boson at the next vertex and repeat the same procedure for the next vertices. This will cancel the terms from \fref{eq:sti-int-n} but leaves new terms of the same form at the next vertices. This process can be iterated, until the external particles are reached. The remaining terms are the contact terms of the STI for the Green's function with the ghost line going through the diagram without interaction: 
\begin{equation}
\parbox{35\unitlength}{
  \fmfframe(2,3)(2,1){
      \begin{fmfgraph*}(35,20)
       \fmftop{x}
       \fmfbottomn{n}{6}
        \fmfleft{li}
       \fmfright{ri}
       \fmfleft{li}
       \fmfright{ri}
       \fmf{phantom,tension=8}{li,i}
       \fmf{phantom,tension=8}{ri,l}  
       \fmf{phantom,tension=6}{x,i}
       \fmf{phantom,tension=6}{x,l}        
       \fmf{phantom}{i,l}       
        \begin{fmffor}{i}{1}{1}{2}
         \fmf{plain,tension=3}{i,n[i]}
       \end{fmffor} 
       \fmf{plain,tension=3}{l,n5}
       \fmf{plain,tension=6}{l,y}
        \fmf{zigzag,tension=6}{y,n6}
       \fmf{plain,tension=1}{i,n}
       \fmf{plain,tension=1}{n,l}        
        \fmffreeze
        \begin{fmffor}{i}{3}{1}{4}
         \fmf{plain}{j,n[i]}
       \end{fmffor} 
        \fmf{plain}{n,j} 
      \fmfdot{n}
       \fmfv{decor.shape=square,decor.filled=empty,decor.size=5}{y}
      \fmffreeze
      \fmf{ghost,left=0.5}{x,y}
      \fmfv{d.sh=circle,d.f=30,d.si=12}{i}
      \fmfv{d.sh=circle,d.f=30,d.si=12}{j}
      \fmfv{d.sh=circle,d.f=30,d.si=12}{l}
      \end{fmfgraph*}}}
\end{equation}
Clearly, the procedure discussed above corresponds to gauge-flipping the external gauge boson through the original diagram. 

The `second order' cancellations because of the ghost terms in the STIs for the gauge boson vertices are more complicated. Again, they lead to the prescription to flip also all internal gauge bosons and give rise to the contact terms where the ghosts interact with the remaining particles. 

Combining the Goldstone and gauge boson diagrams, the diagrams in question look like
\begin{equation}
\parbox{40\unitlength}{
      \begin{fmfgraph*}(40,20)
       \fmftop{x}
       \fmfbottomn{n}{8}
       \fmfleft{li}
       \fmfright{ri}
       \fmf{phantom,tension=8}{li,i}
       \fmf{phantom,tension=8}{ri,l}  
       \fmf{phantom,tension=6}{x,i}
       \fmf{phantom,tension=6}{x,l}        
       \fmf{phantom}{i,l}
        \begin{fmffor}{i}{1}{1}{2}
         \fmf{plain,tension=3}{i,n[i]}
       \end{fmffor}        
       \begin{fmffor}{i}{7}{1}{8}
         \fmf{plain,tension=3}{l,n[i]}
       \end{fmffor}
       \fmf{plain,tension=4}{n,y}
       \fmf{phantom,tension=2}{y,i}
       \fmf{plain,tension=4}{n,l}        
        \fmffreeze
       \fmf{ghost,tension=6}{x,y}
        \begin{fmffor}{i}{5}{1}{6}
         \fmf{plain}{j,n[i]}
       \end{fmffor} 
        \fmf{plain}{n,j} 
      \fmf{ghost,tension=2}{y,k}
       \fmf{double,tension=2}{k,i}
      \fmfdot{n,y,k}
      \fmfv{d.sh=circle,d.f=30,d.si=12pt}{i}
      \fmfv{d.sh=circle,d.f=30,d.si=12pt}{j}
      \fmfv{d.sh=circle,d.f=30,d.si=12pt}{l}
      \end{fmfgraph*}}
\end{equation}
As in the case of the 5 point function in \fref{eq:5-internal-flips}, we have to add the appropriate diagrams so we can use a STI for the subamplitudes. We will proceed by induction and assume that that the $N-1$ particle groves satisfy the STI \eqref{eq:sti-graph} in the sense of \fref{def:sti}. Applying the gauge flips to the subamplitude connected to the double line, we can use the STI and obtain:
\begin{multline}
\parbox{40\unitlength}{
      \begin{fmfgraph*}(40,20)
       \fmftop{x}
       \fmfbottomn{n}{8}
       \fmfleft{li}
       \fmfright{ri}
       \fmf{phantom,tension=8}{li,i}
       \fmf{phantom,tension=8}{ri,l}  
       \fmf{phantom,tension=6}{x,i}
       \fmf{phantom,tension=6}{x,l}        
       \fmf{phantom}{i,l}
        \begin{fmffor}{i}{1}{1}{2}
         \fmf{plain,tension=3}{i,n[i]}
       \end{fmffor}        
       \begin{fmffor}{i}{7}{1}{8}
         \fmf{plain,tension=3}{l,n[i]}
       \end{fmffor}
       \fmf{plain,tension=4}{n,y}
       \fmf{phantom,tension=2}{y,i}
       \fmf{plain,tension=4}{n,l}        
        \fmffreeze
       \fmf{ghost,tension=6}{x,y}
        \begin{fmffor}{i}{5}{1}{6}
         \fmf{plain}{j,n[i]}
       \end{fmffor} 
        \fmf{plain}{n,j} 
      \fmf{ghost,tension=2}{y,k}
       \fmf{double,tension=2}{k,i}
      \fmfdot{n,y,k}
      \fmfv{d.sh=circle,d.f=30,d.si=12pt}{i}
      \fmfv{d.sh=circle,d.f=30,d.si=12pt}{j}
      \fmfv{d.sh=circle,d.f=30,d.si=12pt}{l}
      \end{fmfgraph*}} 
\xrightarrow{T_4^G}
\parbox{40\unitlength}{
      \begin{fmfgraph*}(40,20)
       \fmftop{x}
       \fmfbottomn{n}{8}
       \fmfleft{li}
       \fmfright{ri}
       \fmf{phantom,tension=8}{li,i}
       \fmf{phantom,tension=8}{ri,l}  
       \fmf{phantom,tension=6}{x,i}
       \fmf{phantom,tension=6}{x,l}        
       \fmf{phantom}{i,l}
        \begin{fmffor}{i}{1}{1}{2}
         \fmf{plain,tension=3}{i,n[i]}
       \end{fmffor}        
       \begin{fmffor}{i}{7}{1}{8}
         \fmf{plain,tension=3}{l,n[i]}
       \end{fmffor}
       \fmf{plain,tension=4}{n,y}
       \fmf{phantom,tension=2}{y,i}
       \fmf{plain,tension=4}{n,l}        
        \fmffreeze
       \fmf{ghost,tension=6}{x,y}
        \begin{fmffor}{i}{5}{1}{6}
         \fmf{plain}{j,n[i]}
       \end{fmffor} 
        \fmf{plain}{n,j} 
      \fmf{ghost,tension=2}{y,k}
       \fmf{double,tension=2}{k,i}
      \fmfdot{n,y,k}
      \fmfv{d.sh=circle,d.f=empty,d.si=12pt}{i}
      \fmfv{d.sh=circle,d.f=30,d.si=12pt}{j}
      \fmfv{d.sh=circle,d.f=30,d.si=12pt}{l}
      \end{fmfgraph*}} \\
=\quad \sum_{\phi_i}\parbox{35\unitlength}{
  \fmfframe(2,3)(2,1){
      \begin{fmfgraph*}(35,20)
       \fmftop{x}
       \fmfbottomn{n}{6}
        \fmfleft{li}
       \fmfright{ri}
       \fmfleft{li}
       \fmfright{ri}
       \fmf{phantom,tension=8}{li,i}
       \fmf{phantom,tension=8}{ri,l}  
       \fmf{phantom,tension=6}{x,i}
       \fmf{phantom,tension=6}{x,l}        
       \fmf{phantom}{i,l}       
        \begin{fmffor}{i}{5}{1}{6}
         \fmf{plain,tension=3}{l,n[i]}
       \end{fmffor} 
       \fmf{plain,tension=3}{i,n2}
       \fmf{plain,tension=6}{i,y}
        \fmf{zigzag,tension=6}{y,n1}
       \fmf{ghost,tension=1}{n,i}
       \fmf{plain,tension=1}{n,l}        
        \fmffreeze
        \fmf{ghost,right}{i,y}
        \begin{fmffor}{i}{3}{1}{4}
         \fmf{plain}{j,n[i]}
       \end{fmffor} 
        \fmf{plain}{n,j} 
      \fmfdot{n}
       \fmfv{decor.shape=square,decor.filled=empty,decor.size=5}{y}
      \fmffreeze
      \fmf{ghost}{x,n}
      \fmfv{d.sh=diamond,d.f=empty,d.si=15}{i}
      \fmfv{d.sh=circle,d.f=30,d.si=12}{j}
      \fmfv{d.sh=circle,d.f=30,d.si=12}{l}
      \end{fmfgraph*}}}
\end{multline}
These are contact terms in the STI with internal ghost interactions. Of course now we have to flip the external gauge boson also  through the new diagrams. 

Since by assumption the contact terms of the subamplitude are those generated by the formal mapping defined in \fref{sec:groves-def}, we see that the contact terms are those corresponding to the diagrams
\begin{equation}
\parbox{40\unitlength}{
      \begin{fmfgraph*}(40,20)
       \fmftop{x}
       \fmfbottomn{n}{8}
       \fmfleft{li}
       \fmfright{ri}
       \fmf{phantom,tension=6}{li,i}
       \fmf{phantom,tension=8}{ri,l}  
       \fmf{phantom,tension=6}{x,i}
       \fmf{phantom,tension=6}{x,l}        
       \fmf{phantom}{i,l}
        \begin{fmffor}{i}{1}{1}{2}
         \fmf{plain,tension=3}{i,n[i]}
       \end{fmffor}        
       \begin{fmffor}{i}{7}{1}{8}
         \fmf{plain,tension=3}{l,n[i]}
       \end{fmffor}
       \fmf{plain,tension=4}{n,y}
       \fmf{phantom,tension=2}{y,i}
       \fmf{plain,tension=4}{n,l}        
        \fmffreeze
       \fmf{photon,tension=6}{x,y}
        \begin{fmffor}{i}{5}{1}{6}
         \fmf{plain}{j,n[i]}
       \end{fmffor} 
        \fmf{plain}{n,j} 
      \fmf{photon}{y,i}
      \fmfdot{n,y}
      \fmfv{d.sh=circle,d.f=empty,d.si=12pt}{i}
      \fmfv{d.sh=circle,d.f=30,d.si=12pt}{j}
      \fmfv{d.sh=circle,d.f=30,d.si=12pt}{l}
      \end{fmfgraph*}} 
\end{equation}
This shows that for the set of diagrams obtained by applying the gauge flips, the contact terms that appear by contracting an external gauge boson  are indeed the same ones that are assigned to every diagram by the mapping $\mathscr{F}$ defined in \fref{sec:groves-def} and therefore the STI is satisfied. This finishes our proof. 

It should be clear, that the sets of diagrams connected by gauge flips are indeed the \emph{minimal} gauge invariance classes since by construction an omission of a diagram would lead to an violation of a Ward Identity.
\part{Reconstruction of Feynman rules}\label{part:reconstruction}
\chapter{Lagrangian of a spontaneously broken gauge theory}\label{chap:ssb-lag}\chaptermark{Lagrangian}
We have seen in the previous part that the gauge invariance of the Lagrangian implies the Ward- and Slavnov Taylor identities of the Green's functions and irreducible vertices of the theory. We now take the opposite point of view and investigate if we can reconstruct a gauge invariant Lagrangian from a given set of input parameters by demanding that the Ward Identities are satisfied for a \emph{finite} set of tree level scattering amplitudes. 

Before turning to the proof we must, however,  establish a notation for the Lagrangian and clarify the relations among the coupling constants in a spontaneously broken gauge theory. In \fref{app:general-fr} we introduce the general renormalizable Lagrangian that contains the particle spectrum of a spontaneously broken gauge theory but without imposing gauge invariance and spontaneous symmetry breaking (SSB). We are going to use this Lagrangian to evaluate the Ward identities in \fref{chap:reconstruction}. 

In this chapter we establish the relations among the coupling constants in the case of spontaneous symmetry breaking in order to compare with the results from the calculation of the Ward identities in \fref{chap:reconstruction}. 

\section{Field content and symmetries}\label{sec:ssb-fields}
\subsection{Gauge fields}
We want to parametrize an Lagrangian of a gauge theory with a local symmetry group $G$ that is spontaneously broken down to a subgroup $H$ (see eg. \cite{Weinberg:1995}). Therefore we introduce gauge fields $W_{\alpha\mu}$ that transform under infinitesimal $G$ gauge transformations as
\begin{equation}
\delta W_{a\mu}=\frac{1}{g_a}\dmu \omega_a - f^{abc}\omega_b W_{c\mu}
\end{equation}
Here we allow for different gauge couplings $g_a$ to include the possibility of nonsimple groups $G$ like in the Standard Model. 

The $f^{abc}$ are the structure constants of the  Lie algebra of $G$ that satisfy the Jacobi Identity
\begin{equation}\label{eq:jacobi}
f^{abe}f^{cde}+f^{cae}f^{bde}+f^{ade}f^{bce}=0
\end{equation}
We work in a representation in which the structure constants are totally antisymmetric. 
\subsection{Scalar fields}
To describe spontaneous symmetry breaking we introduce a multiplet of real-valued scalar fields containing Higgs bosons $H$ and Goldstone bosons $\phi$: 
\begin{equation}\label{eq:scalar-explicit}
\phi_A=\begin{pmatrix} \phi_\alpha \\ v_i+ H_i \end{pmatrix}
\end{equation}
transforming under a (in general reducible) representation of the symmetry group
\begin{equation}\label{eq:scalar-gauge}
  \delta \phi_A =- \omega_a T^a_{AB}\phi_B
\end{equation}
This parametrization is no restriction on the scalar sector because we can always split complex fields into real and imaginary parts and we can always combine several irreducible representations into one reducible representation. 

According to Goldstone's Theorem \cite{Goldstone:1961} there is a massless Goldstone boson for every broken generator of the symmetry. Therefore the indices of the Goldstone bosons run over the broken generators and we will also use greek letters from the beginning of the alphabet for gauge bosons corresponding to broken generators. 

We choose to collect physical scalars that are not connected to the mechanism of spontaneous symmetry breaking (like scalar partners of fermions in supersymmetric theories) also in $H_i$ without introducing a separate notation. It will be understood that the vacuum expectation value of these scalars vanishes and they don't mix with the Goldstone bosons.  

We don't consider the possibility of dynamical symmetry breaking (see
e.g. \cite{Weinberg:1995}) without fundamental Higgs fields. The description
of dynamical symmetry breaking in terms of an effective Lagrangian without
reference to the details of the underlying dynamics requires the introduction
of nonlinearly realized symmetries
\cite{Coleman:1969,Feruglio:1993,Dobado:1997jx}
\nocite{Alvarez-Gaume:1985} that leads to nonrenormalizable interactions. It would be interesting to extend the present discussion to that case but this is beyond the scope of this work. 

Since the split in Higgs and Goldstone bosons in \fref{eq:scalar-explicit} has to be sensible  we demand that the vacuum expectation value of $\phi$ is in the Higgs direction:
\begin{equation}\label{eq:vev}
\vev{\phi_A}\equiv\phi_{0A}=\begin{pmatrix} 0\\ v_i \end{pmatrix}
\end{equation}
The generators $T^a_{AB}$ (that are chosen as real antisymmetric matrices) satisfy the Lie algebra
\begin{equation}\label{eq:ssb-kommutator}
[T^a,T^b]=f^{abc}T^c
\end{equation}

We follow Llewellyn-Smith \cite{LlewellynSmith:1973}  and split the generators in sub-matrices:
\begin{equation}\label{eq:ssb-gen}
T^c_{AB}=\begin{pmatrix} t^c_{\alpha\beta}& u^c_{\alpha j} \\ -u^c_{i\beta} &
  T^c_{ij}\end{pmatrix}\equiv 
\begin{pmatrix}
  t^c & u^c \\ -(u^c)^T & T^c
\end{pmatrix}
\end{equation}
The component  form of \fref{eq:ssb-kommutator} is given by
\begin{subequations}\label{subeq:ssb-kommutator}
\begin{align}
  [t^a,t^b]-[u^a,(u^b)^T]&=f^{abc}t^c \label {eq:phiphi-kommutator}\\
  [t^a,u^b]+[u^a,T^b]&=f^{abc}u^c \label{eq:hphi-kommutator}\\
  \bigl(t^a_{\gamma\beta}u^b_{\delta k}+u^a_{\gamma j}T^b_{jk}-(t^b_{\gamma\delta}u^a_{\delta k}+u^b_{\gamma j}T^a_{jk}\bigr)&=f^{abc}u^c_{\gamma k}\bigr)\nonumber\\
   [T^a,T^b]-[(u^{a})^T,u^b]=f^{abc}T^c
\end{align}
\end{subequations}
so neither the $T^a_{ij}$ nor the $t^a_{\beta\gamma}$ form a subalgebra by
themselves. 
\subsection{Fermions}\label{sec:ssb-fermions}
As matter fields  we introduce chiral fermions $\psi_R$ and $\psi_L$ transforming under possibly  different representations of the symmetry group:
\begin{equation}\label{eq:fermion-gauge}
\begin{aligned}
  \delta \psi_{Li}&= \ii\omega_a \tau^a_{Lij}\psi_i\\
  \delta \psi_{Ri}&= \ii\omega_a \tau^\alpha_{Rij} \psi_i
\end{aligned}  
\end{equation}
with the commutation relations
\begin{equation}\label{eq:fermion-lie}
[\tau_{L/R}^a, \tau_{L/R}^b]_{ij}=\ii f^{abc}\tau_{L/Rij}^c
\end{equation}
Again the representations are in general reducible. If we consider only massive (Dirac) fermions, every left-handed fermion  must have a right-handed partner to generate a Dirac mass term. To obtain a consistent quantum field theory, anomaly free representations must be chosen but this will play no role in our discussion. 

As an alternative notation we sometimes use vector and axial-vector couplings defined by
\begin{equation} \label{eq:va-coupling}
g_V=\frac{1}{2}(\tau_L+\tau_R) \qquad
  g_A=\frac{1}{2}(\tau_R-\tau_L)
\end{equation}
\subsection{Majorana Fermions}
A Majorana spinor is defined to be equal to its charge conjugate (following the convention of \cite{Bailin:1994}):\nocite{Weinberg:2000}
\begin{equation}\label{eq:majorana}
\psi_M=\ii\gamma^2 \psi^*_M
\end{equation}
The Majorana condition \eqref{eq:majorana} must  commute with gauge transformations, i.e. 
\begin{equation}
(\Delta \psi_{Mi})^*\overset{!}{=} \ii \gamma^2 \Delta\psi_{Mi} 
\end{equation}
(note that $(\ii \gamma^2)^2=1$). From this condition  we get a constraint on the gauge transformations:
\begin{equation}
(\Delta \psi_{Mi})^* 
=(\ii\gamma^2)\ii(-g_{Vji}^a+g_{Aji}^a\gamma^5)\psi_{Mi}\overset{!}{=}(\ii\gamma^2)\ii(g_{Vij}^a+g_{Aij}^a\gamma^5)\psi_{Mi}
\end{equation}
where we have used  $ g_{V/A}^*=g_{V/A}^T$ and the reality of $\gamma^5$. 

Thus for Majorana fermions the vector coupling must be antisymmetric while the axial vector coupling must be symmetric.

Because of these symmetry properties in  the case of an abelian symmetry only an axial vector coupling is allowed while in for a nonabelian symmetry only a vector coupling is allowed\footnote{The representation matrices of a Lie algebra cannot be totally symmetric since this is inconsistent with the commutation relations.}.
\subsection{Unbroken Symmetries}\label{sec:unbroken}
The symmetry group $G$ might contain an unbroken subgroup $H$ that leaves the vacuum $\phi_0$ invariant and therefore their generators, denoted by $L$, annihilate $\phi_0$:
\begin{equation}\label{eq:annihilate}
L^a_{AB}\phi_{0B}=0
\end{equation} 
In our parametrization the unbroken  subgroup is generated by matrices 
\begin{equation}
L^a=
\begin{pmatrix}
 t^a & h^a \\ -(h^a)^T & T^a
\end{pmatrix}
\end{equation}
that satisfy 
besides \eqref{eq:ssb-condition} the condition 
\begin{equation} \label{eq:unbroken-condition}
  h^c_{\alpha i} v_i=0
\end{equation}
 Using the fact that the unbroken generators form a subalgebra and annihilate the vacuum  \eqref{eq:annihilate} we show  in \fref{app:massless}  that
\begin{equation}
h^a_{i\beta}=0
\end{equation}
so the representation of the unbroken subgroup reduces to 
\begin{equation}
L^a=
\begin{pmatrix}
 t^a & 0 \\ 0 & T^a
\end{pmatrix}
\end{equation}
which shows that Goldstone and Higgs bosons transform under ordinary representations of the unbroken subgroup. This fact is well known from the formalism of nonlinear representations of broken symmetries \cite{Coleman:1969}, that we review in \fref{app:nonlinear}. It turns out that these representation can be reduced further. In \fref{app:massless} it is derived that massless gauge bosons couple only to particles of the same mass. 
This is also clear on general grounds since---by definition---no internal symmetry can connect particles of different mass. 
To avoid a cumbersome notation distinguishing between massive and massless gauge bosons we extend the definitions of the couplings of the Goldstone bosons to unbroken indices:
\begin{equation}\label{eq:massless-definitions}
\begin{aligned}
t^{c}_{ab}&=
        \begin{cases} t^{c}_{\alpha \beta}, & m_a\wedge m_b \neq 0 \\
                0, &  m_a\vee m_b = 0
         \end{cases}\\
u^b_{a i}&=
        \begin{cases}
                u^{b}_{\alpha i},&  m_a\wedge m_b \neq 0 \\
                0, &  m_a\vee m_b = 0
        \end{cases}
\end{aligned}
\end{equation}
\section{Lagrangian}\label{sec:ssb-lag}
The Yang-Mills Lagrangian for the gauge fields is
\begin{equation}
\mathscr{L}_{YM}=-\frac{1}{4}F^{\mu\nu}_{a}F_{a\mu\nu} 
\end{equation}
with the field strength tensor 
\begin{equation}
F_{a\mu\nu}=\dmu A_{a\nu}-\partial_\nu A_{a\mu} +g_af^{abc}A_{b\nu}A_{c\nu}
\end{equation}
The Lagrangian for the scalar fields has the form 
\begin{equation}\label{eq:ssb-lagrangian}
 \mathscr{L}_{\phi}=\frac{1}{2}D_\mu\phi_AD^\mu\phi_A-V(\phi)
\end{equation}
with the covariant derivative of the scalars
\begin{equation}\label{eq:phi-der}
D_\mu\phi_A=\dmu\phi_A+ g_a T_{AB}^a W_{a\mu}\phi_B
\end{equation}
and a potential satisfying the gauge invariance condition 
\begin{equation}\label{eq:potential-gauge}
  \frac{\partial V(\phi)}{\partial \phi_A}T^a_{AB}\phi_B=0
\end{equation}
We  parametrize the scalar potential as 
\begin{equation}\label{eq:scalar-potential}
  V(\phi)=\frac{g_2^{AB}}{2}\phi^A\phi^B+\frac{g_3^{ABC}}{3!}\phi^A\phi^B\phi^C+\frac{g_4^{ABCD}}{4!}\phi^A\phi^B\phi^C\phi^D
\end{equation}
The constraints arising on the coefficients $g_2$ and $g_4$ from the symmetry condition \eqref{eq:potential-gauge} are summarized in \fref{sec:ssb-conditions}. 

The Lagrangian for the fermions is is parametrized as
\begin{equation}\label{eq:ferm-lag}
  \mathscr{L}_f=\ii\bar\psi_i\fmslash D\psi_i +\bar\psi_i \phi_A \bigl(X_{ij}^A(\tfrac{1-\gamma^5}{2})+{X_{
    ij}^A}^\dagger(\tfrac{1+\gamma^5}{2})\bigr)\psi_j
\end{equation}
with the covariant derivative of the fermions
\begin{equation}
 D_\mu\psi_i= \dmu\psi_i -\ii  g_a W_{a\mu}(\tau^a_{Lij}(\tfrac{1-\gamma^5}{2})+\tau_{Rij}^a(\tfrac{1+\gamma^5}{2}))\psi_j
\end{equation}
For the fermion-scalar coupling to be invariant under the gauge transformations \eqref{eq:scalar-gauge} and \eqref{eq:fermion-gauge}, it must satisfy the transformation law
\begin{equation}\label{eq:yukawa-tensor}
  -\ii \tau^{a}_{Rij}X_{jk}^A+iX_{ij}^A\tau^a_{Ljk}
  =X_{ik}^BT^a_{BA}
\end{equation}
that is it transforms as a mixed tensor with one index in the scalar representation, one index in the left- and and one index in the right-handed fermion representation. 

To do Feynman diagram calculations with this Lagrangian, the usual gauge fixing has to be performed. The gauge fixing and ghost terms in the notation used here are given in \fref{app:brs-ssb}. 

The Lagrangian introduced in this section can be compared with the general parametrization introduced in  \fref{app:general-fr} and we can identify the values the general parameters take in the case of spontanteous symmetry breaking . 

In the Yang-Mills Lagrangian \eqref{eq:ssb-lagrangian}  the quartic gauge boson interaction $g_{W^4}$ of the general Lagrangian \eqref{eq:ssb-lag} is given by\footnote{From now on we suppress the coupling constants in the relations among the coupling constants.} 
\begin{equation}\label{eq:gw4}
g_{W^4}^{abcd}=f^{abe}f^{cde}-f^{ade}f^{bce}
\end{equation}
The notation for the couplings of fermions to gauge bosons is unchanged while the Yukawa couplings are 
\begin{equation}
g_{H ij}^k=X^k_{ij}\qquad g_{\phi ij}^a=X^a_{ij}
\end{equation}
The coupling of scalars to gauge bosons is given by the square of the covariant derivative
\begin{align}\label{eq:scalar-derivative}
  \frac{1}{2}D_\mu\phi_A D^\mu\phi_A  
&= \frac{1}{2}\dmu \phi_A\partial^\mu \phi_A\\
 &+\frac{1}{2}
  T_{AB}^aW^\mu_a \phi_B\overleftrightarrow\dmu  \phi_A - \frac{1}{4}\phi_A\{ T^a,T^b \}_{AB}\phi_BW_{a\mu}W^\mu_b
\end{align}
Inserting the parametrization \eqref{eq:ssb-gen} we find that  $T_{ij}^a$ and $t_{ab}^c$ agree with the Lagrangian
\eqref{eq:ssb-lag}
The 2 scalar-2 gauge boson coupling is given by the anticommutator of representation matrices:
\begin{equation}\label{eq:2s2w}
  g_{\phi^2 W^2}^{ABcd}=-\{T^c,T^a\}_{AB}
\end{equation}
Using \fref{eq:hphiw-coupling}, the 2 Higgs 2 gauge boson coupling is given by:
\begin{equation}\label{eq:2h2w}
\{T^a_{ik},T^b_{kj}\}-\{\tfrac{g_{HWW}^{a}}{2m_{W_c}},\tfrac{g_{HWW}^{b}}{2m_{W_c}}
\}_{ij}=-g_{H^2 W^2}^{abij}
\end{equation} 
The cubic scalar gauge boson  couplings originate from the contraction of the anticommutator $\{T^\alpha,T^b\}$ with a vacuum expectation value $\phi_0$:
\begin{equation}\label{eq:phiww-def}
 g_{\phi WW}^{Abc}=g_{\phi^2 W^2}^{ABbc}\phi_{0B}
\end{equation}
In the notation introduced in \fref{eq:massless-definitions} they are given by:
\begin{subequations}\label{subeq:scalar-gauge} 
\begin{align}
 g_{H\phi W}^{iab}&=-  u^b_{ai}\label{eq:ghphiw}\\ 
 g_{\phi WW}^{abc}&=(m_b t^c_{ba}+m_c t^b_{ca})\label{eq:gphiww}\\
  g_{HWW}^{iab}&=(m_a u^b_{a i}+m_b u^a_{b i})\label{eq:ghww}
\end{align}
\end{subequations}
The triple scalar couplings can be expressed through the terms in the scalar potential by
\begin{equation}\label{eq:cubic-higgs}
 g_{\Phi^3}^{ABC}=- g_{3}^{ABC}-g_4^{iABC}v_i
\end{equation}

\section{Symmetry conditions and implications of SSB}\label{sec:ssb-conditions}
The Lagrangian written down in \fref{sec:ssb-lag} is subject to the  constraints from gauge invariance given in \fref{eq:potential-gauge} and \eqref{eq:yukawa-tensor} that we have not exploited yet. 

Furthermore we must ensure that the scalar potential induces spontanteous symmetry breaking  and our  split in Higgs and Goldstone bosons is sensible. It will turn out that the implementation of these conditions allow to express the couplings of the Goldstone bosons by couplings of the physical particles. 

In order that the vev \eqref{eq:vev} is the true ground state of the theory, it must be the minimum of the scalar potential. This condition is analyzed in \fref{app:ssb-relations} and leads to \fref{eq:higgs-mass-sr-app}:
\begin{equation}\label{eq:higgs-mass-sr}
m_{H_i}^2v_i=-\frac{1}{2}g_{H^3}^{ijk}v_jv_k+\frac{1}{6}g_{H^4}^{ijkl}v_jv_kv_l
\end{equation}
This are the so called `Higgs mass sum rules' \cite{Langacker:1984}. These relations are not connected to the gauge invariance of the scalar potential and therefore cannot be verified by the Ward Identities.  
  
Since  our parametrization of the scalar fields should  identify the unphysical Goldstone bosons correctly,  the gauge transformation of the vev must be in the Goldstone boson direction so we demand
\begin{subequations}\label{subeq:ssb-conditions}
\begin{equation}\label{eq:ssb-condition}
T^c_{ij}v_j=0
\end{equation}
We assume that the mass matrix of the gauge bosons obtained by inserting the parametrization of the scalar field \eqref{eq:scalar-explicit} into the Lagrangian \eqref{eq:ssb-lagrangian} has been diagonalized:
\begin{align}
(u^a_{\gamma i}v_i)(u^b_{\gamma j}v_j)=m_a^2\delta_{a,b}
\end{align}
The case  $(u^c_{\alpha i}v_i)=0$ corresponds to the unbroken subgroup $H$  that we have discussed in \fref{sec:unbroken}. Since the  Goldstone bosons correspond to the broken generators their number is equal to the number of massive gauge bosons. 
Therefore the projection  of  the matrix  $(u^b_{\alpha i}v_i)$ onto the subspace of massive gauge bosons 
 $u^\beta_{\alpha i}v_i$ is a \emph{square} matrix that can be chosen diagonal if the mass matrix is diagonalized:
 \begin{align}\label{eq:diag_mass}
u^\beta_{\gamma j}v_j=m_\beta\delta_{\beta,\gamma}
\end{align}
We demand also that the fermions get their masses from the coupling to the scalars:
\begin{equation*}
\bar\psi_i \phi_{0A} \bigl(X_{ij}^A(\tfrac{1-\gamma^5}{2})+{X_{
    ij}^A}^\dagger(\tfrac{1+\gamma^5}{2})\bigr)\psi_j=-m_i\bar\psi_i \psi_i
\end{equation*}
This implies\footnote{For vectorlike symmetries with $\tau_R=\tau_L$ also an explicit mass term for the fermions
\begin{equation*}
\mathscr{L}_{f,m}=-\tilde m_i\bar\psi_i\psi
\end{equation*}
is allowed. In this case, \fref{eq:yukawa-ssb-condition} gets modified to 
\begin{equation*}
  X_{ij}^A\phi_{0A}= {X_{ij}^A}^\dagger\phi_{0A} =-\delta_{ij}(m_i-\tilde m_i)
\end{equation*}
so that in any case $m_i$ is the physical mass of the fermion.}
\begin{equation}\label{eq:yukawa-ssb-condition}
  X_{ij}^A\phi_{0A} = {X_{ij}^A}^\dagger\phi_{0A} =-\delta_{ij}m_i
\end{equation}
\end{subequations}
Together with the Lie algebra of the representation matrices and the invariance of the Yukawa couplings and the scalar potential, this summarizes our symmetry conditions that we will now use to eliminate the Goldstone boson couplings.

Let us first turn to the scalar gauge boson sector. Consistency relations on the generators \eqref{eq:ssb-gen} can be obtained by acting on the vacuum expectation value $\phi_0$  with a  commutator of two generators in the representation of the scalars:
\begin{equation*}
 \left([T^a,T^b]\right)\phi_0=f^{abc}\begin{pmatrix} m_c \\ 0\end{pmatrix}
\end{equation*}
The resulting equations are used in \fref{app:ssb-relations} to express the couplings of the Goldstone bosons to the gauge bosons by structure constants and gauge boson masses:
\begin{subequations}\label{subeq:triple-phi-couplings}
\begin{align}
 t^b_{ac}&=\frac{1}{2 m_a m_c}f^{bac}(m_b^2-m_a^2-m_c^2)\label{eq:2phi-w-coupling}\\
 g_{\phi WW}^{abc}&=(m_b t^c_{ba}+m_c t^b_{ca}) =\frac{1}{m_a}f^{abc}(m_b^2-m_c^2)\label{eq:phi-2w-coupling}
\end{align}
One also obtains the symmetry relation
\begin{equation*}
m_\alpha u^\beta_{i\alpha}=m_\beta u^\alpha_{i\beta}
\end{equation*}
that allows to simplify \fref{eq:ghww} to
\begin{equation}\label{eq:hphiw-coupling}
 m_a  g_{H\phi W}^{iab}\equiv - m_a u^b_{ia}=-\frac{1}{2}g_{HWW}^{iab}
\end{equation}
Similarly one can derive the fermion-Goldstone boson  coupling by contracting the transformation law of the Yukawa couplings \eqref{eq:yukawa-tensor} with the vacuum expectation value  $\phi_{0A}$ and using the condition \eqref{eq:yukawa-ssb-condition}
 \begin{equation}\label{eq:f-phi-coupling}
  m_a g_{\phi Lij}^a \equiv m_a X^a_{ij}=\ii m_j \tau^{a}_{Rij}-\ii m_i\tau^{a}_{Lij}
 \end{equation}
Note that the coupling of  vectorlike fermions with $\tau_L=\tau_R$ to Goldstone bosons vanishes. 
The relations obtained up to now reproduce the results obtained in \cite{LlewellynSmith:1973} from tree level unitarity. 
 
The coupling of the Goldstone bosons to scalar particles can be obtained from the invariance of the scalar potential \eqref{eq:potential-gauge} (see \fref{subeq:triple-scalar}):
\begin{align}
  m_{W_a}g_{\phi H^2}^{a ij}& =T^a_{ij}(m_i^2-m_j^2)\label{eq:phi2h-coupling}\\
m_{W_a} m_{W_b} g_{\phi^2 H}^{ab i}&=-\frac{m_i^2}{2}g_{HWW}^{iab}\label{eq:2phih-coupling}\\
    g_{\phi^3}^{abc}&\equiv 0
 \end{align}
\end{subequations}
Also, all quartic scalar couplings except $g_{H^4}$ can be expressed in terms of the  cubic couplings 
\begin{equation}\label{eq:quartic-scalar}
 m_{W_\alpha}g^{ABC\alpha}_{\Phi^4}+g^{DBC}_{\Phi^3}T^\alpha_{DA}+g^{ADC}_{\Phi^3}T^a_{DB}
+g^{ABD}_{\Phi^3}T^a_{DC}=0
\end{equation}
(see \eqref{eq:gold-higgs-4} for the explicit expressions  for the Higgs and Goldstone boson couplings.) 
Again we can eliminate the cubic Goldstone boson couplings from this equation and express everything in terms of the input parameters. 

Furthermore, the identity \footnote {Formally this is a `Super Jacobi Identity' \cite{Weinberg:2000}}
\begin{equation}\label{eq:super-jacobi}
 0= \lbrack A,\{B,C\} \rbrack +\{\lbrack C, A \rbrack,B\} -\{ \lbrack A,B\rbrack C\}
\end{equation}
can be used together with \fref{eq:2s2w} to obtain the relations (see \eqref{eq:2s-jac})
\begin{subequations} \label{eq:2s-jac_komp}
\begin{align}
m_{W_a}g_{\phi^2W^2}^{abcd}&=f^{eca}g_{\phi
  WW}^{bed}+f^{eda}g_{\phi WW}^{bce}-t^a_{eb}g_{\phi WW}^{ecd}-g_{H\phi
  W}^{iba} g_{HWW}^{icd} \label{eq:2phi-jac} \\
m_{W_a}g^{abci}_{H\phi
  W^2}&=f^{eba}g_{HWW}^{iec}+f^{eca}g^{ieb}_{HWW}+g^{iae}_{H\phi W}g^{ebc}_{\phi
  WW} -T_{ji}^ag^{jbc}_{HWW}\label{eq:hphi-jac}
\end{align}
\end{subequations}
\section{Input parameters and dependent parameters}\label{sec:ssb-summary}
We can summarize the  results from \fref{sec:ssb-conditions} by dividing the parameters of the general Lagrangian from \fref{app:general-fr} into input parameters and dependent parameters. This is not a \emph{minimal} set of input parameters since we don't attempt to solve the relations among the input parameters themselves imposed by the Lie algebra structure, the Jacobi Identities and other symmetry relations. Such a solution doesn't seem very useful in the general setting of our setup and it is well known that in practice there are several equivalent schemes that can be used according to convenience. We will come back to this issue in the electroweak Standard Model in \fref{chap:running}. 

The input parameters can be put in three groups. The structure constants and the fermion-gauge couplings are determined by the gauge group and the fermion representations alone:
\begin{subequations}\label{eq:ssb-input}
\begin{equation}
\begin{aligned}
&f^{abc}\\
&\tau_{L/R ij}^a
\end{aligned}
\end{equation}
These structure constants must satisfy the Jacobi Identity \eqref{eq:jacobi} and the representation matrices the Lie algebra \eqref{eq:fermion-gauge}

The second group consists of the couplings of the Higgs bosons to gauge bosons and fermions. In the notation of \fref{app:general-fr} these are given by    
\begin{equation}
\begin{aligned}
&g_{Hij}^h\\
&g_{HWW}^{iab}\\
&T^a_{ij}
\end{aligned}
\end{equation}
It seems strange that the Higgs-Gauge boson couplings are regarded as input parameters since in the Standard Model they are determined by the gauge boson masses. In more general Higgs models one can give, however, only consistency relations on these couplings. The matrices $g_{HWW}$ and $T$ are connected by the relation \eqref{eq:ssb-kommutator} while the couplings to the fermions must satisfy \eqref{eq:yukawa-tensor}. 

The Higgs self coupling is parametrized by the two matrices 
\begin{equation}g_{H^3}^{ijk} \quad,\quad g_{H^4}^{ijkl}
\end{equation}
\end{subequations}
 They are connected to the Higgs masses by the sum rules \eqref{eq:higgs-mass-sr}. However, these sum rules are \emph{not} connected to gauge invariance since they follow from the minimization of the scalar potential. Thus they \emph{cannot} be checked by the Ward Identities. 

All other couplings can be expressed by these input parameters. In \fref{sec:ssb-conditions} we have seen that all Goldstone boson couplings can be expressed by these parameters via \fref{subeq:triple-phi-couplings} and \eqref{eq:2s-jac_komp}. Furthermore, all quartic couplings of physical particles except the quartic Higgs selfcoupling are dependent parameters. 

To summarize, we have shown how all parameters in a spontaneously broken gauge theory can expressed through the set of input-parameters \eqref{eq:ssb-input}, using the relations \eqref{subeq:triple-phi-couplings}, \eqref{eq:2s-jac_komp} and \eqref{eq:quartic-scalar}. In  \fref{chap:reconstruction} we show how these relations can also be obtained by imposing the Ward-identities on the scattering amplitudes calculated using the general Lagrangian \eqref{eq:ssb-lag} without using the constraints from gauge invariance and spontaneous symmetry breaking. 

\section{Example: Standard model}\label{sec:sm}
As a first example to illustrate our notation and for later applications, we briefly review the electroweak Standard Model 

We first discuss  the gauge sector of the Standard Model that consists---before symmetry breaking---of $SU(2)$ gauge bosons $W^a_\mu$ and a $U(1)$ gauge boson $B_\mu$. After spontanteous symmetry breaking , the mass eigenstates are given by\begin{align}
W_\mu^\pm&=\frac{1}{\sqrt 2}(W^1_\mu\mp iW^2_\mu)\\
Z_\mu&=\cos\theta_w W^3_\mu-\sin\theta_w B_\mu\\
A_\mu&=\sin\theta_w W^3_\mu+\cos\theta_w B_\mu
\end{align}
On tree level,the `Weinberg angle'  $\theta_w$ is given by 
\begin{equation}\label{eq:cw-on-shell}
\cos\theta_w=\frac{m_W}{m_Z}
\end{equation}
The electromagnetic coupling is given in terms of the Weinberg angle and the coupling constants $g$ and $g'$ by 
\begin{equation}
  e=\sin\theta_w g=\cos\theta_w g' 
\end{equation}
and the electromagnetic charge can be expressed by the Gell-Mann Nishijima relation 
\begin{equation}
  Q=\frac{Y}{2}+T^3
\end{equation}
in terms of the third component of the weak isospin $T^3$ and the hypercharge $Y$ that determines the coupling to the $U(1)$ gauge boson $B_\mu$. 
The triple gauge boson couplings follow from the term
\begin{multline}
\mathscr{L}_{W^3}=- g\epsilon^{abc}W^\mu_b W^\nu_c\dmu W_{\nu a}\\
=\ii g W^+_{\mu} W^-_{\nu} (\cos\theta_w\dmu Z_\nu +\sin\theta_w \dmu A_\nu)+\text{permutations}
\end{multline}
The covariant derivative acting on the fermions is given by
\begin{multline}\label{eq:sm-covariant}
D_\mu =\Bigl(\dmu -\ii e Q A_\mu-\ii \tfrac{g}{\sqrt 2} (W_\mu^+ T^++W_\mu^- T^-) \tfrac{(1-\gamma^5)}{2}\\
-\ii \tfrac{g}{\cos\theta_w}Z\left[T^3\tfrac{(1-\gamma^5)}{2}-Q\sin^2\theta_w\right]\Bigr)
\end{multline}

We now turn to the scalar sector that is normally parametrized as a complex scalar $SU(2)$ doublet field
\begin{equation}\label{eq:complex_higgs}  
\begin{aligned}
\phi=\frac{1}{\sqrt 2}\left((v+H)-
\ii\vec\phi\cdot\vec\sigma\right)\begin{pmatrix}0\\ 1\end{pmatrix}
&=\frac{1}{\sqrt
  2}\begin{pmatrix} -\ii(\phi_1-\ii\phi_2) \\ \left((v+H)+\ii\phi^0)\right)\end{pmatrix}\\
 \delta \phi &=\ii \omega_a \frac{\sigma^a}{2}\phi
\end{aligned}
\end{equation}
However, it is well known \cite{Peskin:1995} that one can rewrite this as a vector of real fields
\begin{equation}
\phi=\begin{pmatrix}\phi_1\\ \phi_2\\ \phi_3 \\ v+H\end{pmatrix} \quad \Delta \phi =- \omega_a T^a\phi
\end{equation}
where the \emph{real} representation matrices $T^a$ can be obtained by applying the matrices $-\ii \frac{\sigma}{2}$ to the complex representation of $\phi$ \eqref{eq:complex_higgs} and identifying the transformation laws of the real components. 

Finally we write down the quark sector of the electroweak Standard Model in our notation. The left-handed quarks transform under a irreducible spinor representation of $SU(2)$ while we assemble the right handed quarks in a direct sum of trivial representations of $SU(2)$:
\begin{equation}
Q_L=\begin{pmatrix}u_L\\d_L \end{pmatrix}\qquad \delta Q_L=\ii \omega_a\frac{\sigma^a}{2}Q_L \qquad  Q_R=\begin{pmatrix}u_R\\d_R \end{pmatrix}\qquad \delta Q_R=0
\end{equation}
If we ignore quark mixing the interaction between scalars and quarks can then be written in matrix form
\begin{equation}
\begin{aligned}
-\mathscr{L}_{H q\bar q}&=\lambda _d \bar Q_{L}\cdot\phi
\,d_R+\lambda_u\epsilon^{ab} \bar Q_{La}\phi_b^\dagger u_R +h.c.\\
&=\bar Q_L\begin{pmatrix}\lambda_d(\tfrac{1}{\sqrt 2}(v+H)-\ii\phi_0)& -\ii\lambda_d \phi^+\\
                        -\ii\lambda_u\phi^- & \lambda_u\frac{1}{\sqrt 2}((v+H)+\ii\phi_0) 
        \end{pmatrix} Q_R+\text{ h.c.}\\
&=\bar Q_L \left[\frac{1}{\sqrt 2}\lambda_d ((v+H)-\ii \phi_0 \sigma^3)-\ii \lambda_u (\phi^+\sigma^++\phi^-\sigma^-)\right]Q_R+\text{ h.c.}
\end{aligned}
\end{equation}
In this example the gauge invariance is of course obvious in the original parametrization in the first line so we don't perform the rather tedious exercise of checking the transformation law \eqref{eq:yukawa-tensor} in the notation of the last line.
\section{Example: SUSY Yang-Mills}
As a second example that serves to demonstrate that the above setup also includes supersymmetric field theories, we transcribe the supersymmetric Yang-Mills theory \cite{Weinberg:2000,Bailin:1994} to our notation. In the conventions of \cite{ReuterOhl:2002,Reuter:2002} the SYM Lagrangian (after eliminating auxiliary fields) reads\footnote{This is a simplified version where the fermion field $\Psi$ is a Majorana fermion. For Dirac fermions one must include another sfermion scalar field in the Lagrangian. }
\begin{multline}
 \mathscr{L}_{SYM}=-\frac{1}{4}F^{\mu\nu}_{a}F_{a\mu\nu} + \frac{\ii}{2} \bar\lambda_a(\fmslash D\lambda)_a+i\bar\Psi(\fmslash D-m)\Psi+(D_\mu\phi)^\dagger D^\mu\phi-m^2\phi^\dagger \phi\\
 -y^{\alpha\beta\gamma}\phi_\alpha\bar \Psi_\beta(\tfrac{1-\gamma^5}{2})\Psi_\gamma-{y^*}^{\alpha\beta\gamma}\phi_\alpha\bar \Psi_\beta(\tfrac{1+\gamma^5}{2})\Psi_\gamma\\
-\sqrt{2}g \left[ \bar \lambda_a\phi^\dagger T^a(\tfrac{1-\gamma^5}{2})\Psi+\bar \Psi T^a\phi(\tfrac{1+\gamma^5}{2})\lambda^a\right]-\frac{g}{2}\sum_a|(\phi^\dagger T^a\phi)|^2+V(\phi)
\end{multline}
Here $\lambda$ is the gaugino field, a Majorana fermion transforming in the adjoint representation of the gauge group;  $\phi$ the sfermion field, a complex scalar in the fundamental representation. The field strength tensor and the covariant derivative are defined as usual.  The potential $V$ is determined by the matrices $y^{\alpha\beta\gamma}$ and the mass matrices of the fermions but we won't need the details here. It can be shown that the gauge symmetry might be broken by a vev of the sfermions but the supersymmetry remains unbroken \cite{Weinberg:2000}
 
We see that this is indeed a special case of \eqref{eq:ssb-lagrangian} and \eqref{eq:ferm-lag} with the identifications
\begin{equation}
\begin{aligned}
\psi_i&=\begin{pmatrix}\Psi_\alpha\\\lambda_a\end{pmatrix}\\
\tau^a_{Lij}&=\tau^a_{Rij}=\begin{pmatrix}T^a_{\beta\gamma} &0 \\ 0& (-\ii) f^{abc}\end{pmatrix} \\
T^a_{AB}&=\ii T^a_{\beta\gamma}\\
X^A_{ij}&=\begin{pmatrix}y^{\alpha\beta\gamma} & 0 \\ - {\scriptstyle \sqrt 2g}\, T^b_{\alpha\gamma}&0 \end{pmatrix}\\
\end{aligned}
\end{equation}
It should be noted that gauge invariance alone forces the gaugino-fermion-sfermion coupling to be proportional to the representation matrix $T$. Indeed, after some index shuffling, the transformation law of the Yukawa couplings \fref{eq:yukawa-tensor} turns into the Lie algebra of the representation matrices
\begin{equation}
[T^a,T^b]=\ii f^{abc}T^c
\end{equation}
and the transformation law of the Yukawa coupling:
\begin{equation}
- T^a_{\beta\gamma}y^{\alpha\gamma\lambda}+ y^{\alpha\beta\gamma}T^a_{\gamma\lambda}=T^a_{\alpha\delta}y^{\delta\beta\lambda}
\end{equation}

The \emph{coefficient} of the $\bar\lambda \phi \Psi$ term can, however, only be determined by supersymmetry. Therefore we will be able to apply our method of verifying the Feynman rules also to supersymmetric theories, but of course we cannot hope to check the supersymmetry of the Lagrangian. To do this the BRS transformations and therefore the STIs have to be generalized to include SUSY transformations \cite{Maggiore:1996,ReuterOhl:2002,Reuter:2002}. 

\chapter{Reconstruction of the Feynman rules from the Ward-Identities}\label{chap:reconstruction}\chaptermark{Reconstruction of the Feynman rules}

After we have clarified the structure of a spontaneously broken gauge theory in \fref{chap:ssb-lag} we  are now in a position to demonstrate---using the tools of \fref{part:tools}---that the Ward-Identities \eqref{eq:chanowitz} for 3 and 4 particle tree-level matrix elements with up to 4 contractions are sufficient to reconstruct the Feynman rules of a spontaneously broken gauge theory (apart from the quartic Higgs coupling).

  From the results of \fref{sec:skeleton} it is evident that the \emph{complete set} of Ward Identities for 3 and 4 point functions is equivalent to the complete set of STIs that in turn is known \cite{Piguet:1981} to determine the Lagrangian. The nontrivial fact that we establish in this chapter is that the \emph{limited} set of Ward Identities without external Goldstone bosons is sufficient to reconstruct the Feynman rules. The calculations are done in the language of Ward Identities for scattering amplitudes that appears to be more intuitive, the use of the STIs for irreducible vertices yields the same results and can provide important cross-checks.

As a first step, we establish in \fref{sec:reconstruct_3point} that Ward Identities for the 3 point matrix elements allow to express the triple Goldstone boson couplings in terms of the input parameters in the same way as in \fref{sec:ssb-summary}. We then consider the Ward Identities for 4 point scattering amplitudes with one contraction that yield consistency relations among the coupling constants of the physical particles like the Lie algebra structure, the invariance of the Higgs-Yukawa coupling or the expression for the $WWHH$ coupling. Finally we show that the conditions from  the Ward Identities with more than one contraction determine the quartic couplings involving Goldstone bosons and give consistency relations on the triple Goldstone boson couplings. In this process at every stage the results of the previous steps have to be used. 
We will give some examples for the calculations and list all the resulting conditions while the details of the complete calculations can be found in \fref{app:wi}.

\section{Cubic Goldstone boson couplings}\label{sec:reconstruct_3point}
In \fref{sec:ssb-summary} we have derived \fref{subeq:triple-phi-couplings}  that express the cubic Goldstone boson couplings by the input parameters. While this derivation used only the Yang Mills form of the Lagrangian and basic properties of spontaneous symmetry breaking, we will now re-derive this relations from the Ward Identities alone.
\subsection{Couplings of one Goldstone boson}\label{sec:3point-1phi}
The couplings of one Goldstone boson to two physical particles are  determined by the Ward Identity with one unphysical gauge boson. As an example we consider the Ward Identity for the $WWH$ vertex. 
\begin{equation*}
\parbox{15mm}{
\begin{fmfchar*}(15,15)
  \fmfleft{A1,A2} \fmfright{A3} \fmf{double}{A1,a} \fmf{dashes}{A2,a}
   \fmf{photon}{A3,a} \fmfblob{10}{a}
\end{fmfchar*}}=0 
\end{equation*}
or explicitly 
\begin{equation}\label{eq:2wh-wi}
-\ii  p_a^{\mu_a}\epsilon_b^{\mu_b}\me_{\mu_a\mu_b}(W_a 
 W_bH_i)=m_{W_a}
 \epsilon_b^{\mu_b}\me_{\mu_b}(\phi_a  W_b H_i  )
\end{equation}
Inserting the Feynman rules \eqref{eq:fr-hww} and \eqref {eq:fr-hphiw} from \fref{app:feynman} gives
\begin{equation}
g_{HWW}^{iab} (p_a\cdot\epsilon_b)=m_{W_a}g_{H\phi W}^{iab}\epsilon_b\cdot (p_i-p_a)
\end{equation}
so  for on-shell gauge bosons with polarization vectors satisfying $\epsilon_b\cdot p_b=0$ and using momentum conservation we recover \fref{eq:hphiw-coupling}:
\begin{equation}\label{eq:ghphiw-wi}
g_{H\phi W}^{iab}=-\frac{1}{2m_{W_a}}g_{HWW}^{iab}
\end{equation}
As an important remark, we note that  as long as we contract the $HWW$ vertex with a polarization vector orthogonal to $p_b$, the vertex will satisfy the \emph{WI}
\begin{equation}\label{eq:hww-sti-contr}
\ii p_{a \mu} \epsilon_{b\nu}\onepi{ W_a^{\mu}(p_a)
  W_b^{\nu}(p_b)H_i(p_i)}+m_{W_a}\epsilon_{b\nu}\onepi{\phi_a(p)W_b^{\nu}(p_b)H_i(p_i)}=0
\end{equation}
even if the Higgs is off-shell. This will prove very useful in explicit calculations. This property follows also from the STI for the $HWW$-vertex \eqref{eq:hww-sti}. 
\begin{subequations}\label{subeq:triple-phi-wi1}
Similarly the other relations in \eqref{subeq:triple-phi-couplings} for the couplings of one Goldstone boson can be derived (see \fref{app:3wi})   :
\begin{align}
\parbox{15mm}{
\begin{fmfchar*}(15,15)
  \fmfleft{A1,A2} \fmfright{A3} \fmf{double}{A1,a} \fmf{fermion}{A2,a}
   \fmf{fermion}{a,A3} \fmfblob{10}{a}
\end{fmfchar*}}=0 &\qquad \Rightarrow 
  g_{\phi ij}^a=-\frac{i}{m_{W_a}}(m_{f_i}\tau_{L ij}^a-m_{f_j}\tau_{R ij}^a)\\
\parbox{15mm}{
\begin{fmfchar*}(15,15)
  \fmfleft{A1,A2} \fmfright{A3} \fmf{double}{A1,a} \fmf{dashes}{A2,a}
   \fmf{dashes}{a,A3} \fmfblob{10}{a}
\end{fmfchar*}}=0 &\qquad \Rightarrow g_{\phi H^2}^{aij}=\frac{1}{m_a}T^a_{ij}(m_i^2-m_j^2) \\
\parbox{15mm}{
\begin{fmfchar*}(15,15)
  \fmfleft{A1,A2} \fmfright{A3} \fmf{double}{A1,a} \fmf{photon}{A2,a}
   \fmf{photon}{a,A3} \fmfblob{10}{a}
\end{fmfchar*}}=0 &\qquad \Rightarrow  g_{\phi WW}^{abc}=\frac{1}{m_a}f^{abc}(m_b^2-m_c^2)
\end{align}
\end{subequations}
An important further result is that the Ward Identity for the three gauge boson vertex implies not only the Goldstone boson-gauge boson coupling but also the total antisymmetry of the coupling constants $f^{abc}$ (see the discussion after \fref{eq:phi-2w-app} in \fref{app:3w-wi}.) 
\subsection{Couplings of 2 or 3  Goldstone bosons}\label{sec:3point-2phi}
To get the relations for the couplings of 2 Goldstone bosons to one physical particle, one has to consider the Ward Identity \eqref{eq:chanowitz} with two unphysical gauge bosons. Using again the $HWW$ amplitude as an example,  the Ward Identity 
\begin{equation*}
\parbox{15mm}{
\begin{fmfchar*}(15,15)
  \fmfleft{A1,A2} \fmfright{A3} \fmf{double}{A1,a} \fmf{dashes}{A2,a}
   \fmf{double}{A3,a} \fmfblob{10}{a}
\end{fmfchar*}}=0 
\end{equation*}
reads  
\begin{multline}\label{eq:wwh-2pow}
p_a^\mu p_b^\nu\me_{\mu\nu}(H_i  W_aW_b)=m_{W_a}m_{W_b}\me(H_i 
\phi_a\phi_b)\\
+im_{W_a}p_b^\mu\me_\mu(H_i 
\phi_a W_b)+\ii m_{W_b}p_a^\mu\me_\mu(H_i 
 W_a\phi_b )
\end{multline}
Inserting the Feynman rules from \fref{app:feynman} gives
\begin{multline}
\ii g_{HWW}^{iab}(p_a\cdot p_b)=\ii m_{W_a}m_{W_b}g^{abi}_{\phi^2H}\\
+\ii m_{W_a}g^{iab}_{H\phi W}(p_b\cdot(p_i-p_a))+\ii m_{W_b}g^{iba}_{H\phi W}(p_a\cdot(p_i-p_b))
\end{multline}
Using the expression \eqref{eq:hphiw-coupling} for $g_{H\phi W}$ (that we have reproduced from the Ward Identity in \fref{eq:ghphiw-wi})  and the symmetry of $g_{HWW}$ we find that this Ward Identity is equivalent to \fref{eq:2phih-coupling}:
\begin{subequations}\label{subeq:triple-phi-wi2}
\begin{equation}\label{eq:gphihh}
g^{abi}_{\phi^2H}=\frac{1}{2m_{W_a}m_{W_b}}g_{HWW}^{iab}(p_b+p_b)\cdot
p_i=-\frac{m_{H_i}^2}{2m_{W_a}m_{W_b}}g_{HWW}^{iab}
\end{equation}
Similarly the remaining relations in \fref{subeq:triple-phi-couplings} can be derived (see \fref{app:3wi}): 
\begin{align}
\parbox{15mm}{
\begin{fmfchar*}(15,15)
  \fmfleft{A1,A2} \fmfright{A3} \fmf{double}{A1,a} \fmf{photon}{A2,a}
   \fmf{double}{A3,a} \fmfblob{10}{a}
\end{fmfchar*}}=0 &\qquad \Rightarrow  m_a m_c t^b_{a c}=\frac{1}{2}f^{b ac}(m_b^2-m_a^2-m_c^2)\\
\parbox{15mm}{
\begin{fmfchar*}(15,15)
  \fmfleft{A1,A2} \fmfright{A3} \fmf{double}{A1,a} \fmf{double}{A2,a}
   \fmf{double}{A3,a} \fmfblob{10}{a}
\end{fmfchar*}}=0 &\qquad \Rightarrow
 g_{\phi^3}^{abc}=0
\end{align}
\end{subequations}
Therefore we have succeeded in determining the cubic Goldstone boson couplings from the Ward Identities for 3-particle matrix elements. As an important additional result, it turns out, that the STIs for the three point vertices are satisfied \emph{automatically} if the corresponding Ward Identities are satisfied. We check this explicitly in  \fref{app:3wi}. This property will be very useful in the calculations of the 4 point STIs since we can use the connection between the Ward Identities and the STIs discussed in \fref{sec:skeleton}. 
\section{Gauge invariance of physical couplings}
After we have completed the discussion of the cubic Goldstone boson couplings and the 3 point STIs we now consider the Lie algebra structure of the couplings of the physical particles , i.e. the Higgs bosons, gauge bosons and fermions. 

The calculations are done in unitarity gauge, keeping only the external Goldstone bosons appearing in combination with the unphysical gauge bosons, after the prescription given in \fref{app:uni-sti}. We don't loose information if we work in unitarity gauge since, according to section \ref{sec:gpi}, the 4 particle scattering amplitudes in $R_\xi$ gauge agree automatically with those in unitarity gauge if all 3-particle matrix elements satisfy the Ward Identities . Therefore the internal Goldstone bosons appearing in $R_\xi$ gauge drop out of the calculations `trivially' and don't lead to additional conditions. 

\subsection{Example: $WWHH$ Ward identity}\label{sec:hhww-wi}
As a first example, we will evaluate the  Ward Identity for the $HHWW$ amplitude
\begin{equation}
\me(\mathcal{D}_a W_bH_iH_j)\quad=\parbox{20mm}{
\begin{fmfchar*}(20,20)
  \fmfleft{A1,A2} \fmfright{A3,A4} 
  \fmf{double}{A2,a} 
  \fmf{dashes,label=$H_i$,label.side=right}{A3,a}
  \fmf{photon,label=$W_b$}{A1,a} 
  \fmf{dashes,label=$H_j$,label.side=left}{A4,a}
  \fmfv{decor.shape=circle,decor.filled=empty,decor.size=25}{a}
\end{fmfchar*}}\quad=0
\end{equation}
and show that it reproduces the
 definition of the quartic scalar- gauge boson coupling \eqref{eq:2s2w} and
 the Higgs components of the Lie algebra
\eqref{eq:ssb-kommutator}. 

We first consider the Higgs exchange diagrams. Using the STI \eqref{eq:hww-sti}
\begin{equation}\label{eq:hww-sti-main}
- \onepi{ \mathcal{D}_a(p_a) W_b^\mu(p_b)H_i(p_i)}=\frac{1}{2} g^{iab}_{HW W}p_b^\mu  
\end{equation}
 we see that the $s$-channel diagram satisfies the Ward Identity by itself:
\begin{equation}\label{eq:hhww-higgs}
\parbox{20mm}{
\begin{fmfchar*}(20,15)
  \fmftop{A1,A2} \fmfbottom{A3,A4} \fmf{double}{A1,a} \fmf{dashes}{A2,b}
  \fmf{dashes}{a,b} \fmf{photon}{A3,a} \fmf{dashes}{A4,b}
 \fmfdot{a}
  \fmfdot{b} 
\end{fmfchar*}}=0
\end{equation}
Using the  STI for the  $HHW$ vertex \eqref{eq:hhw-sti}
\begin{equation}
- \onepi{\mathcal{D}_a(p_a)H_i(p_i)H_j(p_j)} =T^a_{ij}\left[D_{H_i}(p_i)^{-1}-D_{H_j}(p_j)^{-1}\right]
 \end{equation}
 gives for the  $t$ and $u$-channel diagrams
\begin{multline}\label{eq:hhaa-tu}
\parbox{15mm}{
\begin{fmfchar*}(15,15)
  \fmftop{A1,A2} \fmfbottom{A3,A4} \fmf{double,label=$B_a$,label.side=right}{A1,a} \fmf{dashes,label=$H_i$,label.side=left}{A2,a}
  \fmf{dashes}{a,b} \fmf{photon,label=$W_b$}{A3,b} \fmf{dashes,label=$H_j$,label.side=right}{A4,b}
 \fmfdot{a}
  \fmfdot{b}  
\end{fmfchar*}}+\parbox{15mm}{
\begin{fmfchar*}(15,15)
  \fmftop{A1,A2} \fmfbottom{A3,A4} \fmf{double}{A1,a} \fmf{phantom}{A2,a}
  \fmf{dashes}{a,b} \fmf{photon}{A3,b} \fmf{phantom}{A4,b}
  \fmffreeze \fmf{dashes}{A4,a} \fmf{dashes}{A2,b}
  \fmfdot{a}
  \fmfdot{b}
\end{fmfchar*}}
\begin{aligned}=& T^a_{ik}T^b_{kj}(-p_b-2p_j)\cdot\epsilon_{b}+ T^a_{jk}T^b_{ki}(-p_b-2p_i)\cdot\epsilon_{b}\\
&=-2T^a_{ik}T^b_{kj}(p_j\cdot\epsilon_{b})-2 T^a_{jk}T^b_{ki}(p_i\cdot\epsilon_{b})
\end{aligned}
\end{multline}
Turning to the $t$- and
$u$-channel gauge boson exchange diagrams we can use the STI \eqref{eq:hww-sti-main} and the relation
\begin{equation}\label{eq:contr-prop}
q_\mu \left(g^{\mu\nu}-\frac{q^\mu q^\nu}{m^2}\right)\frac{-\ii}{q^2-m^2}=\frac{\ii}{m^2}q^\nu
\end{equation}
to find
\begin{equation}\label{eq:2a2h-t}
 \parbox{15mm}{
\begin{fmfchar*}(15,15)
  \fmftop{A1,A2} \fmfbottom{A3,A4} \fmf{double,label=$B_a$,label.side=right}{A1,a} \fmf{dashes,label=$H_i$,label.side=left}{A2,a}
  \fmf{photon}{a,b} \fmf{photon,label=$W_b$,label.side=left}{A3,b} \fmf{dashes,label=$H_j$,label.side=right}{A4,b}
 \fmfdot{a}
  \fmfdot{b}  
\end{fmfchar*}}\quad+
\parbox{15mm}{
\begin{fmfchar*}(15,15)
  \fmftop{A1,A2} \fmfbottom{A3,A4} \fmf{double}{A1,a} \fmf{phantom}{A2,a}
  \fmf{photon}{a,b} \fmf{photon}{A3,b} \fmf{phantom}{A4,b} \fmffreeze
  \fmf{dashes}{A4,a} \fmf{dashes}{A2,b}
  \fmfdot{a}
  \fmfdot{b}
\end{fmfchar*}}
\begin{aligned}\quad
=&
(-\ii)\ii \tfrac{1}{2m_{W_d}^2}g_{HWW}^{iac} g_{HWW}^{jcb}(p_j\cdot
\epsilon_b)\\
&+ (-\ii)\ii\tfrac{1}{2m_{W_d}^2}g_{HWW}^{jac} g_{HWW}^{icb}(p_i\cdot
\epsilon_b)
\end{aligned}
\end{equation}
Using the  STI \eqref{eq:www-sti}
\begin{equation}
\onepi{\mathcal{D}_a^\mu(p_a)W_b^\nu(p_b)W_c^\rho(p_c)}=f^{abc}\left[D_{W_b}(p_b)^{-1}-D_{W_c}(p_c)^{-1}\right]
\end{equation}
we  obtain the $s$-channel diagram: 
\begin{equation}\label{eq:hhaa-s}
\parbox{15mm}{
\begin{fmfchar*}(15,15)
  \fmftop{A1,A2} \fmfbottom{A3,A4} \fmf{double}{A1,a} \fmf{dashes}{A2,b}
  \fmf{photon}{a,b} \fmf{photon}{A3,a} \fmf{dashes}{A4,b}
 \fmfdot{a}
  \fmfdot{b} 
\end{fmfchar*}}=-f^{abc}T^c_{ij}\epsilon_b\cdot
    (p_i-p_j)
\end{equation}

The only diagram with a quartic vertex is:
\begin{equation}\label{eq:hhaa-kon}
  \parbox{15mm}{  \begin{fmfchar*}(15,15)
  \fmftop{A1,A2} \fmfbottom{A3,A4} \fmf{photon}{A1,a} \fmf{dashes}{A2,a}
   \fmf{photon}{A3,a} \fmf{dashes}{A4,a}
  \fmfdot{a}
  \fmfv{decor.shape=square,decor.filled=full,decor.size=3}{A1}
\end{fmfchar*}}=(-\ii)\ii g_{H^2 W^2}^{abij}(p_a\cdot\epsilon_b)
\end{equation}

Adding up the diagrams, we find
\begin{multline}
\left[-2T^a_{ik}T^b_{kj}+(\frac{1}{2m_{W_c}^2})g_{HWW}^{iac} g_{HWW}^{jcb}\right](p_j\cdot
\epsilon_b)\\
\left[-2 T^b_{ik}T^a_{kj}+(\frac{1}{2m_{W_c}^2})g_{HWW}^{jac} g_{HWW}^{icb}\right](p_i\cdot
\epsilon_b)\\
=g_{H^2 W^2}^{abij}\epsilon_b\cdot(p_i+p_j)+f^{abc}T^c_{ij}\epsilon_b\cdot
    (p_i-p_j)
\end{multline}
Splitting the product of the coupling matrices into commutators and anticommutators we find 
by matching coefficients
\begin{subequations}\label{eq:2h2w-wi}
\begin{align}
[T^a,T^b]_{ij}-[\tfrac{g_{HWW}^{a}}{2m_{W_c}},\tfrac{g_{HWW}^{b}}{2m_{W_c}}]_{ij}&=f^{abc}T^c_{ij}\\
\{T^a_{ik},T^b_{kj}\}-\{\tfrac{g_{HWW}^{a}}{2m_{W_c}},\tfrac{g_{HWW}^{b}}{2m_{W_c}}
\}_{ij}&=-g_{H^2 W^2}^{abij}
\end{align}
\end{subequations}
This are just one of the  commutation relations \eqref{eq:ssb-kommutator} and
the definition of the $HHWW$ coupling according to \fref{eq:2s2w}. The same
relations can also be obtained from the STI for the irreducible $2W2H$
vertex\eqref{eq:2s2w-1cov_a}, in agreement with the general result from \fref{sec:skeleton}. 

\subsection{Gauge boson and fermion couplings}
Having discussed one example for the evaluation of the Ward Identities, we will now quote the results of the remaining identities. The details of the calculations can be found in \fref{app:4wi}.

The Ward Identity for the $4W$ amplitude with one unphysical $W$ gives the Jacobi Identity for the structure constants and the quartic gauge coupling (see \fref{app:wwww-wi}):
\begin{multline}\label{eq:4w-wi}
  \parbox{20mm}{
\begin{fmfchar*}(20,20)
  \fmfleft{A1,A2} \fmfright{A3,A4} \fmf{double}{A1,a} \fmf{photon}{A2,a}
   \fmf{photon}{A3,a} \fmf{photon}{A4,a}
   \fmfv{decor.shape=circle,decor.filled=empty,decor.size=25}{a}
\end{fmfchar*}}\Rightarrow 
\begin{cases}
f^{abe}f^{cde}+f^{cae}f^{bde}+f^{ade}f^{bce}=0\\
g_{W^4}^{abcd}=-f^{ace}f^{bde}-2f^{ade}f^{bce}=f^{abe}f^{cde}-f^{ade}f^{bce}
 \end{cases}
\end{multline}
As in an unbroken gauge theory, the Ward Identity for 2 fermions and 2 gauge bosons gives the Lie algebra of the generators of the fermion representations (see \fref{app:ffww-wi}):
\begin{equation}\label{eq:2f2w-wi}
  \parbox{20mm}{
\begin{fmfchar*}(20,20)
  \fmfleft{f1,f2} \fmfright{A,H} \fmf{fermion,label=$f_j$}{f1,a}
  \fmf{fermion,label=$f_i$, label.side=left}{a,f2}
  \fmf{double}{A,a} \fmf{photon,label=$W_b$,label.side=left}{H,a}     
 \fmfv{decor.shape=circle,decor.filled=empty,decor.size=25}{a}   
\end{fmfchar*}}
\Rightarrow \begin{cases}[\tau_{L}^a, \tau_{L}^b]_{ij}-if^{abc}\tau_{Lij}^c&=0\\
[\tau_{R}^a, \tau_{R}^b]_{ij}-if^{abc}\tau_{Rij}^c&=0\end{cases}
\end{equation}
\subsection{Symmetry of Higgs-Yukawa couplings}
The Higgs-Yukawa coupling must satisfy the transformation law \eqref{eq:yukawa-tensor} that expresses the invariance under global transformations. 

This relation can be obtained from the  Ward Identity for 2 fermions, one gauge boson and one Higgs boson \eqref{eq:ffwh-wi}:
\begin{multline}\label{eq:2fhw-wi}
  \parbox{20mm}{
\begin{fmfchar*}(20,20)
  \fmfleft{f1,f2} \fmfright{A,H} \fmf{fermion,label=$f_j$}{f1,a}
  \fmf{fermion,label=$f_i$, label.side=left}{a,f2}
  \fmf{double}{A,a} \fmf{dashes,label=$H_h$,label.side=left}{H,a}
 \fmfv{decor.shape=circle,decor.filled=empty,decor.size=25}{a}
\end{fmfchar*}}\\
\Rightarrow
0=-\ii\frac{g_{HWW}^{hba}}{2m_{W_a}}  g_{\phi ij}^a-i(g_{H ij}^h
T^b_{hk})-g_{H il}^k
\tau_{Llj}^b+\tau_{Ril}^bg_{H lj}^k
\end{multline}
This is indeed the $A=h$ component of the transformation law \eqref{eq:yukawa-tensor}. 

Note that for vectorlike fermions the first term on the right hand side vanishes and \fref{eq:2fhw-wi} becomes the condition for the global invariance of the Higgs Yukawa coupling. 
\section{Goldstone boson couplings}
\subsection{Quartic Goldstone boson -gauge boson couplings}
The quartic Goldstone  Goldstone boson -gauge boson couplings are on one hand determined by the anticommutator relations \eqref{eq:2s2w}. The Higgs component of this anticommutator was reproduced by the $HHWW$ identity, so  the remaining components follow from the similar identities with external Goldstone bosons, where ghost terms have to be taken into account. 

This follows from the STI for the 2 gauge boson -2 scalar vertex \eqref{eq:2s2w-1cov_a}, according to the analysis in \fref{sec:skeleton}.
\begin{equation}
\parbox{20mm}{
\begin{fmfchar*}(20,20)
  \fmfleft{A1,A2} \fmfright{A3,A4} 
  \fmf{double}{A2,a} 
  \fmf{dashes,label=$H_i$,label.side=right}{A3,a}
  \fmf{photon,label=$W_b$}{A1,a} 
  \fmf{dashes,label=$\phi_c$,label.side=left}{A4,a}
  \fmfv{decor.shape=circle,decor.filled=empty,decor.size=25}{a}
\end{fmfchar*}}\quad=\quad
\parbox{20mm}{\begin{fmfchar*}(20,20)
  \fmfleft{A1,A2} \fmfright{A3,A4} 
  \fmf{ghost,label=$\bar c_b$}{A2,a} 
  \fmf{dashes,label=$H_i$,label.side=right}{A3,a}
  \fmf{photon,label=$W_b$}{A1,a} 
  \fmf{ghost,label=$c_c$,label.side=right}{a,A4}
  \fmfv{decor.shape=circle,decor.filled=empty,decor.size=25}{a}
 \fmfv{decor.shape=cross,decor.size=5}{A4}
\end{fmfchar*}} \qquad  \Rightarrow g_{\phi H W^2}^{ciab}=\{T^a,T^b\}_{ci}
\end{equation}
and similar for the component with 2 Goldstone bosons:
\begin{equation}
g_{\phi^2 W^2}^{abcd}=\{T^c,T^d\}_{ab}
\end{equation}

Since we want to avoid the use of ghost terms in our reconstruction of the Feynman rules, we cannot use these identities. Fortunately, the quartic gauge-Goldstone boson couplings are also \emph{uniquely} determined by the Jacobi-like identities  \eqref{eq:2s-jac_komp}, that follow from Ward Identities without external Goldstone bosons as we will demonstrate now.

The Ward Identity for the  $3W H$ amplitude with one unphysical gauge boson results in \eqref{eq:s3-s4}: 
\begin{equation}\label{eq:hphi-jac-exp}
 \parbox{20mm}{
\begin{fmfchar*}(20,20)
  \fmfleft{A1,A2} \fmfright{A3,A4} \fmf{double}{A1,a} \fmf{dashes,label=$H_i$,label.side=right}{A2,a} \fmf{photon,label=$W_b$}{A3,a} \fmf{photon,label=$W_c$}{A4,a}
   \fmfv{decor.shape=circle,decor.filled=empty,decor.size=25}{a}
\end{fmfchar*}}
\Rightarrow
\begin{aligned} m_{W_a}g_{H\phi W^2}^{abci}&=-g_{HWW}^{icd}f^{abd}-g_{HWW}^{ibd}f^{acd}\\
&-\frac{g_{HWW}^{iad}}{2m_{W_d}}g^{dbc}_{\phi WW}+T^a_{ij}g_{HWW}^{jbc}
\end{aligned}
\end{equation}
and this reproduces \eqref{eq:hphi-jac}. The same result follows from
the STI for the irreducible $\phi HWW$ vertex \eqref{eq:2s2w-1cov-b}

To avoid the use of external Goldstone bosons, we cannot use the Ward Identity for the $\phi 3W$ amplitude with one contraction. Instead we show that the same result can
be obtained from the $4W$ Ward Identity with 2  contractions where also the
$2W 2\phi$ vertex contributes. The result of the calculation is given by \fref{eq:2pow4w}: 
\begin{equation}\label{eq:2phi-jac-exp}
 \parbox{20mm}{
\begin{fmfchar*}(20,20)
  \fmfleft{A1,A2} \fmfright{A3,A4} \fmf{double}{A1,a} \fmf{photon,label=$W_c$,label.side=right}{A2,a} \fmf{double}{A3,a} \fmf{photon,label=$W_d$}{A4,a}
\fmfv{decor.shape=circle,decor.filled=empty,decor.size=25}{a}
\end{fmfchar*}}
 \Rightarrow 
\begin{aligned}
&m_{W_a}m_{W_b}g_{\phi^2 W^2}^{abcd}-\frac{1}{2}g_{HWW}^{iab}g_{HWW}^{icd}
=  m_{W_b}g_{\phi WW}^{ecd}t^a_{be}\\
&+ f^{ace}f^{dbe}(m_{W_d}^2-m_{W_e}^2)- f^{cbe}f^{dae}(m_{W_c}^2-m_{W_e}^2))
\end{aligned}
\end{equation}
Indeed this is the explicit form of the relation  \eqref{eq:2phi-jac}.
\subsection{Lie algebra structure of the triple Goldstone boson couplings}\label{sec:phi-lie}
Apart from the relation \eqref{eq:2h2w-wi} for the Higgs-gauge boson coupling,
the Lie algebra of the scalar generators \eqref{eq:ssb-kommutator} also
implies  two  relations for the Goldstone boson -gauge boson couplings
\eqref{eq:phiphi-kommutator} and \eqref{eq:hphi-kommutator}. Since these couplings involve more than one Goldstone boson, we have to consider Ward identities for amplitudes with several unphysical gauge bosons. 

We begin with the $3WH$ Ward Identity with two unphysical gauge bosons. The calculation in \fref{app:app:3wh-2cov} gives
\begin{equation}\label{eq:3wh-wi}
 \parbox{20mm}{
\begin{fmfchar*}(20,20)
  \fmfleft{A1,A2} \fmfright{A3,A4} \fmf{double}{A1,a} \fmf{double}{A2,a}
   \fmf{dashes}{A3,a} \fmf{photon}{A4,a}
     \fmfv{decor.shape=circle,decor.filled=empty,decor.size=25}{a}
\end{fmfchar*}}=
\begin{aligned}
 m_{W_b} g_{H\phi W^2}^{ibac}&=\frac{1}{2} g_{HWW}^{ice}
   f^{abe}\left(1-\frac{m_{W_a}^2-m_{W_b}^2}{m_{W_e}^2}\right)\\
& +\frac{1}{2}
   f^{ace}g_{HWW}^{ibe}-g_{HWW}^{jab} T^c_{ji}
\end{aligned}
\end{equation}
The same relation is obtained from the STI for the irreducible vertices with two contractions \eqref{eq:4point-2cov} as we show in \fref{app:4sti-2cov-exp}.

We cannot yet identify this equation with a relation obtained from the Yang Mills structure of the Lagrangian in \fref{chap:ssb-lag}. However, we can eliminate the quartic coupling using the result from the Ward Identity \eqref{eq:hphi-jac-exp}
and the explicit expression for $g_{\phi WW}^{dac}$. This gives us the condition
\begin{multline}\label{eq:3wh-wi2}
-\frac{1}{2m_{W_e}^2} g_{HWW}^{ice}
   f^{bae}(m_{W_b}^2-m_{W_a}^2-m_{W_e}^2)+\frac{1}{2m_{W_e}^2}g^{ieb}_{HWW}f^{cae}(m_{W_c}^2-m_{W_a}^2-m_{W_e}^2
   )\\
-g_{HWW}^{jab} T^c_{ji}+T_{ji}^bg_{H WW}^{jac}=-f^{bce}g_{H WW}^{iae}
\end{multline}
which can be easily seen to be the explicit form of \eqref{eq:hphi-kommutator}.

To derive the remaining component \eqref{eq:phiphi-kommutator}, we calculate the Ward Identity for the 4 gauge boson amplitude with 3 unphysical gauge bosons and get \eqref{eq:3pow4w}:
\begin{equation}\label{eq:4w-wi3}
  \parbox{20mm}{
\begin{fmfchar*}(20,20)
  \fmfleft{A1,A2} \fmfright{A3,A4} \fmf{double}{A1,a} \fmf{double}{A2,a}
   \fmf{double}{A3,a} \fmf{photon}{A4,a}
 \fmfv{decor.shape=circle,decor.filled=empty,decor.size=25}{a}
\end{fmfchar*}}\Rightarrow
\begin{aligned} 
&f^{ace}f^{ebd}(2m_{W_e}^2-m_{W_a}^2-m_{W_b}^2-m_{W_c}^2-m_{W_d}^2)=\\
&\Biggl\{-f^{aed}f^{cbe}\left[m_{W_e}^2+\frac{(m_{W_a}^2-m_{W_d}^2)(m_{W_c}^2-m_{W_b}^2)}{m_{W_e}^2}\right]\\
&+g_{HWW}^{ibc} g_{HWW}^{iad}\Biggr\}\quad- \quad
   ( a\leftrightarrow c)
\end{aligned}
\end{equation}
It requires some algebraic manipulations and use of the Jacobi Identity to see that this is indeed the same as \eqref{eq:phiphi-kommutator} (see \fref{app:4w-3co}). 
\subsection{Global invariance of Goldstone boson Yukawa couplings}
The Ward Identity for two fermions and 2 unphysical gauge bosons yields \eqref{eq:gphif-cond}
\begin{equation}\label{eq:2f2w-wi2}
  \parbox{20mm}{
\begin{fmfchar*}(20,20)
  \fmfleft{f1,f2} \fmfright{A,H} \fmf{fermion,label=$f_j$}{f1,a}
  \fmf{fermion,label=$f_i$, label.side=left}{a,f2}
  \fmf{double}{A,a} \fmf{double}{H,a}     
 \fmfv{decor.shape=circle,decor.filled=empty,decor.size=25}{a} 
\end{fmfchar*}}
\Rightarrow \ii g_{\phi il}^b\tau_{Llj}^a  -\ii \tau_{Ril}^ag_{\phi lj}^b= g_{\phi ij}^c t^a_{cb}-g_{H ij}^k \frac{g_{HWW}^{kab}}{2m_{W_b}}
\end{equation}
This relation can be identified as the transformation laws of the Yukawa couplings \eqref{eq:yukawa-tensor}. 
\subsection{Scalar potential}
The coupling of three Higgs bosons and one Goldstone boson is determined by the
$W3H$ Ward Identity \eqref{eq:3hw-wi}:
\begin{equation}
\parbox{20mm}{
\begin{fmfchar*}(20,20)
  \fmfleft{A1,A2} \fmfright{A3,A4} \fmf{double}{A1,a} \fmf{dashes,label=$H_i$,label.side=right}{A2,a}
  \fmf{dashes,label=$H_k$}{A3,a} \fmf{dashes,label=$H_j$,label.side=left}{A4,a}
 \fmfv{decor.shape=circle,decor.filled=empty,decor.size=25}{a}
\end{fmfchar*}}\Rightarrow
\begin{aligned}
m_{W_b}g_{\phi H^3}^{aijk}&=g_{H^3}^{jkl}T^a_{il}- g_{\phi H^2}^{bjk}\frac{g_{HWW}^{iab}}{2m_{W_b}}+ g_{H^3}^{ikl}T^a_{jl}- g_{\phi H^2}^{bik}\frac{g_{HWW}^{jab}}{2m_{W_b}}\\
&+ g_{H^3}^{ijl}T^a_{kl}- g_{\phi H^2}^{bij}\frac{g_{HWW}^{kab}}{2m_{W_b}} 
\end{aligned}
\end{equation}
The condition for the $2W2\phi$ coupling is obtained form the Ward Identity for the $2W2H$ amplitude with 2 contractions \eqref{eq:h2phi2-app}:
\begin{multline}
\parbox{20mm}{
\begin{fmfchar*}(20,20)
  \fmfleft{A1,A2} \fmfright{A3,A4} \fmf{double}{A1,a} \fmf{dashes,label=$H_i$,label.side=right}{A2,a}
  \fmf{double}{A3,a} \fmf{dashes,label=$H_j$,label.side=left}{A4,a}
 \fmfv{decor.shape=circle,decor.filled=empty,decor.size=25}{a}
\end{fmfchar*}}
\Rightarrow 
\begin{aligned}
&m_{W_a}m_{W_b}g_{\phi^2H^2}^{ijab}= m_{W_a}t^b_{ca}g_{\phi H^2}^{cij}+\frac{1}{2}g_{HWW}^{kab}g_{H^3}^{ijk}\\
&+\ii T^b_{ki}T^a_{kj}\left(m_{H_i}^2-m_{H_k}^2\right)
+\ii\frac{m_{H_i}^2}{4m_{W_c}^2} g_{HWW}^{ibc}g_{HWW}^{jac}\\
&+\ii T^b_{kj}T^a_{ki}\left(m_{H_j}^2-m_{H_k}^2\right) +\ii \frac{m_{H_j}^2}{4 m_{W_c}^2}g_{HWW}^{jbc}g_{HWW}^{jac}
\end{aligned}
\end{multline}
Inserting the relations for the triple scalar couplings from \eqref{subeq:triple-phi-wi1} and \eqref{subeq:triple-phi-wi2} , we see that this is the same as one component of the invariance condition of the scalar potential \eqref{eq:quartic-scalar} (see \eqref{eq:h2phi2-inv}). 

The relation for $g_{\phi^3 H}$ following from \eqref{eq:quartic-scalar} can be derived by the Ward Identity for the $3WH$ amplitude with three unphysical gauge bosons \eqref{eq:3wh-3cov}:
\begin{multline}
 \parbox{20mm}{
\begin{fmfchar*}(20,20)
  \fmfleft{A1,A2} \fmfright{A3,A4} \fmf{double}{A1,a} \fmf{dashes,label=$H_i$,label.side=right}{A2,a} \fmf{double}{A3,a} \fmf{double}{A4,a}
   \fmfv{decor.shape=circle,decor.filled=empty,decor.size=25}{a}
\end{fmfchar*}}\Rightarrow
\begin{aligned}
&m_{W_a}g_{H\phi^3}^{iabc}=-\frac{m_j^2}{2m_{W_b}m_{W_c}}T^a_{ij} g_{HWW}^{jcb}\\
&-\frac{m_{H_i}^2}{2m_{W_e} m_{W_c}} g_{HWW}^{ice}t^a_{be}+\frac{g_{HWW}^{jac}}{2m_{W_b}m_{W_c}}T^b_{ij}(m_{H_i}^2-m_{H_j}^2 ) \\
&- \frac{m_{H_i}^2}{2m_{W_b}m_{W_e}} g_{HWW}^{ibe}t^a_{ce}+\frac{g_{HWW}^{jcb}}{2m_{W_b}m_{W_c}}  T^c_{ij}(m_{H_i}^2-m_{H_j}^2)
\end{aligned}
\end{multline}
This is indeed equivalent to \eqref{eq:gphi3h}.

Finally, the Ward Identity for the 4 gauge boson amplitude with 4 unphysical gauge bosons gives the condition for the 4 Goldstone boson coupling:
\begin{multline}\label{eq:4w-4cov}
 \parbox{20mm}{
\begin{fmfchar*}(20,20)
  \fmfleft{A1,A2} \fmfright{A3,A4} \fmf{double}{A1,a} \fmf{double}{A2,a} \fmf{double}{A3,a} \fmf{double}{A4,a}
 \fmfv{decor.shape=circle,decor.filled=empty,decor.size=25}{a}
\end{fmfchar*}}\Rightarrow
\begin{aligned}
& m_{W_a}m_{W_b}m_{W_c} m_{W_d} g_{\phi^4}^{abcd}=-\frac{m_{H_i}^2}{4}g^{dia}_{HWW}g_{HWW}^{ibc}\\
&-\frac{m_{H_i}^2}{4}g^{ibc}_{HWW}g_{HWW}^{iac}-\frac{m_{H_i}^2}{4}g^{icd}_{HWW}g_{HWW}^{iab}
\end{aligned}
\end{multline}
\section{Summary}
To conclude our discussion of the reconstruction of the Feynman rules, we compare the results of the Ward Identities with the relations among the coupling constants in a spontaneously broken gauge theory from \fref{sec:ssb-summary} and discuss the relation to the reconstruction of the Feynman rules from tree level unitarity. 

Comparing to the results from \fref{chap:ssb-lag}, we see that all the relations related to gauge invariance can be reproduced by our limited set of Ward identities  except the definitions for the quartic Goldstone-gauge boson couplings from the anticommutator \eqref{eq:2s2w} and the gauge invariance condition of the quartic Higgs coupling resulting from \fref{eq:potential-gauge}. However, the quartic gauge-Goldstone boson couplings are uniquely determined by the `Jacobi Identities' \eqref{eq:2s-jac_komp} that are reproduced by the Ward identities.  It is obvious that the 4 point Ward identities are not sufficient to reproduce
   the invariance conditions of the quartic Higgs coupling so one has to consider a 5 point Ward Identity to verify this Feynman rules. In \fref{chap:ssb-lag} we have also given relations like the `Higgs mass sum rules' \eqref{eq:higgs-mass-sr} that follow from the minimization of the scalar potential. Since these relations are not related to gauge invariance, they cannot be verified by the Ward Identities. 

A complementary approach for the reconstruction of the Feynman rules  is the imposition of unitarity bounds \cite{LlewellynSmith:1973,Cornwall:1973,Horejsi:1994} that we have sketched in \fref{sec:unitarity}. The leading violations of tree level unitarity  are removed if the gauge boson interactions are of the Yang Mills type. In our approach the same conditions have been derived using the Ward identities \eqref{eq:4w-wi} and \eqref{eq:2f2w-wi}. To cancel subleading divergences, the couplings of the Higgs particles have to satisfy certain constraints, the so called 'Higgs sum rules' \cite{Gunion:1991}. As it turns out, these sum-rules are just the Lie algebra
\eqref{eq:ssb-kommutator}, the Yukawa transformation law \eqref{eq:yukawa-tensor} and the quartic gauge boson-Higgs coupling \eqref{eq:2h2w} after the identifications of the triple gauge boson couplings from \fref{subeq:triple-phi-couplings}. Like in the reconstruction of the Feynman rules from the Ward Identities, 5 point functions has to be considered to obtain relations for the quartic Higgs coupling \cite{Cornwall:1973,Horejsi:1994}.

From \cite{LlewellynSmith:1973,Cornwall:1973,Gunion:1991} one observes that the results of the Ward Identities with one unphysical gauge boson correspond to the cancellation of the \emph{leading} divergences while the result of the Ward Identities with several unphysical gauge bosons corresponds to the cancellation of the \emph{subleading} divergences. As example, consider the transformation law of the Goldstone boson-Yukawa couplings from \fref{eq:yukawa-tensor}. In \cite{LlewellynSmith:1973,Gunion:1991} it arises from the cancellation of the subleading unitarity violations in the $f\bar fWW$ amplitude  while in our approach it is reproduced from the $f\bar f WW$ Ward Identity with 2 unphysical gauge bosons \eqref{eq:2f2w-wi2}. 
\chapter{Gauge checks in O'Mega}\label{chap:implementation}
The results of \fref{chap:reconstruction} have given us a finite set of Ward Identities that have to be checked to ensure the correct implementation of the Feynman rules of a spontaneously broken gauge theory. We now discuss the implementation of this identities in the matrix element generator \OMEGA. Some aspects of the architecture of \OMEGA, relevant for the implementation of the Ward Identities are sketched in \fref{sec:omega}. The implementation of the Ward Identities is discussed in \fref{sec:omega-wi}. In \fref{sec:omega-sti} we describe the implementation of STIs that are relevant for checks of SUSY \cite{Reuter:2002,ReuterOhl:2002} and for future applications in loop calculations \cite{Kilian:loops}.  
\section{Architecture of O'Mega}\label{sec:omega}
The matrix element generator \omegacite~is especially suited for processes with a large number of external particles, because it reduces the factorial growth of calculational effort with the number of external particles to an exponential growth.

The complete
electroweak Standard Model and the MSSM have been implemented. The implementation of interfering
color amplitudes is currently being completed.  At the moment \OMEGA~generates the amplitude for a scattering process as a Fortran90/95 function, if needed other programming languages can be added as target languages.The possibility to calculate loop amplitudes from \OMEGA~tree-level matrix elements via the Feynman loop theorem \cite{Feynman:1963} is currently being studied \cite{Kilian:loops}.

In the \OMEGA~algorithm, the amplitudes are constructed from subamplitudes with one particle off its mass shell, the so called  \emph{one particle off shell wave functions (1POWs).} The 1POWs are obtained
from connected Green's functions by applying the LSZ reduction formula
to all but one external line while the remaining line is kept off the
mass shell
\begin{multline}
  \Phi(p_1+\ldots+p_n-q_1-\ldots-q_m) = \\
    \Braket{\phi(q_1),\ldots,\phi(q_m);\text{out}|\Phi(x)
           |\phi(p_1),\ldots,\phi(p_n);\text{in}}\,.
\end{multline}
 For example the 1POW 
\begin{equation}
A_\mu(p_1+p_2+p_3)\equiv \braket{A_\mu (p_2),\Out| A_\mu(x)|e^-(p_1),e^+(p_3),\In}
\end{equation}
consists of two diagrams:
\begin{equation}\label{eq:bremsstrahlung}
\parbox{32mm}{
\fmfframe(6,7)(9,6){%
\begin{fmfchar*}(25,20)
\fmftop{t}
\fmfbottomn{b}{3}
\fmf{photon,tension=3}{v2,t}
\fmf{fermion}{b1,v2,b3}
\fmf{photon}{b2,v2}
\fmfv{d.sh=circle,d.f=empty,d.size=20pt}{v2}
\fmflabel{$e^-(p_1)$}{b1}
\fmflabel{$e^+(p_3)$}{b3}
\fmflabel{$A_\mu(p_2)$}{b2}
\fmflabel{$A_\mu(p_1+p_2+p_3)$}{t}
\end{fmfchar*}}}=
\parbox{20mm}{
\begin{fmfchar*}(20,15)
\fmftop{t}
\fmfbottomn{b}{3}
\fmf{photon,tension=2}{v2,t}
\fmf{fermion,tension=2}{b1,v1}
\fmf{plain,tension=2}{v1,v2}
\fmf{fermion}{v2,b3}
\fmffreeze
\fmf{photon}{b2,v1}
\fmfdot{v1,v2}
\end{fmfchar*}}
\qquad 
\parbox{20mm}{
\begin{fmfchar*}(20,15)
\fmftop{t}
\fmfbottomn{b}{3}
\fmf{photon,tension=2}{v1,t}
\fmf{fermion}{b1,v1}
\fmf{plain,tension=2}{v1,v2}
\fmf{fermion,tension=2}{v2,b3}
\fmffreeze
\fmf{photon}{b2,v2}
\fmfdot{v1,v2}
\end{fmfchar*}}
\end{equation}

At tree-level, the set of all 1POWs for a given set of external
momenta can be constructed recursively
\begin{equation}
\label{eq:recursive-1POW}
  \parbox{22\unitlength}{%
    \fmfframe(2,3)(2,1){%
      \begin{fmfgraph*}(17,15)
       \fmflabel{$x$}{x}
       \fmftop{x}
       \fmfbottomn{n}{6}
       \fmf{plain,tension=6}{x,n}
       \fmfv{d.sh=circle,d.f=empty,d.si=30pt,l=$n$,l.d=0}{n}
       \begin{fmffor}{i}{1}{1}{6}
         \fmf{plain}{n,n[i]}
       \end{fmffor}
      \end{fmfgraph*}}} = 
  \sum_{k+l=n}
  \parbox{32\unitlength}{%
    \fmfframe(2,3)(2,1){%
      \begin{fmfgraph*}(27,15)
       \fmflabel{$x$}{x}
       \fmftop{x}
       \fmfbottomn{n}{6}
       \fmf{plain,tension=8}{x,n}
       \fmf{plain,tension=4}{n,k}
       \fmf{plain,tension=4}{n,l}
       \fmfv{d.sh=circle,d.f=empty,d.si=20pt,l=$k$,l.d=0}{k}
       \fmfv{d.sh=circle,d.f=empty,d.si=20pt,l=$l$,l.d=0}{l}
       \fmffixed{(30pt,0pt)}{k,l}
       \begin{fmffor}{i}{1}{1}{4}
         \fmf{plain}{k,n[i]}
       \end{fmffor}
       \begin{fmffor}{i}{5}{1}{6}
         \fmf{plain}{l,n[i]}
       \end{fmffor}
       \fmfdot{n}
      \end{fmfgraph*}}}\,,
\end{equation}
where the sum extends over all partitions of the set of $n$~momenta.
This recursion will terminate at the external wave functions.

The amplitude for a given process can now be obtained by summing over products of 1POWs. In a theory
with only cubic couplings this is expressed as
\begin{equation}
\label{eq:keystones}
  \me = 
      \sum_{k,l,m=1}^{P(n)}
        K^{3}_{\Phi_k\Phi_l\Phi_m}(p_k,p_l,p_m)
        \Phi_k(p_k)\Phi_l(p_l)\Phi_m(p_m)
\end{equation}
where $P(n)=2^{n-1}-1$ is the number of distinct momenta, that can be formed from $n$ external momenta. The generalization to  theories with quartic vertices is obvious. The quantities $K^{3}_{\Phi_k\Phi_l\Phi_m}$ are called \emph{Keystones}. The nontrivial problem of the determination of the keystones without double counting of diagrams has been solved in the algorithm of\omegacite~. Using this construction, Feynman diagrams are not generated separately and 1POWs occurring more than once in the amplitude can be factorized.

As an example we consider the process $e^+ e^- \to \mu^+\mu^- \gamma$ that can be decomposed into 1POWs according to 
\begin{equation}
\parbox{20mm}{
\begin{fmfchar*}(20,20)
\fmftopn{t}{3}
\fmfbottomn{b}{2}
\fmf{fermion}{b1,v,t1}
\fmf{fermion}{b2,v,t3}
\fmffreeze
\fmf{photon}{t2,v}
\fmfv{d.sh=circle,d.f=empty,d.size=20}{v}
\end{fmfchar*}}\, =\quad
\parbox{30mm}{
\fmfframe(5,0)(0,0){
\begin{fmfchar*}(25,20)
\fmftopn{t}{3}
\fmfbottomn{b}{2}
\fmf{fermion}{b1,v1,t1}
\fmf{fermion}{b2,v2,t3}
\fmf{photon}{v1,v2}
\fmffreeze
\fmf{photon}{t2,v2}
\fmfdot{v1}
\fmflabel{$K_{e^-e^+ A}$}{v1}
\fmfv{d.sh=circle,d.f=empty,d.size=20pt}{v2}
\end{fmfchar*}}}
\, +\,
\parbox{25mm}{
\begin{fmfchar*}(25,20)
\fmftopn{t}{3}
\fmfbottomn{b}{2}
\fmf{fermion}{b1,v1,t1}
\fmf{fermion}{b2,v2,t3}
\fmf{photon}{v1,v2}
\fmffreeze
\fmf{photon}{t2,v1}
\fmfdot{v2}
\fmflabel{$K_{\mu^-\mu^+ A}$}{v2}
\fmfv{d.sh=circle,d.f=empty,d.size=20pt}{v1}
\end{fmfchar*}}
\end{equation}
As illustrated in this simple example,
the factorization of diagrams into a sum over a product of 1POWs leads to the re-use of code that is needed more than once. This procedure is also known as  \emph{common subexpression elimination}.

The recursive definition of the 1POWs in \fref{eq:recursive-1POW}
allows to construct the amplitude in terms  of the 1POWs using ~(\ref{eq:keystones})
\emph{directly}, without having to construct and
factorize the Feynman diagrams explicitly.

The amplitude can now be constructed in the following way. Starting from the external particles, the \emph{tower} of all 1POWs up to a given height is constructed and \emph{all} possible  keystones for the process under consideration are determined. Now all 1POWs not appearing in the keystones are eliminated and finally the amplitude is constructed using \eqref{eq:keystones}. By construction, the resulting expression contains no more
redundancies and can be translated to a numerical expression.

\section{Implementation of Ward identities}\label{sec:omega-wi}
An advantage of the expression of the amplitude in terms of 1POWs---apart from the efficiency of the resulting code---is the fact that they are precisely the objects satisfying the simple Ward Identities \eqref{eq:gf-wi} 
\begin{equation}
 -\ii p_\mu W^\mu(p_1+\ldots+p_n-q_1-\ldots-q_m) =m_W\phi(p_1+\ldots+p_n-q_1-\ldots-q_m)
\end{equation}
Therefore \OMEGA~allows to check gauge invariance for every gauge boson occurring in the amplitude. This feature is enabled by calling \OMEGA~in the gauge checking mode
\begin{verbatim}
        f90_SM4.bin -warning:g e+ e- W+ W- Z     
\end{verbatim}
In gauge checking mode the complete tower is generated in order to allow gauge checks for all occurring gauge bosons. 

To use the Ward Identities in numerical calculations, appropriate criteria have to be determined to distinguish irrelevant numerical fluctuations from serious gauge invariance violations. More specifically, problems arise for very small wavefunctions, where small numerical fluctuations can be large relative to the absolute value of the wave functions. Such cases have to be discarded, in order not to generate spurious warnings. Two criteria have been found, that are sensitive to violations of gauge invariance and give comparable results: 
 
 \begin{enumerate}
  \item comparison with the euclidean norm $\|\cdot \|$ of the gauge boson wavefunction:
    \begin{equation}
|\ii\tfrac{p^\mu}{m}W_\mu +\phi|\leq \|\tfrac{p}{m}\|\, \||W\| \epsilon
\end{equation}
where the threshold is chosen as
\begin{equation*}
  \epsilon=3500\epsilon_M
\end{equation*}
where  $\epsilon_M\sim 10^{-16}$ is the machine precision and we demand that the violation of the Ward Identity is larger than a certain value
\begin{equation*}
 |\ii\tfrac{p^\mu}{m}W_\mu +\phi|  > 10\epsilon_M
\end{equation*}
\item comparison with a projection on the physical part of the gauge boson wavefunction:
 \begin{equation}
|\ii\tfrac{p^\mu}{m}W_\mu +\phi|\leq \|(g^{\mu\nu}-\tfrac{p^\mu p^\nu}{m^2})W_\nu\| \epsilon
\end{equation}
with the thresholds
\begin{equation*}
  \epsilon =5000\epsilon_M\qquad |\ii\tfrac{p^\mu}{m}W_\mu +\phi| > 100\epsilon_M
\end{equation*}
 \end{enumerate}
The numerical values have been chosen as low as possible without generating spurious warnings.

To be able to check all the Feynman rules already from 4 point functions we need the Ward Identities with several contractions  \eqref{eq:chanowitz} like 
\begin{equation}\label{eq:chanowitz-app}
\me(\mathcal{D}_a\mathcal{D}_a \Phi_i\Phi_j)=0
\end{equation}
as we have demonstrated in \fref{chap:reconstruction}. Therefore it is useful to enable the checking of this identity in \OMEGA. However, the use of 1POWs as building blocks for the amplitude means we cannot check the Ward Identities with several contractions for \emph{internal} gauge bosons. Nevertheless it is possible to check this identity for \emph{external} gauge bosons. This makes it necessary to compute also the Goldstone boson amplitudes corresponding to the physical relevant amplitude. Therefore we have implemented the automatic generation of the necessary amplitudes and of a Fortran function that sums up the different terms of \fref{eq:chanowitz-app}.

The automatic generation of the Ward Identity  is enabled by the commando line option \texttt{-warning:w}. The call
\begin{verbatim}
        f90_SM4.bin -warning:w e+ e- W+ W- Z     
\end{verbatim}
will generate the matrix elements for the required processes
\begin{equation}
 e^+ e^- \to  W^+ W^- Z \Rightarrow 
\begin{cases}  e^+ e^- &\to W^+ W^- \phi^0\\
               e^+ e^-&\to W^+ \phi^- \phi^0\\
                      &\dots \\
              e^+ e^- &\to \phi^+ \phi^- \phi^0
\end{cases}
\end{equation}
and a function that sums the amplitudes with the required phases. This subroutine is automatically called whenever the amplitude is evaluated. This allows checking the Ward Identities parallel to the calculation of the physical amplitude.

As a criterion for the violation of the Ward Identity we take (as in \cite{Reuter:2002} in the case of SUSY STIs) 
\begin{equation}\label{eq:sti-viol}
\frac{|\sum_i \me_i|}{\sum_i |\me_i|}\leq \epsilon
\end{equation}
with 
\begin{equation}
\epsilon=10^7\times \epsilon_M\qquad \text{and}\qquad |\sum_i \me_i|\geq 1000 \epsilon_M
\end{equation}
Applications of the Ward Identities in \OMEGA~are described in \fref{chap:ssb_groves} and \fref{chap:running}. As an example of the efficiency of the gauge checks, we consider the sensitivity of the Ward Identities to the coupling $g_{H^2Z^2}$ in the Standard Model. If this coupling is used with the wrong sign, the Ward Identity for the external $Z$ boson in the amplitude $e^+ e^- \to HHZ$ is violated with 
\begin{equation}
\frac{|\ii\tfrac{p^\mu}{m}W_\mu +\phi|}{ \|\tfrac{p}{m}\|\, \||W\| }\sim 10^{12}\epsilon
\end{equation}
The Ward Identities are sensitive to a relative error of 
\begin{equation}
\frac{\delta g_{H^2Z^2}}{g_{H^2Z^2}} \sim 3\times 10^{-8}
\end{equation}
Also numerical instabilities in the expressions of the polarization spinors have been discovered using the Ward Identities. 
\section{Implementation of Slavnov-Taylor identities}\label{sec:omega-sti}
According to the results of \fref{part:reconstruction}, it is not necessary to consider the STIs for off-shell Green's functions to obtain gauge checks verifying the Feynman rules in ordinary gauge theories. Also on-shell matrixelements are the natural quantities appearing in numerical tree level calculations so the on-shell identities are sufficient for the purposes of this work. 

However, since the minimal supersymmetric Standard Model has  already been implemented in \OMEGA, one has also to consider the appropriate consistency checks \cite{ReuterOhl:2002,Reuter:2002} that involve off-shell STIs following from the full SUSY-BRS transformations \cite{Maggiore:1996}. Furthermore, work is under way  to calculate loop amplitudes from \OMEGA~tree-level matrix elements \cite{Kilian:loops} and here off-shell STIs are needed for gauge checks. Because of these reasons it is useful to provide the necessary infrastructure for gauge checks involving off-shell STIs in \OMEGA~and the implementation described in this section has already been used in \cite{Reuter:2002}. 

To generate the contact terms the STI \eqref{eq:lsz-wi-boson}, we introduce additional sources in the Lagrangian---just as in \fref{eq:antifields} for the derivation of the ZJ-equation---that couple to the BRS transformations of the physical fields:
    \begin{equation}
      \mathscr{L}_{BRS}= (-\ii)\sum_\Phi \Phi^* \deltabrs \Phi 
    \end{equation}
The sources $\Phi^*$ have to be treated as nonpropagating particles.
We have chosen to include the inverse propagators appearing in front of  the contact terms in the STI \eqref{eq:lsz-wi-boson} in the external wavefunctions of the BRS sources:
\begin{equation}
 \begin{aligned}
    H^* &: (-\ii)(p^2-m_H^2)\\
    W^*_\mu &: \ii(p^2-m_W^2)\epsilon_\mu\\
    \psi^* &: (-\ii) (\fmslash p-m)u(p)
  \end{aligned}
\end{equation}
For fermions, particles and antiparticles have to be distinguished according to \fref{eq:ferm-wf}. 
 In this way, the \OMEGA~scattering amplitude for the `process'
\begin{equation}\label{eq:antifield-call}
  \bar c(k)\Phi_1(p_1)\dots\Phi_j^*(p_j)\dots\Phi_n(p_n)
\end{equation}
is precisely a contact term from the STI \eqref{eq:lsz-wi-boson}:
\begin{equation}
D_{\Phi_j}^{-1}(p_j) \me( \bar c(k)\Phi_1(p_1)\dots(c\Delta\Phi_j)(p_j)\dots \Phi_n(p_n))
\end{equation}
For physical polarization vectors with $\epsilon\cdot p=0$  the terms of the form \eqref{eq:antifield-call} are sufficient. For gauge checks involving the unphysical modes of gauge bosons, one has to include terms of the form \eqref{eq:lsz-wi-gauge} arising from the inhomogeneous BRS transformation of the gauge bosons to check the STIs. This could be done by explicitly considering the amplitude 
\begin{equation}\label{eq:sti-ghost-term}
-\frac{\ii}{\xi}p_b^\nu\me(c_a(p)\Phi_1(p_1)\dots\bar c_b(p_b)\dots\Phi_n(p_n))
\end{equation}
or by introducing another auxiliary field $\chi_W$ \cite{Reuter:2002} with the Feynman rules
\begin{equation}
\mathscr{L}_\chi=-\frac{1}{\xi}W^* \chi_W \dmu c
\end{equation}
The additional auxiliary field is necessary to convert this vertex into a cubic vertex so it can be included in \OMEGA. The matrix element \eqref{eq:sti-ghost-term} is then generated as the amplitude for
\begin{equation}
 \bar c(k) \chi_{W_a}\Phi_1(p_1)\dots {W_a^*}^\mu(p_a)\dots\Phi_n(p_n)
\end{equation}
For calculations in unitarity gauge, the ghost Feynman rules have to be modified according to the results of \fref{app:uni-sti}.

The construction sketched in this section has the advantage that for  checks of the STIs, one only has to add up  amplitudes of the form \eqref{eq:antifield-call} and no contraction of group indices or multiplication with inverse propagators has to be done by hand outside of the amplitude. Also the automatic generation of the contact terms should be straightforward, similar to the case of the Ward Identity with several contractions discussed in \fref{sec:omega-wi}. As numerical criterion for the violation of the STIs we can use again \fref{eq:sti-viol}.
\part{Consistency of realisitic calculations: selection and resummation of diagrams}\label{part:resummation} 
\chapter{Forests and groves in spontaneously broken gauge theories}\label{chap:ssb_groves}\chaptermark{Forests and groves}
The prescription for the construction of gauge flips in spontaneously broken gauge theories given in \cite{Boos:1999} is to treat Higgs bosons like gauge bosons. As discussed already in \fref{sec:groves}, the characteristic $HWW$ vertex prevents the assignment of  conserved charges to the Higgs bosons, so it is plausible that no new groves should appear. However, no explicit proof for the treatment of Higgs bosons was given in \cite{Boos:1999} and we find it worthwhile to investigate this problem based on the results of \fref{sec:gi-classes}. We will find that for the usual linear realization of the symmetries of the Standard Model the brief discussion given above is essentially correct. However, in the case of nonlinear realized symmetries \cite{Coleman:1969} additional groves appear that are discussed in \fref{sec:nl-groves}. These grove are also consistent (i.e. they satisfy the Ward Identities) in unitarity gauge. We give numerical checks of these results for processes with up to 8 external particles. 

\section{Definition of gauge flips}\label{sec:sti-flips}
In order to apply the formalism of flips and groves sketched in \fref{sec:groves} to spontaneously broken gauge theories, we have to define the elementary flips of the theory. According to \cite{Boos:1999} and the results of \fref{sec:gi-classes}, the gauge flips are given by the minimal sets of 4 point diagrams satisfying the STIs. In spontaneously broken gauge theories there is a peculiar property of Higgs exchange diagrams that  could lead to confusion in the definition of the gauge flips. We first treat the case of the amplitude $\bar f f \to WW$ in some detail, before extending the discussion to the remaining elementary flips. In \fref{sec:nl-flips} we will discuss theories with nonlinear realized symmetries \cite{Coleman:1969} where the gauge flips can be simplified and additional groves appear. This has also interesting implications for unitarity gauge as will be discussed in \fref{sec:nl-flips}. 
\subsection{$\bar f f \to WW$}\label{sec:2f2w-flips}
To see the origin of the subtleties in the definition of the gauge flips, we recall that in the calculations of Ward Identities often Higgs exchange diagrams satisfy the Ward Identity by themselves. We have met one example in the $HHWW$ Ward Identity in \fref{sec:hhww-wi} (see \fref{eq:hhww-higgs}). The reason for this is that a simple Ward Identity for the $HWW$ vertex is valid even if the Higgs is off shell (see the discussion around \fref{eq:hww-sti-contr}). 
 
The same happens in the Ward Identity for  $\bar f f \to W W$ (see \fref{app:ffww-wi}) where the Higgs exchange diagram
\begin{equation*}
\parbox{15mm}{
\begin{fmfchar*}(15,15)
\fmfleft{a,b}
\fmfright{f1,f2}
\fmf{fermion}{a,fwf}
\fmf{fermion}{fwf,b}
\fmf{dashes}{fwf,www}
\fmf{photon}{www,f1}
\fmf{photon}{www,f2}
\fmfdot{www,fwf}
\end{fmfchar*}}
\end{equation*}
 gives no contribution.

Considering the STI for the amplitude where all external particles are off-shell,  using the STI \eqref{eq:hww-sti-main} for the $HWW$ vertex and the Feynman rules from \fref{app:general-fr} we find that the Higgs diagram in the $\bar f f W W$ amplitude reproduces a ghost diagram from the STI :
\begin{equation} 
\parbox{15mm}{
\begin{fmfchar*}(15,15)
  \fmfleft{f1,f2} \fmfright{A,H} \fmf{fermion}{f1,a1}
  \fmf{fermion}{a1,f2}
  \fmf{dashes}{a1,a2}
  \fmf{photon}{H,a2} \fmf{double}{a2,A}
\fmfdot{a1,a2}  
\end{fmfchar*}}\quad =\quad
\parbox{15mm}{
\begin{fmfchar*}(15,15)
  \fmfleft{f1,f2} \fmfright{A,H} \fmf{fermion}{f1,a1}
  \fmf{fermion}{a1,f2}
  \fmf{dashes}{a1,a2}
  \fmf{ghost}{A,a2} \fmf{ghost}{a2,H}
\fmfdot{a1,a2}  
    \fmfv{decor.shape=square,decor.filled=full,decor.size=5}{H}
\end{fmfchar*}}
\end{equation}
According to the definitions in \fref{sec:groves-def}, this means that this diagram satisfies the STI by itself. Therefore we would conclude that the elementary gauge flips in a spontaneously broken gauge theory are still defined by \eqref{eq:gauge_flips2}:
\begin{equation}
\left\{
\parbox{15mm}{
\begin{fmfchar*}(15,15)
\fmfleft{a,b}
\fmfright{f1,f2}
\fmf{fermion}{a,fwf1}
\fmf{fermion}{fwf1,fwf2}
\fmf{fermion}{fwf2,b}
\fmf{photon}{fwf1,f1}
\fmf{photon}{fwf2,f2}
\fmfdot{fwf1,fwf2}
\end{fmfchar*}}\quad,\quad
\parbox{15mm}{
\begin{fmfchar*}(15,15)
\fmfleft{a,b}
\fmfright{f1,f2}
\fmf{fermion}{a,fwf1}
\fmf{fermion}{fwf1,fwf2}
\fmf{fermion}{fwf2,b}
\fmf{phantom}{fwf2,f2}
\fmf{phantom}{fwf1,f1}
\fmffreeze
\fmf{photon}{fwf2,f1}
\fmf{photon}{fwf1,f2}
\fmfdot{fwf1,fwf2}
\end{fmfchar*}}\quad,\quad 
\parbox{15mm}{
\begin{fmfchar*}(15,15)
\fmfleft{a,b}
\fmfright{f1,f2}
\fmf{fermion}{a,fwf}
\fmf{fermion}{fwf,b}
\fmf{photon}{fwf,www}
\fmf{photon}{www,f1}
\fmf{photon}{www,f2}
\fmfdot{www,fwf}
\end{fmfchar*}}\right \}
\end{equation}
while it seems that the Higgs exchange diagram
\begin{equation*}
\parbox{15mm}{
\begin{fmfchar*}(15,15)
\fmfleft{a,b}
\fmfright{f1,f2}
\fmf{fermion}{a,fwf}
\fmf{fermion}{fwf,b}
\fmf{dashes}{fwf,www}
\fmf{photon}{www,f1}
\fmf{photon}{www,f2}
\fmfdot{www,fwf}
\end{fmfchar*}}
\end{equation*}
has to be regarded as a new kind of flip that might lead to new groves.
However, as discussed in \fref{sec:flip-def}, we have to make sure that the set of Goldstone boson diagrams corresponding to  the elementary gauge flips also satisfies the STIs by themselves. As we will now show, this forces us to include the Higgs exchange diagram in the gauge flips.

Considering the Higgs exchange diagram in the $\bar f f \to W \phi$ amplitude,  we see, using the STI for the $WH\phi$ vertex \eqref{eq:phihw-sti},  that in this case one gets additional contributions: 
\begin{equation}
\parbox{15mm}{
\begin{fmfchar*}(15,15)
  \fmfleft{f1,f2} \fmfright{A,H} \fmf{fermion}{f1,a1}
  \fmf{fermion}{a1,f2}
  \fmf{dashes,label=$H$}{a1,a2}
  \fmf{dashes,label=$\phi$}{H,a2} \fmf{double}{a2,A}
\fmfdot{a1,a2}  
\end{fmfchar*}}\quad = \quad 
\parbox{15mm}{
\begin{fmfchar*}(15,15)
  \fmfleft{f1,f2} \fmfright{A,H} \fmf{fermion}{f1,a1}
  \fmf{fermion}{a1,f2}
  \fmf{dashes}{a1,a2}
  \fmf{ghost}{A,a2} \fmf{ghost}{a2,H}
\fmfdot{a1,a2}  
    \fmfv{decor.shape=cross,decor.size=5}{H}
\end{fmfchar*}}\quad +\quad
\parbox{15mm}{
\begin{fmfchar*}(15,15)
  \fmfleft{f1,f2} \fmfright{A,H} \fmf{fermion}{f1,a,f2}
  \fmf{dashes}{H,i,a} 
  \fmf{phantom}{a,A}
\fmffreeze
\fmf{ghost}{A,i}
\fmfdot{a}
\fmfv{decor.shape=square,decor.filled=empty,decor.size=5}{i}
\end{fmfchar*}}\quad +\quad
\parbox{15mm}{
\begin{fmfchar*}(15,15)
  \fmfleft{f1,f2} \fmfright{A,H} \fmf{fermion}{f1,a1}
  \fmf{fermion}{a1,f2}
  \fmf{dashes}{a1,a2}
  \fmf{ghost}{A,a2} \fmf{zigzag}{a2,H}
\fmfdot{a1}  
\fmfv{decor.shape=square,decor.filled=empty,decor.size=5}{a2}
\end{fmfchar*}}
\end{equation}
The first and the last diagram are contact terms required by the STI. To cancel the second diagram, we are forced to add two more diagrams:
\begin{equation*}
\parbox{15mm}{
\begin{fmfchar*}(15,15)
\fmfleft{a,b}
\fmfright{f1,f2}
\fmf{fermion}{a,fwf1}
\fmf{fermion}{fwf1,fwf2}
\fmf{fermion}{fwf2,b}
\fmf{double}{fwf1,f1}
\fmf{dashes}{fwf2,f2}
\fmfdot{fwf1,fwf2}
\end{fmfchar*}}\quad ,\quad
\parbox{15mm}{
\begin{fmfchar*}(15,15)
\fmfleft{a,b}
\fmfright{f1,f2}
\fmf{fermion}{a,fwf1}
\fmf{fermion}{fwf1,fwf2}
\fmf{fermion}{fwf2,b}
\fmf{phantom}{fwf2,f2}
\fmf{phantom}{fwf1,f1}
\fmffreeze
\fmf{double}{fwf2,f1}
\fmf{dashes}{fwf1,f2}
\fmfdot{fwf1,fwf2}
\end{fmfchar*}}
\end{equation*}
According to the general analysis from \fref{chap:diagrammatica}, these terms provide the remaining diagrams, so a cancellation because of the STI 
\begin{equation*}
0=\quad
\parbox{15mm}{
\begin{fmfchar*}(15,15)
  \fmfleft{f1,f2} \fmfright{A,H} \fmf{fermion}{f1,a,f2}
  \fmf{dashes}{H,i,a} 
  \fmf{phantom}{a,A}
\fmffreeze
\fmf{ghost}{A,i}
\fmfdot{a}
\fmfv{decor.shape=square,decor.filled=empty,decor.size=5}{i}
\end{fmfchar*}}\quad +\quad
 \parbox{15mm}{
\begin{fmfchar*}(15,15)
  \fmfleft{f1,f2} \fmfright{A,H} \fmf{fermion}{f1,a,i,f2}
  \fmf{dashes}{H,a} 
  \fmf{phantom}{a,A}
\fmffreeze
\fmf{ghost,left=0.5}{A,i}
\fmfdot{a}
\fmfv{decor.shape=square,decor.filled=empty,decor.size=5}{i}
\end{fmfchar*}}\quad +\quad
\parbox{15mm}{
\begin{fmfchar*}(15,15)
  \fmfleft{f1,f2} \fmfright{A,H} \fmf{fermion}{f1,i,a,f2}
  \fmf{dashes}{H,a} 
  \fmf{phantom}{a,A}
\fmffreeze
\fmf{ghost}{A,i}
\fmfdot{a}
\fmfv{decor.shape=square,decor.filled=empty,decor.size=5}{i}
\end{fmfchar*}}
\end{equation*}
takes place. Interestingly, in this case the gauge boson exchange diagram satisfies the Ward Identity by itself.

Therefore, our definition in \fref{sec:flip-def} forces us  to include two diagrams
\begin{equation*}
 \parbox{15mm}{
\begin{fmfchar*}(15,15)
\fmfleft{a,b}
\fmfright{f1,f2}
\fmf{fermion}{a,fwf1}
\fmf{fermion}{fwf1,fwf2}
\fmf{fermion}{fwf2,b}
\fmf{photon}{fwf1,f1}
\fmf{photon}{fwf2,f2}
\fmfdot{fwf1,fwf2}
\end{fmfchar*}}\quad ,\quad
\parbox{15mm}{
\begin{fmfchar*}(15,15)
\fmfleft{a,b}
\fmfright{f1,f2}
\fmf{fermion}{a,fwf1}
\fmf{fermion}{fwf1,fwf2}
\fmf{fermion}{fwf2,b}
\fmf{phantom}{fwf2,f2}
\fmf{phantom}{fwf1,f1}
\fmffreeze
\fmf{photon}{fwf2,f1}
\fmf{photon}{fwf1,f2}
\fmfdot{fwf1,fwf2}
\end{fmfchar*}}
\end{equation*}
in the gauge flips for $f\bar f \to WW$. But then also the diagram with a triple gauge boson vertex has to be included and finally we obtain the correct set of flips:
\begin{equation}\label{ffww-flips}
\{ t_{4}^{G,8},\dots,t_{4}^{G,11}\}=
\left\{
\parbox{15mm}{
\begin{fmfchar*}(15,15)
\fmfleft{a,b}
\fmfright{f1,f2}
\fmf{fermion}{a,fwf1}
\fmf{fermion}{fwf1,fwf2}
\fmf{fermion}{fwf2,b}
\fmf{photon}{fwf1,f1}
\fmf{photon}{fwf2,f2}
\fmfdot{fwf1,fwf2}
\end{fmfchar*}}\,,\,
\parbox{15mm}{
\begin{fmfchar*}(15,15)
\fmfleft{a,b}
\fmfright{f1,f2}
\fmf{fermion}{a,fwf1}
\fmf{fermion}{fwf1,fwf2}
\fmf{fermion}{fwf2,b}
\fmf{phantom}{fwf2,f2}
\fmf{phantom}{fwf1,f1}
\fmffreeze
\fmf{photon}{fwf2,f1}
\fmf{photon}{fwf1,f2}
\fmfdot{fwf1,fwf2}
\end{fmfchar*}}\,,\, 
\parbox{15mm}{
\begin{fmfchar*}(15,15)
\fmfleft{a,b}
\fmfright{f1,f2}
\fmf{fermion}{a,fwf}
\fmf{fermion}{fwf,b}
\fmf{photon}{fwf,www}
\fmf{photon}{www,f1}
\fmf{photon}{www,f2}
\fmfdot{www,fwf}
\end{fmfchar*}}\, ,\,
\parbox{15mm}{
\begin{fmfchar*}(15,15)
\fmfright{a,b}
\fmfleft{f1,f2}
\fmf{photon}{a,fwf}
\fmf{photon}{fwf,b}
\fmf{dashes}{fwf,www}
\fmf{fermion}{www,f1}
\fmf{fermion}{f2,www}
\fmfdot{www}
\fmfdot{fwf}
\end{fmfchar*}}\right \}
\end{equation}
\subsection{Elementary flips in spontaneously broken gauge theories}\label{sec:ssb-flips}
The discussion in \fref{sec:2f2w-flips} can easily be generalized to the other elementary gauge flips since the argument did only depend on the  structure of the STIs for the $WWH$ and $W\phi H$ vertices. Considering the remaining vertices, we only used the fact that they satisfy the appropriate STIs. Therefore the conclusions apply also for the other elementary flips with at least two gauge bosons and thus we have to include the Higgs exchange diagram in \emph{all} gauge flips. For example, the Higgs exchange diagrams have to be included in the gauge flips for 4 gauge bosons.  All flips including new elementary gauge flips for gauge-Higgs boson 4 point functions are displayed  in \fref{app:flips}. 

As another example, we discuss the flips for the 4 point amplitude for $\bar f f \to W H$. Here the internal Higgs bosons contribute to the Ward Identities (see \fref{app:ffwh-wi}) , so no confusion can arise and it is clear that Higgs exchange diagrams have to be included in the gauge flips: 
\begin{equation}\label{eq:h_flips_1}
\{ t_{4,H_1}^{G,1},\dots, t_{4,H_1}^{G,5} \}=
\left\{
\parbox{15mm}{
\begin{fmfchar*}(15,15)
\fmfleft{a,b}
\fmfright{f1,f2}
\fmf{fermion}{a,fwf1}
\fmf{fermion}{fwf1,fwf2}
\fmf{fermion}{fwf2,b}
\fmf{photon}{fwf1,f1}
\fmf{dashes}{fwf2,f2}
\fmfdot{fwf1,fwf2}
\end{fmfchar*}}\, ,\,
\parbox{15mm}{
\begin{fmfchar*}(15,15)
\fmfleft{a,b}
\fmfright{f1,f2}
\fmf{fermion}{a,fwf1}
\fmf{fermion}{fwf1,fwf2}
\fmf{fermion}{fwf2,b}
\fmf{phantom}{fwf2,f2}
\fmf{phantom}{fwf1,f1}
\fmffreeze
\fmf{photon}{fwf2,f1}
\fmf{dashes}{fwf1,f2}
\fmfdot{fwf1,fwf2}
\end{fmfchar*}}\, ,\, 
\parbox{15mm}{
\begin{fmfchar*}(15,15)
\fmfleft{a,b}
\fmfright{f1,f2}
\fmf{fermion}{a,fwf}
\fmf{fermion}{fwf,b}
\fmf{photon}{fwf,www}
\fmf{photon}{www,f1}
\fmf{dashes}{www,f2}
\fmfdot{www,fwf}
\end{fmfchar*}}\, , \,
\parbox{15mm}{
\begin{fmfchar*}(15,15)
\fmfleft{a,b}
\fmfright{f1,f2}
\fmf{fermion}{a,fwf}
\fmf{fermion}{fwf,b}
\fmf{dashes}{fwf,www}
\fmf{photon}{www,f1}
\fmf{dashes}{www,f2}
\fmfdot{www,fwf}
\end{fmfchar*}}
\right \} 
\end{equation}
4 particle diagrams with only external Higgs bosons or fermions are always gauge parameter independent by themselves, so we have to introduce another class of flips, that plays a role similar to the flavor flips and  might be called `Higgs flips'. They consist of all diagrams contributing to the $\bar f f \to HH$ amplitude  and to the $4H$ amplitude and are given in \fref{subeq:higgs_flips}.

The \emph{forest} for a given set of external diagrams is now defined as the set of diagrams connected by flavor,  Higgs and gauge flips while the definition of the groves remains as before. 

\subsection{Flips for nonlinear realizations of the symmetry}\label{sec:nl-flips}
Apart from the linear parametrization of the scalar fields \eqref{eq:scalar-explicit} used in the parametrization of \fref{chap:ssb-lag}, one can also introduce a nonlinear realization of the symmetry \cite{Coleman:1969} as is reviewed in \fref{app:nonlinear}. In this parametrization, the Higgs bosons are not connected to the symmetry breaking mechanism but merely additional matter particles.

The appearance of new groves in nonlinear realizations is very plausible if one considers the nonlinearly realized electroweak Standard Model \cite{Feruglio:1993,Dobado:1997jx}. Here the Higgs boson transforms trivially under gauge transformations and can be removed from the theory. The trivial transformation law  implies that  the diagrams without Higgs bosons are gauge invariant by themselves, in contrast to the linear parametrization. Therefore one can simplify the elementary gauge flips be omitting the internal Higgs bosons.

It is not obvious that these rather general arguments carry over to theories with a more complicated  Higgs sector where the Higgs bosons transform nontrivial under the unbroken subgroup. We show in \fref{app:nl-sti} for a general nonlinear realized symmetry that the STI for the $WH\phi$ vertex becomes trivial at tree level. Therefore, according to the discussion in \fref{sec:2f2w-flips}, the Higgs exchange diagrams have not to be included in the gauge flips without external Higgs bosons. One can introduce a new class of flips that  consist of the diagrams not needed for the gauge flips, i.e. 
\begin{equation}
\left\{
\parbox{15mm}{
\begin{fmfchar*}(15,15)
\fmfleft{a,b}
\fmfright{f1,f2}
\fmf{photon}{a,fwf1}
\fmf{dashes}{fwf1,fwf2}
\fmf{photon}{fwf2,b}
\fmf{photon}{fwf1,f1}
\fmf{photon}{fwf2,f2}
\fmfdot{fwf1}
\fmfdot{fwf2}
\end{fmfchar*}}\quad,\quad
\parbox{15mm}{
\begin{fmfchar*}(15,15)
\fmfleft{a,b}
\fmfright{f1,f2}
\fmf{photon}{a,fwf1}
\fmf{dashes}{fwf1,fwf2}
\fmf{photon}{fwf2,b}
\fmf{phantom}{fwf2,f2}
\fmf{phantom}{fwf1,f1}
\fmffreeze
\fmf{photon}{fwf2,f1}
\fmf{photon}{fwf1,f2}
\fmfdot{fwf1}
\fmfdot{fwf2}
\end{fmfchar*}}\quad , \quad 
\parbox{15mm}{
\begin{fmfchar*}(15,15)
\fmfleft{a,b}
\fmfright{f1,f2}
\fmf{photon}{a,fwf}
\fmf{photon}{fwf,b}
\fmf{dashes}{fwf,www}
\fmf{photon}{www,f1}
\fmf{photon}{www,f2}
\fmfdot{fwf}
\fmfdot{www}
\end{fmfchar*}}
\right \}
\end{equation}
for the $4 W$ function and
\begin{equation}\label{eq:higgs-flip2}
\parbox{15mm}{
\begin{fmfchar*}(15,15)
\fmfleft{a,b}
\fmfright{f1,f2}
\fmf{photon}{a,fwf}
\fmf{photon}{fwf,b}
\fmf{dashes}{fwf,www}
\fmf{fermion}{www,f1}
\fmf{fermion}{f2,www}
\fmfdot{www}
\fmfdot{fwf}
\end{fmfchar*}}
\end{equation}
for the $2$ fermion- 2 gauge boson function. These  `Higgs exchange flips' have to be included in the definition of the forests while 
the groves are still defined as the sets of diagrams connected by gauge flips only.

 As we discuss in \fref{app:nl-sti}, the simplification only effects the $WWH$ vertex while the STI for the $WHH$ vertex is similar to the linear realization. Thus diagrams with internal Higgs bosons can not be neglected if a $WHH$ vertex appears. This affects flips with external Higgs bosons in theories with general Higgs sectors.

In the flips of  \fref{eq:h_flips_1} the internal Higgs appears with a $WHH$ vertex so it has to be included in the gauge flips. In contrast, in the flips for the $WWHH$ subdiagrams \eqref{eq:hhww-flips} the diagram 
\begin{equation*}
\parbox{15mm}{
\begin{fmfchar*}(15,15)
\fmfleft{a,b}
\fmfright{f1,f2}
\fmf{photon}{a,i1}
\fmf{photon}{i1,b}
\fmf{dashes}{i1,i2}
\fmf{dashes}{i2,f1}
\fmf{dashes}{i2,f2}
\fmfdot{i1,i2}
\end{fmfchar*}}
\end{equation*}
includes the $HWW$ vertex and has not to be included in the gauge flips for nonlinear realizations of the symmetry.
All the relevant flips for nonlinear realizations are again listed in \fref{app:flips}.

The reduced number of gauge flips has also interesting consequences for unitarity gauge. In \fref{app:uni-sti} we have defined unitarity gauge as 
 the limit $\xi\to\infty$ of the linear realized theory in $R_\xi$ gauge. Since it can  be equivalently obtained from 
 nonlinear realized symmetries by transforming the Goldstone bosons away \cite{Grosse-Knetter:1993},
it follows that the groves obtained from the reduced sets of flips are also consistent in unitarity gauge, i.e. they satisfy the appropriate Ward Identities. Of course it is not sensible to speak of `gauge invariance classes' in a fixed gauge but this result at least indicates that no numerical instabilities due to violations of Ward Identities appear. 
\section{Structure of the groves}
\subsection{Linear parametrization}
\begin{figure}[htbp]
\begin{multline*}
\parbox{22mm}{
\begin{fmfgraph*}(22,17)
    \fmfleft{i1,i2}
    \fmfright{o3,o5,o4}
  \fmf{fermion}{i1,v1}
  \fmfdot{v1}
  \fmf{fermion}{v1,v4}
  \fmfdot{v1,v4}
  \fmf{photon}{v16,v4}
  \fmfdot{v4,v16}
  \fmfdot{v16}
  \fmf{fermion}{v16,i2}
  \fmfdot{v16}
  \fmf{fermion,tension=0.5}{o4,v16}
  \fmfdot{v4}
  \fmf{fermion,tension=0.5}{v4,o5}
  \fmfdot{v1}
  \fmf{dashes,tension=0.5}{v1,o3}
  \end{fmfgraph*}}
\quad\parbox{22mm}{
  \begin{fmfgraph*}(22,17)
    \fmfleft{i1,i2}
    \fmfright{o3,o5,o4}
  \fmf{fermion}{i1,v1}
  \fmfdot{v1}
  \fmf{photon}{v4,v1}
  \fmfdot{v1,v4}
  \fmfdot{v4}
  \fmf{fermion}{v4,i2}
  \fmfdot{v4}
  \fmf{fermion,tension=0.5}{o4,v4}
  \fmf{fermion}{v1,v5}
  \fmfdot{v1,v5}
  \fmfdot{v5}
  \fmf{fermion,tension=0.5}{v5,o5}
  \fmfdot{v5}
  \fmf{dashes,tension=0.5}{v5,o3}
  \end{fmfgraph*}}
\quad\parbox{22mm}{
  \begin{fmfgraph*}(22,17)
    \fmfleft{i1,i2}
    \fmfright{o5,o3,o4}
  \fmf{fermion}{i1,v1}
  \fmfdot{v1}
  \fmf{photon}{v4,v1}
  \fmfdot{v1,v4}
  \fmfdot{v4}
  \fmf{fermion}{v4,i2}
  \fmf{fermion}{v17,v4}
  \fmfdot{v4,v17}
  \fmfdot{v17}
  \fmf{fermion,tension=0.5}{o4,v17}
  \fmfdot{v17}
  \fmf{dashes,tension=0.5}{v17,o3}
  \fmfdot{v1}
  \fmf{fermion,tension=0.5}{v1,o5}
  \end{fmfgraph*}}
\quad\parbox{22mm}{
  \begin{fmfgraph*}(22,17)
    \fmfleft{i1,i2}
    \fmfright{o5,o4,o3}
  \fmf{fermion}{i1,v1}
  \fmfdot{v1}
  \fmf{photon}{v4,v1}
  \fmfdot{v1,v4}
  \fmf{fermion}{v4,v16}
  \fmfdot{v4,v16}
  \fmfdot{v16}
  \fmf{fermion}{v16,i2}
  \fmfdot{v16}
  \fmf{dashes,tension=0.5}{v16,o3}
  \fmfdot{v4}
  \fmf{fermion,tension=0.5}{o4,v4}
  \fmfdot{v1}
  \fmf{fermion,tension=0.5}{v1,o5}
  \end{fmfgraph*}}\\
\quad\parbox{22mm}{
  \begin{fmfgraph*}(22,17)
    \fmfleft{i1,i2}
    \fmfright{o5,o3,o4}
  \fmf{fermion}{i1,v1}
  \fmfdot{v1}
  \fmf{photon}{v4,v1}
  \fmfdot{v1,v4}
  \fmf{photon}{v16,v4}
  \fmfdot{v4,v16}
  \fmfdot{v16}
  \fmf{fermion}{v16,i2}
  \fmfdot{v16}
  \fmf{fermion,tension=0.5}{o4,v16}
  \fmfdot{v4}
  \fmf{dashes,tension=0.5}{v4,o3}
  \fmfdot{v1}
  \fmf{fermion,tension=0.5}{v1,o5}
  \end{fmfgraph*}}
\quad\parbox{22mm}{
  \begin{fmfgraph*}(22,17)
    \fmfleft{i1,i2}
    \fmfright{o5,o3,o4}
  \fmf{fermion}{i1,v1}
  \fmfdot{v1}
  \fmf{photon}{v4,v1}
  \fmfdot{v1,v4}
  \fmf{dashes}{v4,v16}
  \fmfdot{v4,v16}
  \fmfdot{v16}
  \fmf{fermion}{v16,i2}
  \fmfdot{v16}
  \fmf{fermion,tension=0.5}{o4,v16}
  \fmfdot{v4}
  \fmf{dashes,tension=0.5}{v4,o3}
  \fmfdot{v1}
  \fmf{fermion,tension=0.5}{v1,o5}
  \end{fmfgraph*}}
\quad\parbox{22mm}{
  \begin{fmfgraph*}(22,17)
    \fmfleft{i1,i2}
    \fmfright{o5,o3,o4}
  \fmf{fermion}{i1,v1}
  \fmfdot{v1}
  \fmf{dashes}{v1,v4}
  \fmfdot{v1,v4}
  \fmf{photon}{v16,v4}
  \fmfdot{v4,v16}
  \fmfdot{v16}
  \fmf{fermion}{v16,i2}
  \fmfdot{v16}
  \fmf{fermion,tension=0.5}{o4,v16}
  \fmfdot{v4}
  \fmf{dashes,tension=0.5}{v4,o3}
  \fmfdot{v1}
  \fmf{fermion,tension=0.5}{v1,o5}
  \end{fmfgraph*}}
\end{multline*}
 \caption{$t$-channel grove for $\bar f f \to \bar f f H$ in a spontaneously broken gauge theory }
\label{fig:4qh-grove}
\end{figure}
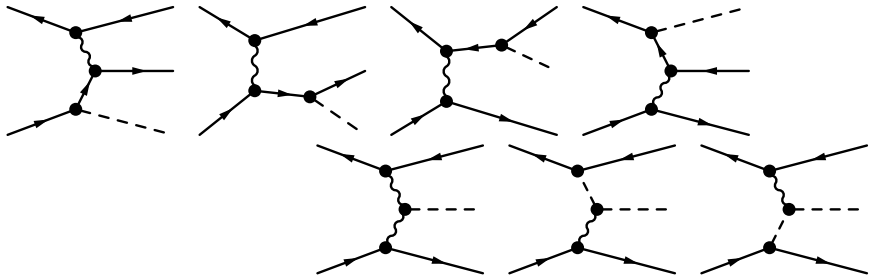

To analyze the groves in spontaneously broken gauge theories, we have implemented a spontaneously broken,  nonabelian gauge theory in the program \texttt{bocages} \cite{Ohl:bocages}. As an example, we consider the amplitude for the process $\bar f f \to f\bar f H$. The only new features compared to the QCD example $\bar q q \to \bar q q g$ from \fref{eq:qcd-grove} are single diagram groves that consist of diagrams without gauge bosons.
The remaining groves are similar to the groves in QCD with the external gluon replaced by a Higgs boson. An example is 
  shown in \fref{fig:4qh-grove}. 
Similarly, in the process $\bar f f \to \bar f f W$ there appear only two  groves  like in the case of QCD discussed in \fref{sec:qcd-groves} they include, of course, additional diagrams involving Higgs bosons. 

Apart from the one-diagram groves, the Higgs flips \eqref{subeq:higgs_flips} don't lead to additional groves. If there is \emph{one} gauge boson in a diagram, the gauge flips can always be used to flip to diagrams with more than one internal gauge boson. For example the diagrams   
\begin{equation*}
\parbox{20mm}{
  \begin{fmfgraph*}(20,15)
    \fmfleft{i1,i2}
    \fmfright{o5,o3,o4}
    \fmf{fermion}{i1,v1}
  \fmfdot{v1}
  \fmf{dashes}{v1,v4}
  \fmfdot{v1,v4}
  \fmf{photon}{v16,v4}
  \fmfdot{v4,v16}
  \fmfdot{v16}
  \fmf{fermion}{v16,i2}
  \fmfdot{v16}
  \fmf{fermion,tension=0.5}{o4,v16}
  \fmfdot{v4}
  \fmf{dashes,tension=0.5}{o3,v4}
  \fmfdot{v1}
  \fmf{fermion,tension=0.5}{v1,o5}
  \end{fmfgraph*}}\quad\leftrightarrow\quad
  \parbox{20mm}{
\begin{fmfgraph*}(20,15)
    \fmfleft{i1,i2}
    \fmfright{o5,o3,o4}
    \fmf{fermion}{i1,v1}
  \fmfdot{v1}
  \fmf{photon}{v1,v4}
  \fmfdot{v1,v4}
  \fmf{photon}{v16,v4}
  \fmfdot{v4,v16}
  \fmfdot{v16}
  \fmf{fermion}{v16,i2}
  \fmfdot{v16}
  \fmf{fermion,tension=0.5}{o4,v16}
  \fmfdot{v4}
  \fmf{dashes,tension=0.5}{o3,v4}
  \fmfdot{v1}
  \fmf{fermion,tension=0.5}{v1,o5}
  \end{fmfgraph*}}
\end{equation*}
are  connected by a gauge flip from \fref{eq:h_flips_1}. 

It turns out that this is the generic structure: the only new groves compared to the case of unbroken gauge theories consist of one diagram each,  where all internal particles are Higgs bosons. The remaining diagrams fall in the same groves that were discussed in \fref{sec:groves}. 

Since the Higgs-fermion Yukawa couplings are proportional to the fermion masses, the coupling of the Higgs boson to light fermions is usually set to zero in practical calculations. 
Of course this is a  consistent approximation if  the masses of light fermions are also set to zero. From \fref{fig:4qh-grove} we see that the set of diagrams obtained by neglecting the coupling to light fermions  in general does \emph{not} correspond to a gauge invariant subset if the fermion masses are not set to zero. In general, the numerical instabilities caused by this inconsistency are negligible but there can be some small corners in phase space where they become relevant. 
\subsection{Nonlinear realizations}\label{sec:nl-groves}
In the case of nonlinear realized symmetries, the gauge flips simplify as we have shown in \fref{sec:nl-flips} so there is a more interesting structure of the groves than in the case of linearly parametrized scalar sectors.

As a first example, we consider the process $\bar f f \to \bar f f W$. Let us discuss the case of a single Higgs boson in a nonlinear parametrization  first. In this case the Higgs boson is a singlet under the unbroken symmetry group and there there is no $HHW$ vertex. We see from the gauge flips in \fref{subeq:h_flips} that in this case the gauge flips `conserve' Higgs number in the sense that only external Higgs bosons appear. Therefore gauge flips cannot change the number of Higgs bosons and the groves can be classified according to the number of internal Higgs bosons. 

\begin{figure}[htbp]
\begin{equation*}
\parbox{22mm}{
  \begin{fmfgraph*}(22,17)
    \fmfleft{i1,i2}
    \fmfright{o5,o3,o4}
 \fmf{fermion}{i1,v1}
    \fmfdot{v1}
  \fmf{dashes}{v1,v4}
  \fmfdot{v1,v4}
  \fmfdot{v4}
  \fmf{fermion}{v4,i2}
  \fmf{fermion}{v17,v4}
  \fmfdot{v4,v17}
  \fmfdot{v17}
  \fmf{fermion,tension=0.5}{o4,v17}
  \fmfdot{v17}
  \fmf{photon,tension=0.5}{o3,v17}
  \fmfdot{v1}
  \fmf{fermion,tension=0.5}{v1,o5}
  \end{fmfgraph*}}\quad
\parbox{22mm}{
  \begin{fmfgraph*}(22,17)
    \fmfleft{i1,i2}
    \fmfright{o5,o4,o3}
    \fmf{fermion}{i1,v1}
  \fmfdot{v1}
  \fmf{dashes}{v1,v4}
  \fmfdot{v1,v4}
  \fmf{fermion}{v4,v16}
  \fmfdot{v4,v16}
  \fmfdot{v16}
  \fmf{fermion}{v16,i2}
  \fmfdot{v16}
  \fmf{photon,tension=0.5}{o3,v16}
  \fmfdot{v4}
  \fmf{fermion,tension=0.5}{o4,v4}
  \fmfdot{v1}
  \fmf{fermion,tension=0.5}{v1,o5}
  \end{fmfgraph*}}\quad
\parbox{22mm}{%
  \begin{fmfgraph*}(22,17)
    \fmfleft{i1,i2}
    \fmfright{o5,o3,o4}
    \fmf{fermion}{i1,v1}
  \fmfdot{v1}
  \fmf{dashes}{v1,v4}
  \fmfdot{v1,v4}
  \fmf{photon}{v16,v4}
  \fmfdot{v4,v16}
  \fmfdot{v16}
  \fmf{fermion}{v16,i2}
  \fmfdot{v16}
  \fmf{fermion,tension=0.5}{o4,v16}
  \fmfdot{v4}
  \fmf{photon,tension=0.5}{o3,v4}
  \fmfdot{v1}
  \fmf{fermion,tension=0.5}{v1,o5}
  \end{fmfgraph*}}
 \end{equation*}
 \caption{Mixed grove for $\bar f f \to \bar f f W$ for a singlet Higgs }
\label{fig:4qw-grove-uni}
\end{figure}
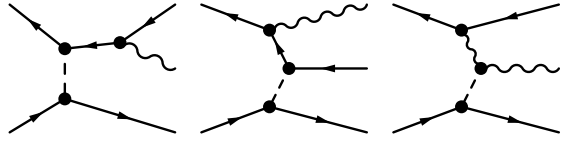

We find that the amplitude for  $\bar f f \to \bar f f W$ can be decomposed into 6 groves instead of the 2 appearing in unbroken gauge theories and in the linear parametrization of spontaneously broken gauge theories. Two groves are  `gauge groves' consisting of  5 diagrams without internal Higgs boson that look exactly like in QCD  \eqref{eq:qcd-grove}. The remaining  groves are `mixed groves' with one internal Higgs boson. An example of a mixed grove is given in \fref{fig:4qw-grove-uni}. 
Here all diagrams are proportional to  the coupling of the Higgs to one fermion pair, but this is not always the case as we will see below. 
In this example, two mixed groves correspond to one gauge grove. This additional structure arises because there is no gauge flip between the two diagrams
\begin{equation}\label{eq:no-flip}
\parbox{20mm}{ 
 \begin{fmfgraph*}(20,15)
    \fmfleft{i1,i2}
    \fmfright{o5,o3,o4}
  \fmf{fermion}{i1,v1}
  \fmfdot{v1}
  \fmf{photon}{v4,v1}
  \fmfdot{v1,v4}
  \fmf{dashes}{v4,v16}
  \fmfdot{v4,v16}
  \fmfdot{v16}
  \fmf{fermion}{v16,i2}
  \fmfdot{v16}
  \fmf{fermion,tension=0.5}{o4,v16}
  \fmfdot{v4}
  \fmf{photon,tension=0.5}{o3,v4}
  \fmfdot{v1}
  \fmf{fermion,tension=0.5}{v1,o5}
  \end{fmfgraph*}}\quad \nleftrightarrow\quad
\parbox{20mm}{ 
 \begin{fmfgraph*}(20,15)
    \fmfleft{i1,i2}
    \fmfright{o5,o3,o4}
  \fmf{fermion}{i1,v1}
  \fmfdot{v1}
  \fmf{dashes}{v4,v1}
  \fmfdot{v1,v4}
  \fmf{photon}{v4,v16}
  \fmfdot{v4,v16}
  \fmfdot{v16}
  \fmf{fermion}{v16,i2}
  \fmfdot{v16}
  \fmf{fermion,tension=0.5}{o4,v16}
  \fmfdot{v4}
  \fmf{photon,tension=0.5}{o3,v4}
  \fmfdot{v1}
  \fmf{fermion,tension=0.5}{v1,o5}
  \end{fmfgraph*}} 
\end{equation}

In a nonlinear realization of a more complicated Higgs sector, the Higgs bosons transform according to a linear representation of the unbroken subgroup and therefore couple to the massless gauge boson through $HHW$ vertices. This enables an indirect flip between the two diagrams from \eqref{eq:no-flip} since the flips  \eqref{eq:h_flips_1} have to be included in the gauge flips also in the nonlinear parametrization:
\begin{equation}\label{eq:indirect-flip}
\parbox{20mm}{ 
 \begin{fmfgraph*}(20,15)
    \fmfleft{i1,i2}
    \fmfright{o5,o3,o4}
  \fmf{fermion}{i1,v1}
  \fmfdot{v1}
  \fmf{photon}{v4,v1}
  \fmfdot{v1,v4}
  \fmf{dashes}{v4,v16}
  \fmfdot{v4,v16}
  \fmfdot{v16}
  \fmf{fermion}{v16,i2}
  \fmfdot{v16}
  \fmf{fermion,tension=0.5}{o4,v16}
  \fmfdot{v4}
  \fmf{photon,tension=0.5}{o3,v4}
  \fmfdot{v1}
  \fmf{fermion,tension=0.5}{v1,o5}
  \end{fmfgraph*}}\quad \leftrightarrow\quad
\parbox{20mm}{ 
 \begin{fmfgraph*}(20,15)
    \fmfleft{i1,i2}
    \fmfright{o5,o3,o4}
  \fmf{fermion}{i1,v1}
  \fmfdot{v1}
  \fmf{dashes}{v4,v1}
  \fmfdot{v1,v4}
  \fmf{dashes}{v4,v16}
  \fmfdot{v4,v16}
  \fmfdot{v16}
  \fmf{fermion}{v16,i2}
  \fmfdot{v16}
  \fmf{fermion,tension=0.5}{o4,v16}
  \fmfdot{v4}
  \fmf{photon,tension=0.5}{o3,v4}
  \fmfdot{v1}
  \fmf{fermion,tension=0.5}{v1,o5}
  \end{fmfgraph*}}\quad \leftrightarrow\quad
\parbox{20mm}{ 
 \begin{fmfgraph*}(20,15)
    \fmfleft{i1,i2}
    \fmfright{o5,o3,o4}
  \fmf{fermion}{i1,v1}
  \fmfdot{v1}
  \fmf{dashes}{v4,v1}
  \fmfdot{v1,v4}
  \fmf{photon}{v4,v16}
  \fmfdot{v4,v16}
  \fmfdot{v16}
  \fmf{fermion}{v16,i2}
  \fmfdot{v16}
  \fmf{fermion,tension=0.5}{o4,v16}
  \fmfdot{v4}
  \fmf{photon,tension=0.5}{o3,v4}
  \fmfdot{v1}
  \fmf{fermion,tension=0.5}{v1,o5}
  \end{fmfgraph*}} 
\end{equation}
We see that the `Higgs conservation' in the gauge flips breaks down. Therefore the appearance of Higgs bosons charged under the unbroken subgroup reduces the number of groves. 
One finds that mixed groves are still present, however, in general only \emph{one} mixed grove corresponds to a gauge grove and they don't contain a fixed number of Higgs bosons.

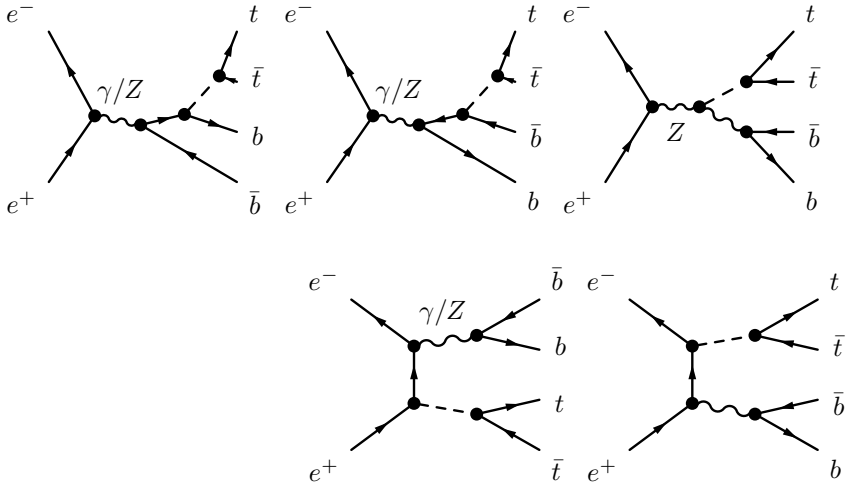
\begin{figure}[htbp]
\begin{center}
\begin{multline*}
\fmfframe(6,7)(6,6){
  \begin{fmfgraph*}(25,20)
    \fmfleft{i1,i2}
    \fmfright{o5,o4,o3,o6}
  \fmflabel{$e^+$}{i1}
  \fmf{fermion}{i1,v1}
  \fmfdot{v1}
  \fmflabel{$e^-$}{i2}
  \fmfdot{v1}
  \fmf{fermion}{v1,i2}
  \fmf{photon,label=$\gamma/Z$,tension=2}{v5,v1}
  \fmfdot{v1,v5}
  \fmf{fermion}{v5,v20}
  \fmfdot{v5,v20}
  \fmf{dashes,tension=0.5}{v20,v80}
  \fmfdot{v20,v80}
  \fmflabel{$t$}{o6}
  \fmfdot{v80}
  \fmf{fermion,tension=0.5}{v80,o6}
  \fmflabel{$\bar t$}{o3}
  \fmfdot{v80}
  \fmf{fermion,tension=0.5}{o3,v80}
  \fmflabel{$b$}{o4}
  \fmfdot{v20}
  \fmf{fermion,tension=0.5}{v20,o4}
  \fmflabel{$\bar b$}{o5}
  \fmfdot{v5}
  \fmf{fermion,tension=0.5}{o5,v5}
  \end{fmfgraph*}}
\fmfframe(6,7)(6,6){%
  \begin{fmfgraph*}(25,20)
    \fmfleft{i1,i2}
    \fmfright{o4,o5,o3,o6}
  \fmflabel{$e^+$}{i1}
  \fmf{fermion}{i1,v1}
  \fmfdot{v1}
  \fmflabel{$e^-$}{i2}
  \fmfdot{v1}
  \fmf{fermion}{v1,i2}
  \fmf{photon,label=$\gamma/Z$,tension=2}{v5,v1}
  \fmfdot{v1,v5}
  \fmf{fermion}{v20,v5}
  \fmfdot{v5,v20}
  \fmf{dashes,tension=0.5}{v20,v80}
  \fmfdot{v20,v80}
  \fmflabel{$t$}{o6}
  \fmfdot{v80}
  \fmf{fermion,tension=0.5}{v80,o6}
  \fmflabel{$\bar t$}{o3}
  \fmfdot{v80}
  \fmf{fermion,tension=0.5}{o3,v80}
  \fmflabel{$\bar b$}{o5}
  \fmfdot{v20}
  \fmf{fermion,tension=0.5}{o5,v20}
  \fmflabel{$b$}{o4}
  \fmfdot{v5}
  \fmf{fermion,tension=0.5}{v5,o4}
  \end{fmfgraph*}}
\fmfframe(6,7)(6,6){%
  \begin{fmfgraph*}(25,20)
    \fmfleft{i1,i2}
    \fmfright{o4,o5,o3,o6}
  \fmflabel{$e^+$}{i1}
  \fmf{fermion}{i1,v1}
  \fmfdot{v1}
  \fmflabel{$e^-$}{i2}
  \fmfdot{v1}
  \fmf{fermion}{v1,i2}
  \fmf{photon,label=$Z$,tension=2}{v5,v1}
  \fmfdot{v1,v5}
  \fmf{dashes}{v5,v20}
  \fmfdot{v5,v20}
  \fmflabel{$t$}{o6}
  \fmfdot{v20}
  \fmf{fermion,tension=0.5}{v20,o6}
  \fmflabel{$\bar t$}{o3}
  \fmfdot{v20}
  \fmf{fermion,tension=0.5}{o3,v20}
  \fmf{photon}{v21,v5}
  \fmfdot{v5,v21}
  \fmflabel{$\bar b$}{o5}
  \fmfdot{v21}
  \fmf{fermion,tension=0.5}{o5,v21}
  \fmflabel{$b$}{o4}
  \fmfdot{v21}
  \fmf{fermion,tension=0.5}{v21,o4}
  \end{fmfgraph*}}\\
\fmfframe(6,7)(6,6){%
  \begin{fmfgraph*}(25,20)
    \fmfleft{i1,i2}
    \fmfright{o3,o6,o4,o5}
    \fmflabel{$e^+$}{i1}
  \fmf{fermion}{i1,v1}
  \fmfdot{v1}
  \fmf{fermion}{v1,v4}
  \fmfdot{v1,v4}
  \fmflabel{$e^-$}{i2}
  \fmfdot{v4}
  \fmf{fermion}{v4,i2}
  \fmf{photon,label=$\gamma /Z$}{v17,v4}
  \fmfdot{v4,v17}
  \fmflabel{$\bar b$}{o5}
  \fmfdot{v17}
  \fmf{fermion,tension=0.5}{o5,v17}
  \fmflabel{$b$}{o4}
  \fmfdot{v17}
  \fmf{fermion,tension=0.5}{v17,o4}
  \fmf{dashes}{v1,v5}
  \fmfdot{v1,v5}
  \fmflabel{$t$}{o6}
  \fmfdot{v5}
  \fmf{fermion,tension=0.5}{v5,o6}
  \fmflabel{$\bar t$}{o3}
  \fmfdot{v5}
  \fmf{fermion,tension=0.5}{o3,v5}
  \end{fmfgraph*}}
\fmfframe(6,7)(6,6){%
  \begin{fmfgraph*}(25,20)
    \fmfleft{i1,i2}
    \fmfright{o4,o5,o3,o6}
   \fmflabel{$e^+$}{i1}
  \fmf{fermion}{i1,v1}
  \fmfdot{v1}
  \fmf{fermion}{v1,v4}
  \fmfdot{v1,v4}
  \fmflabel{$e^-$}{i2}
  \fmfdot{v4}
  \fmf{fermion}{v4,i2}
  \fmf{dashes}{v4,v17}
  \fmfdot{v4,v17}
  \fmflabel{$t$}{o6}
  \fmfdot{v17}
  \fmf{fermion,tension=0.5}{v17,o6}
  \fmflabel{$\bar t$}{o3}
  \fmfdot{v17}
  \fmf{fermion,tension=0.5}{o3,v17}
  \fmf{photon}{v5,v1}
  \fmfdot{v1,v5}
  \fmflabel{$\bar b$}{o5}
  \fmfdot{v5}
  \fmf{fermion,tension=0.5}{o5,v5}
  \fmflabel{$b$}{o4}
  \fmfdot{v5}
  \fmf{fermion,tension=0.5}{v5,o4}
  \end{fmfgraph*}}
\end{multline*}
\end{center}
\caption{Mixed grove in the $e^+e- \to b\bar b t\bar t$ amplitude} \label{fig:6f-grove}
\end{figure}

The same structure is found in amplitudes with more external particles. Let us first consider the 6-fermion amplitude. For a single Higgs boson, 18 mixed groves  are obtained by inserting a fermion-antifermion pair via a Higgs boson in all possible places in the  gauge groves of the $4f$ amplitude. An application to the process $e^+ e^- \to b\bar b t \bar t$ that is relevant for the measurement of the top Yukawa coupling is shown in \fref{fig:6f-grove}. Again all the diagrams in the mixed groves are proportional to  the coupling of the Higgs to one fermion pair. For a general higgs sector, because of the additional flips as in \fref{eq:indirect-flip} only 6 mixed groves remain, corresponding to the gauge groves.

\begin{figure}[htbp]
\begin{center}
\begin{equation*}
\parbox{25mm}{
  \begin{fmfgraph*}(25,20)
    \fmfleft{i1,i2}
    \fmfright{o6,o4,o3,o5}
  \fmf{fermion}{i1,v1}
  \fmfdot{v1}
  \fmf{dashes}{v1,v4}
  \fmfdot{v1,v4}
  \fmf{photon}{v16,v4}
  \fmfdot{v4,v16}
  \fmf{dashes}{v16,v64}
  \fmfdot{v16,v64}
  \fmfdot{v64}
  \fmf{fermion}{v64,i2}
  \fmfdot{v64}
  \fmf{fermion,tension=0.5}{o5,v64}
  \fmfdot{v16}
  \fmf{photon,tension=0.5}{o3,v16}
  \fmfdot{v4}
  \fmf{photon,tension=0.5}{o4,v4}
  \fmfdot{v1}
  \fmf{fermion,tension=0.5}{v1,o6}
  \end{fmfgraph*}}\quad
\parbox{25mm}{
  \begin{fmfgraph*}(25,20)
    \fmfleft{i1,i2}
    \fmfright{o6,o3,o4,o5}
  \fmf{fermion}{i1,v1}
  \fmfdot{v1}
  \fmf{dashes}{v1,v4}
  \fmfdot{v1,v4}
  \fmf{photon}{v16,v4}
  \fmfdot{v4,v16}
  \fmf{dashes}{v16,v64}
  \fmfdot{v16,v64}
  \fmfdot{v64}
  \fmf{fermion}{v64,i2}
  \fmfdot{v64}
  \fmf{fermion,tension=0.5}{o5,v64}
  \fmfdot{v16}
  \fmf{phantom,tension=0.5}{o4,v16}
  \fmfdot{v4}
  \fmf{phantom,tension=0.5}{o3,v4}
  \fmfdot{v1}
  \fmf{fermion,tension=0.5}{v1,o6}
  \fmffreeze
  \fmf{photon,tension=0.5}{o3,v16}
  \fmfdot{v4}
  \fmf{photon,tension=0.5}{o4,v4}
  \end{fmfgraph*}}\quad
\parbox{25mm}{
  \begin{fmfgraph*}(25,20)
    \fmfleft{i1,i2}
    \fmfright{o6,o3,o4,o5}
  \fmf{fermion}{i1,v1}
  \fmfdot{v1}
  \fmf{dashes}{v1,v4}
  \fmfdot{v1,v4}
  \fmf{dashes}{v4,v16}
  \fmfdot{v4,v16}
  \fmfdot{v16}
  \fmf{fermion}{v16,i2}
  \fmfdot{v16}
  \fmf{fermion,tension=0.5}{o5,v16}
  \fmfdot{v4}
  \fmf{photon,tension=0.5}{o4,v4}
  \fmfdot{v4}
  \fmf{photon,tension=0.5}{o3,v4}
  \fmfdot{v1}
  \fmf{fermion,tension=0.5}{v1,o6}
  \end{fmfgraph*}}
\end{equation*}
\end{center}
\caption{Mixed grove with 2 Higgs bosons in the 4 fermion 2 gauge boson amplitude}\label{fig:2hgrove}
\end{figure}
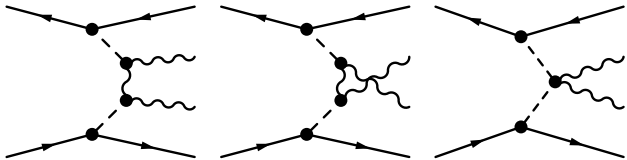

 As a final  example we discuss the amplitude $\bar f f\to\bar f f WW$. For a single Higgs boson, the Higgs number conservation leads to the appearance of 2 mixed groves with 2 Higgs and one gauge boson. One example is shown in \fref{fig:2hgrove}.

\begin{figure}[htbp]
\begin{center}
\begin{align*}
&\parbox{25mm}{
  \begin{fmfgraph*}(25,20)
    \fmfleft{i1,i2}
    \fmfright{o5,o3,o4,o6}
  \fmf{fermion}{i1,v1}
  \fmfdot{v1}
  \fmfdot{v1}
  \fmf{fermion}{v1,i2}
  \fmf{photon,tension=2.0}{v5,v1}
  \fmfdot{v1,v5}
  \fmf{fermion}{v5,v20}
  \fmfdot{v5,v20}
  \fmfdot{v20}
  \fmf{fermion,tension=0.5}{v20,o6}
  \fmf{dashes,tension=0.5}{v20,v81}
  \fmfdot{v20,v81}
  \fmfdot{v81}
  \fmf{photon,tension=0.5}{o4,v81}
  \fmfdot{v81}
  \fmf{photon,tension=0.5}{o3,v81}
  \fmfdot{v5}
  \fmf{fermion,tension=0.5}{o5,v5}
  \end{fmfgraph*}}
\qquad
\parbox{25mm}{
  \begin{fmfgraph*}(25,20)
    \fmfleft{i1,i2}
    \fmfright{o3,o4,o5,o6}
  \fmf{fermion}{i1,v1}
  \fmfdot{v1}
  \fmfdot{v1}
  \fmf{fermion}{v1,i2}
  \fmf{photon,tension=2.0}{v5,v1}
  \fmfdot{v1,v5}
  \fmfdot{v5}
  \fmf{fermion,tension=0.5}{v5,o6}
  \fmf{fermion}{v21,v5}
  \fmfdot{v5,v21}
  \fmfdot{v21}
  \fmf{fermion,tension=0.5}{o5,v21}
  \fmf{dashes,tension=0.5}{v21,v85}
  \fmfdot{v21,v85}
  \fmfdot{v85}
  \fmf{photon,tension=0.5}{o4,v85}
  \fmfdot{v85}
  \fmf{photon,tension=0.5}{o3,v85}
  \end{fmfgraph*}}
\qquad \parbox{25mm}{
  \begin{fmfgraph*}(25,20)
    \fmfleft{i1,i2}
    \fmfright{o3,o4,o5,o6}
  \fmf{fermion}{i1,v1}
  \fmfdot{v1}
  \fmfdot{v1}
  \fmf{fermion}{v1,i2}
  \fmf{photon,tension=2.0}{v5,v1}
  \fmfdot{v1,v5}
  \fmf{photon}{v20,v5}
  \fmfdot{v5,v20}
   \fmfdot{v20}
  \fmf{fermion,tension=0.5}{v20,o6}
  \fmfdot{v20}
  \fmf{fermion,tension=0.5}{o5,v20}
  \fmf{dashes}{v5,v21}
  \fmfdot{v5,v21}
  \fmfdot{v21}
  \fmf{photon,tension=0.5}{o4,v21}
   \fmfdot{v21}
  \fmf{photon,tension=0.5}{o3,v21}
  \end{fmfgraph*}}\\
&\parbox{25mm}{
  \begin{fmfgraph*}(25,20)
    \fmfleft{i1,i2}
    \fmfright{o3,o4,o5,o6}
  \fmf{fermion}{i1,v1}
  \fmfdot{v1}
  \fmf{fermion}{v1,v4}
  \fmfdot{v1,v4}
  \fmfdot{v4}
  \fmf{fermion}{v4,i2}
  \fmf{photon}{v17,v4}
  \fmfdot{v4,v17}
  \fmfdot{v17}
  \fmf{fermion,tension=0.5}{v17,o6}
  \fmfdot{v17}
  \fmf{fermion,tension=0.5}{o5,v17}
  \fmf{dashes}{v1,v5}
  \fmfdot{v1,v5}
  \fmfdot{v5}
  \fmf{photon,tension=0.5}{o4,v5}
  \fmfdot{v5}
  \fmf{photon,tension=0.5}{o3,v5}
  \end{fmfgraph*}}
\qquad \parbox{25mm}{
  \begin{fmfgraph*}(25,20)
    \fmfleft{i1,i2}
    \fmfright{o5,o6,o3,o4}
  \fmf{fermion}{i1,v1}
  \fmfdot{v1}
  \fmf{fermion}{v1,v4}
  \fmfdot{v1,v4}
  \fmfdot{v4}
  \fmf{fermion}{v4,i2}
  \fmf{dashes}{v4,v17}
  \fmfdot{v4,v17}
  \fmfdot{v17}
  \fmf{photon,tension=0.5}{o4,v17}
  \fmfdot{v17}
  \fmf{photon,tension=0.5}{o3,v17}
  \fmf{photon}{v5,v1}
  \fmfdot{v1,v5}
  \fmfdot{v5}
  \fmf{fermion,tension=0.5}{v5,o6}
  \fmfdot{v5}
  \fmf{fermion,tension=0.5}{o5,v5}
  \end{fmfgraph*}}
\end{align*}
\end{center}
\caption{Mixed grove in the 4 fermion 2 gauge boson amplitude} \label{fig:4f2w-grove}
\end{figure}
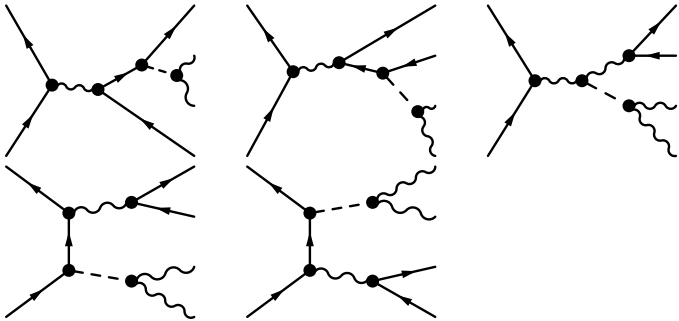

Furthermore, apart from  2 gauge groves (and several one diagram groves), there are  10 mixed groves with one Higgs boson. 
In  the example shown in  \fref{fig:4f2w-grove}, th grove is obtained by inserting the external gauge boson pair via the Higgs boson into the $\bar f f \to \bar f f$ amplitude. Therefore all diagrams of this grove are proportional to the gauge boson-Higgs coupling.
 However, in general it is not true that all mixed groves are proportional to one Higgs coupling. An example is provided by the three diagrams
\begin{equation}
\parbox{25mm}{
  \begin{fmfgraph*}(25,20)
    \fmfleft{i1,i2}
    \fmfright{o3,o6,o5,o4}
  \fmf{fermion}{i1,v1}
  \fmfdot{v1}
  \fmf{fermion}{v1,v4}
  \fmfdot{v1,v4}
  \fmf{dashes}{v4,v16}
  \fmfdot{v4,v16}
  \fmf{fermion}{v16,v64}
  \fmfdot{v16,v64}
  \fmfdot{v64}
  \fmf{fermion}{v64,i2}
  \fmfdot{v64}
  \fmf{photon,tension=0.5}{o4,v64}
  \fmfdot{v16}
  \fmf{fermion,tension=0.5}{o5,v16}
  \fmfdot{v4}
  \fmf{fermion,tension=0.5}{v4,o6}
  \fmfdot{v1}
  \fmf{photon,tension=0.5}{o3,v1}
  \end{fmfgraph*}}\leftrightarrow
\parbox{25mm}{
  \begin{fmfgraph*}(25,20)
    \fmfleft{i1,i2}
    \fmfright{o3,o6,o4,o5}
  \fmf{fermion}{i1,v1}
  \fmfdot{v1}
  \fmf{fermion}{v1,v4}
  \fmfdot{v1,v4}
  \fmf{dashes}{v4,v16}
  \fmfdot{v4,v16}
  \fmf{photon}{v64,v16}
  \fmfdot{v16,v64}
   \fmfdot{v64}
  \fmf{fermion}{v64,i2}
   \fmfdot{v64}
  \fmf{fermion,tension=0.5}{o5,v64}
   \fmfdot{v16}
  \fmf{photon,tension=0.5}{o4,v16}
   \fmfdot{v4}
  \fmf{fermion,tension=0.5}{v4,o6}
   \fmfdot{v1}
  \fmf{photon,tension=0.5}{o3,v1}
  \end{fmfgraph*}} \leftrightarrow
\parbox{25mm}{
  \begin{fmfgraph*}(25,20)
    \fmfleft{i1,i2}
    \fmfright{o6,o3,o4,o5}
  \fmf{fermion}{i1,v1}
  \fmfdot{v1}
  \fmf{photon}{v4,v1}
  \fmfdot{v1,v4}
  \fmf{dashes}{v4,v16}
  \fmfdot{v4,v16}
  \fmf{photon}{v64,v16}
  \fmfdot{v16,v64}
  \fmfdot{v64}
  \fmf{fermion}{v64,i2}
  \fmfdot{v64}
  \fmf{fermion,tension=0.5}{o5,v64}
  \fmfdot{v16}
  \fmf{photon,tension=0.5}{o4,v16}
  \fmfdot{v4}
  \fmf{photon,tension=0.5}{o3,v4}
  \fmfdot{v1}
  \fmf{fermion,tension=0.5}{v1,o6}
  \end{fmfgraph*}}
\end{equation}
that are connected by flips from \fref{eq:h_flips_1} and that have no Higgs coupling in common. Again, in a more general Higgs sector with $HHW$ vertices the `Higgs conservation law' breaks down and only 2 mixed groves remain. 

\subsection{Numerical checks}
We have checked the consistency of the groves discussed in the previous sections in the Standard Model numerically. We have generated the  matrix elements  for processes with up to 6 external fermions using \omegacite. We use the following criteria:  
\begin{itemize}
\item comparison of unitarity and $R_\xi$ gauge. The error quoted is the relative difference of the squared matrix elements:
\begin{equation}
\delta_{\text{rel}}=\frac{|\me(U)|^2-|\me(R_\xi)|^2}{|\me(U)|^2+|\me(R_\xi)|^2}
\end{equation}
\item violation of the Ward Identities for the internal (and external gauge bosons). The error quoted is the quantity 
\begin{equation}
\delta_{\text{rel}}=\frac{|\ii\tfrac{p^\mu}{m}W_\mu +\phi|}{\|\tfrac{p}{m}\|\, \||W\|}
\end{equation}
where $\|x\|$ denotes the \emph{euclidean} norm of Lorentz-vectors. (see \fref{sec:omega-wi})
\end{itemize}
In \fref{tab:g_groves} we show the results for the `gauge groves', obtained by setting all Higgs couplings to zero. As expected, this is consistent in unitarity gauge, i.e. all Ward Identities are satisfied. In $R_\xi$ gauge, the Ward Identities for Green's functions with 4 external particles are satisfied while the Ward Identities with 5 external particles are violated badly. This has to be expected, since according to the discussion in \fref{sec:2f2w-flips} the 4 particle gauge boson exchange diagrams satisfy the Ward Identity, but the 4 point diagrams with external Goldstone bosons that appear in the 5 point functions violate the Ward Identities. Starting from the 6 point functions, these violations of the 5 point Ward Identities cause inconsistencies between the matrix elements in unitarity gauge and $R_\xi$ gauge.
\begin{table}[htbp]
  \begin{center}
    \begin{tabular}{c|c|c|c}
     Process           &$U \leftrightarrow R_\xi$ &  WI ($U$)&   WI ($R_\xi$)  \\\hline
 $\bar b b \to W^+ W^-$ & $\surd$ &  $\surd$ &  $\surd$ \\
$\bar d d \to t \bar b W^-$ & $\surd$& $\surd$& $\delta_{\text{rel}}=\mathcal{O}(1)$ \\
$\bar d d \to t \bar t t\bar t$ & $\delta_{\text{rel}}=\mathcal{O}(10^{-3})$& $\surd$& $\delta_{\text{rel}}=\mathcal{O}(1)$\\
$\bar d d \to t \bar t b\bar b$ & $\delta_{\text{rel}}=\mathcal{O}(10^{-8})$& $\surd$& $\delta_{\text{rel}}=\mathcal{O}(1)$\\
$\bar d d \to t \bar b b\bar b W^-$ & $\delta_{\text{rel}}=\mathcal{O}(10^{-6})$& $\surd$& $\delta_{\text{rel}}=\mathcal{O}(1)$\\
$\bar d d \to b\bar b b\bar b W^- W^+$ & $\delta_{\text{rel}}=\mathcal{O}(10^{-5})$& $\surd$& $\delta_{\text{rel}}=\mathcal{O}(1)$
    \end{tabular}\caption{Consistency of the gauge groves}\label{tab:g_groves}
     \end{center}
\end{table}
Similar results are obtained for the `mixed groves' that can be  obtained by setting all internal Higgs boson but one to zero. 

\chapter{Finite width effects}\label{chap:width}
In \fref{sec:intro-widths} we have seen that the use of a naive Breit-Wigner propagator for unstable gauge bosons violates gauge invariance. We will review  this problem in more detail in this chapter and study several suggested solutions numerically. In \fref{sec:qft-widths} we will discuss briefly that even a careful field theoretic treatment of unstable gauge bosons in nonabelian gauge theories leads to violations of gauge invariance. 

In \fref{sec:width-schemes} we describe several simple schemes for the description of finite widths effects that have been introduced in the literature and their properties with respect to gauge invariance. 

In \fref{sec:widths-comparison} we compare these schemes numerically at the matrix element level and in \fref{sec:widths-numeric} we present our numerical results for cross sections in these schemes obtained with \omegacite~  and WHIZARD \cite{Kilian:WHIZARD} in the single $W$ production process.
\section{Dyson summation and violation of Ward Identities}\label{sec:qft-widths}
In quantum field theory, finite widths of unstable particles arise from imaginary parts of self energies. If one writes the full inverse propagator as
\begin{equation}\label{eq:dyson}
-\Gamma_{\Phi\Phi}(p^2)=D_\Phi^{-1}(p^2)\equiv {D^0_\Phi}^{-1}(p^2) - \ii \Pi_\Phi (p^2)
\end{equation}
with the self energy $\Pi_\Phi$ and the free propagator $D^0_\Phi$, then the propagator is given by
\begin{equation}
D_\Phi=\frac{D_\Phi^0}{1-\ii \Pi_\Phi D_\Phi^0}=D_\Phi^0\sum_{i=0}^\infty(\ii\Pi_\Phi D_\Phi^0)^n
\end{equation}
Therefore the solution of the so called \emph{Dyson equation} \eqref{eq:dyson} corresponds to a \emph{resummation} of Feynman diagrams:
\begin{equation*}
\parbox{14mm}{
\begin{fmfchar*}(14,14)
\fmfleft{l}
\fmfright{r}
\fmf{photon}{l,i,r}
\fmfblob{15pt}{i}
\end{fmfchar*}}\, =\,
\parbox{10mm}{
\begin{fmfchar*}(10,10)
\fmfleft{l}
\fmfright{r}
\fmf{photon}{l,r}
\end{fmfchar*}}\, +\,
\parbox{14mm}{
\begin{fmfchar*}(14,14)
\fmfleft{l}
\fmfright{r}
\fmf{photon}{l,i,r}
\fmfv{d.sh=circle,d.f=empty,d.size=15pt,l=$\Pi$,l.d=0}{i}
\end{fmfchar*}}\, +\,
\parbox{25mm}{
\begin{fmfchar*}(25,14)
\fmfleft{l}
\fmfright{r}
\fmf{photon}{l,i,j,r}
\fmfv{d.sh=circle,d.f=empty,d.size=15pt,l=$\Pi$,l.d=0}{i,j}
\end{fmfchar*}}\quad +\dots
\end{equation*}
For the simple case of a scalar particle we obtain as resummed propagator
\begin{equation}
D_\phi=\frac{\ii}{p^2-m^2+\Pi(p^2)}
\end{equation}
Defining the physical mass $M^2$ of the particle as the real part of the pole of the propagator 
\begin{equation}
M^2=m^2-\Re\Pi(M^2)
\end{equation}
and the width as
\begin{equation}
\Gamma=-\frac{1}{M}\Im\Pi(M^2)
\end{equation}
we can approximate the resummed propagator near the pole by a Breit-Wigner propagator  
\begin{equation}
D_\phi\sim\frac{\ii}{p^2-M^2-\ii M\Gamma}
\end{equation}

In the case of massive gauge bosons, the solution of the Dyson equation  is more complicated because of the Lorentz structure of the propagator and the self energy. 

Introducing the decomposition of the self energy into a transverse  and a longitudinal part, 
\begin{equation}
 \Pi_W^{\mu\nu}(q^2)=\Pi_{W}^T(q^2) \left(g^{\mu\nu}-\frac{q^\mu
  q^\nu}{q^2}\right)+\Pi_W^L(q^2) \frac{q^\mu q^\nu}{q^2}
\end{equation}
the resummed propagator in unitarity gauge  is given by: 
\begin{multline}\label{eq:width-prop}
  D_W=D^0_W\Bigl[ (\ii \Pi_W)D_W^0+(\ii\Pi_W)D_W^0(\ii\Pi_W)D_W^0+\dots\Bigr]\\
=\frac{(-\ii)}{q^2-m^2-\Pi_W^T}\left[g^{\mu\nu}-\frac{q^\mu q^\nu}{q^2}\frac{q^2+\Pi_W^L-\Pi_W^T}{m^2+\Pi_W^L}\right]
\end{multline}
In $R_\xi$ gauge, the gauge-Goldstone boson mixing by loop effects has to be taken into account and the Dyson equation becomes a matrix equation (see e.g. \cite{Boehm:2001}).

The  resummation of self energies takes a special class of Feynman diagram into account at \emph{all} orders perturbation theory, while neglecting other diagrams. This mixing of  different orders of perturbation theory violates gauge invariance. We have seen already in \fref{sec:intro-widths} that the Breit-Wigner propagator of a gauge boson  violates the STI for the triple gauge boson vertex and this is also true for the solution of the Dyson equation \eqref{eq:width-prop}. It is even true that the STI are violated if all radiative corrections  corrections in \emph{fixed order} are included. Because of the nonlinearity of the STI for the effective action \eqref{eq:sti}, the vertices calculated from the effective action to $n$ loop order  don't satisfy simple Ward Identities. Therefore the Green's functions calculated with the effective vertices and resummed propagators at the $n$-loop level violate the STIs at the next order of perturbation theory.

\section{Simple schemes for finite width effects}\label{sec:width-schemes}
We now turn to the description of simple schemes that have been proposed in the literature to include finite width effects  in tree level calculations on a effective level \cite{LopezCastro:1991,Denner:1999,Baur:1991,Baur:1995,Beenakker:1999}. We have implemented all these schemes in \OMEGA. A numerical comparison is given in \fref{sec:widths-comparison} and \fref{sec:widths-numeric}. Finally we will give a very brief overview of schemes that have been proposed  to introduce Dyson-resummed propagators without violations of gauge invariance in loop calculations \cite{Denner:1996,Beenakker:1997,Aeppli:1994rs,Papavassiliou:1998}.
\subsection{Step width}
Apart from the violation of gauge invariance, 
another unphysical feature of the propagator \eqref{eq:uni-width} is the use of a width
even for spacelike momenta $q^2<0$. This is unphysical because the self energy cannot develop an imaginary part 
below the threshold for $W$ decay. As a simple way to take this into account, the so called \emph{step width scheme}  proposes to replace
the fixed width in \eqref{eq:uni-width} by a step function
\begin{equation}
  m\Gamma \to m\Gamma\theta(q^2)
\end{equation}
Unfortunately, this doesn't improve gauge invariance since the Ward Identity is
still violated. In fact the use of the step function turns out to worsen problems with gauge invariance. For small scattering  angles of the electron in single $W$ production, disastrous results are obtained (see \fref{sec:widths-comparison}). However, it was found recently that the step with scheme seems to lead to consistent results in $6$ fermion production \cite{Dittmaier:2002}.
\subsection{$U(1)$ restoring schemes}
To restore $U(1)$ gauge invariance while keeping finite width effects several
schemes have been proposed:
\begin{description}
\item[Fixed width scheme] The simplest proposal is the so called fixed width scheme
\cite{LopezCastro:1991} i.e. the replacement $M_W^2 \to
M_W^2-\ii M_W\Gamma_W$ everywhere in the propagator:
\begin{equation}\label{eq:cm-propagator}
  D_W^{\mu\rho}=\frac{-\ii}{q^2-M_W^2+\ii M_W\Gamma_W}\left(g^{\mu\nu}-\frac{q^\mu q^\nu}{M_W^2-\ii
  M_W\Gamma_W} \right) 
\end{equation}
This satisfies the QED Ward Identity if a constant width is used, the use of step width still violates gauge invariance. $SU(2)$ gauge invariance is violated in any case, nevertheless stable  numerical results were obtained in single $W$ \cite{Beenakker:1997} and 6 fermion production \cite{Dittmaier:2002} even for high energies.
\item[Running-Width-Scheme] 
If one keeps only the imaginary part of the self-energies and takes the
fermions in the loop as massless one can approximate \cite{Baur:1995}
\begin{align}\label{eq:running-width}
  \Im \Pi_T =-\Gamma_W \frac{q^2}{M_W}\equiv- q^2\gamma_W \qquad \Im \Pi_L=0 
\end{align}
Plugging this into the resummed propagator \eqref{eq:width-prop} yields 
\begin{equation}\label{eq:zepp-prop}
 \ii D_W^{\mu\nu}=\frac{1}{q^2-m^2+\ii q^2 \gamma}\left( g^{\mu\nu}-\frac{q^\mu
  q^\nu}{m^2}(1+\ii \gamma)\right)
\end{equation}
The QED Ward Identity can now be satisfied if the  $WW\gamma$-vertex  is
modified by a factor $(1+\ii\gamma)$. This factor can also be derived from the
explicit calculation of the fermion triangle vertex correction to the
$WW\gamma$-vertex \cite{Baur:1995}. The use of the width for non-timelike momenta cannot be avoided in order to keep electromagnetic gauge invariance.

The running width scheme violates $SU(2)$ and is known to lead to results wrong by several orders of magnitudes at high energies \cite{Beenakker:1997,Dittmaier:2002}.
\end{description}
\subsection{$SU(2)$ restoring schemes}
\begin{description}
\item[Overall  Scheme]
A rather ad hoc proposal is the so called \emph{overall factor} scheme \cite{Baur:1991}. (sometimes also called \emph{fudge factor} scheme). The Feynman diagrams  are evaluated with propagators with vanishing widths, so gauge invariance is manifest. Then the \emph{complete}
matrix element is multiplied by a factor
\begin{equation}
  \frac{p^2-m^2}{p^2-m^2+\ii m\Gamma}
\end{equation}
for every propagator that can become resonant. This scheme treats the resonant diagrams correctly, non-resonant diagrams are not treated properly, however.

\item[Complex Mass Scheme]
A more sophisticated version of the fixed width scheme, that ensures also $SU(2)$  gauge invariance, changes the Feynman rules by replacing the gauge boson masses by \emph{complex masses}
\begin{equation} 
M_{W,Z}^2 \to M_{W,Z}^2-\ii M_{W,Z}\Gamma_{W,Z}
\end{equation}
everywhere in the theory \cite{Denner:1999}. This implies for example the use of a complex Weinberg angle
\begin{equation}
\cos\theta_w^2=\frac{M_{W}^2-\ii M_{W}\Gamma_{W}}{M_{Z}^2-\ii M_{Z}\Gamma_{Z}}
\end{equation}
Similarly, the widths of top quarks \cite{Heide:2000} and Higgs bosons can be taken into account. This scheme is manifestly gauge invariant, however the physical content of the complex Feynman rules is rather unclear. 

\item[Effective (nonlocal) Lagrangian] This method \cite{Beenakker:1999} gives a prescription to modify the vertices of the theory so $SU(2)$ gauge invariance is satisfied also for running widths. To describe running widths in a manifestly gauge invariant way, nonlocal terms involving Wilson lines
  \begin{equation}
    U(x,y)=\text{P} e^{-\ii g\int_x^y A^a_\mu T_a (\omega)d\omega^\mu}
  \end{equation}
are added to the Lagrangian. 

The selfenergy of gauge bosons is introduced by a term
\begin{equation}
  S_{\text{NL}}=\frac{1}{2}\int d^4x d^4 \Sigma_W(x-y)\Tr[U(y,x)T_aF^a_{\mu\nu}(x)U(x,y)T_b{F^{b}}^{\mu\nu}(y)]
\end{equation}
The derivation of the resulting Feynman rules is rather lengthy and we quote the results from \cite{Beenakker:1999}.
The gauge boson propagator is given by
\begin{equation}
 \ii D_W^{\mu\nu}=\frac{1}{q^2-m^2+q^2\Sigma(q^2)}\left( g^{\mu\nu}-\frac{q^\mu
  q^\nu}{m^2}\left(1+\Sigma(q^2) \right)\right)
\end{equation}
and the addition to the triple gauge boson vertex becomes
\begin{equation}\label{eq:non-local-vertex}
\sum_{\text{Permutations}}\epsilon_{ijk}  \Sigma(p_i^2)\left(\frac{1}{2}A^{\mu_i,\mu_j\mu_k}(q_i)+\frac{q_i^{\mu_i}-q_j^{\mu_i}}{q_i^2-q_j^2}T^{\mu_i\mu_j}(q_i,q_j)\right)
\end{equation}
with
\begin{align}\label{eq:t-a-def}
  T^{\mu\nu}(p,q)&=(p\cdot q)g^{\mu\nu}-p^\nu q^\mu\\
  A^{\mu,\nu\rho}(p)&=g^{\mu\nu}p^\rho-g^{\mu\rho}p^\nu 
\end{align}
 To apply this scheme to six fermion production processes, one has also to  modify the quartic gauge boson vertices \cite{Beenakker:1999}. 
In our implementation in O'Mega, we use 
\begin{equation}
\Sigma_W(q^2)=\ii\frac{\Gamma_W}{M_W}\theta(q^2)
\end{equation}
so---apart from the step function---the propagator is the same as in the running width scheme given in \fref{eq:zepp-prop}. 

Recently the nonlocal vertex scheme was matched to the full fermion loop scheme \cite{Beenakker:2003}. It was found that for low energies the scheme underestimates the cross section compared to the full fermion loop scheme while for large energies problems with unitarity appear. A modification was proposed to overcome this problems by adding an additional solution of the Ward Identities to the the vertices in order to yield better agreement with the full fermion loop results. 
\end{description}
We remark that is inconsistent to use gauge bosons as external on-shell particles in the complex mass scheme and the nonlocal vertices scheme. To satisfy the Ward Identities, the gauge bosons must be on the `complex mass shell' $p^2=M_W^2+\ii M_W\Gamma_W$ in the complex mass scheme and   $p^2=\frac{m_{W_a}^2}{1+\ii\gamma_W}$ in the nonlocal vertex scheme, respectively. 
\subsection{Loop schemes}
In  \emph{background field gauge}  on can exploit the fact  that the effective action satisfies a simple linear Ward Identity 
order by order in perturbation theory \cite{Denner:1996,Boehm:2001} and that therefore Dyson summation doesn't violate the Ward Identities. However, to use this scheme in practical calculations, one has to calculate \emph{all} diagrams with  a given number of loops to obtain a gauge invariant result. 

The \emph{fermion loop scheme} \cite{Beenakker:1997} proposes to include all \emph{fermionic} 1 loop diagrams. The gauge invariance of this scheme can be derived in different ways. A simple way is to use the fact that one can freely introduce additional fermion families without destroying gauge invariance. This argument is similar to the flavor selection rules discussed in \fref{sec:groves}. The gauge invariance follows also from the background field method \cite{Denner:1996} and from the application of the formalism of gauge flips to loop diagrams \cite{Ondreka:2003}. The fermion loop scheme has been applied to calculations for 4 fermions in the final state \cite{Beenakker:1997,Accomando:1999,Passarino:2000} but is insufficient for 6 fermion production at large energies \cite{Dittmaier:2002} where also boson loops have to be taken into account. 

Both the background field method and the fermion loop scheme require the calculation of loop diagrams and thus go beyond the present capabilities of O'Mega. Therefore they will not be considered further in this work. We mention that there are other schemes like the \emph{pole-scheme} \cite{Aeppli:1994rs}, and the \emph{pinch technique}  \cite{Papavassiliou:1998} that we don't consider here for the same reason.
\section{Numerical results for single $W$ production}

\begin{figure}[htbp]
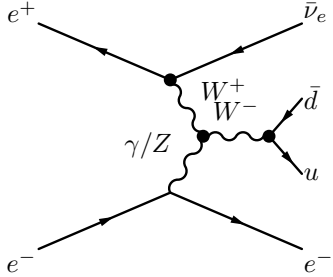

\begin{equation*}
\parbox{35mm}{
\begin{fmfchar*}(35,30)
\fmfleft{a,b}
\fmfright{f1,f3,f4,f2}
\fmfv{label=$e^- $,la.di=1}{a,f1}
\fmfv{label=$e^+ $,la.di=0.8}{b}
\fmfv{label=$\bar\nu_e$,la.di=0.8}{f2}
\fmfv{label=$\bar d$,la.di=0.8}{f4}
\fmfv{label=$u$,la.di=0.8}{f3}
\fmf{fermion}{a,fvf,f1}
\fmf{fermion}{f2,wp,b}
\fmf{phantom}{fvf,wp}
\fmffreeze
\fmf{photon, label=$\gamma/Z$,label.side=left}{fvf,v}
\fmf{photon, label=$W^+$}{wp,v}
\fmf{photon, label=$W^-$}{wm,v}
\fmf{fermion}{wm,f3}
\fmf{fermion}{f4,wm}
\fmfdot{v,wp,wm}
\end{fmfchar*}}
\end{equation*}
\caption{Single $W$ production}\label{fig:single_w}
\end{figure}

The violation of gauge invariance for finite widths  plays an important role for  single $W$ production in the process $e \bar e \to e \bar \nu_e  u\bar
d$\cite{Argyres:1995}, where we get a contribution from the Feynman diagram displayed in \fref{fig:single_w}. 

For small scattering angles of the electron, the photon propagator 
\begin{equation*}
D_\gamma \propto \frac{1}{1-\cos\theta_e^2}
\end{equation*}
becomes large and a tiny violation of gauge invariance is blown up with possible dramatical consequences for the numerical stability. 
\subsection{Comparison of Matrix elements}\label{sec:widths-comparison}
To investigate the violation of gauge invariance and compare the different schemes discussed in \fref{sec:width-schemes} in the process $e \bar e \to e \bar \nu_e  u\bar
d$, we have written a simple event generator that evaluates the complete scattering matrix element generated by O'Mega, using momentum configurations where the single $W$ diagram of \fref{fig:single_w} dominates and where the violation of $U(1)$ gauge invariance becomes relevant. This allows to obtain a qualitative comparison of the different schemes on the matrix element level and to discuss the behavior of the matrix elements as a function of the scattering angle and the center of mass energy. For the quantitative comparison of the cross sections in \fref{sec:widths-numeric} we will use the phase space generator WHIZARD \cite{Kilian:WHIZARD}.

Our program allows to give the center of mass energy and the scattering angle of the electron and generates the remaining variables  according to random distributions. The squared momentum $p_W^2$ of the $W^-$ is generated with a Lorentz distribution 
\begin{equation}
  P(p^2)=\mathcal{N}\frac{1}{(p^2-m_W^2)^2+m_W^2\Gamma^2_W}
\end{equation}

In the following, we take the complex mass scheme as a reference scheme since it is manifestly gauge invariant and numerically stable. 
To make comparisons with the other schemes, we generate a sample of phase space points $\{\Pi_i(s,\theta_e)\}$ with a given center of mass energy and electron scattering angle.

The expectation value of an observable in a sample of $N$ events is given by 
\begin{equation}
\braket{\mathcal O(s,\theta_e)}=\frac{1}{N}\sum_i \mathcal{O}(\Pi_i(s,\theta_e)) 
\end{equation}
We measure the deviation of a given scheme $S$ to the complex mass scheme using the observable   
\begin{equation}\label{eq:def-delta}
\Delta_{S-CM}(s,\theta_e)\equiv \frac{\Braket{|\me_{S}^2-\me_{CM}^2|}}{\braket{\me_{CM}^2}}
\end{equation}
For clarity of the figures, we don't include the errorbars for the uncertaintiy of the numerical integration that is responsible for the fluctuations.
\begin{figure}[htbp]
  \begin{center}
 \includegraphics[width=\graphwidth,angle=270]{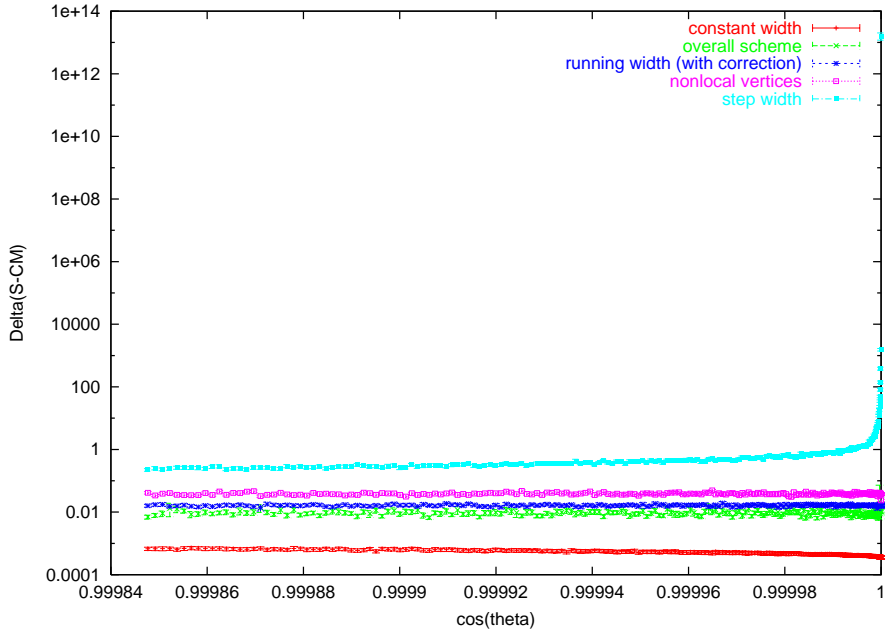}
\caption{Comparison of different schemes with the complex mass scheme in the process $e \bar e \to e \bar \nu_e  u\bar
d$ according to \fref{eq:def-delta} at
    $\sqrt s =800$ GeV for small angles}
    \label{fig:diff_05}
  \end{center}
\end{figure}
From \fref{fig:diff_05} we can see that at $800$ GeV and for small scattering angles the complex mass and the fixed width scheme agree at the per mille level while the overall scheme, the timelike scheme with correction factor for the $\gamma WW$ vertex and the nonlocal vertex scheme agree at the per cent level. 

The step width scheme shows deviations ranging from $\mathcal{O}(1)$ for scattering angles $\theta_e \sim 1^\circ$ to catastrophic divergences for $\theta_e\to 0$.
 \begin{figure}[htbp]
  \begin{center}
\includegraphics[width=\graphwidth,angle=270]{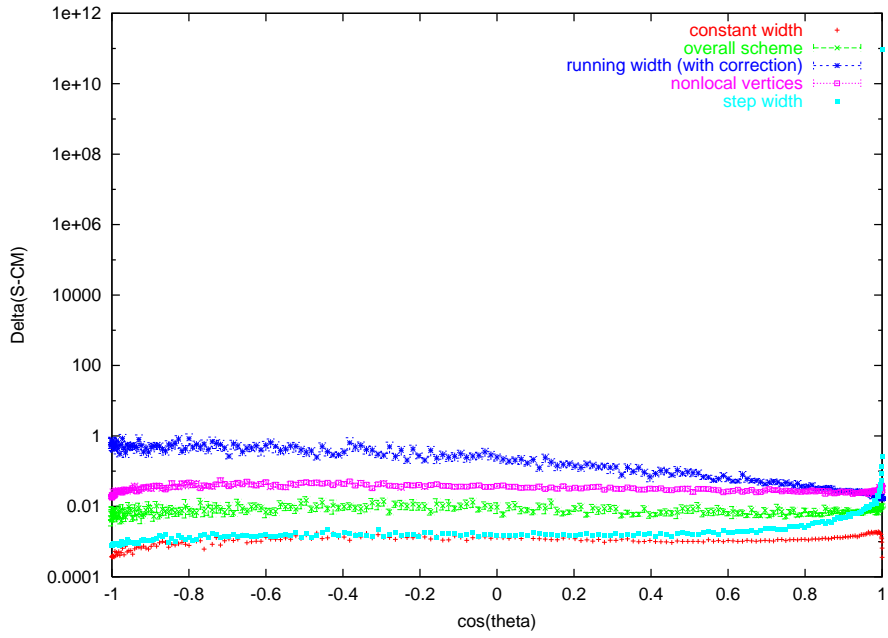}
   \caption{Comparison of different schemes with the complex mass scheme in the process $e \bar e \to e \bar \nu_e  u\bar
d$ according to \fref{eq:def-delta} at
    $\sqrt s =800$ GeV for $-180^\circ<\theta <180^\circ$}
    \label{fig:diff_180}
  \end{center}
\end{figure}

At large scattering angles, shown in \fref{fig:diff_180} we obtain a different picture: while deviations of the nonlocal vertex and the overall scheme from the complex mass scheme are approximately independent of the scattering angle, the step width scheme shares the good behavior of the fixed width scheme at large scattering angles---contrary to the behavior at small angles. 

In contrast, the running width scheme shows deviations of $\mathcal{O}(1)$ at large scattering angles while it is consistent for small angles. 
 
Note that at large scattering angles, our sampling of the phase space might not be realistic, since  other diagrams than the single $W$ production become important. However, the observations made in this section are supported by the numerical results for the cross sections and angular distributions for unweighted events obtained with WHIZARD discussed in \fref{sec:widths-numeric}

\begin{figure}[htbp]
 \begin{center}
    \includegraphics[width=\graphwidth,angle=270]{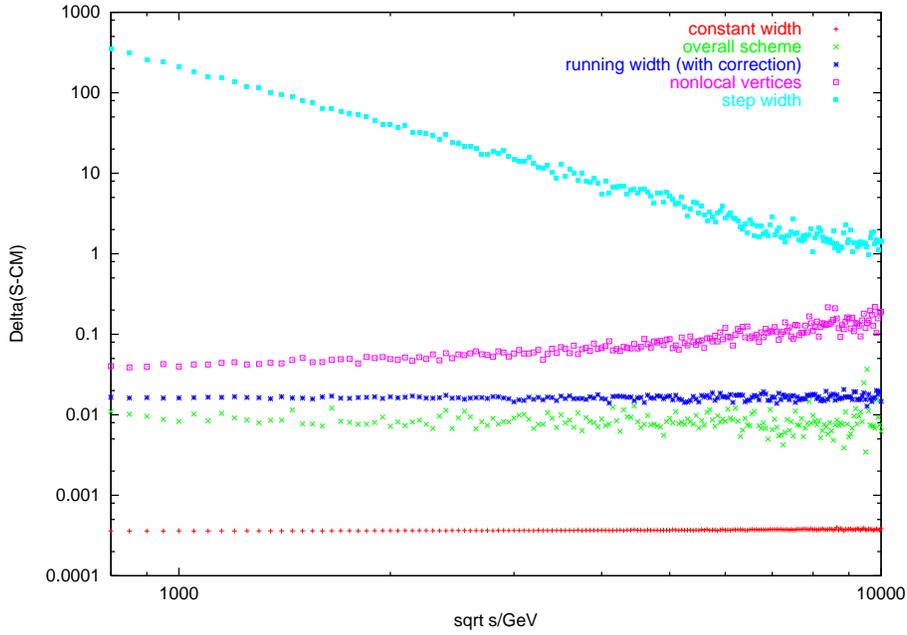}
 \caption{Comparison of different schemes in the process $e^- e^+ \to e^-\bar\nu_e u\bar d$ with the complex mass scheme as a function of $\sqrt s$ at  $\theta=0.01^\circ$}
    \label{fig:roots_angle01}
 \end{center}
\end{figure}

\begin{figure}[htbp]
 \begin{center}
    \includegraphics[width=\graphwidth,angle=270]{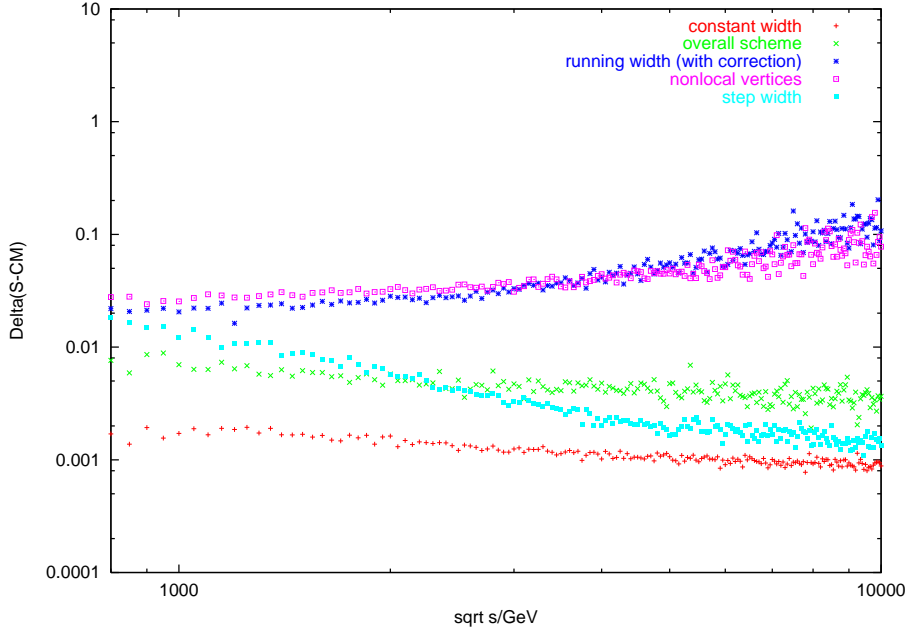}
    \caption{Comparison of different schemes in the process $e^- e^+ \to e^-\bar\nu_e u\bar d$ with the complex mass scheme as a function of $\sqrt s$ at
      $\theta=10^\circ$}
    \label{fig:roots_angle10}
  \end{center}
\end{figure}

Considering the behavior as a function of energy, shown in \fref{fig:roots_angle01}  at $0.01^\circ$ and in \ref{fig:roots_angle10} at $10^\circ$, we see that the step width scheme shows \emph{increasing} agreement with the complex mass scheme while the nonlocal vertices scheme becomes \emph{more inconsistent}. 
The behavior of the step width scheme can be understood from the fact that the violation of the Ward Identity gets enhanced by a factor 
\begin{equation}
  \frac{1}{E^2(1-\cos\theta)}
\end{equation}
from the photon propagator, so one can expect dropping inconsistencies with growing energy. 

The deviations of the running width scheme with vertex corrections from the complex mass scheme are approximately independent of the energy for a small scattering angle but show a similar energy dependence as the nonlocal vertex scheme for larger angles. 

To summarize our discussion, we have found that  for all scattering angles and center of mass energies the complex mass and the fixed width scheme agree at the per mille level while the overall scheme shows agreement at the per cent level. 

In contrast the  timelike scheme is totally unreliable for small angles while the inconsistencies of the running width scheme grow with energy. The nonlocal vertex scheme shows deviations  growing with the energy from $\mathcal{O}(1\%)$ to  $\mathcal{O}(10\%)$.
\subsection{Results for cross sections}\label{sec:widths-numeric}
To obtain a quantitative comparison among the different schemes and to compare with existing calculations in the literature, we have performed a phase space generation of the \OMEGA~matrix element using the phase space and event generator WHIZARD \cite{Kilian:WHIZARD}.

In the case of single $W$ production, apart from the simple tree level schemes,  numerical results have also been obtained in the full fermion loop scheme \cite{Beenakker:1997,Accomando:1999,Passarino:2000}. We compare our results to those of \cite{Beenakker:1997} for the case of large scattering angles and to those of \cite{Passarino:2000} for small scattering angles. 
  
To compare with the results of \cite{Beenakker:1997}, we use the so called `canonical LEP cuts' from \cite{Bardin:1997}, i.e.
\begin{subequations}
  \begin{align}
    10^\circ< \theta_e &<180^\circ\\
    E_e&> 1 \text{ GeV}\\
    E_u,E_d &> 3 \text{ GeV}
  \end{align}
\end{subequations}
and the same input parameters as \cite{Beenakker:1997}. Our results are shown in \fref{tab:lacc20}. 
\begin{table}[htbp]
  \begin{center}
    \begin{tabular}{|l|c|c|c|c|c|}
     \hline $\sqrt s$  & 200 GeV & 500 GeV & 1 TeV & 2 TeV & 10 TeV \\ \hline
     FW &0.709 (1)&0.3736 (7)&0.2725 (6)&0.1992 (4)&0.03162 (9)\\ \hline 
     CM &0.707 (1)&0.3729 (7)&0.2712 (6)&0.2001 (4)&0.03148 (7)\\ \hline 
     RW &0.708 (1)&0.3740(7)&0.2772 (6)&0.2139 (5)&0.3850(4)\\  \hline 
     NV &0.709 (1)&0.3735 (7)&0.2719 (6)&0.2005 (5)&0.0441 (1)\\ \hline \hline 
     FW  \cite{Beenakker:1997}&0.7062 (8)&0.3734 (11)&0.2709 (17)&0.1965 (19)
     &0.0305 (6)\\\hline  
     RW  \cite{Beenakker:1997}&0.7045 (10)&0.3736 (11)&0.2763 (17)&0.2176 (19)&
     0.5261 (10)\\\hline  
     MFL \cite{Beenakker:1997} &0.7060 (6)&0.3713 (11)&0.2705 (17)&
     0.2000 (19) &0.0306 (6)\\ \hline
     FL\cite{Beenakker:1997} &0.7177 (7)&0.3776 (11)&0.2797 (17)&
     0.2076 (17)&0.0326 (6)\\\hline
    \end{tabular}
    \caption{Cross section (pb) of $e^- e^+ \to e^-\bar\nu_e u\bar d$ for $\theta_e>10^\circ$}
    \label{tab:lacc20}
  \end{center}
\end{table}
We obtain a reasonable agreement with existing results in the fixed with and the running width scheme. At small energies, all schemes agree with the complex mass schemes, while at $10$ TeV the running width scheme is wrong by an order of magnitude. The nonlocal vertex scheme shows shows discrepancies of the order of $30\%$. 

Comparison with the fermion loop scheme and the minimal fermion loop scheme (MFL) that considers only imaginary parts of the loop diagrams, shows that no effective scheme is able to approximate the full fermion loop scheme while the MFL gives comparable results comparable to the fixed width scheme. Similar results have been obtained in \cite{Accomando:1999}.

In \ref{tab:lacc20_A} we show our results for a scattering angle
\begin{equation}
   0.1^\circ< \theta_e <180^\circ
\end{equation}
while the remaining cuts and input parameters are the same as before.
\begin{table}[htbp]
  \begin{center}
    \begin{tabular}{|l|c|c|c|c|c|}
     \hline $\sqrt s$  & 200 GeV & 500 GeV & 1 TeV & 2 TeV & 10 TeV \\ \hline
     FW &0.790 (2)&0.803 (2)&1.103 (3)&1.496 (3)&2.103 (5)\\ \hline 
     CM &0.791 (2)&0.805 (2)&1.108 (3)&1.494 (4)&2.094 (5)\\ \hline 
     RW &0.789 (2)&0.804 (2)&1.107 (2)&1.504 (4)
     &2.468 (6)\\  \hline
     NV &0.788 (2)&0.801 (2)&1.109 (3)&1.498 (5)&2.289 (7)\\ \hline 
     TL &1.538 (3)&1.218 (3)&1.276 (4)&1.554 (5)&2.101 (6)\\\hline 
    \end{tabular}
    \caption{Cross section (pb) of $e^- e^+ \to e^-\bar\nu_e u\bar d$ for $\theta_e>0.1^\circ$}
    \label{tab:lacc20_A}
  \end{center}
\end{table}

To compare to the results for the fermion loop scheme at small angles \cite{Passarino:2000},  we use the input parameters and the cuts from  \cite{Passarino:2000} i.e.
\begin{subequations}
  \begin{align}
    \cos\theta_e&>0.997 \\
    M(ud)&>45 \text{ GeV}
  \end{align}
\end{subequations}
The results are shown in \fref{tab:sacc20}. 
\begin{table}[htbp]
\begin{center}
  \begin{tabular}{|l|c|c|c|c|c|}
\hline  $\sqrt s$  & 200 GeV & 500 GeV & 1 TeV & 2 TeV & 10 TeV \\ \hline
     Fixed Width &0.1161 (9)  & 0.881 (5)  & 2.03 (1)  & 3.59 (2) & \\ \hline 
     Complex Mass &0.118 (1)& 0.867 (4) & 2.03 (1) & 3.68 (7) & 4.2 (4) \\
     \hline 
     Running Width &0.1173 (7)& 0.871 (4)&2.018(8) & 3.61(4)& 7.4(2) \\  \hline 
     Nonlocal Vertices&0.1178 (8)&0.895 (7)&2.07 (4)&3.64 (3) &6.6(3) \\\hline\hline 
     Fixed Width \cite{Passarino:2000} & 0.11977 (67) & - & - & - &-\\ \hline
     FL \cite{Passarino:2000} & 0.11367 (8) & - & - &
     - & - \\ \hline
\end{tabular}
  \caption{Cross section (pb) of $e^- e^+ \to e^-\bar\nu_e u\bar d$ for $\cos\theta_e>0.997$}
  \label{tab:sacc20}
\end{center}
\end{table}
From tables \ref{tab:lacc20}, \ref{tab:lacc20_A} and \fref{tab:sacc20} we see a qualitative agreement with the results of \fref{sec:widths-comparison}: 
The step width scheme is wrong by $O(1)$ at small energies and scattering angles and gets better for growing energies. In contrast, the running width scheme shows drastic deviations for large scattering angles and high energies. 

The nonlocal vertex scheme shows discrepancies for high energies that are  $\sim 10\%- 50\%$ and therefore don't depend drastically on the scattering angle. This is also in agreement with the results from \fref{sec:widths-comparison} and confirms the violations of unitarity reported in \cite{Beenakker:2003}.

\chapter{Effective coupling constants}\label{chap:running}
\section{Running coupling constants and effective Weinberg angle}\label{sec:sinthw}
In \fref{sec:qft-widths} we have considered the imaginary part of the self-energy to include finite width effects via Dyson summation. We will now discuss the effects of the \emph{real} part of the self energies and vacuum polarizations, that leads to running coupling constants and masses. 

In QED, due to the Ward Identity the vacuum polarization is purely transverse
\begin{equation}
 \Pi_\gamma^{\mu\nu}(q^2)= q^2\Pi_{\gamma}(q^2) \left(g^{\mu\nu}-q^\mu q^\nu\right)
\end{equation}
and the radiative corrections to the photon propagator can be taken into account by defining the \emph{running coupling constant}
\begin{equation}
e^2(q^2)=e^2\frac{1}{(1-\Pi_\gamma(q^2))}
\end{equation}
In nonabelian gauge theories, the resummation of the vacuum polarization is not sufficient for the definition of a gauge independent running coupling constant. This is possible using the Callan-Symanzik equation (see e.g. \cite{Peskin:1995}) provided a suitable renormalization scheme is used \cite{Caswell:1974,Piguet:1985} but we will not go into this any further. 

A gauge invariant prescription to include the running coupling constants is to replace 
\begin{equation}
g\to g (Q^2)
\end{equation}
with the same $Q^2$ \emph{everywhere} in the scattering amplitude
where $Q^2$ is a typical  scale  of the order of the momentum invariants in the scattering process (see e.g. \cite{Peskin:1995}). This so called `improved perturbation theory' allows to absorb large logarithmic corrections from loop diagrams in tree level amplitudes. 

If one introduces running coupling constants directly in the Feynman rules, the coupling appears at different vertices with different momenta and this disturbs the gauge cancellations that involve different Feynman diagrams. As in the case of finite width effects, a gauge invariant prescription for running coupling constants is given by the fermion loop scheme \cite{Beenakker:1997}. Furthermore, it is consistent to take the different momentum scales in different groups of Feynman diagrams into account by using different values of $g(Q^2)$ in different gauge invariance classes of diagrams. 

In theories which contain several gauge groups with independent coupling constants, additional problems occur. As an example, we consider the electroweak Standard Model that we have briefly reviewed in \fref{sec:sm}. 
 
\begin{figure}[htbp]
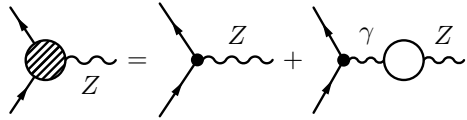

\begin{equation*}
\parbox{14mm}{
\begin{fmfchar*}(14,14)
\fmfleft{l1,l2}
\fmfright{r}
\fmf{fermion}{l1,i,l2}
\fmf{photon,label=$Z$}{i,r}
\fmfblob{15pt}{i}
\end{fmfchar*}}\, =\,
\parbox{15mm}{
\begin{fmfchar*}(15,15)
\fmfleft{l1,l2}
\fmfright{r}
\fmf{fermion}{l1,i,l2}
\fmfdot{i}
\fmf{photon,label=$Z$}{r,i}
\end{fmfchar*}}\, +\,
\parbox{20mm}{
\begin{fmfchar*}(20,14)
\fmfleft{l1,l2}
\fmfright{r}
\fmf{fermion}{l1,i,l2}
\fmf{photon,label=$Z$}{r,m}
\fmf{photon,label=$\gamma$}{m,i}
\fmfdot{i}
\fmfv{d.sh=circle,d.f=empty,d.size=15pt}{m}
\end{fmfchar*}}
\end{equation*}
\caption{Effective fermion-$Z$ vertex}\label{fig:z-gamma-mix}
\end{figure}

The Weinberg angle enters several observables that can be used to determine its value in an experiment. For example, \fref{eq:cw-on-shell} allows to determine the so called `on-shell' value of $\sin\theta_w$ from the measured gauge boson masses \cite{PDG}
\begin{equation}\label{eq:sw-os}
\sin^2\theta_w^{OS}=1-\frac{m_W^2}{m_Z^2}=0.22278\pm 0.00036
\end{equation}
Alternatively, in \fref{eq:sm-covariant} the Weinberg angle enters the $Z$-boson fermion coupling 
\begin{equation}\label{eq:ferm_sinw}
\ii \tfrac{g}{\cos\theta_w}Z\left[T^3\tfrac{(1-\gamma^5)}{2}-Q\sin^2\theta_w\right]
\end{equation}
in a form that allows for a measurement from forward-backward asymmetries. A third possibility is the expression in terms of the Fermi constant, determined from the muon lifetime:
\begin{equation}\label{eq:gf}
\frac{G_F}{\sqrt 2}=\frac{g^2}{8 m_W^2}=\frac{e^2\sin^2\theta_w\cos^2\theta_w}{8m_Z}
\end{equation}
On tree level, all these definitions agree, while the inclusion of radiative corrections leads to deviations from these expressions \cite{Peskin:1995,Hollik:1993,PDG}. The corrections introduce a dependence on the top quark and Higgs mass and therefore on unknown or not precisely measured quantities. For example the definition from the $Z$-fermion vertex \eqref{eq:ferm_sinw} receives radiative corrections from photon-$Z$ mixing as shown in \fref{fig:z-gamma-mix}. 
The theoretical prediction for the effective Weinberg angle entering the asymmetries \cite{PDG}:
\begin{equation}\label{eq:sw-mix} 
\sin^2\theta_w^*=0.23143\pm 0.00015
\end{equation}
therefore differs from other definitions. 

Several schemes for input parameters in the Standard Model can be used \cite{Hollik:1993,PDG}. The \emph{on-shell} scheme maintains the relation \eqref{eq:sw-os} in all orders perturbation theory while the remaining expressions are modified. Other schemes based on \fref{eq:ferm_sinw} or \eqref{eq:gf} lead to a violation of this identity. These violations are parametrized by the so called  $\rho$ parameter:
\begin{equation}
\cos\theta_w\frac{m_Z}{m_W}\equiv\rho^{-\frac{1}{2}}
\end{equation}
The advantage of schemes different from the on shell scheme is that they allow the determination of $\sin\theta_w$ from input parameters that can be measured with greater precision than the $W$ mass. Currently, the most accurate determination \cite{PDG} 
\begin{equation}\label{eq:sw-pdg}
\sin^2\theta_w^{G_F}=0.23105\mp 0.00008
\end{equation}
comes from \fref{eq:gf}. However, it is inconsistent to use this value together with the measured value of the gauge boson masses in tree level calculations since gauge cancellations rely on the relation \eqref{eq:sw-os}. We will demonstrate this in the following sections. If one uses the experimental value \eqref{eq:sw-pdg} as input parameter together with the $Z$ mass, one has to treat the $W$-mass as dependent parameter, different from its on-shell value. 

We list some consistent tree-level schemes of input-and dependent parameters:

The on-shell scheme:
\begin{equation}\label{eq:os-scheme}
e,m_W,m_Z \rightarrow \begin{cases} \cos\theta_w=\frac{m_W}{m_Z}&\\
                                               g=\frac{e}{\sin\theta_w}& 
                      \end{cases}
\end{equation}
This scheme is currently implemented in \omegacite.

The $G_F$ scheme:
\begin{equation}\label{eq:gf-scheme}
G_F,m_W,m_Z \rightarrow \begin{cases} \cos\theta_w=\frac{m_W}{m_Z}&\\
                                       e=\sin\theta_w M_W\sqrt {\sqrt 2 G_F}&\\
                                       g=2\sqrt{ \sqrt 2 G_F} m_W &
                      \end{cases}
\end{equation}
This is the standard option of WHIZARD \cite{Kilian:WHIZARD}.

To use $\sin^2\theta_w$ from \eqref{eq:sw-pdg} as input, one has to eliminate $m_W$ from the input parameter set, e.g. by 
\begin{equation}\label{eq:sw-scheme}
e,\sin^2\theta_w ,m_Z \rightarrow \begin{cases} m_W=\cos\theta_w{m_Z}&\\
                                                g=\frac{e}{\sin\theta_w}& 
                      \end{cases}
\end{equation}
\section{Cubic vertices involving Goldstone bosons}
In this section and the next, we will investigate the consequences of an inconsistent scheme of input parameters, for example using the Weinberg angle  as additional \emph{free} input parameter in \eqref{eq:os-scheme}:
\begin{equation}\label{eq:inconsistent}
e,\sin^2\theta_w\neq \left(1-\frac{m_W^2}{m_Z^2}\right),m_W, m_Z, g=\frac{e}{\sin\theta}
\end{equation}
Here $g$ could also  be determined from $m_W$ and $G_F$ as in \eqref{eq:gf-scheme}, the important point is that we treat both $\sin^2\theta_w$ and $m_W$ as independent input parameters. 

We will first turn to the case of the Goldstone boson couplings. According to the results of \fref{part:reconstruction}, they are dependent parameters completely determined from the input parameters. As we have seen in \fref{sec:sinthw}, in the Standard Model we have the situation that the `input parameters' in the sense of \fref{chap:ssb-lag} are not independent. 

We will derive a parametrization of the cubic couplings  involving Goldstone bosons in the Standard Model that respects the Ward Identities of the three point functions even if the relation \eqref{eq:sw-os} is violated. In \fref{sec:sw-ward} we will see that a similar analysis of the 4 point Ward Identities leads to a complete elimination of one of the redundant input parameters. 
  
According to \fref{subeq:triple-phi-couplings}, we can obtain the couplings of Goldstone bosons to gauge bosons from the rules 
\begin{equation}
f^{abc}\partial_\mu W^{a\nu}W^b_\mu W^{c\nu} \to 
\begin{cases}-\frac{1}{2m_{W_a}}f^{abc}(m_{W_b}^2-m_{W_c}^2)\phi^aW^{b\mu}W^c_\mu\\
f^{abc}\frac{1}{2m_{W_b}m_{W_c}}(m_{W_a}^2-m_{W_b}^2-m_{W_c}^2)W_a^\mu(\phi_b\overleftrightarrow \dmu\phi_c)
\end{cases}
\end{equation}
We now apply this to the $WWZ$ coupling in the Standard Model:
\begin{multline}
\mathscr{L}_{ZWW}=-\ii g\cos\theta_w\Bigl[\dmu W^{+\nu}(W^{\mu -}Z^\nu-W^{\nu -}Z^\mu) -\dmu
 W^{-}_\nu(W^{\nu +}Z^\mu- W^{\mu +}Z^\nu) \\
+\dmu Z_\nu(W^{+\mu}
 W^{-\nu}-W^{-\mu}W^{+\nu})\Bigr]\\
\end{multline}
and find the couplings of the Goldstone bosons that satisfy the 3 point Ward Identities:
\begin{equation}\label{eq:wphia-lag}
\begin{aligned}
\mathscr{L}_{\phi^\pm W^\mp Z}&=-\ii\frac{g\cos\theta_w}{m_W}(m_W^2-m_Z^2) \Bigl[\phi^+W^{-\mu}Z_\mu
-\phi^-W^{+\mu} Z_\mu\Bigr]\\
\mathscr{L}_{\phi^0 \phi^\mp W^\pm}&
=\ii\frac{g\cos\theta_w}{2}\frac{m_Z}{m_W} \left[
  W^{+\mu}(\phi^-\overleftrightarrow\dmu \phi^0)- 
  W^{-\mu}(\phi^+\overleftrightarrow\dmu \phi^0)\right]\\
\mathscr{L}_{\phi^\pm \phi^\mp Z}&=(-i\frac{g\cos\theta_w}{2})\frac{m_Z^2-2m_W^2}{m_W^2} Z^\mu(\phi^+\overleftrightarrow\dmu \phi^-)
\end{aligned}
\end{equation}
This is not the parametrization of the Feynman rules  usually given in the literature. For example, the first line of \fref{eq:wphia-lag} is given in \cite{Cheng:1984} as\footnote{To compare with \cite{Cheng:1984}, note that we use a different convention for the phases in the Goldstone boson sector to simplify the Ward Identities. Our rules can be obtained from \cite{Cheng:1984} by replacing their $\phi^\pm$ by $\mp \ii \phi^\pm$. The Feynman rules for the Standard Model with our parametrization of the Goldstone bosons have been given in \cite{Schwinn:2000}. }.
\begin{subequations}\label{eq:wphia-lag-wrong}
\begin{equation}
\mathscr{L}_{\phi^\pm W^\mp Z}=-\ii g\sin^2\theta_w m_Z \Bigl[\phi^+W^{-\mu}Z_\mu
-\phi^-W^{+\mu} Z_\mu\Bigr]
\end{equation}
that agrees with \fref{eq:wphia-lag} only for $\cos\theta_w=\frac{m_W}{m_Z}$. Similarly the rest of \fref{eq:wphia-lag} in the form from \cite{Cheng:1984} are
\begin{equation}
\begin{aligned}
\mathscr{L}_{\phi^0 \phi^\mp
W^\pm}&=\ii\frac{g}{2}W^{+\mu}(\phi^-\admu\phi^0)-\ii\frac{g}{2}W^{-\mu}(\phi^+\admu\phi^0)\\
\mathscr{L}_{\phi^\pm \phi^\mp Z}&=-i\frac{g}{\cos\theta_w}\gz
  Z^\mu(\phi^-\admu\phi^+)
\end{aligned}
\end{equation}
\end{subequations}
and agree with our version obtained from the Ward Identities only for the on-shell value of $\sin\theta_w$. 
To study the violation of gauge invariance caused by wrong expressions for the Goldstone boson couplings numerically, we have investigated the process $W^+Z\to W^+ Z$ in the Standard Model. We have generated an isotropic phase space and averaged the difference of $R_\xi$ gauge to unitarity gauge over the sample, analogous  to \fref{eq:def-delta}. 

\begin{figure}[htbp]
 \begin{center}
     \includegraphics[width=\graphwidth,angle=270]{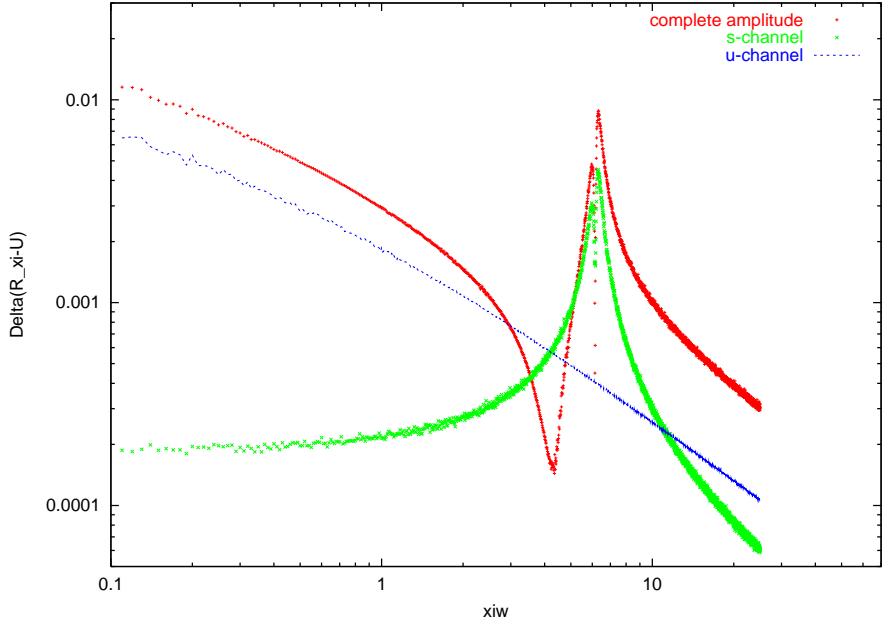}
    \caption{Comparison of $R_\xi$ and unitarity gauge for $W^+ Z \to W^+ Z$ at $\sqrt s=200$ GeV}
    \label{fig:wzwz_xiw}
  \end{center}
\end{figure}

In \fref{fig:wzwz_xiw} we show the results obtained with the Goldstone boson coupling in terms of the Weinberg angle from \fref{eq:wphia-lag-wrong} as a function of $\xi_W$ at  $\sqrt s =200$ GeV and using the effective Weinberg angle as determined from \fref{eq:sw-mix}. 
Since the violations of gauge invariance get multiplied by the unphysical part of the propagator
\begin{equation}
\frac{i}{p^2-\xi_W M_W^2}
\end{equation}
we expect a `resonance' at the unphysical pole in the $s$-channel for 
\begin{equation}
\xi_W^{\text{res}}=\left(\frac{s}{M_W}\right)^2 \approx\left(\frac{200}{80}\right)^2 \approx 6.5
\end{equation}
Because of this pole, we have to take the gauge boson width into account. For our numerical comparison, we choose the fixed width scheme. The violations of gauge invariance caused by this scheme have already been discussed in \fref{chap:width}. Since the momentum transfer in the $u$-channel is spacelike, we don't expect a pole but instead a $\xi$ dependence of the squared amplitude that becomes $\propto \frac{1}{\xi^2}$ for large $\xi$.

The expected behavior can be seen clearly in  \fref{fig:wzwz_xiw} where we show the contribution from the $s$- and $u$-channel diagrams alone together with the complete amplitude including $t$-channel Higgs-exchange and the quartic gauge boson interaction.

We see that errors at the order of $1\%$ can arise at the unphysical pole  and also for $\xi_W\sim 0$ where the errors in the $u$-channel diagram become large.

\begin{figure}[htbp]
 \begin{center}
     \includegraphics[width=\graphwidth,angle=270]{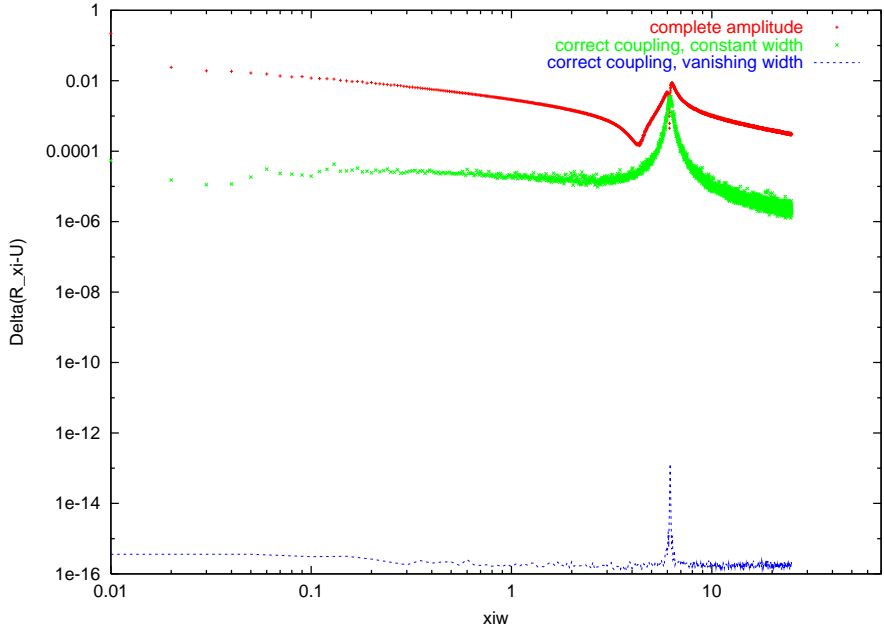}
    \caption{Comparison of $R_\xi$ and unitarity gauge for $W^+ Z\to W^+ Z$ at $\sqrt s=200$ GeV as a function of $\xi_W$.}
    \label{fig:wzwz_sinw_width}
  \end{center}
\end{figure}

As shown in \fref{fig:wzwz_sinw_width}, using the expressions of the Goldstone boson couplings determined from the Ward Identities \eqref{eq:wphia-lag} instead, we find perfect agreement if the gauge boson widths are set to zero. We know from the discussion in \fref{sec:width-schemes} that the fixed width scheme violates $SU(2)$ gauge invariance, and indeed we find a similar error at the unphysical $s$-channel pole as in the case of the wrong Goldstone boson couplings. The violations of gauge invariance for small $\xi_W$ in the $t$-channel are only of the order $10^{-5}$, however, so for practical purposes the fixed width scheme allows stable calculations over a wide range of the gauge parameter. 

\section{Gauge boson couplings}\label{sec:sw-ward}
We have seen, that in the case of a running Weinberg angle the expressions of the triple Goldstone boson couplings in terms of the (constant) gauge boson masses obtained as solutions of the Ward Identities for the cubic vertices remain consistent. 

We will now show that a consistent `deformation' of the Feynman rules is not possible if the Ward Identities for 4 particle functions are taken into account. Instead, in the solution of the Ward Identity the Weinberg angle is eliminated from the set of input parameters.  
\subsection{Conditions from Ward Identities}\label{sec:ward-cond}
We will `reconstruct' coupling constants from the set of input parameters given in \fref{eq:inconsistent}. We use the Lie algebra of the fermion couplings and the invariance of the Yukawa couplings, that are  equivalent to  the set of Ward Identities  obtained from the $f\bar f \to WW$ Ward Identity with one contraction \eqref{eq:2f2w-wi},with two contractions \eqref{eq:2f2w-wi2} and and the $\bar f f \to W H$ Ward Identity \eqref{eq:2fhw-wi}. This set of Ward Identities can be solved in terms of the parameters
\begin{equation}\label{eq:input}
\begin{aligned}
g_{CC}&=\frac{g}{\sqrt 2}\\
g^L_{A\bar ee}&=g^R_{A\bar ee}=-e\\
g_{\gamma WW}&=ie\\
\end{aligned}
\end{equation}
To evaluate the Ward Identities in the Standard Model, we use form of the generators from \fref{eq:sm-covariant}.
We begin our analysis by evaluating the Ward Identity \eqref{eq:2f2w-wi} for the amplitude $e^+ e^-\to W^+ W^-$ in the Standard Model and obtain
\begin{equation}
\begin{aligned}\label{eq:fermion-lie-sm}
-\ii g_{ZWW} g^L_{Z\bar e e}+\ii g_{\gamma WW} e&=g_{CC}^2\qquad \\
-\ii g_{ZWW} g^R_{Z\bar e
  e}+\ii g_{\gamma WW} e&=0
\end{aligned}
\end{equation}
where we have used the explicit form of the triple gauge boson couplings.
Similarly, the Ward Identity with two contractions \eqref{eq:2f2w-wi2} yields:
\begin{equation}\label{eq:yukawa-gold}
\begin{aligned}
g_{CC}^2 &=-\frac{1}{2m_e}g_{HW^+W^-}g_{H\bar e
  e}-\ii g_{WWZ}g^A_{Z\bar e e }\\
m_e(g^A_{Z\bar ee})^2&=-\frac{1}{8}g_{HZZ}g_{H\bar
  ee}
\end{aligned}
\end{equation}
Finally Ward Identities for $e^+ e^- \to Z H$ Ward Identity \eqref{eq:2fhw-wi} reads:
\begin{equation}\label{eq:yukawa-higgs}
- g_{Hee}=\frac{m_e}{2m_W^2}g_{HWW}=\frac{m_e}{2m_Z^2}g_{HZZ}
\end{equation}
Both equations of \fref{eq:fermion-lie-sm} together can be turned into a condition for the axial part of the electron $Z$-coupling:
\begin{equation}\label{eq:gcc-kommutator}
\frac{1}{2}g_{CC}^2=-\ii g_{ZWW} g^A_{Z\bar ee}
\end{equation}
The Ward Identities \eqref{eq:yukawa-gold} and \eqref{eq:yukawa-higgs} together can be used to derive the relations
\begin{equation}
\begin{aligned}
(g^A_{Z\bar ee})^2&=\frac{m_Z^2}{4m_e^2}g^2_{H\bar
  ee}\\
g_{CC}^2&=\frac{4m_W^2}{m_Z^2}(g^A_{Z\bar ee})^2 -\ii g_{WWZ}g^A_{Z\bar e e
  }
\end{aligned}
\end{equation}
Combining the last equation with the result from the Lie algebra \eqref{eq:gcc-kommutator} finally results in
\begin{equation}\label{eq:axial-z}
(g^{A}_{Z\bar ee})^2=\frac{m_Z^2}{8m_W^2}g_{CC}^2
\end{equation}
This fixes the $Z$-fermion coupling (up to a sign) and shows that the introduction of a running Weinberg angle is inconsistent.

Therefore we see that \emph{all} coupling constants appearing in the Ward Identities considered in this analysis can be expressed in terms of the parameters \eqref{eq:input} and the masses. 
Tracing our steps back we find\footnote{In the cases where the Ward Identities determine the coupling constants only up to a sign, we choose the same sign as in the Standard Model}:
\begin{equation}
\begin{aligned}
\ii g_{ZWW}&=-\frac{g_{CC}^2}{2g^A_{Z\bar ee}}= \frac{m_W}{m_Z}g\\
g_{H\bar ee}&=-\frac{m_e}{2m_Z}g^A_{Z\bar ee}=-\frac{m_e}{ m_W}g\\
g_{HWW}&=-\frac{2 m_W^2}{m_e}g_{Hee}= g m_W\\
g_{HZZ}&=-2\frac{m_Z^2}{m_e} g_{H\bar e e}=\frac{m_Z^2}{m_W}g\\
g_{Z\bar e e}^R&=\frac{g_{\gamma WW}g_{\gamma\bar e e}}{g_{ZWW}}=\frac{e^2}{g}\frac{m_Z}{m_W} \\
g_{Z\bar e e}^L&=2 g_{Z\bar e e}^A+g_{Z\bar e e}^R=- g\frac{m_Z}{m_W}\left[\frac{1}{2}-\frac{e^2}{g^2}\right]
\end{aligned}
\end{equation}
These are indeed the Standard Model values with an on-shell Weinberg angle. To fix the ratio $e/g$ to the on-shell value
\begin{equation}
\frac{e^2}{g^2}=\sin^2\theta_w=1-\frac{m_W^2}{m_Z^2}
\end{equation}
one has to consider additional identities like the Ward Identity for the $e^-\bar \nu_e \to W^- Z$ amplitude that we have not included in our analysis. 
\subsection{Modification of Feynman rules}
We have seen that the solution of the set of Ward Identities used in \fref{sec:ward-cond} doesn't allow the use of an effective Weinberg angle that doesn't satisfy the relation \eqref{eq:sw-os}\footnote{That doesn't mean one cannot use the Weinberg angle as input parameter. It is consistent to use, for example, the measured value \eqref{eq:sw-pdg} if the $W$ mass is regarded as derived parameter according to \fref{eq:sw-os}.}. However, it is possible to introduce effective couplings in such a way that some Ward Identities are still satisfied. Although, according to the famous statement of M.Veltman, working in a scheme that is  `a little bit gauge invariant' is as impossible being `a little bit pregnant', in tree level calculations for amplitudes with few external particles such a modification of the Feynman rules can yield stable numerical results. 

As a first  example consider the $e^+ e^- \to W^+W^-$ Ward Identity \eqref{eq:fermion-lie-sm}.
If we use a running Weinberg angle in the $ZWW$ coupling
\begin{equation}
g_{ZWW}=g\cos\theta_w(p^2)
\end{equation}
we see that \fref{eq:fermion-lie-sm} alone allows a running of the Weinberg angle, keeping $g_{CC}$ fixed, provided  we run also the axial $Z$ coupling. 
 according to 
\begin{equation}\label{eq:axial-z-run}
g^A_{Z\bar ee}=\ii \frac{g}{4\cos\theta_w(p^2)}
\end{equation}
However, we have seen that this becomes  inconsistent if the remaining Ward Identities are taken  into account. Furthermore, \fref{eq:fermion-lie-sm} shows that a modification of the vectorial part of the $Z$ coupling by an effective Weinberg angle to include the $W-Z$ mixing \eqref{eq:sw-mix} results in a violation of the Ward Identity.

Similarly we can try to satisfy the Ward Identity \eqref{eq:yukawa-gold} alone by modifying the Higgs-gauge boson couplings. Inserting the values read from the Standard Model Lagrangian, we find that the Weinberg angle cancels in the identity for $e^+e^-\to W^+W^-$ while the identity for $e^+e^-\to W^+W^-$ gives the condition
\begin{equation}
m_e \frac{g^2}{\cos^2\theta_w}=\frac{gm_e}{m_W}g_{HZZ}
\end{equation}
This is satisfied in the standard parametrization \cite{Cheng:1984}
\begin{equation}
g_{HZZ}=\frac{g m_Z}{\cos\theta_w}
\end{equation}
only for an on shell value of $\cos\theta_w$ but if we can use Higgs-$Z$ coupling in the form 
\begin{equation}\label{eq:hzz-mod}
g_{HZZ}=\frac{gm_Z^2}{m_W}
\end{equation}
the Ward Identity will be satisfied for an arbitrary value of $\sin\theta_w$. Of course by continuing such redefinition, we would arrive at the theory with the on-shell value of the Weinberg angle, but it is an interesting question, what can be achieved by performing this redefinition for only some of the couplings. Therefore we have investigated the modifications of the Feynman rules numerically.

We will first discuss the violations of gauge invariance caused by the use of an effective Weinberg angle \eqref{eq:sw-mix} in the fermion-$Z$ coupling and the on-shell value \eqref{eq:sw-os} in the remaining Feynman rules. As we can see from the numerical results in \fref{tab:sw-eff}, significant deviations between unitarity and $R_\xi$ gauge occur already for light particles in the initial state. 
\begin{table}[htbp]
  \begin{center}
    \begin{tabular}{c|c|c|c}
     Process           &$U \leftrightarrow R_\xi(\xi=1)$ &  WI  \\\hline
 $\bar e^+ e^- \to W^+ W^-$& $\surd$ & $\mathcal{O}(10^{-1})$ \\
 $e^+ e^- \to W^+ W^- Z$&$\mathcal{O}(10^{-2})$  &$\mathcal{O}(10^{-2})$ \\
 $e^+ e^- \to b \bar t W^+$&$\mathcal{O}(10^{-1})$  &$\mathcal{O}(10^{-2})$ \\
 $e^+ e^- \to b \bar t u \bar d$&$\mathcal{O}(10^{-1})$&$\mathcal{O}(10^{-2})$
    \end{tabular}\caption{Effective Weinberg angle in the fermion-$Z$ couplings}\label{tab:sw-eff}
     \end{center}
\end{table}

\begin{table}[htbp]
  \begin{center}
    \begin{tabular}{c||c|c||c|c|}
     Process           &$U \leftrightarrow R_\xi$ & corrected& WI&corrected \\\hline
 $\bar b b \to Z Z$& $\surd$ &  $\surd$& $\surd$ & $\surd$  \\
 $\bar b b \to Z H$& $\surd$ &$\surd$&  $\mathcal{O}(10^{-2})$ &$\surd$ \\
$\bar b b \to b \bar b H$& ${O}(10^{-4}) $& $\surd$ &$\mathcal{O}(10^{-3})$  &   $\surd$ \\
$\mu^- \mu^+  \to t \bar t H$&$\mathcal{O}(10^{-7})$  & $\surd$& $\mathcal{O}(10^{-2})$ & $\surd$ \\
 $d\bar d \to t\bar t t \bar t$& $\mathcal{O}(10^{-6})$&$\mathcal{O}(10^{-6})$&$\mathcal{O}(10^{-3})$&$\mathcal{O}(10^{-3})$\\
 $b\bar b \to t\bar t t \bar t$& $\mathcal{O}(10^{-2})$& $\mathcal{O}(10^{-2})$&$\mathcal{O}(10^{-3})$&$\mathcal{O}(10^{-3})$
    \end{tabular}\caption{Effects of corrected $HZZ$ coupling}\label{tab:sw-wrong}
     \end{center}
\end{table}
We next turn to the case of a Weinberg angle different from the on-shell value \eqref{eq:sw-os}. The numerical results in \fref{tab:sw-wrong} show the behavior expected from the discussion in \fref{sec:ward-cond}. Wee see that the 4 point Ward Identities are violated for the $\bar f f \to ZH$ amplitudes and the   $\bar f f \to ZZ$ amplitude with 2 contractions. The violations of the 4 point Ward Identities cause errors in the scattering amplitudes starting from the 5 point functions. This is indeed what happens, however in the examples considered, the numerical effects are small for realistic processes \footnote{In polarized amplitudes the relative errors in `forbidden' amplitudes are considerable larger}. 

As we have seen in \fref{sec:ward-cond}, the modification of the $HZZ$ Feynman rules can save the Ward Identity for $\bar f f\to ZH$. From the numerical results obtained with the modification from \fref{eq:hzz-mod} we find that this is indeed the case and inconsistencies in the amplitudes can be delayed until 6 point functions are considered.

\addtocounter{chapter}{1}
\chapter*{Summary and Outlook}
\markboth{SUMMARY AND OUTLOOK}{}
\addcontentsline{toc}{part}{Summary and Outlook}
Predictions for scattering processes with a large number of external particles, that will become important at future colliders, make automatized calculations of scattering cross sections indispensable. In \fref{chap:intro} we have reviewed the importance of maintaining gauge invariance in numerical calculations and the need for automatized gauge checks  that allow to verify both the consistency of the Feynman rules and the numerical stability of the code used in the calculation. 

In \fref{part:reconstruction}, we have identified simple, model independent, tools to verify the gauge invariance of the Lagrangian used in a numerical calculation. We found that the Ward Identities for on shell 4 point amplitudes are sufficient to reconstruct the Feynman rules of a spontaneously broken gauge theory, apart from the quartic Higgs coupling that requires the use of a 5 point Ward Identity. However, the Ward Identities with several momentum contractions have to be used to reconstruct the Goldstone boson couplings. The connection between the Ward Identities for Green's functions and the STIs for irreducible vertices---that are known to be sufficient for the reconstruction of the Feynman rules---has been clarified in \fref{chap:diagrammatica}. In the analysis, we have used new identities for vertex functions with several momentum contractions that we have derived in \fref{chap:sti}. Additional work is required to determine  a minimal set of identities to verify the supersymmetry of SUSY particle physics models, building on the results of \cite{Reuter:2002,ReuterOhl:2002}. 

The identities necessary for the gauge checks have been implemented in the matrix element generator \OMEGA, as described in \fref{sec:omega} and have been used in debugging the implementation of the complete Feynman rules of the Standard Model in $R_\xi$ gauge. The infrastructure for the implementation of STIs for off shell amplitudes that are relevant for checks of supersymmetry \cite{Reuter:2002,ReuterOhl:2002} and in loop calculations \cite{Kilian:loops} has also been provided.

In \fref{chap:diagrammatica} we have given a new proof for the formalism of flips \cite{Boos:1999} for the determination of gauge invariant subsets of Feynman diagrams (groves). Our proof clarifies the precise definition of gauge flips in spontaneously broken gauge theories that has been applied to the classification of the gauge invariance classes in \fref{chap:ssb_groves}. We found new gauge invariance classes in theories with a nonlinearly realized scalar sector. In this case the groves in theories with only neutral Higgs bosons can be classified according to the number of internal Higgs boson lines. These results are also relevant for calculations in unitarity gauge. In theories with a linear realized scalar sector in $R_\xi$ gauge, no additional groves compared to the unbroken case exist. The applications of gauge flips to loop diagrams is currently being studied \cite{Ondreka:2003} and the extension of our proof of \fref{chap:diagrammatica} to loop diagrams, using the Feynman tree theorem \cite{Feynman:1963} in the formulation of \cite{Kilian:loops}, is under investigation.  

In realistic calculations, the need to include finite width effects and the introduction of  running coupling constants violates gauge invariance. In \fref{chap:width} we have reviewed schemes that have been proposed to treat unstable gauge bosons in effective tree level calculations and that we have implemented in \OMEGA. We have investigated these schemes numerically in the single $W$ production process $e^-e^+\to e^- \bar \nu_e u\bar d$, both at the matrixelement and at the cross section level. We have obtained results consistent with existing literature and find that the fixed width \cite{LopezCastro:1991} and the complex mass scheme \cite{Denner:1999} give consistent results while the timelike and the running width scheme \cite{Baur:1995} are unreliable at large energies. We can also confirm problems of the nonlocal vertex scheme \cite{Beenakker:1999} with unitarity that recently have been reported in \cite{Beenakker:2003}. Results for 6 fermion production processes and comparison with \cite{Dittmaier:2002} will be given in future work.

Problems with gauge invariance arise also from an inconsistent introduction of running coupling constants. We have discussed this issue in \fref{chap:running} for the example of the Weinberg angle in the Standard Model, using the constraints from  the Ward Identities obtained in \fref{part:reconstruction}. We find that discrepancies between unitarity and $R_\xi$ gauge  at the per cent level can arise from a inconsistent use of the Weinberg angle in the triple couplings involving Goldstone bosons and in the neutral current couplings. The inconsistencies can be reduced in  the considered examples if an on-shell value of the Weinberg angle is used in the Goldstone boson couplings and in the Higgs-Z couplings.

\section*{Acknowledgments}
I am grateful to  Prof. Manakos for the possibility to write this thesis in his working group.
Special thanks go to Dr.Thorsten Ohl for his advice, the encouragement of good ideas (and dis-encouragement of bad ideas) and technical support. I also thank the other members of the Darmstadt high energy physics working group, especially David Ondreka for arising my interest in groves in spontaneously broken gauge theories and useful discussions on programming questions,  and Dr.J\"urgen Reuter for his cooperation in the implementation of STIs in \OMEGA. 

I like to thank Prof. R\"uckl for making it possible to finish this work in W\"urzburg. I am grateful for the kind reception I received from his working group and that of  Prof. Fraas. I like to mention especially Federico v.d.Pahlen and Olaf Kittel who shared their office with me and Dr.Heinrich P\"as who made sure I got my coffee breaks.

Finally I thank my parents for their support throughout my studies.

\vspace{1cm}

\noindent This work has been supported by the Bundesministerium f\"ur Bildung und Forschung Germany, (05HT1RDA/6).
\appendix
\part{Appendix}
\chapter{BRS symmetry}\label{app:swi}
In this appendix we briefly sketch the formalism of BRS invariance  \cite{Becchi:1975} and set up our  notation. The general formalism is reviewed in \fref{app:brs}, the application to the spontaneously broken gauge theory introduced in \fref{chap:ssb-lag} is given in \fref{app:brs-ssb}. A more detailed discussion can be found in standard textbooks \cite{Weinberg:1995,Kugo:1997,Boehm:2001}.
In \fref{app:brs-graph} we introduce our graphical notation for BRS transformations and STIs.  

\section{BRS formalism}\label{app:brs}
The BRS transformations are generated by a fermionic, hermitian operator $Q$, the BRS charge. We write the transformations of a general field  $\Psi$ as
\begin{equation}
\Delta_{\text{BRS}}\Psi=\epsilon\deltabrs\Psi\equiv\lbrack\ii\epsilon Q,\Psi\rbrack 
\end{equation}
where $\epsilon$ is a Grassmann number.

The  BRS transformations are chosen so the BRS-charge is \emph{nilpotent}:
\begin{equation}
  Q^2=0
\end{equation}
This is a crucial property that allows to decompose the Hilbert space of the theory into physical and unphysical states and to define BRS invariant gauge fixing terms as we will see below. In gauge theories the BRS transformation of physical fields is obtained by replacing the gauge-transformation parameter by the product between $\epsilon$ and a ghost field. The transformation of the ghost is chosen in an appropriate way to make the BRS transformation nilpotent. 
 
Because of the nilpotency of $Q$, states that are obtained by applying $Q$ to another arbitrary state (so called `BRS exact states') have vanishing norm: 
\begin{equation}
\ket{\psi}= Q\ket{\eta}:\quad \braket{\psi|\psi}=0\qquad  \forall \ket{\eta}
\end{equation}
States that are annihilated by the BRS charge  are called `BRS closed'. They are orthogonal to the exact states:
\begin{equation}
\braket{\psi|\phi}=\braket{\eta|Q|\phi}=0 \qquad \forall \ket{\psi}=Q\ket{\eta} \quad ,\quad Q\ket{\phi}=0
\end{equation}
Therefore we can decompose the Hilbert space into orthogonal subspaces. Because of the nilpotency of $Q$, a closed state stays closed if one adds an arbitrary exact state. 

One can show (see e.g. \cite{Kugo:1997}) that provided the BRS closed states have positive norm, it is consistent to define the \emph{physical} states of the theory as closed states modulo exact states:
\begin{equation}\label{eq:q-cohomology} 
\begin{aligned}
  Q \ket{\psi_{\text{phys}}}&=0\\
    \ket{\psi_{\text{phys}}}& \sim  \ket{\psi_{\text{phys}}}+Q\ket{\eta}
\end{aligned}
\end{equation}
In mathematical terms, this is the \emph{cohomology} of the operator $Q$. We will see in \fref{app:brs-ssb} that in a spontaneously broken gauge theory this condition eliminates the unphysical components of the gauge fields, the Goldstone bosons and the ghosts from the spectrum.  

Gauge fixing of the lagrangian can be performed by adding the BRS transformation of an \emph{arbitrary} functional $\mathcal{F}$  of the fields with ghost number $-1$ to the classical lagrangian $\mathscr{L}_0$ 
\begin{equation}\label{eq:gauge-fix-funct}
  \mathscr{L}=\mathscr{L}_0 +\deltabrs\mathcal{F}\lbrack\Psi(x)\rbrack
\end{equation}
This construction ensures that matrix elements between physical states are independent of the gauge fixing:
\begin{equation}
 \braket{\phi_{\text{phys}}|\mathscr{L}|\psi_{\text{phys}}}= \braket{\phi_{\text{phys}}|\mathscr{L}_0|\psi_{\text{phys}}}
\end{equation}
We can derive the general \emph{Slavnov Taylor Identities} of the theory by sandwiching the commutator (or anticommutator) of an arbitrary products of fields with the BRS-charge between physical fields:
\begin{multline}\label{eq:gf-sti-app}
0=\braket{\phi_{\text{phys}}|\tprod{[\ii Q,\Psi_1\Psi_2\dots\Psi_n]_{\pm}}|\psi_{\text{phys}}}\\
=\sum_i(-)^{s(i)}\braket{\phi_{\text{phys}}|\tprod{\Psi_1\dots\deltabrs\Psi_i\dots \Psi_n}|\psi_{\text{phys}}}
\end{multline}
\section{Application to a spontaneously broken gauge theory}\label{app:brs-ssb}
\sectionmark{Application to a SBGT}
We now apply the general BRS formalism to the spontaneously broken gauge theory discussed in \fref{chap:ssb-lag}. In the parametrization of the fields introduced in \fref{sec:ssb-fields}, the BRS transformations are given by
\begin{subequations}\label{eq:ssb-brs}
  \begin{align}
   \deltabrs \psi_{Li}&= \ii c_a \tau^{a}_{Lij}\psi_{Lj}\\
   \deltabrs \psi_{Ri}&= \ii c_a \tau^{a}_{Rij} \psi_{Lj}\\
   \deltabrs \phi_A &=- c_a T^a_{AB}\phi_B\\
   \deltabrs W_{\mu a}&=\dmu c_a+f^{abc}W_{\mu b}c_c \label{eq:gauge-brs} \\
   \deltabrs c_a&= -\frac{1}{2}f^{abc}c_b c_c\\
   \deltabrs\bar c_a&= B_a\\
   \deltabrs B_a&=0
 \end{align}
\end{subequations}
The transformations of the physical fields are gauge transformations with $\omega$ replaced by $c$ and the transformation of ghost and antighost are chosen so that $Q$ is nilpotent. 

For the Higgs and Goldstone bosons the explicit transformation laws in the parametrization  \eqref{eq:ssb-gen} are given by
\begin{equation}\label{eq:brs-scalar}
\begin{aligned}
\deltabrs H_i&=c_c (u^c_{ib}\phi_b -T^c_{ij}H_j)\\
\deltabrs \phi_a&=-m_ac_a-c_c(t^c_{ab}\phi_b+ u^c_{ai}H_i) 
\end{aligned}
\end{equation}
We will see that the inhomogeneous term $m_ac_a$ in the transformation of the Goldstone bosons ensures that they are not part of the physical spectrum. 

The BRS transformations of asymptotic fields receive only contributions from the terms in \eqref{eq:ssb-brs} linear in the fields \cite{Boehm:2001,Kugo:1997}. Therefore the only asymptotic fields transforming nontrivially are:
\begin{eqnarray*}
\deltabrs  W^\mu_{\In/\Out} &=&\dmu c_{\In/\Out}\\
\deltabrs \phi_{\In/\Out}&=&-m_{W}c_{\In/\Out} \\
\deltabrs \bar c_{\In/\Out}&=&B_{\In/\Out}
\end{eqnarray*}
According to the definition of physical states given in \fref{eq:q-cohomology}, the physical spectrum consists only of fermions, physical scalars and three components of the gauge bosons. The scalar mode of the gauge bosons with polarization $\propto p_\mu$, the Goldstone bosons and the antighost are eliminated from the physical spectrum because they are not annihilated by Q while the ghost fields and the auxiliary field $B$ are zero-norm states in the image of $Q$. 

To perform gauge fixing, we have to chose a gauge fixing functional $\mathcal{F}[\Psi]$ in \fref{eq:gauge-fix-funct}. The Faddeev-Popov lagrangian \cite{Faddeev:1967}, originally derived from the path integral, is reproduced by the choice
\begin{equation}
  \mathcal{F}= \bar c_a(G_a+\frac{\xi}{2}B_a)
\end{equation}
that results in the lagrangian
\begin{subequations}\label{eq:gauge-fix}
\begin{equation}
  \mathscr{L}=\mathscr{L}_0+\mathscr{L}_{GF}+\mathscr{L}_{FP}
\end{equation}
with 
\begin{equation}
\mathscr{L}_{GF}=B_aG_a+\frac{\xi}{2}B_a^2
\end{equation}
and
\begin{equation}
 \mathcal{L}_{FP}=-\bar c_a (\deltabrs G_a)   
\end{equation}
\end{subequations}

The explicit form of the gauge fixing function is given in a linear  $R_\xi$-gauge by
\begin{equation}\label{eq:rxi-gf}
G_a=(\dmu W^\mu_a-\xi m_{W_a}\phi_a)  
\end{equation}
The equation of motion for the auxiliary field $B$ resulting from \eqref{eq:gauge-fix}
\begin{equation}\label{eq:b-eom-app}
B_a=-\frac{1}{\xi}G_a=-\frac{1}{\xi}(\dmu W_a^\mu-\xi m_{W_a}\phi_a)
\end{equation} 
allows to write the gauge fixing lagrangian as
\begin{equation}
  \mathscr{L}_{GF}=-\frac{1}{2\xi}G_a^2
\end{equation}
Using the parametrization \eqref{eq:ssb-lagrangian} we find
\begin{multline}
  \deltabrs G_a=\dmu (\deltabrs W_a^\mu)-\xi m_{W_a}\deltabrs \phi_a\\
  =\dmu(\partial^\mu c_a+f^{abc}W^\mu_bc_c)+\xi m_{W_a}^2c_a+\xi m_{W_a}c_c(t^c_{ab}\phi_b +u^c_{ai}H_i) 
\end{multline}
so, after an integration by parts, the ghost lagrangian becomes
\begin{equation}\label{eq:ghost-lag}
  \mathscr{L}_{FP}=\dmu\bar c_a D^\mu c_a -\xi m_{W_a}^2\bar c_ac_a -\xi m_{W_a}\bar c_a(t^c_{ab}\phi_b +u^c_{ai}H_i)c_c 
\end{equation}
and the Feynman rules for the ghosts are given by
\begin{subequations}\label{subeq:ghost-fr}
  \begin{align}
    \bar c_aW^\mu_b c_c&: f^{abc} p_a^\mu \\
    \bar c_a \phi _b c_c&: -\ii \xi m_{W_a} t^c_{ab} \\
    \bar c_a H_i c_c&: -\ii \xi m_{W_a} u^c_{ai}=-\frac{\ii}{2}\xi g_{HWW}^{iac}
  \end{align}
\end{subequations}
The propagators resulting from \fref{eq:gauge-fix} are

\begin{align}
D_{W\mu\nu}(p)&\equiv\int d^4x d^4y\,e^{-\ii p(x-y)}\greensfunc{W_\mu(x)
  W_\nu(y)}\nonumber\\
&=\frac{-\ii}{p^2-m_w^2}\left(g_{\mu\nu}-(1-\xi)\dfrac{p_\mu p_\nu}{p^2-\xi
  m_w^2}\right) \label{eq:rxi-prop}\\
D_{c}(p)&\equiv\int d^4x d^4y\,e^{-\ii p(x-y)}\greensfunc{c(x)
  \bar c(y)}=\frac{i}{p^2-\xi m_w^2}\label{eq:ghost-prop}
\end{align}
\section{Graphical notation}\label{app:brs-graph}
To represent STIs diagrammatically, we will introduce the following graphical notation for the homogeneous part of the BRS-transformations of the fields:
\begin{equation}\label{eq:brs-graph}
\deltabrs \Phi_i = T^a_{ij}c_a\Phi_j: \qquad
\parbox{15mm}{
\begin{fmfchar*}(15,15)
\fmfright{g1}
\fmftop{g2}
\fmfbottom{b}
\fmf{plain,label=$\Phi_j$}{g1,g2}
\fmf{ghost,label=$c_a$}{b,g1}
\fmfv{decor.shape=square,decor.filled=empty,decor.size=5,label=$T^a_{ij}$}{g1}
\end{fmfchar*}}
\end{equation}
The contraction with a momentum is denoted by a black square. Therefore the BRS transformation of a gauge boson looks like
\begin{equation}\label{eq:brs-gauge-graph}
\deltabrs W^a_\mu = \dmu c^a+f^{abc}W_b c_c: \quad
\parbox{20mm}{
\fmfframe(0,0)(5,0){
\begin{fmfchar*}(15,15)
\fmfright{g1}
\fmfleft{g2}
\fmf{ghost,label=$c^a$,la.si=left}{g2,g1}
\fmfv{decor.shape=square,decor.filled=full,decor.size=5,label=$\ii p^\mu$}{g1}
\end{fmfchar*}}}
\quad+
\parbox{20mm}{
\fmfframe(0,0)(5,0){
\begin{fmfchar*}(15,15)
\fmfright{g1}
\fmftop{g2}
\fmfbottom{b}
\fmf{photon,label=$W_b$}{g1,g2}
\fmf{ghost,label=$c_c$}{b,g1}
\fmfv{decor.shape=square,decor.filled=empty,decor.size=5,label=$f^{abc}$}{g1}
\end{fmfchar*}}}
\end{equation}
The multiplication with a gauge boson mass will be denoted by a cross so the transformation of a Goldstone boson is written as
\begin{multline}\label{eq:brs-gold-graph}
\deltabrs \phi_a=-m_{W_a}c_a-c_c(t^c_{ab}\phi_b+ u^c_{ai}H_i):\\
\parbox{25mm}{
\fmfframe(0,0)(10,0){
\begin{fmfchar*}(15,15)
\fmfright{g1}
\fmfleft{g2}
\fmf{ghost,label=$c^a$,la.si=left}{g2,g1}
\fmfv{decor.shape=cross,decor.size=5,label=$(-m_{W_a})$}{g1}
\end{fmfchar*}}}
\qquad+
\parbox{25mm}{
\fmfframe(0,0)(10,0){
\begin{fmfchar*}(15,15)
\fmfright{g1}
\fmftop{g2}
\fmfbottom{b}
\fmf{dashes,label=$\phi_b$}{g1,g2}
\fmf{ghost,label=$c_c$}{b,g1}
\fmfv{decor.shape=square,decor.filled=empty,decor.size=5,label=$-t^{c}_{ab}$}{g1}
\end{fmfchar*}}}\quad+
\parbox{25mm}{
\fmfframe(0,0)(10,0){
\begin{fmfchar*}(15,15)
\fmfright{g1}
\fmftop{g2}
\fmfbottom{b}
\fmf{dashes,label=$H_i$}{g1,g2}
\fmf{ghost,label=$c_c$}{b,g1}
\fmfv{decor.shape=square,decor.filled=empty,decor.size=5,label=$-u^{c}_{ai}$}{g1}
\end{fmfchar*}}}
\end{multline}

Because of the nonlinearity of the BRS transformations, these transformations receive radiative corrections. The insertion of a BRS transformed gauge field in a Green's function therefore is represented as
\begin{center}
\begin{equation*}
\braket{0|\Tprod{W_\mu\bar c \deltabrs W_\nu}|0}= \quad 
\parbox{15mm}{
\begin{fmfchar*}(15,15)
\fmfleft{g1,g2}
\fmfright{b}
\fmf{ghost,label=$c$}{y,g2}
\fmf{ghost}{g1,y}
\fmf{photon,label=$W$}{y,b}
\fmfblob{10}{y}
\fmfv{decor.shape=square,decor.filled=full,decor.size=5}{g1}
\end{fmfchar*}}
\quad+\quad
\parbox{15mm}{
\begin{fmfchar*}(15,15)
\fmfleft{g1,g2}
\fmfright{b}
\fmf{photon}{b,m}
\fmf{ghost}{m,g2}
\fmf{ghost,right=0.5}{g1,m}
\fmffreeze
\fmf{photon,right=0.5}{m,g1}
\fmfv{decor.shape=square,decor.filled=empty,decor.size=5}{g1}
\fmfv{decor.shape=diamond,decor.filled=empty,d.si=13}{m}
\end{fmfchar*}}
\end{equation*}
 \end{center}
The second term consists of the tree level contribution from \fref{eq:brs-gauge-graph} plus loop corrections. The tree level contribution is a disconnected diagram, therefore we denote the Green's functions with insertions of BRS-transformed fields by diamond-shaped blobs to distinguish them from connected Green's functions. In the above example, the tree level diagram and a one loop contribution look like:
 \begin{center}
\begin{equation*}
\parbox{15mm}{
\begin{fmfchar*}(15,15)
\fmfleft{g1,g2}
\fmfright{b}
\fmf{photon}{b,m}
\fmf{ghost}{m,g2}
\fmf{ghost,right=0.5}{g1,m}
\fmffreeze
\fmf{photon,right=0.5}{m,g1}
\fmfv{decor.shape=square,decor.filled=empty,decor.size=5}{g1}
\fmfv{decor.shape=diamond,d.filled=empty,d.si=13}{m}
\end{fmfchar*}}
\quad =\quad
\parbox{15mm}{
\begin{fmfchar*}(15,15)
\fmfleft{g1,g2}
\fmfright{b}
\fmf{photon}{b,g1}
\fmf{ghost}{g1,g2}
\fmfv{decor.shape=square,decor.filled=empty,decor.size=5}{g1}
\end{fmfchar*}}
\quad+\quad
\parbox{15mm}{
\begin{fmfchar*}(15,20)
\fmfleft{g1,g2}
\fmfright{b}
\fmf{phantom}{g1,m1}
\fmf{phantom}{g2,m1}
\fmf{photon,tension=2.5}{b,m1}
\fmffreeze
\fmf{phantom,tension=2.5}{g1,c}
\fmf{ghost,right=0.5}{c,m1}
\fmf{ghost,right=0.5}{m1,m2}
\fmf{ghost,tension=2.5}{m2,g2}
\fmf{photon}{m2,c}
\fmfv{decor.shape=square,decor.filled=empty,decor.size=5}{c}
\fmfdot{m1}
\fmfdot{m2}
\end{fmfchar*}}
+\dots
\end{equation*}
 \end{center}
For tree level applications it is sufficient to keep only the first term. 

The Nakanishi-Lautrup field is represented by a double line. We will use the notation 
\begin{equation*}
\Greensfunc{\Phi_1\dots\Phi_n B}= \quad 
\parbox{15mm}{
\begin{fmfchar*}(15,15)
\fmfleftn{p}{4}
\fmfright{b}
\fmfpolyn{empty,smooth}{y}{5}
\fmf{double,label=$B$}{y1,b}
\fmf{plain,label=$\Phi_1$,label.side=right}{p1,y5}
\fmf{plain}{p2,y4}
\fmf{plain}{p3,y3}
\fmf{plain,label=$\Phi_n$,label.side=left}{p4,y2}
\end{fmfchar*}}=
 \quad 
\parbox{15mm}{\begin{fmfchar*}(15,15)
\fmfleftn{p}{4}
\fmfright{b}
\fmfpolyn{empty,smooth}{y}{5}
\fmf{double}{b,i}
\fmf{photon,label=$W$}{y1,i}
\fmf{plain}{p1,y5}
\fmf{plain}{p2,y4}
\fmf{plain}{p3,y3}
\fmf{plain}{p4,y2}
\fmfdot{i}
\end{fmfchar*}}
\quad+\quad 
\parbox{15mm}{\begin{fmfchar*}(15,15)
\fmfleftn{p}{4}
\fmfright{b}
\fmfpolyn{empty,smooth}{y}{5}
\fmf{double}{i,b}
\fmf{dashes,label=$\phi$}{y1,i}
\fmf{plain}{p1,y5}
\fmf{plain}{p2,y4}
\fmf{plain}{p3,y3}
\fmf{plain}{p4,y2}
\fmfdot{i}
\end{fmfchar*}}
\end{equation*}
for the insertion of the auxiliary field $B$ into Green's functions. The bilinear Feynman rules for the coupling to gauge- and Goldstone bosons follow from the equation of motion \eqref{eq:b-eom-app}:
\begin{equation}
\begin{aligned}
\parbox{15mm}{\begin{fmfchar*}(15,15)
\fmfleft{p}
\fmfright{b}
\fmf{double}{i,b}
\fmf{dashes,label=$\phi$}{p,i}
\fmfdot{i}
\end{fmfchar*}}
\quad &= m\\
\parbox{15mm}{\begin{fmfchar*}(15,15)
\fmfleft{p}
\fmfright{b}
\fmf{double}{i,b}
\fmf{photon,label=$W$}{p,i}
\fmfdot{i}
\end{fmfchar*}}
\quad &= -\frac{1}{\xi}\ii p^\mu 
\end{aligned}
\end{equation}

As an example for the  graphical representation of a STI, we consider the STI for the $W\bar \psi\psi$ vertex: 
\begin{equation*}
\begin{array}{rccl}
0=\greensfunc{\lbrack \ii Q, \bar c\bar\psi \psi }=&
\greensfunc{B \bar\psi \psi }&-\greensfunc{\bar c c \Delta \bar \psi\psi
  }&+\greensfunc{\bar c\bar\psi c \Delta \psi}\vspace{5mm}\\
=&
\parbox{15mm}{
\begin{fmfchar*}(15,15)
\fmfleft{l}
\fmfright{r1,r2}
\fmf{double}{l,v}
\fmf{fermion}{r1,v}
\fmf{fermion}{v,r2}
\fmfblob{10}{v}
\end{fmfchar*}}
&+\parbox{15mm}{\begin{fmfchar*}(15,15)
\fmfleft{l}
\fmfright{r1,r2}
\fmf{fermion}{i,r2}
\fmf{ghost}{l,i}
\fmf{ghost,right=0.5}{i,r1}
\fmffreeze
\fmf{fermion,right=0.5}{r1,i}
\fmfv{decor.shape=diamond,d.filled=empty,d.si=13}{i}
\fmfv{decor.shape=square,decor.filled=empty,decor.size=5}{r1}
\end{fmfchar*}}
&+\parbox{15mm}{\begin{fmfchar*}(15,15)
\fmfleft{l}
\fmfright{r1,r2}
\fmf{ghost}{l,i}
\fmf{fermion}{r1,i}
\fmf{ghost}{l,i}
\fmf{ghost,left=0.5}{i,r2}
\fmffreeze
\fmfv{decor.shape=diamond,d.filled=empty,d.si=13}{i}
\fmf{fermion,right=0.5}{i,r2}
\fmfv{decor.shape=square,decor.filled=empty,decor.size=5}{r2}
\end{fmfchar*}}
\end{array}
\end{equation*}
Beginning with the 4 point functions, the structure of the `contact terms' with the insertion of the operator $\bar c \Delta\Phi$ gets more complicated, since the interactions of the ghosts with gauge, Higgs and Goldstone bosons from the lagrangian \eqref{eq:ghost-lag} have to be taken into account even on tree level (see \fref{eq:4-contact} in \fref{chap:sti}).

\chapter{Nonlinear realizations of symmetries}\label{app:nonlinear}
In \fref{app:ccwz} we review the theory of nonlinear realizations of symmetries
 that is useful for constructing effective field theory descriptions of spontaneously broken symmetries. In \fref{app:nl-sti} we derive the STIs relevant for the application to gauge flips for nonlinear realized scalar sectors in \fref{chap:ssb_groves}.
\section{General setup}\label{app:ccwz}
We will sketch the formalism of nonlinear realized symmetries
\cite{Coleman:1969} 
(for reviews see \cite{Feruglio:1993,Dobado:1997jx,Weinberg:1995,Kugo:1997}). A description of the formalism in the language of differential geometry is given in \cite{Alvarez-Gaume:1985}.
We consider a symmetry group $G$ that is spontaneously broken down to a subgroup $H$. Keeping the index convention introduced in \fref{chap:ssb-lag}, we denote the generators of $G$ by $T^a$, the generators of $H$ by $L^a$ and the the generators  that are not in
$H$ as $V^\alpha$.

Since $H$ is a subgroup of $G$, the unbroken generators must form an subalgebra:
\begin{subequations}
\begin{equation}\label{eq:l-comm}
[L^a,L^b]=f^{abc}L^c
\end{equation}
This shows that the structure constants with one unbroken index vanish. This implies that the broken generators carry a representation of the unbroken subalgebra:
\begin{equation}\label{eq:broken-unbroken-commutator}
 [L^a,V^\beta]= f^{a\beta\gamma}V^\gamma
\end{equation}
In general we must assume 
\begin{equation}
 [V^\alpha,V^\beta]= f^{\alpha\beta c}L^c+\ii f^{\alpha\beta\gamma}V^\gamma
\end{equation}
\end{subequations}
while for chiral symmetries the commutator of $V$s is only a linear combination of $L$s.

According to Goldstone's theorem, for every broken generator $V^\alpha$ there is a Goldstone boson $\phi_\alpha$. The goldstone bosons can be used to
parametrize the coset space $G/H$ by introducing the so called canonical representation
\begin{equation}
U=e^{\frac{\ii}{f}\phi_\alpha V^\alpha}\in G/H
\end{equation}
 One can show, that the $\phi$ transform nonlinearly under $G$ but
linearly under $H$ \cite{Coleman:1969}. 
The transformation of the $\phi$ is defined as is defined as follows: multiplication of $U$ by an arbitrary group element 
\begin{equation}
\mathcal{G}=e^{\ii\omega_aT^a}
\end{equation}
results in another element of $G$ that can be written as:
\begin{equation}\label{eq:nl-trans}
\mathcal{G}U=e^{\frac{\ii}{f}\phi'_\alpha(\phi,\omega) V^\alpha}e^{\ii\alpha_b(\phi,\omega)L^b} \equiv U'(\phi,\omega) \mathcal{H}(\phi,\omega)\in G
\end{equation}
Here the result of the transformation has been written as the product of an element of the coset space in the canonical representation and a `compensating' $H$ transformation $\mathcal{H}$. The nonlinear transformations of $\phi$ are defined as
\begin{equation}\label{eq:nl-dphi}
\phi_\alpha\to \phi'_\alpha(\phi,\omega)
\end{equation}
We will now introduce matter fields $\Phi$, carrying a representation of the unbroken subgroup $H$. We can lift this representation to a nonlinear realization of $G$ by introducing the transformation 
\begin{equation}
\Phi= U\Psi
\end{equation}
$G$ transformations are realized on the fields $\Psi$  as linear, $\phi$ dependent transformations of the unbroken subgroup $H$. To see this, we act with $\mathcal{G}\in G$ on the original field $\Phi$ and insert the definition of $\Psi$:
\begin{equation}
\mathcal{G}\Phi=\mathcal{G}U\Psi =U'(\phi,\omega) \mathcal{H}(\phi,\omega)\Psi
 \end{equation}
where $U'\in G/H$ and $\mathcal{H}(\phi,\omega)\in H$ are defined as in \eqref{eq:nl-trans}. Therefore we can take the transformed field as 
\begin{equation}\label{eq:nl-dpsi}
\Psi'=\mathcal{H}(\phi,\omega)\Psi
\end{equation}

Because of the field dependent transformations, the derivatives of $U$ and $\Psi$ have no simple transformation law:
\begin{equation}
\begin{aligned}
\dmu (\mathcal{G}U)&=(\dmu U')\mathcal{H}+U'(\dmu \mathcal{H})\\
\dmu \Psi'&= \mathcal{H}\dmu \Psi+(\dmu \mathcal{H})\Psi
\end{aligned}
\end{equation}
 and one has to construct covariant derivatives to write down kinetic terms for the fields.

The object 
 \begin{equation}
\mathcal{D}_\mu= U^\dagger\dmu U
 \end{equation}
transforms according to 
\begin{equation}
\mathcal{D}_\mu\to  \mathcal{H}^\dagger (\mathcal{D'}_\mu)\mathcal{H}+\mathcal{H}^\dagger\dmu \mathcal{H}
\end{equation}
where $\mathcal{D'}$ is made out of the $U'$. Because of the commutation relation \eqref{eq:l-comm}, the inhomogeneous last term is an element of $H$ so the component of $\mathcal{D}_\mu$ in $G/H$
\begin{equation}
 \mathcal{U}_\mu^\alpha \equiv \Tr[\mathcal{D}_\mu V^\alpha] 
\end{equation}
 has a simple transformation law:
\begin{equation}
 \mathcal{U'}_\mu=\mathcal{H}\mathcal{U'}_\mu \mathcal{H}^\dagger
\end{equation}
where we have defined $\mathcal{U}_\mu\equiv \mathcal{U}_\mu^\alpha V^\alpha$.
This object serves as a covariant derivative of the Goldstone bosons and can be used to construct a kinetic term:
\begin{equation}
\mathcal{L}_{\phi}=f^2\tr[\mathcal{U}^{\dagger}_\mu\mathcal{U}^{\mu}]=f^2\tr[\dmu U^\dagger \partial^\mu U]
\end{equation}
The component along $H$:
\begin{equation}
 \mathcal{E}_\mu^a \equiv \Tr[\mathcal{D}_\mu L^a] 
\end{equation}
transforms similar to a gauge field:
\begin{equation}
 \mathcal{E'}_\mu=\mathcal{H}\mathcal{E}_\mu \mathcal{H}^\dagger-\dmu\mathcal{H}\mathcal{H}^\dagger
\end{equation}
and can be used to make a covariant derivative for the matter fields $\Psi$: 
\begin{equation}
D_\mu\Psi = (\dmu + \mathcal{E}_\mu^aL^a)\Psi
\end{equation}

The construction for local transformations is similar, except one has to covariantize also with respect to the gauge symmetry \cite{Kugo:1997,Weinberg:1995,Coleman:1969,Feruglio:1993,Dobado:1997jx}.
\section{STIs for nonlinearly realized symmetries}\label{app:nl-sti}
\subsection{Abelian toy model}
It is instructive to  check the  rather  general arguments given in \fref{sec:nl-flips} in a more concrete way using the STIs. We will discuss the toy model of an spontaneously broken abelian  gauge theory first. We start  with a complex  scalar
field in a linear parametrization:
\begin{equation}
\Phi=\frac{1}{\sqrt 2}\left ((v+H)+\ii \phi \right)
\end{equation}
that transforms under BRS transformations according to 
\begin{equation}\label{eq:l-symm} 
  \begin{aligned}
    \deltabrs H&=g c\phi\\
    \deltabrs\phi&=-m_Ac-gHc
  \end{aligned}
\end{equation}
The cubic interaction terms with the gauge boson arising from the square of
the covariant derivative
\begin{equation}
D_\mu\Phi=\dmu \Phi+\ii g A_\mu \Phi
\end{equation}
and the potential
\begin{equation}
V(\Phi)=\mu^2|\Phi|^2-\lambda|\Phi|^4
\end{equation}
are given by
\begin{equation}\label{eq:siam_int}
\mathcal{L}_{Int}=g(\phi \overleftrightarrow{\partial_\mu} H )A^\mu+gm_AHA^2 -\frac{g m_H^2}{2m_A} (H^3+H\phi^2)+\dots
\end{equation}
where we have used
\begin{equation}
m_A=gv\qquad m_H=\sqrt{2}\mu 
\end{equation} 
and
\begin{equation}
\lambda=\frac{\mu^2}{v^2}=\frac{m_H^2}{2v^2}=\frac{g^2 m_H^2}{2m_A^2}
\end{equation}

To obtain the nonlinear realization, 
now we parameterize the scalar fields  as
\begin{equation}
\Phi=(v+H(x))e^{\ii\phi(x)/v}
\end{equation}
In accordance with the general theory, the reparametriezd Higgs bosons
transform trivial under the symmetry and the Goldstone boson transforms inhomogeneously:
\begin{equation}\label{eq:nl-symm} 
  \begin{aligned}
    \deltabrs H&=0\\
    \deltabrs\phi&=-m_Ac
  \end{aligned}
\end{equation}
Because of this `shift symmetry', only derivatives of $\phi$ can enter the
lagrangian and indeed, from the square of the covariant derivative we find
this time
\begin{equation}
\mathcal{L}_{Int}=  gm_A H
A^2 +2g HA^\mu\dmu\phi+\frac{1}{v}H(\dmu \phi)^2 +\dots
\end{equation}

We will now discuss the STI for the $W\phi H$ vertex that is relevant for the
definition of the gauge flips in spontaneously broken gauge theories as we have seen in \fref{sec:sti-flips}.
In the case of the nonlinear realization, the STI is simply
\begin{equation}\label{eq:phihw-nl}
\ii p_{a \mu_a} \onepi{ A^{\mu}(p_a)
  \phi(p_b)H(k_i)}+m_{A}\onepi{\phi(p)\phi(p_b)H(k_i)}=0
\end{equation}
since the transformations \fref{eq:nl-symm} are trivial. Note that in the nonlinear parametrization the Higgs boson decouples from the ghosts so no ghost term appears in \fref{eq:phihw-nl}. From the Feynman rules we can
verify this identity :
\begin{equation}
\begin{aligned}
 \ii p_{a \mu_a} \onepi{ A^{\mu}(p_a)
  \phi(p_b)H(k_i)}&=2\ii g (p_a\cdot p_b)\\
m_{A}\onepi{\phi(p)\phi(p_b)H(k_i)}&=-2\ii g(p_a\cdot p_b)
\end{aligned}
\end{equation}
where the factor $2$ in the last line is a symmetry factor. 

Therefore, a Ward Identity like identity is valid also for off shell Higgs and Goldstone
bosons. This ensures that the diagram
\begin{equation}\label{eq:3wphi-diag}
\parbox{15mm}{
\begin{fmfchar*}(15,15)
\fmfleft{a,b}
\fmfright{f1,f2}
\fmf{photon}{a,fwf1}
\fmf{dashes,label=$H$}{fwf1,fwf2}
\fmf{photon}{fwf2,b}
\fmf{dashes,label=$\phi$}{fwf1,f1}
\fmf{photon}{fwf2,f2}
\fmfdot{fwf1}
\fmfdot{fwf2}
\end{fmfchar*}}
\end{equation}
satisfies the Ward Identity by itself. 

This is in contrast to the linearly realized symmetry that  transforms the Higgs and Goldstone
bosons into one another. In this case  we have a nontrivial STI :
\begin{multline}\label{eq:phihw-lin}
\ii p_{a \mu_a} \onepi{ W_a^{\mu}(p_a)
  \phi_b^{\nu}(p_b)H_i(k_i)}+m_{W_a}\onepi{\phi_a(p)\phi_b(p_b)H_i(k_i)}\\
= -g\onepi{H(p_a+p_b)H(k_i)}+g \onepi{\phi_b^\mu(p_b)\phi_b(-p_b)}
\end{multline}
Indeed, inserting the Feynman rules gives 
\begin{multline}
\ii p_{a \mu_a} \onepi{ W_a^{\mu}(p_a)
  \phi_b^{\nu}(p_b)H_i(k_i)}+m_{W_a}\onepi{\phi_a(p)\phi_b(p_b)H_i(k_i)}\\
= -\ii gp_a\cdot (k_i-p_b)-\ii gm_H^2=\ii g\left[(k_i^2-m_H^2)-p_b^2\right]
\end{multline}
\subsection{General Symmetries}
For the derivation of the BRS transformations in theories with nonlinear symmetries \cite{Dobado:1997jx}, we consider the  transformations \eqref{eq:nl-dphi} and \eqref{eq:nl-dpsi} for infinitesimal parameters $\omega=\epsilon$. In linear order  of $\epsilon$ we can write
\begin{equation}
\begin{aligned}
U'(\phi,\epsilon)&=1+\ii \mathcal{K}_\alpha^b (\phi)\epsilon_b V^\alpha +\mathcal{O}(\epsilon^2) \\
\mathcal{H}(\phi,\epsilon)&=1+\ii\Omega_a^b(\phi) \epsilon_b L^a  +\mathcal{O}(\epsilon^2)
\end{aligned}
\end{equation}
In the language of differential geometry, the quantities $\mathcal{K}^a_\beta$ are Killing vector fields, since they generate isometries in the coset space $G/H$. The object $\Omega^a=\Omega_b^aL^b$ is known as $H$-compensator \cite{Alvarez-Gaume:1985}.

The BRS transformations are now given by 
\begin{equation}
\begin{aligned}
\deltabrs \Psi&=-\ii c_a \Omega^a(\phi) \Psi\\
\deltabrs \phi_\alpha&=f c_b \mathcal{K}_\alpha^b (\phi) 
\end{aligned}
\end{equation}
We will still use the linear gauge fixing \eqref{eq:rxi-gf} instead of a gauge fixing function in terms of the $U$ \cite{Dobado:1997jx}. Note that the gauge fixing term in the nonlinear parametrization contains no Higgs-ghost interaction since the Higgs don't appear in the BRS transformation of the Goldstone bosons.
The STIs for Green's functions depend nonlinear on the GBs through the 
Killing vectors and Compensators, e.g.
\begin{multline}\label{eq:phihw-sti-nl}
\ii p_{a \mu_a} \greensfunc{ W_a^{\mu}(p_a)
  \phi_\beta^{\nu}(p_b)H_i(k_i)}+m_{W_a}\greensfunc{\phi_a(p)\phi_\beta(p_b)H_i(k_i)}\\
= f\greensfunc{(\mathcal{K}^a_\beta(\phi))(p_a+p_b)H_i(k_i)}-\ii \greensfunc{\phi_\beta(p_b)(\Omega^\alpha_{ij}(\phi)H_j)(-p_b)}
\end{multline}
Diagrammatically, this can be written as
\begin{equation}
\parbox{20mm}{
\begin{fmfchar*}(15,15)
\fmfleft{l}
\fmfright{r1,r2}
\fmfpolyn{smooth,filled=shaded}{g}{3}
\fmf{double}{l,g1}
\fmf{plain,label=$H$}{r1,g2}
\fmf{dashes,label=$\phi$}{r2,g3}
\end{fmfchar*}}
=\qquad \parbox{25mm}{
\fmfframe(0,5)(0,5){
\begin{fmfchar*}(15,15)
\fmfleft{l}
\fmfright{r1,r2}
\fmf{plain}{r1,i}
\fmf{ghost}{l,i}
\fmf{dashes}{i,r2}
\fmffreeze
\fmf{ghost,left=0.75}{i,r2}
\fmf{dashes,right=0.5}{i,r2}
\fmf{dashes,right=1}{i,r2}
\fmfv{decor.shape=diamond,decor.filled=empty,d.si=15}{i}
\fmfv{decor.shape=square,decor.filled=empty,decor.size=5,label=$\mathcal{K}$}{r2}
\end{fmfchar*}}}
\qquad +\qquad \parbox{25mm}{
\fmfframe(0,5)(0,5){
\begin{fmfchar*}(15,15)
\fmfleft{l}
\fmfright{r1,r2}
\fmf{ghost}{l,i}
\fmf{plain}{r1,i}
\fmf{dashes}{i,r2}
\fmffreeze
\fmf{ghost,left=0.75}{i,r1}
\fmf{dashes,right=0.5}{i,r1}
\fmf{dashes,right=1}{i,r1}
\fmfv{decor.shape=diamond,decor.filled=empty,d.si=15}{i}
\fmfv{decor.shape=square,decor.filled=empty,decor.size=5,label=$\Omega$}{r1}
\end{fmfchar*}}}
\end{equation}

We now turn to the STI for the $W\phi H$ vertex that is 
relevant for the discussion of gauge flips.
Expanding the Killing vectors and compensators according to 
\begin{equation}
\begin{aligned}
\mathcal{K}^b_\alpha(\phi)&=g_\alpha^b+t_{\alpha \gamma}^{b}\phi_\gamma+\mathcal{O}(\phi^2) \\
\Omega^a_{ij}(\phi)&= \mathcal{T}^a_{ij}+\mathcal{O}(\phi)
\end{aligned}
\end{equation}
 we find that the STI for the $WH\phi$ vertex is given by:
\begin{multline}
\ii p_{a \mu_a} \onepi{ W_a^{\mu}(p_a)
  \phi_\beta^{\nu}(p_b)H_i(k_i)}+m_{W_a}\onepi{\phi_a(p)\phi_\beta(p_b)H_i(k_i)}\\
= f t^a_{\beta\gamma}\onepi{\phi_\gamma(p_a+p_b)H_i(k_i)}-\ii \mathcal{T}^a_{ij}\onepi{\phi_b(p_b)H_j(-p_b)}\overset{\text{tree level}}{=}0
\end{multline}
On tree level, the right hand side vanishes because there is no Higgs-Goldstone boson mixing on tree level. Therefore the same simplification as in the  abelian model takes place and we can conclude that the diagram \eqref{eq:3wphi-diag} satisfies the Ward Identity by itself also in the nonabelian case. In higher orders perturbation theory the right hand side of \fref{eq:phihw-sti-nl} no longer vanishes since a $\phi-H$ mixing is generated by loop diagrams. 

No simplification compared to the linear case appears for the STI for the $HHW$ vertex that is on tree level 
\begin{multline}
\ii p_{a \mu_a} \onepi{ W_a^{\mu}(p_a)
  H_i(k_i) H_j(k_j)}+m_{W_a}\onepi{\phi_a(p)H_i(k_i)H_j(k_j)}\\
=-\ii \mathcal{T}^a_{ik}\onepi{H_k(-p_j)H_j(-p_j)}-\ii \mathcal{T}^a_{jk}\onepi{H_i(p_i)H_k(-p_i)}
\end{multline}
This identity is similar to the linear case \eqref{eq:hhw-sti}.
\chapter{More on STIs}
In this appendix we discuss some technicalities concerning the  STIs. In \fref{app:amputate},  we review the amputation of contracted gauge boson lines and the ghost terms for external gauge bosons. In \fref{app:uni-sti} we clarify the correct use of the STIs in unitarity gauge and derive the form \eqref{eq:uni-ghosts-main} of the ghost interactions. \Fref{app:2cov-ghosts} contains more details on the derivation of the STI with two contractions in \fref{sec:2cov-sti}.  Finally, \fref{app:sti_examples} collects explicit results for the STIs in the model of \fref{chap:ssb-lag} that will be used in the calculations of Ward Identities. 
\section{STIs for amputated Green's Functions}\label{app:amputate}
\subsection{Definition of amputated Green's Functions}\label{app:amp}
We define amputated Green's functions as a continuation of the $S$-matrix elements off the mass shell:
\begin{multline}\label{eq:lsz-boson}
  \int\prod_{i=1}^l d^4x_i e^{-\ii k_i x_i} \prod_{j=l+1}^n d^4 y_j e^{\ii p_j
    y_j}\Greensfunc{\Phi(y_j)\dots\Phi(x_i)}\\
  \equiv\prod_{i=1}^K D(k_i) \prod_{j=l+1}^n
  D(p_j)\,\me(\Phi(k_1)\dots \Phi(k_l)\to \Phi(p_{l+1})\dots \Phi(p_n))
\end{multline}
If all external particles are on-shell, this becomes the LSZ formula (for this formulation see e.g. \cite{Peskin:1995}) \footnote{Since our main interest is in tree level calculations we suppress field renormalization constants}. 

Using the  convention to treat the operators $(c\Delta\Phi)$ as insertions that are not amputated,  the contact terms on the right hand side of \fref{eq:local-wi} are turned into (for simplicity we treat all momenta as incoming)
\begin{multline}\label{eq:contact-amp}
 \int d^4x e^{-\ii k x}\prod_{i=1}^n d^4 y_i e^{-\ii p_i
    y_j}\Greensfunc{c(y_j)\bar c(x) \Phi_1(y_1)\dots\Delta\Phi_j(y_j)\dots\Phi_n(y_n)}\\
=D_c(k)\prod_{i=1,i\neq j}^n
    D_{\Phi_i}(p_i)\me(\bar c(k)\Phi_1(p_1)\dots (c\Delta\Phi_j)(p_j)\dots \Phi_{n}(p_n))
\end{multline}

The amputation of fermions must be discussed a little more carefully since we have to distinguish particles and antiparticles. The corresponding LSZ formula are \cite{Itzykson:1980}: 
\begin{equation}\label{eq:lsz-ferm}
\begin{aligned}
\Braket{\Out|d^\dagger|\In}&=\ii\int d^4x \bar v
e^{-\ii kx}(\ii \fmslash\partial-m)\Braket{\Out|\psi|\In}\\
\Braket{\Out|b|\In}&=-\ii\int d^4x \bar u
e^{\ii kx}(\ii \fmslash\partial-m)\Braket{\Out|\psi|\In}\\
\Braket{\Out|d|\In}&=\ii\int d^4x \Braket{\Out|\bar \psi|\In}(-\ii \fmslash{\overleftarrow\partial}-m)v e^{\ii kx}\\
\Braket{\Out|b^\dagger|\In}&=-\ii\int d^4x \Braket{\Out|\bar \psi|\In}(-\ii \fmslash{\overleftarrow\partial}-m)u e^{-\ii kx}
\end{aligned}
\end{equation}
Therefore an amputated Green's function with the insertion of the BRS transformation of a fermion becomes
\begin{multline}
(-\ii )^2\int d^4x  d^4 y_1d^4 y_2 \, e^{-\ii (k x-p_1y_1)} \\
 \bar u(p_1)(\ii \fmslash\partial_{y_1}-m)
\Greensfunc{c(y_2)\bar c(x) \dots\Psi(y_1)\dots \Delta\bar\Psi(y_2)}
(-\ii \fmslash{\overleftarrow\partial}_{y_2}-m)u(p_2)e^{-\ii p_2 y_2} \\
=(-\ii) D_c(k) \bar
 u(p_1)\me\left(\bar c(x)\psi(p_1)\dots (c\Delta\bar\psi)(p_2+k)\right)(\fmslash p_2-m)u(p_2)
\end{multline}
This can be regarded as the prescription to compute the matrix element with the `polarization spinor'
$(-\ii) (\fmslash p_2-m)u(p_2)$ for the transformed particle $\Delta\bar\psi$. 

Similar considerations apply for the remaining cases in \fref{eq:lsz-ferm} so we have to apply the replacements
\begin{equation}\label{eq:ferm-wf}
\begin{aligned}
  u(p)&\to (-\ii) (\fmslash p-m)u(p)\\
  v(p)&\to (-\ii) (\fmslash p+m)v(p)\\
  \bar u(p)&\to (-\ii)\bar u(p)(\fmslash p-m)\\
  \bar v(p)&\to(-\ii)\bar v(p)(\fmslash p+m)
\end{aligned}
\end{equation}
for the calculations of contact terms involving fermions.
\subsection{STI for the gauge boson propagator}
To go in the  STI \eqref{eq:local-wi} from Green's functions to amputated matrix elements, one has to amputate external gauge boson lines contracted with a momentum. This can be done using the (tree level) relations
\begin{equation}\label{eq:prop-rel-app}
\begin{aligned}
k_\mu D^{\mu\nu}_W&=-\xi D_c k^\nu\\
D_\phi&=D_c
\end{aligned}
\end{equation}
On tree level they can be checked from the explicit expressions, but we will sketch the derivation from the STIs that allows the generalization to loop calculations\cite{Yao:1988}. 

\Fref{eq:prop-rel-app} is a consequence of the STI (see \fref{fig:prop-sti})
\begin{multline}\label{eq:gauge-prop-sti}
0 = \deltabrs\greensfunc{W_a^\mu(x)\bar
  c_b(y)}\\
=\greensfunc{\left(D^\mu c(x)\right)_a\bar c_b(y)}+\greensfunc{W_a^\mu(x)B(y)}
\end{multline}
that becomes on tree level
\begin{equation}
\greensfunc{W^\mu_a(x)\partial_\nu W_b^\nu(y)}=\xi\greensfunc{\partial^\mu
  c(x)_a\bar c_b(y)}
\end{equation}
In momentum space, this becomes the desired relation \eqref{eq:prop-rel-app}. 
\begin{figure}[htbp]
\begin{center}
\begin{equation*}
\parbox{20mm}{
\begin{fmfchar*}(15,20)
\fmfleft{l}
\fmfright{r}
\fmf{boson}{l,b}
\fmf{double}{b,r}
\fmfblob{10}{b}
\end{fmfchar*}}
\qquad\qquad 
=-\ii p^\mu \quad \parbox{20mm}{
\begin{fmfchar*}(15,20)
\fmfleft{l}
\fmfright{r}
\fmf{ghost}{r,b}
\fmf{ghost}{b,l}
\fmfblob{10}{b}
\fmfv{decor.shape=square,decor.filled=full,decor.size=5}{l}
\end{fmfchar*}}
-\quad \parbox{20mm}{
\begin{fmfchar*}(15,20)
\fmfleft{l}
\fmfright{r}
\fmf{dots}{l,b}
\fmf{dots}{b,r}
\fmfv{decor.shape=diamond,decor.filled=empty,d.si=15}{b}
\fmffreeze
\fmf{boson,right}{b,l}
\fmfv{decor.shape=square,decor.filled=empty,decor.size=5}{l}
\end{fmfchar*}}
\end{equation*}
\end{center}
\caption{STI for the gauge boson propagator} \label{fig:prop-sti}
\end{figure}
Similarly one can use the STI
\begin{equation}\label{eq:gb-sti}
0=\delta_{\text{BRS}}\greensfunc{\phi_a(x)\bar c_b(y)}
\end{equation}
to show the equivalence of the ghost and Goldstone boson propagators on tree level. The radiative corrections to this simple relations are discussed e.g. in \cite{Boehm:2001,Yao:1988}.
\subsection{Amputation of contracted gauge bosons}
In a spontaneously broken gauge theory in $R_\xi$ gauge, the Ward Identity is given by \fref{eq:gf-wi}
\begin{equation}\label{eq:1gb-wi}
 -\frac{1}{\xi}\braket{\Out|\dmu W^\mu|\In}+ m_W\braket{\Out|\phi|\In}=0
\end{equation}
To obtain a relation for scattering amplitudes, we amputate the gauge boson according to the LSZ formula, we obtain for an incoming $W$ 
\begin{multline}\label{eq:lsz-wi}
\int d^4x e^{-\ii kx}\braket{\Out|\dmu W^\mu(x)|\In}=\ii k_\mu
D_W^{\mu\nu}(k)\me_\nu(\In+W\to\Out )\\
=\xi m_w D_\phi(k)\me(\In\to\Out+\phi) 
\end{multline}
For an outgoing gauge boson the sign has to be changed.
On tree level, the relation \eqref{eq:prop-rel-app} can be used to eliminate the gauge parameter $\xi$ and the propagators from \fref{eq:lsz-wi} so we obtain the identity \eqref{eq:gf-wi}:
\begin{equation}\label{eq:gf-wi-app}
-\ii k_\mu\me^\mu(\In+W\to\Out)
=m_w\me(\In+\phi\to\Out) 
\end{equation}
On loop level, there are correction factors \cite{Yao:1988} that can be computed from the STIs \eqref{eq:gauge-prop-sti} and \eqref{eq:gb-sti}. 
\subsection{Amputation of ghost contributions}\label{app:ghost-amp}
In the STI with external off-shell gauge bosons \eqref{eq:sti-gauge}, there appear additional ghost terms of the form 
\begin{multline}\label{eq:sti-gauge-app}
-\frac{1}{\xi}\Greensfunc{(\dmu W^\mu_a(x)-\xi m_{W_a}\phi_a(x))\cdots W_b^\nu(y_b)\cdots}\\
=\Greensfunc{\partial^\nu c_b(y_b)\bar c_a(x)\cdots}\quad +\text{ contact terms}
\end{multline}
The ghost terms have the same pole structure as the left hand side of the STI so no inverse propagators appear when going to amputated Green's functions. To perform the amputation, we write the right hand side of \eqref{eq:sti-gauge-app} as\footnote{In going from Green's functions to amputated matrix elements,  (incoming) ghosts and antighosts have to be exchanged since ghost field operators create incoming antighosts and vice versa.}
\begin{multline}
\int d^4x e^{-\ii p_a x}\prod_{i=1}^n d^4 y_i e^{-\ii p_i
   y_j}\braket{0|\tprod{\left(\partial_\mu c_b(y_b)\right)\bar c_a(x)\dots \Phi_i(y_i)\dots}|0}\\
=D_c(k)D_c(p_a)(\ii p_{b\mu})\prod_{i=1,i\neq a}^n
   D_{\Phi_i}(p_i)\me(c_a(p_a) \bar c_b(p_b)\dots\Phi_{i}(p_i)\dots)
\end{multline}
The derivation of the STI for amputated Green's functions involves the multiplication with the inverse gauge boson propagator  $D_{W\mu\nu}^{-1}(p_b)$. In the first term this cancels the ghost propagator because of  \fref{eq:prop-rel-app}: 
\begin{equation}
 D_{W\mu\nu}^{-1}(p_b)p_b^\nu D_c(p_b)=-\frac{1}{\xi}p_{b\mu}
\end{equation}
Therefore the amputated version of \fref{eq:sti-gauge-app} is
\begin{multline}
\me^{\mu\nu}(\mathcal{D}_a(p_a)\cdots W_b^\nu(p_b)\cdots)\\
= -\frac{1}{\xi}p_{b\mu}\me(\bar c_b(x_b)\cdots c_a(x_a)\cdots)\quad +\text{ contact terms}
\end{multline}
\section{STIs in unitarity gauge}\label{app:uni-sti}
In unitarity gauge the Goldstone bosons and ghosts decouple from the physical matrix elements, however, on the quantum level there is an additional, divergent term in the lagrangian of the form
\begin{equation}\label{eq:higgs-div}
\Delta\mathscr{L}\propto \delta^4(0)\ln\left(1+\frac{H}{v}\right)
\end{equation}
This form is valid for a single Higgs, we will not need the general expression in this work. Different derivations of this result are reviewed in \cite{Grosse-Knetter:1993}.

Instead of eliminating the ghosts from the theory and using the nonpolynomial Higgs interaction \eqref{eq:higgs-div}, one can also introduce ghost fields even in unitarity gauge. In an abelian spontaneously broken gauge theory the appropriate lagrangian is found to be 
\begin{equation}\label{eq:uni-ghosts}
\mathscr{L}_{FP}^{U}=-m_{W_a}\bar c c-g\bar c H c
\end{equation}
This results in a ghost `propagator' 
\begin{equation}
D_c=\frac{-\ii}{m_W}
\end{equation}
and a ghost-Higgs vertex 
\begin{equation}
\bar c H c\quad :\quad -\ii g
\end{equation}
 It can be shown \cite{Grosse-Knetter:1993} that the resummation of ghosts loops calculated using  this lagrangian leads back to \fref{eq:higgs-div}. We will now show that the  STI \eqref{eq:lsz-wi-boson} is valid also in unitarity gauge if we include in addition to the ghost lagrangian \eqref{eq:uni-ghosts} also a Goldstone boson ghost interaction. 

We define unitarity gauge by taking the limit $\xi\to\infty$ \emph{after} amputating the external propagators. In this way all \emph{internal} Goldstone bosons are removed from the theory but the Ward Identity \eqref{eq:gf-wi} remains valid. 
We show that also the STIs remain valid, if the Higgs-ghost coupling \eqref{eq:uni-ghosts} and a similar Goldstone boson-ghost coupling are used.

As the simplest example, consider a 4 point contact term similar to \fref{eq:4-contact} but with a external Higgs boson:
\begin{multline}
\Greensfunc{c(x_1)\bar c(x)\Delta\psi(x_1)\bar \psi(x_2) H(x_3)}=
\parbox{20mm}{
\begin{fmfchar*}(20,20)
\fmfleft{l1,l2}
\fmfright{r1,r2}
\fmf{fermion}{l1,m,l2}
\fmf{dashes}{r1,m}
\fmf{ghost}{r2,m}
\fmffreeze
\fmf{ghost,right=0.5}{m,l1}
\fmfv{decor.shape=square,decor.filled=empty,decor.size=5}{l1}
\fmfv{decor.shape=diamond,decor.filled=empty,d.si=15}{m}
\end{fmfchar*}}\vspace{20mm}\\
 \overset{\text{Tree level}}{=} \quad
\parbox{20mm}{
\begin{fmfchar*}(20,20)
\fmfleft{l1,l2}
\fmfright{r1,r2}
\fmf{fermion}{l1,m}
\fmf{fermion}{m,l2}
\fmf{dashes}{r1,m}
\fmf{phantom}{r2,m}
\fmffreeze
\fmf{ghost,left=0.5}{r2,l1}
\fmfv{decor.shape=square,decor.filled=empty,decor.size=5}{l1}
\fmfdot{m}
\end{fmfchar*}}\qquad+ \qquad
\parbox{20mm}{
\begin{fmfchar*}(20,20)
\fmfleft{l1,l2}
\fmfright{r1,r2}
\fmf{ghost}{r2,m}
\fmf{ghost}{m,l1}
\fmf{dashes}{r1,m}
\fmf{phantom}{l2,m}
\fmffreeze
\fmf{fermion}{l1,l2}
\fmfv{decor.shape=square,decor.filled=empty,decor.size=5}{l1}
\fmfdot{m}
\end{fmfchar*}}
\end{multline} 
Since the ghosts are amputated \emph{before} sending $\xi\to\infty$,  the first term remains unchanged. In the second term, we have to consider the ghost-Higgs coupling that we write schematically as  \footnote{Use \fref{eq:gauge-fix} with  $u^a_{ib}=g$}
\begin{equation}
\bar c H c\quad :\quad -\ii \xi m_W g\\
\end{equation}
Then we obtain for the combination of the ghost-Higgs vertex and the ghost propagator
\begin{equation}
\frac{\ii}{p^2-\xi m_W^2}(-\ii \xi m_W g)\xrightarrow{\xi\to\infty}\frac{-\ii}{m_W} (-\ii g)
\end{equation}
This is just the same result as the one from the lagrangian \eqref{eq:uni-ghosts}.

Similarly, we obtain for a ghost line, coupled to $N$ Higgs bosons:
\begin{equation}\label{eq:higgs-ladder}
\parbox{35mm}{
\begin{fmfchar*}(35,15)
\fmfleft{l}
\fmfright{r}
\fmfbottomn{b}{6}
\fmftopn{t}{6}
\fmf{ghost}{r,i4,i3,i2,i1,l}
\fmf{dashes}{b6,r}
\fmf{phantom}{r,t6}
\fmf{phantom}{b1,l,t1}
\begin{fmffor}{j}{1}{1}{4}
\fmf{dashes}{b[j+1],i[j]}
\fmf{phantom}{i[j],t[j+1]}
\fmfdot{i[j]}
\end{fmffor}
\fmfdot{r}
\end{fmfchar*}}\quad\xrightarrow{\xi\to\infty}\quad\left(\frac{-\ii}{m_W}\right)^N(-\ii g)^N
\end{equation}

We now consider ghost lines with insertions of gauge bosons. In a linear $R_\xi$ gauge the ghost-gauge boson vertex  is independent of $\xi$. Therefore a ghost line that includes at least one interaction with a gauge boson  drops out if we go to unitarity gauge:
 \begin{equation}
\parbox{35mm}{
\begin{fmfchar*}(35,15)
\fmfleft{l}
\fmfright{r}
\fmfbottomn{b}{6}
\fmftopn{t}{6}
\fmf{ghost}{r,i4,i3,i2,i1,l}
\fmf{dashes}{b6,r}
\fmf{photon}{b5,i4}
\fmf{phantom}{r,t6}
\fmf{phantom}{i4,t5}
\fmf{phantom}{b1,l,t1}
\begin{fmffor}{j}{1}{1}{3}
\fmf{dashes}{b[j+1],i[j]}
\fmf{phantom}{i[j],t[j+1]}
\fmfdot{i[j]}
\end{fmffor}
\fmfdot{r,i4}
\end{fmfchar*}}\quad\xrightarrow{\xi\to\infty}\quad 0
\end{equation}
Thus, the ghosts decouple from the gauge bosons as expected.

If we consider a ghost line with insertions of Goldstone and  Higgs bosons,
the  argument leading to \fref{eq:higgs-ladder} remains valid. In contrast to the case of Higgs bosons, the contributions of \emph{internal} Goldstone bosons vanish in the limit $\xi\to\infty$ because of the $\xi$ dependence of the propagators. However, external Goldstone bosons must be kept, and we have to include the ghost-Goldstone boson vertices in addition to \fref{eq:uni-ghosts}. 

From \fref{eq:higgs-ladder} we see that we can obtain the ghost vertices in unitarity gauge from those in $R_\xi$ gauge: 
\begin{equation}
g_{\bar c \Phi c}^U=-\frac{1}{\xi m_{W}^2}g_{\bar c \Phi c}^{R_\xi}
\end{equation}
 The expression for these couplings in a nonabelian spontaneously broken gauge theory is given in \fref{eq:gauge-fix}. 

Therefore the ghost lagrangian, to be used in calculations of STIs in unitarity gauge, is 
 \begin{equation}\label{eq:uni-ghosts2}
\mathscr{L}_{FP}^U=-m_{W_a}\bar c_a c_a+g_{\bar c H c}^{aib} \bar c_a H_i c_b+g_{\bar c \phi c}^{abc}\bar c_a \phi_b c_c
\end{equation}
Here the ghost couplings on the right hand side are those in unitarity gauge.

We have shown that the ghost lagrangian \eqref{eq:uni-ghosts2} reproduces the contact terms in the limit $\xi\to\infty$, provided we amputate the ghost propagators before taking the limit. 

We still have to  consider the term 
\begin{equation*}
-\frac{\ii}{\xi}p_b^\nu\me(c_a(p) \bar c_b(p_b)\dots)
\end{equation*}
that appears in the STI for Green's functions with  off-shell gauge bosons \eqref{eq:lsz-wi-gauge}.
The only contributions to this matrix element again include a ghost line with $N$ insertions of Higgs bosons, this time with one missing ghost propagator compared to \fref{eq:higgs-ladder}. However, this is compensated by the $\frac{1}{\xi}$ in front of the matrix element, so   in the limit $\xi\to\infty$ we obtain
\begin{equation}
-\frac{\ii}{\xi}p_b^\nu\me_{R_\xi}(c_a(p) \bar c_b(p_b)\dots)\xrightarrow{\xi\to\infty}=-\ii p_b^\nu\me_{U}(c_a(p) \bar c_b(p_b)\dots)
\end{equation}
Up to this modification, the STIs remain valid in unitarity gauge if we use the ghost lagrangian \eqref{eq:uni-ghosts2}.

\section{Ghost terms in the STI with 2 contractions}\label{app:2cov-ghosts}
In the derivation of the STI for three point vertices with 2 contraction we have omitted a few steps that we will discuss in this appendix. Taking the derivative of  \eqref{eq:2cov-sti} with respect to a ghost field $c$ and an  arbitrary field $\Phi$  yields, setting classical fields and BRS-sources to zero:
\begin{multline}\label{eq:2cov-3-complicated}
0=\sum_\Phi\Biggl[\onepi{c_b (p_b)\mathcal{D}_a(p_a)\Phi^\star(p)}\onepi{\Phi(p)\Phi(k)}\\
+\onepi{ c_b (p_b)\Phi^\star(p)}\onepi{ \mathcal{D}_a(p_a)\Phi(p)\Phi(k)}\\
+\onepi{ c_b (p_b)\Phi^\star(p)\Phi(k)}\onepi{ \mathcal{D}_a(p_a)\Phi(p)}\Biggr]\\
+\left[\frac{1}{\xi}p_a^2-m_{W_a}^2\right]\onepi{c_b(p_b)\bar c_a(p_a)\Phi(p)}
+\frac{\delta B}{\delta\Phi}\onepi{c(p_b) \mathcal{D}_a(p_a)\bar c(p)}
\end{multline}
where we have used the identity
\begin{equation}
(\mathcal{D}(x)B(y))f(y) 
=\left[-\frac{1}{\xi}\dmu \partial^\mu+m_{W_a}^2\right]f(x)
\end{equation}
We will now show that the identity
\begin{multline}
\onepi{ c_b (p_b)\Phi^\star(p)\Phi(k)}\onepi{ \mathcal{D}_a(p_a)\Phi(p)}\\
+\left[\tfrac{1}{\xi}p_a^2-m_{W_a}^2\right]\onepi{c_b(p_b)\bar c_a(p_a)\Phi(k)}=0
\end{multline}
holds for $\Phi=W,\phi,H$. This allows a considerable simplification in \fref{eq:2cov-3-complicated}. We will use the parametrization introduced in \fref{chap:ssb-lag}.
We begin by checking the case $\Phi=W$. From 
\begin{equation}
\onepi{ \mathcal{D}_a(p_a)W^\mu_a(p_b+p_c)}=\frac{p_a^\mu}{\xi}(p_a^2-\xi m_{W_a}^2)
\end{equation}
and the ghost Feynman rules \eqref{subeq:ghost-fr}
\begin{equation}
\onepi{c_b(p_b)\bar c_a(p_a)W^\mu_c(p_c)}=f^{acb}p^\mu_a
\end{equation}
we see that indeed the desired combination vanishes:
\begin{equation}
f^{abc}\onepi{ \mathcal{D}_a(p_a)W_a(p_b+p_c)}+\left[\tfrac{1}{\xi}p_a^2-m_{W_a}^2\right]\onepi{c_b(p_b)\bar c_a(p_a)W_c(p_c)}=0
\end{equation}
Similarly, for $\Phi=\phi/H$ we obtain
\begin{equation}
\begin{aligned}
-t^b_{ac}\onepi{\mathcal{D}_a(p_a)\phi_a(p_b+p_c)}+\left[\tfrac{1}{\xi}p_a^2-m_{W_a}^2\right]\onepi{c_b(p_b)\bar c_a(p_a)\phi_c(p_c)}&=0\\
-u^b_{ai}\onepi{\mathcal{D}_a(p_a)\phi_a(p_i+p_b)}+\left[\tfrac{1}{\xi}p_a^2-m_{W_a}^2\right]\onepi{c_b(p_b)\bar c_a(p_a)H_i(p_i)}&=0
\end{aligned}
\end{equation}
which can again be seen from the ghost Feynman rules \eqref{subeq:ghost-fr}.

Therefore we have shown that the identity \eqref{eq:2cov-3-complicated} can be simplified to the form used  in \fref{eq:2cov-3}:
\begin{multline}\label{eq:2cov-3-ghost}
0=\sum_\Phi\Biggl[\onepi{c_b (p_b)\mathcal{D}_a(p_a)\Phi^\star(p)}\onepi{\Phi(p)\Phi(k)}\\
+\onepi{ c_b (p_b)\Phi^\star(p)}\onepi{ \mathcal{D}_a(p_a)\Phi(p)\Phi(k)}
+\frac{\delta B}{\delta\Phi}\onepi{c(p_b)\mathcal{D}_a(p_a)\bar c(p)}\Bigr]
\end{multline}
\section{Explicit form of STIs}\label{app:sti_examples}
\subsection{3 point STIs}
In this appendix we list the explicit form of the STIs for the three point vertices in the model described in \fref{chap:ssb-lag}.  They are obtained by inserting the form of the BRS transformations \eqref{eq:ssb-brs} into \fref{eq:sti-irr-3tree}. We list only the terms contributing on tree level, so mixed 2 point functions among gauge, Higgs and Goldstone bosons that are present in the STIs are not displayed. 
\begin{description}
\item[HWW]:
\begin{multline}\label{eq:hww-sti}
- \onepi{ \mathcal{D}_a(p_a) W_b^\mu(p_b)H_i(p_i)}= \ii \frac{1}{\xi}p_b^\nu\onepi{c_a(p_a)\bar c_b^\mu(p_b)H_i(p_i)}\\
=\frac{1}{2} g^{iab}_{HW W}p_b^\mu  
\end{multline}
\item[WHH]:
\begin{multline}\label{eq:hhw-sti}
 -\onepi{\mathcal{D}_a(p_a)H_i(p_i)H_j(p_j)} \\
=-T^a_{ij}\left[\onepi{H_i(p_i)H_i(p_i+p_a)}  -\onepi{H_j(p_i+p_a)H_j(p_j)}\right]
 \end{multline}
\item[WWW]:
\begin{multline}\label{eq:3w-sti}
-\onepi{\mathcal{D}_a(p_a)W_b^\nu(p_b)W_c^\rho(p_c)}\\
=f^{adb}\onepi{W_d^\nu(p_a+p_b)W_c^\rho(p_c)}+f^{adc}\onepi{W_b^\nu(p_b)W_d^\rho(p_a+p_c)}\\
+\ii \frac{1}{\xi}p_b^\nu \onepi {c_a(p_a)\bar c_b(p_b) W_c(p_c)^\rho}
+\ii\frac{1}{\xi}p_c^\rho \onepi{c_a(p_a)\bar c_c(p_c) W_b^\nu(p_b)}
\end{multline}
Here the two point functions  are in $R_\xi$ gauge. We can simplify this equation, if we note that the two point functions together with the ghost terms give the 2 point function in unitarity gauge: 
\begin{multline}\label{eq:unitarity-2point}
 f^{adc}\onepi{W_b(p_b)W_d(p_a+p_c)}_{R_\xi}+\ii \frac{1}{\xi}p_b^\nu \onepi {c_a(p_a)\bar c_b(p_b)W_c(p_c) }_{R_\xi}\\
= f^{adc}\onepi{W_b(p_b)W_d(p_a+p_c)}_{U}
\end{multline}
as can be seen by inserting the explicit expressions.
\item[W$\bar \psi \psi$]:
\begin{multline}\label{eq:ffw-sti}
-\onepi{\bar\psi_i(p_i)\mathcal{D}_a(p_a)\psi_j(p_j)}=-\ii(g_{Vji}^a-g_{Aji}^a\gamma^5)\onepi{\bar\psi_j(p_i+p_a)\psi_j(p_j)}\\
+\ii \onepi{\bar\psi_i(p_i)\psi_i(p_j+p_a)} (g_{Vij}^a+g_{Aij}^a\gamma^5)\\
=\ii(g_{Vij}^a-g_{Aij}^a\gamma^5)S_F^{-1}(p_j)-\ii S_F^{-1}(-p_i)(g_{Vij}^a+g_{Aij}^a\gamma^5)
 \end{multline} 
Here we have used the transformations in the  vector/axial vector notation of \fref{eq:va-coupling}:
\begin{equation*}
\begin{aligned}
\Delta\psi_i&=\ii (g_{Vij}^a+g_{Aij}^a\gamma^5)\psi_j\\
\Delta \bar \psi_i&=-\ii\bar \psi_j(g_{Vij}^a-g_{Aij}^a\gamma^5)
\end{aligned}
\end{equation*}
According to \fref{eq:fermion-sti} we have to put the transformation of the fermion to the right of the inverse propagator.
Note also that here the sign of the momentum in the propagators is important:
\begin{equation*}
\begin{aligned}
\onepi{\bar\psi_j(p_i+p_a)\psi_j(p_j)}&=-S_F^{-1}(p_j)=\ii (\fmslash p_j-m_j)\\
\onepi{\bar\psi_j(p_i)\psi_j(p_j+p_a)}&=-S_F^{-1}(-p_i)=-\ii (\fmslash p_i+m_i)\\
\end{aligned}
\end{equation*}
\item[WW$\phi$]:
\begin{multline}\label{eq:phiww-sti}
- \onepi{ \mathcal{D}_a(p_a)W_b^{\nu}(p_b)\phi_c(p_c)}\\
= \ii \frac{1}{\xi}p_b^\nu \onepi { c_a(p_a)\bar c_b(p_b)\phi_c(p_c) }+m_{W_c}\onepi {c_a(p_a)  W_b^{\nu}(p_b) \bar c_c(p_c)} 
\end{multline}
\item[WH$\phi$]:
\begin{multline}\label{eq:phihw-sti}
\onepi{ \mathcal{D}_a(p_a)
  \phi_b^{\nu}(p_b)H_i(k_i)}= u^a_{ib}\onepi{H_i(p_a+p_b)H_i(k_i)} \\
-u^a_{bi}\onepi{\phi_b^\mu(p_b)\phi_b(-p_b)} +m_{W_b}\onepi {c_a(p_a) \bar c_b(p_b)H_i(k_i)} 
\end{multline}
\item[W$\phi\phi$]:
\begin{multline}\label{eq:2phiw-sti}
-\onepi{ \mathcal{D}_a(p_a)\phi_b^{\nu}(p_b)\phi_c(p_c)}\\
=- t^a_{cb}\left[\onepi{\phi_c^\mu(p_c)\phi_c(-p_c)}-\onepi{\phi_b^\mu(p_b)\phi_b(-p_b)}\right]\\
+m_{W_b}\onepi {c_a(p_a) \bar c_b(p_b)\phi_c(p_c)} +m_{W_c}\onepi {c_a(p_a) \bar c_c(p_c)\phi_b(p_b)} 
\end{multline}
\end{description}
The STIs for vertices with 2 contractions, obtained from \fref{eq:3point-2cov}, read
\begin{description}
\item[HWW]:
 \begin{equation}\label{eq:hww-sti-2}
\onepi{ \mathcal{D}_a(p_a)\mathcal{D}_b(p_b)H_i(p_i)}
=m_{W_a}u^b_{ia}\onepi{H_i(p_a+p_b)H_i(p_i)}
 \end{equation}
\item[WWW]:
\begin{multline}\label{eq:www-sti2}
\onepi{ \mathcal{D}_a(p_a)\mathcal{D}_b(p_b)W_{c\nu}(p_c))}\\ 
=\ii f^{dab}p_{a\mu}\onepi{W_d^\mu(p_a+p_b)W_{c\nu}(p_c)}
-\ii\frac{1}{\xi} p_{c\nu}\onepi{\mathcal{D}_a(p_a)\bar c_c(p_c)c_b(p_b)} 
 \end{multline}
The gauge boson part of the ghost term converts the gauge boson two point function to unitarity gauge as in \fref{eq:unitarity-2point}. Therefore this identity can also be written as
\begin{multline}
\onepi{ \mathcal{D}_a(p_a)\mathcal{D}_b(p_b)W_{c\nu}(p_c))}\\ 
=\ii f^{dab}p_{a\mu}\onepi{W_d^\mu(p_a+p_b)W_{c\nu}(p_c)}_U
+\ii\frac{m_{W_a}}{\xi} p_{c\nu}\onepi{\phi_a(p_a)\bar c_c(p_c)c_b(p_b)} 
\end{multline}
If we symmetrize the right hand side in  $a$ and $b$ and use the explicit expression for the ghost vertex, we can arrive at the equivalent form
\begin{multline}\label{eq:www-sti2a}
 \onepi{ \mathcal{D}_a(p_a)\mathcal{D}_b^\nu(p_b)W_c(p_c)}\\
=\frac{1}{2}f^{abc}\Bigl[(p_a^\nu-p_b^\nu)
  \left[ (p_c^2-m_{W_c}^2)g_{\nu\rho} -p_{c\nu} p_{c\rho}\right] +(p_a^\nu+p_b^\nu)(m_a^2-m_b^2)\Bigr]
\end{multline}
In the last step we used the explicit expression for
$t^a_{cb}$.  
\end{description}

\subsection{STI for 4 point functions}
We list some of the STIs for 4 point vertices obtained from \eqref{eq:sti-irr-4tree}.
\begin{description}
\item[WW$\phi\phi$]:
\begin{multline}\label{eq:2s2w-1cov_a}
\ii p_{a\mu}\onepi{W_a^\mu(p_a)W_b^\nu(p_b)\phi_C(p_c)\phi_D(p_d)}\\
=f^{eba}\onepi{W_e^\nu(p_a+p_b)\phi_C(p_c)\phi_D(p_d)}\\
-T^a_{EC}\onepi{W_b^\nu(p_b)\phi_E(p_a+p_c)\phi_D(p_d)}\\
-T^a_{ED}\onepi{W_b^\nu(p_b)\phi_C(p_c)\phi_E(p_a+p_d)}
\end{multline}
After inserting the Feynman rules, one can show that this identity reproduces the definition of the quartic scalar- gauge boson coupling \eqref{eq:2s2w} and the Lie algebra \eqref{eq:ssb-kommutator}. 
\item[WWW$\phi$]
\begin{multline}\label{eq:2s2w-1cov-b}
m_{W_a}\onepi{\phi_a(p_a)W_b^\nu(p_b)W_c(p_c)\phi_D(p_d)} \\
=f^{eba}\onepi{W_e^\nu(p_a+p_b)W_c(p_c)\phi_D(p_d)}\\
+f^{eca}\onepi{W_b(p_b)W_e(p_a+p_c)\phi_D(p_d)}\\
-T_{ED}^a\onepi{W_b(p_b)W_c(p_c)\phi_E(p_a+p_d)}
\end{multline}
From this relation follow the `Jacobi Identities'  \eqref{eq:2s-jac_komp}.
\item[WWWW]:
\begin{multline}
\ii p_{a\mu_a}\onepi{W_a^{\mu_a}(p_a)W_b^{\mu_b}(p_b)W_c^{\mu_c}(p_c)W_d^{\mu_d}(p_d)}=\\
f^{aeb}\onepi{W_e^{\mu_b}(p_b+p_a)W_c^{\mu_c}(p_c)W_d^{\mu_d}(p_d)}\\
+f^{aec}\onepi{W_b^{\mu_b}(p_b)W_e^{\mu_b}(p_c+p_a)W_d^{\mu_d}(p_d)}\\
+f^{aed}\onepi{W_b^{\mu_b}(p_b)W_c^{\mu_c}(p_c)W_e^{\mu_d}(p_d+p_a)}
\end{multline}

The conditions on the coupling constants arising from this relation are the Jacobi Identity \eqref{eq:jacobi} for the structure constants and the form of the quartic gauge boson coupling \eqref{eq:gw4}

\end{description}

\subsection{STI for 4 point functions with 2 contractions}\label{app:4sti-2cov-exp}
As an example for the 4 point STI with 2 contractions, we consider the $3WH$ vertex. 
From \fref{eq:4point-2cov} we find
\begin{multline}
\onepi{ \mathcal{D}_a(p_a)\mathcal{D}_b(p_b)W_c(p_c)H_i(k_i)}=\ii f^{dab}p_{a\mu}\onepi{W_{d\nu}(p_a+p_b)W_c^\nu(p_c)H_i(k_i)}\\
-m_{W_a}t^b_{da}\onepi{\phi_d(p_a+p_b)W_c^\nu(p_c)H_i(k_i)}+m_{W_a}u^b_{ja}\onepi{H_j(p_a+p_b)W_c^\nu(p_c)H_i(k_i)}\\
- f^{bdc}\onepi{ \mathcal{D}_a(p_a)W_d^\nu(p_b+p_c)H_i(k_i)}
+T^b_{ji}\onepi{ \mathcal{D}_a(p_a))W_c^\nu(p_c)H_j(p_b+k_i)}\\
+u^b_{di}\onepi{ \mathcal{D}_a(p_a))W_c^\nu(p_c)\phi_d(p_b+k_i)}
\end{multline}
Using the STIs for the $WWH$ vertex \eqref{eq:hww-sti} and the $WW\phi$ vertex \eqref{eq:phiww-sti} and inserting the Feynman rules this becomes
\begin{multline}
(\ii)^2 \left [m_{W_a}g_{H\phi
    W^2}^{iabc} p_b^\nu+m_{W_b} g_{H\phi W^2}^{ibac} p_a^\nu\right]
=(\ii)^2 f^{dab}p_a^\nu g_{HWW}^{icd}\\
 +\frac{m_{W_a}}{2m_{W_d}}t^b_{da}g_{HWW}^{icd}(k_i^\nu-p_a^\nu-p_b^\nu)
+\frac{1}{2}g_{HWW}^{jab}T^c_{ij}(k_i^\nu-p_a^\nu-p_b^\nu)\\
+\frac{1}{2}f^{bdc}g_{HWW}^{iad}(p_b^\nu+p_c^\nu)-\frac{1}{2}T^b_{ji}g_{HWW}^{jac}p_c^\nu\\
+\frac{1}{2}g_{HWW}^{ibd}\left[f^{adc}(p_a^\nu+p_c^\nu)+\frac{m_c}{m_d}t^a_{cd}p_c^\nu\right]
\end{multline}
Elimination of $k_i$ using momentum conservation and contraction with $\epsilon_c$ gives as coefficient of $(\epsilon_c\cdot p_a)$: 
\begin{multline}\label{eq:ghphiw2-2cov}
m_{W_b} g_{H\phi W^2}^{ibac}=f^{dab} g_{HWW}^{icd}+(\frac{m_{W_a}}{m_{W_d}}t^b_{da}g_{HWW}^{icd}+g_{HWW}^{jab}T^c_{ij})+\frac{1}{2}g_{HWW}^{ibd}f^{dac}\\
=\frac{1}{2}f^{bda} g_{HWW}^{icd}\left(1-\frac{m_{W_a}^2-m_{W_b}^2}{m_{W_d}^2}\right)+g_{HWW}^{jab}T^c_{ij}+\frac{1}{2}g_{HWW}^{ibd}f^{dac}
\end{multline}
The same result is obtained from the Ward Identity of the $3WH$ Green's function with 2 contractions \eqref{eq:3wh-wi}.
\chapter{Lagrangian and coupling constants}
In \fref{app:general-fr} we give the conventions for the Lagrangian and the Feynman rules used in the calculation of the Ward Identities in \fref{chap:reconstruction}. Some technical steps that we have omitted in the discussion of the expression of the Goldstone boson couplings in terms of the input parameters are given in \fref{app:ssb-relations}.
\section{Parametrization of the general Lagrangian}\label{app:general-fr}
 \subsection{Parametrization}
We give our parametrization of the general Lagrangian with the particle spectrum of a spontaneously broken gauge theory but \emph{without} imposing gauge invariance. We use the same notation for the fields as in \fref{chap:ssb-lag}, i.e. we denote gauge bosons by $W^a$, Goldstone bosons by $\phi_\alpha$ and all other scalar fields by $H_i$. 

Apart from terms $\propto \epsilon^{\mu\nu\rho\sigma}W_\mu W_\nu W_\rho W_\sigma$, the most general renormalizable Lagrangian for these fields is 
\begin{multline}\label{eq:ssb-lag}
  \mathscr{L}=-\frac{1}{4}(\dmu W_{\nu a}-\partial_\nu W_{\mu a})(\partial^\mu W^\nu_a-\partial^\nu W^\mu_a)+\frac{1}{2}m_a^2W_{a\mu}W_a^{\mu}
  - f^{abc}W_{b\mu} W_{c\nu}\partial^\mu
  W_a^\nu\\
  -\frac{1}{4!} (g_{W^4}^{abcd}g^{\mu\rho}g^{\nu\sigma}+g_{W^4}^{abdc}g^{\mu\sigma}g^{\nu\rho}+g_{W^4}^{ac bd}g^{\mu\nu}g^{\rho\sigma})W_{a\mu}
  W_{b\nu} W_{c\rho} W_{d\sigma}\\
  +\frac{1}{2}(\dmu\phi_\alpha\partial^\mu\phi_\alpha-\xi m_a^2 \phi_\alpha^2)
  +\frac{1}{2}(\dmu H_i\partial^\mu H_i-m^2_{H_i}H_i^2)\\
  -\frac{1}{2}t_{\alpha\beta}^c(\phi_\alpha\overleftrightarrow \dmu\phi_\beta)W_c^\mu+\frac{g_{\phi WW}^{\alpha bc}}{2}\phi_\alpha W_b^\mu
  W_{c\mu}-\frac{1}{2}T^a_{ij}(H_i\overleftrightarrow \dmu H_j)W_a^\mu\\
+g_{H\phi W}^{i\alpha b}(\phi_\alpha\overleftrightarrow \dmu  H_i) W_b^\mu  +\frac{1}{2}g_{HWW}^{iab}H_iW_{a\mu}W_b^\mu+\frac{1}{4}g_{\phi^2 W^2}^{\alpha\beta cd}\phi_\alpha\phi_\beta
  W_{c\mu}W_d^\mu\\
+\frac{1}{4}g_{H^2 W^2}^{ab ij}H_iH_jW_a^\mu
  W_{b\mu}+\frac{1}{2}g_{H\phi W^2}^{\alpha bc i} H_i\phi_\alpha W_{b\mu}W_c^\mu +\frac{1}{2}g_{\phi^2 H}^{\alpha\beta i}\phi_\alpha\phi_\beta H_i\\
+\frac{1}{2}g_{\phi H^2}^{\alpha ij} \phi_\alpha H_iH_j
  +\frac{1}{3!}g_{\phi^3}^{\alpha\beta\gamma}\phi_\alpha\phi_\beta\phi_c
 +\frac{1}{3!}g_{H^3}^{ijk}H_iH_jH_k +\text{quartic scalar interactions}
\end{multline}
Here all matrices are assumed to be symmetric with respect to permutations of identical particles. We have included the symmetry factors explicitly in the Lagrangian so the coupling constants appear in the Feynman rules without any numerical factors. It can be shown that our form of the quartic gauge boson coupling is the most general form that is totally symmetric under simultaneous permutations of Lorentz and group indices. 

The Lagrangian of the fermions is parametrized by 
\begin{multline}\label{eq:ferm-ssb-lag}
\mathscr{L}_f=\ii\bar\psi_i\fmslash\partial\psi +\bar\psi_i\fmslash
W_a(\tau^{a}_{Lij}(\tfrac{1-\gamma^5}{2})+\tau_{Rij}^a(\tfrac{1+\gamma^5}{2})\psi_j \\
+\bar\psi_i \phi_\alpha
(g_{\phi ij}^\alpha(\tfrac{1-\gamma^5}{2})+g_{\phi ij}^{\alpha\dagger}(\tfrac{1+\gamma^5}{2}))\psi_j+\bar\psi_i H_k
(g_{H ij}^k(\tfrac{1-\gamma^5}{2})+ g_{Hij}^{k\dagger} (\tfrac{1+\gamma^5}{2}))\psi_j
\end{multline}
Sometimes scalar and pseudoscalar couplings defined by
\begin{equation} \label{eq:sp-coupling}
g_S=\frac{1}{2}(g_{\phi/H}+g_{\phi/H}^\dagger) \qquad
  g_P=\frac{1}{2}(g_{\phi/H}-g_{\phi/H}^\dagger)
\end{equation}
will prove more useful.
\subsection{Feynman rules}\label{app:feynman}
The Lagrangians  \eqref{eq:ssb-lag} and  \eqref{eq:ferm-ssb-lag} result in the following Feynman rules (all momenta incoming)
\begin{subequations}
\begin{multline}\label{eq:3-fr}
W_aW_bW_c: -\Bigl[f^{abc}p_a^{\mu_b} g^{\mu_a\mu_c}+f^{acb}p_a^{\mu_c} g^{\mu_a\mu_b}\\
+f^{bac}p_b^{\mu_a} g^{\mu_b\mu_c}+f^{bca}p_b^{\mu_c} g^{\mu_a\mu_b}+f^{cab}p_c^{\mu_a} g^{\mu_b\mu_c}+f^{cba}p_c^{\mu_b} g^{\mu_a\mu_c}\Bigr]
\end{multline}
For totally antisymmetric $f^{abc}$ this simplifies to the usual gauge boson three point vertex
\begin{equation}
W_aW_bW_c: f^{abc} C^{\mu_a\mu_b\mu_c}(k_a,k_b,k_c)
\end{equation}
with
\begin{equation}
C^{\mu_a\mu_b\mu_c}(k_a,k_b,k_c)= ( g^{\mu_a\mu_b} (k_a^{\mu_c}-k_b^{\mu_c})
               + g^{\mu_b\mu_c} (k_b^{\mu_a}-k_c^{\mu_a})
               + g^{\mu_c\mu_a} (k_c^{\mu_b}-k_a^{\mu_b}) )
\end{equation}
\begin{eqnarray}
W_aW_bW_cW_d&:&-\ii
(g_{W^4}^{abcd}g^{\mu\rho}g^{\nu\sigma}+g_{W^4}^{abdc}g^{\mu\sigma}g^{\nu\rho}+g_{W^4}^{acbd}g^{\mu\nu}g^{\rho\sigma})\\
\phi_\alpha\phi_\beta W_c&:& t_{\alpha\beta}^c (p_\alpha^\mu-p_\beta^\mu)\\
\phi_\alpha W_b W_c &:& \ii g^{\alpha bc}_{\phi WW}g_{\mu\nu} \label{eq:fr-phiww}\\
H_iH_j W_a&:& T_{ij}^c(p_i^\mu-p_j^\mu)\\
H_i\phi_\alpha W_b&:&g_{H\phi W}^{i\alpha b}(p_i-p_a)\label{eq:fr-hphiw}\\
H_i W_a W_b&:&\ii g_{HWW}^{iab}\label{eq:fr-hww}\\
\phi_\alpha\phi_\beta W_c W_d&:&\ii g_{\phi^2 W^2}^{\alpha\beta cd}\\
H_iH_jW_a^\mu W_{b\mu}&:& \ii g_{H^2 W^2}^{ijab}\\
H_i\phi_\alpha W_bW^c_\mu&:&\ii g_{H\phi W^2}^{\alpha ibc}\\
\phi_\alpha\phi_\beta H_i&:&\ii g_{\phi^2 H}^{\alpha\beta i}\\
\phi_\alpha H_iH_j&:&\ii g_{\phi H^2}^{\alpha ij}\\ 
\phi_\alpha\phi_\beta\phi_c&:&\ii g_{\phi^3}^{\alpha\beta\gamma}\\
H_iH_jH_k&:&\ii g_{H^3}^{ijk}\\
\bar \psi_i\psi_j W_a&:&\ii  \gamma^\mu(\tau^{a}_{Lij}(\tfrac{1-\gamma^5}{2})+ \tau_{Rij}^a(\tfrac{1+\gamma^5}{2})) \\
\bar \psi_i\psi_j \phi_\alpha&:&\ii (g_{\phi ij}^\alpha(\tfrac{1-\gamma^5}{2})+g_{\phi ij}^{\alpha\dagger}(\tfrac{1+\gamma^5}{2})\\
\bar \psi_i \psi_j H_k &:& \ii (g_{H ij}^k(\tfrac{1-\gamma^5}{2})+ g_{H ij}^{k\dagger}(\tfrac{1+\gamma^5}{2}))
\end{eqnarray}
\end{subequations}
\section{Relations among coupling constants}\label{app:ssb-relations}
In this appendix we derive the expressions \eqref{subeq:triple-phi-couplings}, \eqref{eq:2s-jac_komp} and \eqref{eq:quartic-scalar} for the parameters of the Lagrangian \eqref{eq:ssb-lagrangian} in terms of the input parameters \eqref{eq:ssb-input}. These relations can be obtained by exploiting the gauge invariance of the Lagrangian and imposing the conditions \eqref{subeq:ssb-conditions}. For simplicity we consider the case of only massive vector bosons first and then turn to the differences that occur for massless gauge bosons in \fref{app:massless}. 
\subsection{Triple Goldstone boson-gauge boson couplings}\label{app:scalar-gauge-couplings}
We can find constraints on the cubic scalar-gauge boson coupling constants using the commutator algebra \eqref{eq:ssb-kommutator} and the relations \eqref{subeq:ssb-conditions}. 

Multiplying the commutator relation \eqref{eq:ssb-kommutator} with $\phi_0$ gives, when all gauge bosons are massive
\begin{equation}
 \left([T^\alpha,T^\beta]\right)\phi_0=f^{\alpha\beta\gamma}\begin{pmatrix} m_\gamma \\ 0\end{pmatrix}
\end{equation}
and therefore, inserting the parametrization of the generators $T$ \eqref{eq:ssb-gen}:
\begin{align}\label{eq:contr-commutator}
m_\alpha u^\beta_{i\alpha}=m_\beta u^\alpha_{i\beta}\\
t^\alpha_{\gamma\beta}m_\beta-t^\beta_{\gamma\alpha}m_\alpha&\overset{!}{=}f^{\alpha\beta\gamma}m_\gamma \label{eq:t-m-sum-rule}
\end{align}
This can be used to express the matrix $t^\alpha_{\gamma\beta}$ in terms of the  structure constants and gauge boson masses. First we eliminate $t$ from the coupling $ g_{\phi WW}$ \eqref{eq:gphiww}
\begin{equation}
    m_\alpha g_{\phi WW}^{\alpha\beta\gamma}=f^{\alpha\beta\gamma}(m_\beta^2-m_\gamma^2)
\end{equation}
Now we can determine $t^\alpha_{\beta\gamma}$ itself:
\begin{equation}\label{eq:t-solution}
  \begin{aligned}
    (m_\gamma g^{\gamma\beta\alpha}_{\phi WW} -m_\alpha g^{\alpha\beta\gamma}_{\phi WW}) &=2m_\alpha m_\gamma
    t^\beta_{\alpha\gamma}+m_\beta (m_\gamma t^\alpha_{\beta\gamma}-m_\alpha t^\gamma_{\beta\alpha}) \\
& =2m_\alpha m_\gamma
    t^\beta_{\alpha\gamma}+m_\beta^2f^{\alpha\gamma\beta}\\
\Rightarrow \qquad m_\alpha m_\gamma
    t^\beta_{\alpha\gamma}&=\frac{1}{2}f^{\beta\alpha\gamma}(m_\beta^2-m_\alpha^2-m_\gamma^2)
  \end{aligned}
\end{equation}
\subsection{Quartic couplings}
We can derive constraints on the quartic scalar-gauge boson couplings \eqref{eq:2s2w}
\begin{equation}
  g_{\phi^2 W^2}^{ABcd}=-\{T^c,T^a\}_{AB}
\end{equation}
by taking the commutator with a generator $T$ and applying the identity  \eqref{eq:super-jacobi}
\begin{equation}
 0= \lbrack A,\{B,C\} \rbrack +\{\lbrack C, A \rbrack,B\} -\{ \lbrack A,B\rbrack C\}
\end{equation}
This yields
\begin{equation}\label{eq:2s-jac_1}
  T^a_{AB}g^{BCbc}_{\phi^2 W^2}-g^{ABbc}_{\phi^2 W^2}T^a_{BC}=f^{acd}g^{ACbd}_{\phi^2 W^2}+f^{abd}g^{ACcd}_{\phi^2 W^2}
\end{equation}
which is the condition for the interaction 
\[ g^{ABab}_{\phi^2  W^2}\phi_A^T\phi_B W^a W^b\]
to be invariant under \emph{global} transformations. 

Multiplying this equation with $\phi_{0C}$ and using the definition \eqref{eq:phiww-def} for the triple scalar-gauge coupling we get the expression
\begin{equation}\label{eq:2s-jac}
  T^a_{AB}g^{Bbc}_{\phi WW}-m_{W_a}g^{Aabc}_{\phi^2 W^2}=f^{acd}g^{Abd}_{\phi WW}+f^{abd}g^{Acd}_{\phi WW}
\end{equation}
While the global transformation law \eqref{eq:2s-jac_1} leaves an ovearall factor of $g^{ABcd}_{\phi^2 W^2}$ undetermined, the contracted version \eqref{eq:2s-jac} determines the quartic coupling uniquely from cubic couplings. 
\subsection{Scalar potential}\label{app:scalar-gauge}
For spontanteous symmetry breaking , we must demand that the state
\begin{equation}
  \phi_{0A}=
  \begin{pmatrix}
    0\\v_i
  \end{pmatrix}
\end{equation}
is a minimum of the potential \eqref{eq:scalar-potential}. This gives the condition
\begin{equation}
  g_2^{Aj}v_j+\frac{g_3^{Ajk}}{2}v_jv_k+\frac{g_4^{Ajkl}}{3!}v_jv_kv_l=0
\end{equation}
These condition eliminates also the linear terms in the Lagrangian
(`tadpoles').

The mass matrix of the scalars is given by
\begin{equation}
 m_{AB}^2\equiv \frac{\partial^2 V(\phi)}{\partial\phi_A\partial\phi_B}|_{\phi=\phi_0}=g_2^{AB}+g_3^{ABi}v_i+\frac{g_4^{ijAB}}{2}v_iv_j
\end{equation}
Contracting this equation with a vev and inserting the relations from the minimization of the potential, the matrix $g_2$ can be eliminated and we obtain the so called `Higgs mass sum rules' \cite{Langacker:1984}:
\begin{equation}\label{eq:higgs-mass-sr-app}
m_{Ai}^2v_i=\frac{1}{2}g_3^{Aij}v_iv_j+\frac{1}{3}g_4^{Aijk}{2}v_iv_jv_k=-\frac{1}{2}g_{\Phi^3}^{Ajk}v_jv_k+\frac{1}{6}g_{\Phi^4}^{Ajkl}v_jv_kv_l
\end{equation}

We can derive conditions for the scalar couplings  using \eqref{eq:potential-gauge}  
\begin{equation}
  \frac{\partial V(\phi)}{\partial \phi_A}T^\alpha_{AB}\phi_B=0
\end{equation}
Taking one derivative and setting $\phi=\phi_0$ gives
\begin{equation}
  m_{AB}T^\alpha_{BC}\phi_{0C}=0
\end{equation}
From the components we obtain the condition that the masses of the Goldstone boson and the mixing between Goldstone and Higgs bosons vanishes:
\begin{equation}
0=m_{ab}^2 m_{W_b}=m_{ia}^2m_{W_a}
\end{equation}

Taking the derivative of  \eqref{eq:potential-gauge} with respect to two fields  and setting  $\phi=\phi_0$ we find 
\begin{equation}\label{eq:potential-gauge-3}
 g_{\Phi^3}^{ABC}m_{W_\alpha}=m_{B}^2T^\alpha_{BA}+m^2_{A}T^\alpha_{AB}
\end{equation}
where we have defined
\begin{equation}
 g_{\Phi^3}^{ABC}=-\frac{\partial^3 V(\phi)}{\partial\phi_A\partial\phi_B\partial\phi_C}|_{\phi=\phi_0}
\end{equation}
The components of \fref{eq:potential-gauge-3} give the conditions
\begin{subequations}\label{subeq:triple-scalar}
  \begin{align}
    m_{W_\alpha}g_{\phi H^2}^{\alpha ij}&=T^\alpha_{ij}(m_i^2-m_j^2) \label{eq:phi-hh-coupling}\\
    m_{W_\alpha}g_{\phi^2 H}^{\alpha\beta i}&=-m_i^2 u^{\alpha}_{i\beta}\label{eq:2phi-h-coupling}\\
    g_{\phi^3}^{\alpha\beta\gamma}&=0
  \end{align}
\end{subequations}

We can derive relations  that can be used to express the quartic Goldstone boson scalar couplings by cubic coupling by repeating this procedure taking three derivatives of \eqref{eq:potential-gauge}. This gives the equation 
\begin{equation}\label{eq:potential-gauge-4}
 m_{W_\alpha}g^{ABC\alpha}_{\Phi^4}+g^{DBC}_{\Phi^3}T^\alpha_{DA}+g^{ADC}_{\Phi^3}T^a_{DB}\\
+g^{ABD}_{\Phi^3}T^a_{DC}=0
\end{equation}
which reads in component form:
\begin{subequations}\label{eq:gold-higgs-4}
  \begin{align}
   m_\kappa g_{\phi^4}^{\alpha\beta\gamma\kappa}&=u^\kappa_{i\alpha}g_{H\phi^2}^{i\beta\gamma}+u^\kappa_{i\beta}g_{H\phi^2}^{i\alpha\gamma}+u^\delta_{i\gamma}g_{H\phi^2}^{i\alpha\beta}\\
 m_\gamma g_{H\phi^3}^{i\alpha\beta\gamma}&=T^\gamma_{ij}g_{H\phi^2}^{j\alpha\beta}+t^\gamma_{\alpha\kappa}g_{H\phi^2}^{i\kappa\beta}+u^\gamma_{j\alpha}g_{H^2\phi}^{ij\beta}+ t^\gamma_{\beta\kappa}g_{H\phi^2}^{i\kappa\alpha}+u^\gamma_{j\beta}g_{H^2\phi}^{ij\alpha}\label{eq:gphi3h}\\
  m_\beta g_{H^2\phi^2}^{ij\alpha\beta}&=T^\beta_{ik}g_{H^2\phi}^{kj\alpha}-u^\beta_{\gamma i}g_{H
 \phi^2}^{j\gamma\alpha}&\nonumber\\
&+T^\beta_{jk}g_{H^2\phi}^{ki\alpha}-u^\beta_{\gamma j}g_{H
 \phi^2}^{i\gamma\alpha}+ t^\beta_{\gamma\alpha}g_{H^2\phi}^{ij\gamma}+u^\beta_{k\alpha}g_{H^3}^{ijk}\label{eq:h2phi2-inv}\\
  m_\alpha g_{H^3\phi}^{ijk\alpha}&=T^\alpha_{il}g_{H^3}^{ljk}-u^\alpha_{\beta i}g_{ \phi H^2}^{\beta jk}\nonumber\\
&+T^\alpha_{jl}g_{H^3}^{lik}-u^\alpha_{\beta j}g_{
 \phi H^2}^{\beta ik}+T^\alpha_{kl}g_{H^3}^{lij}-u^\alpha_{\beta k}g_{
 \phi H^2}^{\beta ij}
  \end{align}
\end{subequations}
 Taking 4 derivatives of \eqref{eq:potential-gauge} we get a relation that constrains $g_{H^4}$. 
\subsection{Couplings of massless gauge bosons}\label{app:massless}
In this section we show that massless gauge bosons only couple to particles of the same mass. To do this we repeat the analysis of \fref{sec:ssb-summary} but allow for massless gauge bosons. 
\subsubsection{Scalar-gauge boson couplings}
To consider one massless and two massive gauge bosons, we use \fref{eq:broken-unbroken-commutator}
and get instead of \eqref{eq:contr-commutator}:
\begin{equation}\label{eq:contr-unbroken}
f^{a \beta\gamma} m_\gamma =t^a_{\gamma\beta}m_\beta\qquad,\qquad h^a_{i\beta}m_\beta=0
\end{equation}
and so since $m_\beta\neq 0$
\begin{equation}\label{eq:h_is_zero}
h^a_{i\beta}=0
\end{equation}
Using the antisymmetry of the structure constants and the $t^a_{cb}$, instead of the first equation in \eqref{eq:contr-unbroken}, we can derive analogously 
\begin{equation}
f^{a\beta\gamma} m_\beta=t^a_{\gamma\beta}m_\gamma
\end{equation}
Together these equations imply the two possibilities 
\begin{equation}\label{eq:f_massless}
\begin{aligned}
m_\beta=m_\gamma &\quad \Rightarrow \quad f^{a \beta\gamma}=t^a_{\gamma\beta}\neq 0\\
m_\beta\neq m_\gamma&\quad \Rightarrow \quad f^{a \beta\gamma}=t^a_{\gamma\beta}=0
\end{aligned}
\end{equation}
 The coupling $g_{\phi WW}$ is for one or two massless gauge bosons, using the defintion \eqref{eq:gphiww}: 
\begin{equation}
\begin{aligned}
 g_{\phi WW}^{\alpha \beta c}&=m_\beta t^c_{\beta\alpha}=
        \begin{cases} m_\beta  f^{\alpha \beta c},\quad m_\alpha=m_\beta\neq 0\\
                        0, \quad  m_\alpha\neq m_\beta            
         \end{cases}\\
 g_{\phi WW}^{\alpha bc}&=0,\qquad m_b=m_c=0
\end{aligned}
\end{equation}
\subsubsection{Fermion Couplings}
For unbroken generators we get from \fref{eq:yukawa-tensor} instead of \eqref{eq:f-phi-coupling}
\begin{equation}
\ii m_j \tau^{a}_{Rij}=\ii m_i\tau^{a}_{Lij}
\end{equation}
Using the conjugate equation of \eqref{eq:yukawa-tensor} we get instead
\begin{equation}
\ii m_j \tau^{a}_{Lij}=\ii m_i\tau^{a}_{Rij}
\end{equation}
so we must have $m_i=m_j$ for $\tau^{a}_{L/Rij}\neq 0$.
\subsubsection{Scalar potential}
For unbroken generators $T^a$  we get from \eqref{eq:potential-gauge-3}
\begin{equation}
0=m_{B}^2T^a_{BA}+m^2_{A}T^a_{AB}
\end{equation}
This gives in components
\begin{subequations}
\begin{align}
0&=T^a_{ij}(m_j^2-m_i^2)\\
0&=-m_i^2g_{H\phi W}^{i \beta a}
\end{align}
\end{subequations}
The first equation implies
\begin{equation}\label{eq:T_massless}
        T^\alpha_{ij}= 0 \qquad \text{if} \quad m_\alpha =0 \quad \text{and} \quad m_j\neq m_i
\end{equation}
The second equation confirms that $g_{H\phi W}^{i\alpha b}=0$ for $m_\alpha=0$ as we know already from \eqref{eq:h_is_zero}.
\chapter{Explicit form of flips}\label{app:flips}
In this appendix we summarize the flips for spontaneously broken gauge theories, both in the linear and  nonlinear representation of the scalar sector. 
\section{Gauge flips} 
\begin{subequations}\label{subeq:h_flips}

The gauge flips for the 4 gauge boson function are in the linear realization: 
\begin{equation}\label{eq:gauge_flips1}
\{ t_{4}^{G,1}, \dots ,t_{4}^{G,7} \}= \left\{
\begin{aligned}
&\parbox{15mm}{
\begin{fmfchar*}(15,15)
\fmfleft{a,b}
\fmfright{f1,f2}
\fmf{photon}{a,fwf1}
\fmf{photon}{fwf1,fwf2}
\fmf{photon}{fwf2,b}
\fmf{photon}{fwf1,f1}
\fmf{photon}{fwf2,f2}
\fmfdot{fwf1}
\fmfdot{fwf2}
\end{fmfchar*}}\,,\,
\parbox{15mm}{
\begin{fmfchar*}(15,15)
\fmfleft{a,b}
\fmfright{f1,f2}
\fmf{photon}{a,fwf1}
\fmf{photon}{fwf1,fwf2}
\fmf{photon}{fwf2,b}
\fmf{phantom}{fwf2,f2}
\fmf{phantom}{fwf1,f1}
\fmffreeze
\fmf{photon}{fwf2,f1}
\fmf{photon}{fwf1,f2}
\fmfdot{fwf1}
\fmfdot{fwf2}
\end{fmfchar*}}\, ,\, 
\parbox{15mm}{
\begin{fmfchar*}(15,15)
\fmfleft{a,b}
\fmfright{f1,f2}
\fmf{photon}{a,fwf}
\fmf{photon}{fwf,b}
\fmf{photon}{fwf,www}
\fmf{photon}{www,f1}
\fmf{photon}{www,f2}
\fmfdot{fwf}
\fmfdot{www}
\end{fmfchar*}}\, , \, 
\parbox{15mm}{
\begin{fmfchar*}(15,15)
\fmfleft{a,b}
\fmfright{f1,f2}
\fmf{photon}{a,c}
\fmf{photon}{c,b}
\fmf{photon}{c,f1}
\fmf{photon}{c,f2}
\fmfdot{c}
\end{fmfchar*}}\\  &
\parbox{15mm}{
\begin{fmfchar*}(15,15)
\fmfleft{a,b}
\fmfright{f1,f2}
\fmf{photon}{a,fwf1}
\fmf{dashes}{fwf1,fwf2}
\fmf{photon}{fwf2,b}
\fmf{photon}{fwf1,f1}
\fmf{photon}{fwf2,f2}
\fmfdot{fwf1}
\fmfdot{fwf2}
\end{fmfchar*}}\,,\,
\parbox{15mm}{
\begin{fmfchar*}(15,15)
\fmfleft{a,b}
\fmfright{f1,f2}
\fmf{photon}{a,fwf1}
\fmf{dashes}{fwf1,fwf2}
\fmf{photon}{fwf2,b}
\fmf{phantom}{fwf2,f2}
\fmf{phantom}{fwf1,f1}
\fmffreeze
\fmf{photon}{fwf2,f1}
\fmf{photon}{fwf1,f2}
\fmfdot{fwf1}
\fmfdot{fwf2}
\end{fmfchar*}}\, , \, 
\parbox{15mm}{
\begin{fmfchar*}(15,15)
\fmfleft{a,b}
\fmfright{f1,f2}
\fmf{photon}{a,fwf}
\fmf{photon}{fwf,b}
\fmf{dashes}{fwf,www}
\fmf{photon}{www,f1}
\fmf{photon}{www,f2}
\fmfdot{fwf}
\fmfdot{www}
\end{fmfchar*}}
\end{aligned}
\right\}
\end{equation}
For nonlinear realizations, the Higgs exchange diagrams of the last line are not present in the gauge flips. 

The flips for $\bar f f \to WW$ are in the linear representation:
\begin{equation}
\{ t_{4}^{G,8},\dots,t_{4}^{G,11}\}=
\left\{
\parbox{15mm}{
\begin{fmfchar*}(15,15)
\fmfleft{a,b}
\fmfright{f1,f2}
\fmf{fermion}{a,fwf1}
\fmf{fermion}{fwf1,fwf2}
\fmf{fermion}{fwf2,b}
\fmf{photon}{fwf1,f1}
\fmf{photon}{fwf2,f2}
\fmfdot{fwf1,fwf2}
\end{fmfchar*}}\,,\,
\parbox{15mm}{
\begin{fmfchar*}(15,15)
\fmfleft{a,b}
\fmfright{f1,f2}
\fmf{fermion}{a,fwf1}
\fmf{fermion}{fwf1,fwf2}
\fmf{fermion}{fwf2,b}
\fmf{phantom}{fwf2,f2}
\fmf{phantom}{fwf1,f1}
\fmffreeze
\fmf{photon}{fwf2,f1}
\fmf{photon}{fwf1,f2}
\fmfdot{fwf1,fwf2}
\end{fmfchar*}}\,,\, 
\parbox{15mm}{
\begin{fmfchar*}(15,15)
\fmfleft{a,b}
\fmfright{f1,f2}
\fmf{fermion}{a,fwf}
\fmf{fermion}{fwf,b}
\fmf{photon}{fwf,www}
\fmf{photon}{www,f1}
\fmf{photon}{www,f2}
\fmfdot{www,fwf}
\end{fmfchar*}}\, ,\,
\parbox{15mm}{
\begin{fmfchar*}(15,15)
\fmfright{a,b}
\fmfleft{f1,f2}
\fmf{photon}{a,fwf}
\fmf{photon}{fwf,b}
\fmf{dashes}{fwf,www}
\fmf{fermion}{www,f1}
\fmf{fermion}{f2,www}
\fmfdot{www}
\fmfdot{fwf}
\end{fmfchar*}}\right \}
\end{equation}
Again the Higgs exchange diagram
\begin{equation}
\parbox{15mm}{
\begin{fmfchar*}(15,15)
\fmfleft{a,b}
\fmfright{f1,f2}
\fmf{photon}{a,fwf}
\fmf{photon}{fwf,b}
\fmf{dashes}{fwf,www}
\fmf{fermion}{www,f1}
\fmf{fermion}{f2,www}
\fmfdot{www}
\fmfdot{fwf}
\end{fmfchar*}}
\end{equation}
is not present for nonlinear symmetries. 

The gauge $\bar f f \to W H$ flips are for linear and nonlinear realizations:
\begin{equation}
\{ t_{4,H_1}^{G,1},\dots, t_{4,H_1}^{G,5} \}=
\left\{
\parbox{15mm}{
\begin{fmfchar*}(15,15)
\fmfleft{a,b}
\fmfright{f1,f2}
\fmf{fermion}{a,fwf1}
\fmf{fermion}{fwf1,fwf2}
\fmf{fermion}{fwf2,b}
\fmf{photon}{fwf1,f1}
\fmf{dashes}{fwf2,f2}
\fmfdot{fwf1,fwf2}
\end{fmfchar*}}\, ,\,
\parbox{15mm}{
\begin{fmfchar*}(15,15)
\fmfleft{a,b}
\fmfright{f1,f2}
\fmf{fermion}{a,fwf1}
\fmf{fermion}{fwf1,fwf2}
\fmf{fermion}{fwf2,b}
\fmf{phantom}{fwf2,f2}
\fmf{phantom}{fwf1,f1}
\fmffreeze
\fmf{photon}{fwf2,f1}
\fmf{dashes}{fwf1,f2}
\fmfdot{fwf1,fwf2}
\end{fmfchar*}}\, ,\, 
\parbox{15mm}{
\begin{fmfchar*}(15,15)
\fmfleft{a,b}
\fmfright{f1,f2}
\fmf{fermion}{a,fwf}
\fmf{fermion}{fwf,b}
\fmf{photon}{fwf,www}
\fmf{photon}{www,f1}
\fmf{dashes}{www,f2}
\fmfdot{www,fwf}
\end{fmfchar*}}\, , \,
\parbox{15mm}{
\begin{fmfchar*}(15,15)
\fmfleft{a,b}
\fmfright{f1,f2}
\fmf{fermion}{a,fwf}
\fmf{fermion}{fwf,b}
\fmf{dashes}{fwf,www}
\fmf{photon}{www,f1}
\fmf{dashes}{www,f2}
\fmfdot{www,fwf}
\end{fmfchar*}}
\right \} 
\end{equation}
A new feature in spontaneously broken gauge theories are the $3WH$ flips that have the same form in linear and nonlinear realizations:
\begin{equation}
\{ t_{4,H_2}^{G,1}, \dots, t_{4,H_2}^{G,6} \}=
\left\{\begin{aligned}
&\parbox{15mm}{
\begin{fmfchar*}(15,15)
\fmfleft{a,b}
\fmfright{f1,f2}
\fmf{photon}{a,fwf1}
\fmf{photon}{fwf1,fwf2}
\fmf{photon}{fwf2,b}
\fmf{dashes}{fwf1,f1}
\fmf{photon}{fwf2,f2}
\fmfdot{fwf1}
\fmfdot{fwf2}
\end{fmfchar*}}\,,\,
\parbox{15mm}{
\begin{fmfchar*}(15,15)
\fmfleft{a,b}
\fmfright{f1,f2}
\fmf{photon}{a,fwf1}
\fmf{photon}{fwf1,fwf2}
\fmf{photon}{fwf2,b}
\fmf{phantom}{fwf2,f2}
\fmf{phantom}{fwf1,f1}
\fmffreeze
\fmf{dashes}{fwf2,f1}
\fmf{photon}{fwf1,f2}
\fmfdot{fwf1}
\fmfdot{fwf2}
\end{fmfchar*}}\,,\, 
\parbox{15mm}{
\begin{fmfchar*}(15,15)
\fmfleft{a,b}
\fmfright{f1,f2}
\fmf{photon}{a,fwf}
\fmf{photon}{fwf,b}
\fmf{photon}{fwf,www}
\fmf{dashes}{www,f1}
\fmf{photon}{www,f2}
\fmfdot{www}
\fmfdot{fwf}
\end{fmfchar*}}\\
&
\parbox{15mm}{
\begin{fmfchar*}(15,15)
\fmfleft{a,b}
\fmfright{f1,f2}
\fmf{photon}{a,fwf1}
\fmf{dashes}{fwf1,fwf2}
\fmf{photon}{fwf2,b}
\fmf{dashes}{fwf1,f1}
\fmf{photon}{fwf2,f2}
\fmfdot{fwf1}
\fmfdot{fwf2}
\end{fmfchar*}}\,,\,
\parbox{15mm}{
\begin{fmfchar*}(15,15)
\fmfleft{a,b}
\fmfright{f1,f2}
\fmf{photon}{a,fwf1}
\fmf{dashes}{fwf1,fwf2}
\fmf{photon}{fwf2,b}
\fmf{phantom}{fwf2,f2}
\fmf{phantom}{fwf1,f1}
\fmffreeze
\fmf{dashes}{fwf2,f1}
\fmf{photon}{fwf1,f2}
\fmfdot{fwf1}
\fmfdot{fwf2}
\end{fmfchar*}}\,,\, 
\parbox{15mm}{
\begin{fmfchar*}(15,15)
\fmfleft{a,b}
\fmfright{f1,f2}
\fmf{photon}{a,fwf}
\fmf{photon}{fwf,b}
\fmf{dashes}{fwf,www}
\fmf{dashes}{www,f1}
\fmf{photon}{www,f2}
\fmfdot{www}
\fmfdot{fwf}
\end{fmfchar*}}
       \end{aligned}
\right \}
\end{equation}
The $2W2H$ flips are for linear symmetries:
\begin{equation}\label{eq:hhww-flips}
\{ t_{4,H_3}^{G,1}, \dots, t_{4,H_3}^{G,7} \}=
\left\{
\begin{aligned}
&\parbox{15mm}{
\begin{fmfchar*}(15,15)
\fmfleft{a,b}
\fmfright{f1,f2}
\fmf{photon}{a,fwf1}
\fmf{photon}{fwf1,fwf2}
\fmf{photon}{fwf2,b}
\fmf{dashes}{fwf1,f1}
\fmf{dashes}{fwf2,f2}
\fmfdot{fwf1}
\fmfdot{fwf2}
\end{fmfchar*}}\,,\,
\parbox{15mm}{
\begin{fmfchar*}(15,15)
\fmfleft{a,b}
\fmfright{f1,f2}
\fmf{photon}{a,fwf1}
\fmf{photon}{fwf1,fwf2}
\fmf{photon}{fwf2,b}
\fmf{phantom}{fwf2,f2}
\fmf{phantom}{fwf1,f1}
\fmffreeze
\fmf{dashes}{fwf2,f1}
\fmf{dashes}{fwf1,f2}
\fmfdot{fwf1}
\fmfdot{fwf2}
\end{fmfchar*}}\,,\,\parbox{15mm}{
\begin{fmfchar*}(15,15)
\fmfleft{a,b}
\fmfright{f1,f2}
\fmf{photon}{a,fwf1}
\fmf{dashes}{fwf1,fwf2}
\fmf{photon}{fwf2,b}
\fmf{dashes}{fwf1,f1}
\fmf{dashes}{fwf2,f2}
\fmfdot{fwf1}
\fmfdot{fwf2}
\end{fmfchar*}}\, ,\, 
\parbox{15mm}{
\begin{fmfchar*}(15,15)
\fmfleft{a,b}
\fmfright{f1,f2}
\fmf{photon}{a,fwf1}
\fmf{dashes}{fwf1,fwf2}
\fmf{photon}{fwf2,b}
\fmf{phantom}{fwf2,f2}
\fmf{phantom}{fwf1,f1}
\fmffreeze
\fmf{dashes}{fwf2,f1}
\fmf{dashes}{fwf1,f2}
\fmfdot{fwf1}
\fmfdot{fwf2}
\end{fmfchar*}}\\
&\parbox{15mm}{
\begin{fmfchar*}(15,15)
\fmfleft{a,b}
\fmfright{f1,f2}
\fmf{photon}{a,i1}
\fmf{photon}{i1,b}
\fmf{photon}{i1,i2}
\fmf{dashes}{i2,f1}
\fmf{dashes}{i2,f2}
\fmfdot{i1,i2}
\end{fmfchar*}}\, ,\, 
\parbox{15mm}{
\begin{fmfchar*}(15,15)
\fmfleft{a,b}
\fmfright{f1,f2}
\fmf{photon}{a,i1}
\fmf{photon}{i1,b}
\fmf{dashes}{i1,i2}
\fmf{dashes}{i2,f1}
\fmf{dashes}{i2,f2}
\fmfdot{i1,i2}
\end{fmfchar*}}\, ,\, 
\parbox{15mm}{
\begin{fmfchar*}(15,15)
\fmfleft{a,b}
\fmfright{f1,f2}
\fmf{photon}{a,c}
\fmf{photon}{c,b}
\fmf{dashes}{c,f1}
\fmf{dashes}{c,f2}
\fmfdot{c}
\end{fmfchar*}}
\end{aligned}
\right \}
\end{equation}
\end{subequations}
Here the diagram 
\begin{equation*}
\parbox{15mm}{
\begin{fmfchar*}(15,15)
\fmfleft{a,b}
\fmfright{f1,f2}
\fmf{photon}{a,i1}
\fmf{photon}{i1,b}
\fmf{dashes}{i1,i2}
\fmf{dashes}{i2,f1}
\fmf{dashes}{i2,f2}
\fmfdot{i1,i2}
\end{fmfchar*}}
\end{equation*}
is not included in the gauge flips for nonlinear realizations of the symmetry. 

Finally we have  the $3HW$ flips that again have the same form in linear and nonlinear realizations:
\begin{equation}
\{ t_{4,H_4}^{G,1}, \dots, t_{4,H_4}^{G,6} \}=
\left\{\begin{aligned}
&\parbox{15mm}{
\begin{fmfchar*}(15,15)
\fmfleft{a,b}
\fmfright{f1,f2}
\fmf{dashes}{a,fwf1}
\fmf{dashes}{fwf1,fwf2}
\fmf{dashes}{fwf2,b}
\fmf{photon}{fwf1,f1}
\fmf{dashes}{fwf2,f2}
\fmfdot{fwf1}
\fmfdot{fwf2}
\end{fmfchar*}}\,,\,
\parbox{15mm}{
\begin{fmfchar*}(15,15)
\fmfleft{a,b}
\fmfright{f1,f2}
\fmf{dashes}{a,fwf1}
\fmf{dashes}{fwf1,fwf2}
\fmf{dashes}{fwf2,b}
\fmf{phantom}{fwf2,f2}
\fmf{phantom}{fwf1,f1}
\fmffreeze
\fmf{photon}{fwf2,f1}
\fmf{dashes}{fwf1,f2}
\fmfdot{fwf1}
\fmfdot{fwf2}
\end{fmfchar*}}\,,\, 
\parbox{15mm}{
\begin{fmfchar*}(15,15)
\fmfleft{a,b}
\fmfright{f1,f2}
\fmf{dashes}{a,fwf}
\fmf{dashes}{fwf,b}
\fmf{dashes}{fwf,www}
\fmf{photon}{www,f1}
\fmf{dashes}{www,f2}
\fmfdot{www}
\fmfdot{fwf}
\end{fmfchar*}}\\
&
\parbox{15mm}{
\begin{fmfchar*}(15,15)
\fmfleft{a,b}
\fmfright{f1,f2}
\fmf{dashes}{a,fwf1}
\fmf{photon}{fwf1,fwf2}
\fmf{dashes}{fwf2,b}
\fmf{photon}{fwf1,f1}
\fmf{dashes}{fwf2,f2}
\fmfdot{fwf1}
\fmfdot{fwf2}
\end{fmfchar*}}\,,\,
\parbox{15mm}{
\begin{fmfchar*}(15,15)
\fmfleft{a,b}
\fmfright{f1,f2}
\fmf{dashes}{a,fwf1}
\fmf{photon}{fwf1,fwf2}
\fmf{dashes}{fwf2,b}
\fmf{phantom}{fwf2,f2}
\fmf{phantom}{fwf1,f1}
\fmffreeze
\fmf{photon}{fwf2,f1}
\fmf{dashes}{fwf1,f2}
\fmfdot{fwf1}
\fmfdot{fwf2}
\end{fmfchar*}}\,,\, 
\parbox{15mm}{
\begin{fmfchar*}(15,15)
\fmfleft{a,b}
\fmfright{f1,f2}
\fmf{dashes}{a,fwf}
\fmf{dashes}{fwf,b}
\fmf{photon}{fwf,www}
\fmf{photon}{www,f1}
\fmf{dashes}{www,f2}
\fmfdot{www}
\fmfdot{fwf}
\end{fmfchar*}}
       \end{aligned}
\right \}
\end{equation}
\section{Flavor and Higgs flips}
Higgs exchange has to be included in the flavor flips so they are given by
\begin{equation}
\{ t_{4}^{F,1}, \dots, t_{4}^{F,6} \}=
\left\{
\begin{aligned}
\parbox{15mm}{
\begin{fmfchar*}(15,15)
\fmfleft{a,b}
\fmfright{f1,f2}
\fmf{fermion}{a,fwf1}
\fmf{photon}{fwf1,fwf2}
\fmf{fermion}{b,fwf2}
\fmf{fermion}{fwf1,f1}
\fmf{fermion}{fwf2,f2}
\fmfdot{fwf1}
\fmfdot{fwf2}
\end{fmfchar*}}\,,\,
\parbox{15mm}{
\begin{fmfchar*}(15,15)
\fmfleft{a,b}
\fmfright{f1,f2}
\fmf{fermion}{a,fwf1}
\fmf{photon}{fwf1,fwf2}
\fmf{fermion}{b,fwf2}
\fmf{phantom}{fwf2,f2}
\fmf{phantom}{fwf1,f1}
\fmffreeze
\fmf{fermion}{fwf2,f1}
\fmf{fermion}{fwf1,f2}
\fmfdot{fwf1}
\fmfdot{fwf2}
\end{fmfchar*}}\, ,\, 
\parbox{15mm}{
\begin{fmfchar*}(15,15)
\fmfleft{a,b}
\fmfright{f1,f2}
\fmf{fermion}{a,fwf}
\fmf{fermion}{fwf,b}
\fmf{photon}{fwf,www}
\fmf{fermion}{f1,www}
\fmf{fermion}{www,f2}
\fmfdot{fwf}
\fmfdot{www}
\end{fmfchar*}}\\
\parbox{15mm}{
\begin{fmfchar*}(15,15)
\fmfleft{a,b}
\fmfright{f1,f2}
\fmf{fermion}{a,fwf1}
\fmf{dashes}{fwf1,fwf2}
\fmf{fermion}{b,fwf2}
\fmf{fermion}{fwf1,f1}
\fmf{fermion}{fwf2,f2}
\fmfdot{fwf1}
\fmfdot{fwf2}
\end{fmfchar*}}\,,\,
\parbox{15mm}{
\begin{fmfchar*}(15,15)
\fmfleft{a,b}
\fmfright{f1,f2}
\fmf{fermion}{a,fwf1}
\fmf{dashes}{fwf1,fwf2}
\fmf{fermion}{b,fwf2}
\fmf{phantom}{fwf2,f2}
\fmf{phantom}{fwf1,f1}
\fmffreeze
\fmf{fermion}{fwf2,f1}
\fmf{fermion}{fwf1,f2}
\fmfdot{fwf1}
\fmfdot{fwf2}
\end{fmfchar*}}\, ,\, 
\parbox{15mm}{
\begin{fmfchar*}(15,15)
\fmfleft{a,b}
\fmfright{f1,f2}
\fmf{fermion}{a,fwf}
\fmf{fermion}{fwf,b}
\fmf{dashes}{fwf,www}
\fmf{fermion}{f1,www}
\fmf{fermion}{www,f2}
\fmfdot{fwf}
\fmfdot{www}
\end{fmfchar*}}
\end{aligned}
\right \}
\end{equation}
Finally there are `Higgs flips' for diagrams without external gauge
bosons that are gauge parameter independent by themselves:
\begin{subequations}
\label{subeq:higgs_flips}
\begin{multline}
 \{ t_4^{H_1},\dots t_4^{H_4}\} = 
\left\{
\parbox{15\unitlength}{
\begin{fmfgraph}(15,15)
\fmfleft{a,b}
\fmfright{f1,f2}
\fmf{fermion}{a,fwf1}
\fmf{fermion}{fwf1,fwf2}
\fmf{fermion}{fwf2,b}
\fmf{dashes}{fwf1,f1}
\fmf{dashes}{fwf2,f2}
\fmfdot{fwf1,fwf2}
\end{fmfgraph}}\, ,\,
\parbox{15\unitlength}{
\begin{fmfgraph}(15,15)
\fmfleft{a,b}
\fmfright{f1,f2}
\fmf{fermion}{a,fwf1}
\fmf{fermion}{fwf1,fwf2}
\fmf{fermion}{fwf2,b}
\fmf{phantom}{fwf2,f2}
\fmf{phantom}{fwf1,f1}
\fmffreeze
\fmf{dashes}{fwf2,f1}
\fmf{dashes}{fwf1,f2}
\fmfdot{fwf1,fwf2}
\end{fmfgraph}}\, ,\, 
\parbox{15\unitlength}{
\begin{fmfgraph}(15,15)
\fmfleft{a,b}
\fmfright{f1,f2}
\fmf{fermion}{a,fwf}
\fmf{fermion}{fwf,b}
\fmf{photon}{fwf,www}
\fmf{dashes}{www,f1}
\fmf{dashes}{www,f2}
\fmfdot{www,fwf}
\end{fmfgraph}}\, , \,
\parbox{15\unitlength}{
\begin{fmfgraph}(15,15)
\fmfleft{a,b}
\fmfright{f1,f2}
\fmf{fermion}{a,fwf}
\fmf{fermion}{fwf,b}
\fmf{dashes}{fwf,www}
\fmf{dashes}{www,f1}
\fmf{dashes}{www,f2}
\fmfdot{www,fwf}
\end{fmfgraph}}
\right \} 
\end{multline}
and
\begin{multline}
\{t_4^{H,5},\dots,t_4^{H,11}\} = \\
\left\{
\begin{aligned}
&\parbox{15\unitlength}{
\begin{fmfgraph}(15,15)
\fmfleft{a,b}
\fmfright{f1,f2}
\fmf{dashes}{a,fwf1}
\fmf{photon}{fwf1,fwf2}
\fmf{dashes}{fwf2,b}
\fmf{dashes}{fwf1,f1}
\fmf{dashes}{fwf2,f2}
\fmfdot{fwf1}
\fmfdot{fwf2}
\end{fmfgraph}},
\parbox{15\unitlength}{
\begin{fmfgraph}(15,15)
\fmfleft{a,b}
\fmfright{f1,f2}
\fmf{dashes}{a,fwf1}
\fmf{photon}{fwf1,fwf2}
\fmf{dashes}{fwf2,b}
\fmf{phantom}{fwf2,f2}
\fmf{phantom}{fwf1,f1}
\fmffreeze
\fmf{dashes}{fwf2,f1}
\fmf{dashes}{fwf1,f2}
\fmfdot{fwf1}
\fmfdot{fwf2}
\end{fmfgraph}} , 
\parbox{15\unitlength}{
\begin{fmfgraph}(15,15)
\fmfleft{a,b}
\fmfright{f1,f2}
\fmf{dashes}{a,fwf}
\fmf{dashes}{fwf,b}
\fmf{photon}{fwf,www}
\fmf{dashes}{www,f1}
\fmf{dashes}{www,f2}
\fmfdot{fwf}
\fmfdot{www}
\end{fmfgraph}} ,  
\parbox{15\unitlength}{
\begin{fmfgraph}(15,15)
\fmfleft{a,b}
\fmfright{f1,f2}
\fmf{dashes}{a,c}
\fmf{dashes}{c,b}
\fmf{dashes}{c,f1}
\fmf{dashes}{c,f2}
\fmfdot{c}
\end{fmfgraph}}\\
&\parbox{15\unitlength}{
\begin{fmfgraph}(15,15)
\fmfleft{a,b}
\fmfright{f1,f2}
\fmf{dashes}{a,fwf1}
\fmf{dashes}{fwf1,fwf2}
\fmf{dashes}{fwf2,b}
\fmf{dashes}{fwf1,f1}
\fmf{dashes}{fwf2,f2}
\fmfdot{fwf1}
\fmfdot{fwf2}
\end{fmfgraph}},
\parbox{15\unitlength}{
\begin{fmfgraph}(15,15)
\fmfleft{a,b}
\fmfright{f1,f2}
\fmf{dashes}{a,fwf1}
\fmf{dashes}{fwf1,fwf2}
\fmf{dashes}{fwf2,b}
\fmf{phantom}{fwf2,f2}
\fmf{phantom}{fwf1,f1}
\fmffreeze
\fmf{dashes}{fwf2,f1}
\fmf{dashes}{fwf1,f2}
\fmfdot{fwf1}
\fmfdot{fwf2}
\end{fmfgraph}} ,  
\parbox{15\unitlength}{
\begin{fmfgraph}(15,15)
\fmfleft{a,b}
\fmfright{f1,f2}
\fmf{dashes}{a,fwf}
\fmf{dashes}{fwf,b}
\fmf{dashes}{fwf,www}
\fmf{dashes}{www,f1}
\fmf{dashes}{www,f2}
\fmfdot{fwf}
\fmfdot{www}
\end{fmfgraph}}
\end{aligned}
\right \}
\end{multline}
\end{subequations}

\chapter{Calculation of Ward Identities}\label{app:wi}
In this appendix we calculate the Ward Identities that are used in \fref{chap:reconstruction} to show that the coupling constants of a spontaneously broken gauge theory can be reconstructed from the 4 point functions. The three point functions are discussed in \fref{app:3wi} while the 4 point functions are evaluated in \fref{app:4wi} and \fref{app:os_4wi}.
\section{Ward identities for 3 point functions}\label{app:3wi}
In this section we evaluate the remaining Ward Identities for three point functions to obtain the relations \eqref{subeq:triple-phi-wi1} and
 \eqref{subeq:triple-phi-wi2} that allow to express the coupling constants of the Goldstone bosons in terms of the input parameters. We also check that the STIs for the vertices are satisfied which simplifies the calculations of the Ward Identities for the 4 point functions. 
\subsection{3 $W$ Ward Identity}\label{app:3w-wi}
To evaluate  the WI for the 3 gauge boson vertex with one unphysical gauge boson.
\begin{equation*}
\parbox{15mm}{
\begin{fmfchar*}(15,15)
  \fmfleft{A1,A2} \fmfright{A3} \fmf{double}{A1,a} \fmf{photon}{A2,a}
   \fmf{photon}{a,A3} \fmfblob{10}{a}
\end{fmfchar*}}=0
\end{equation*}
i.e. 
\begin{equation}\label{eq:www-wi}
-\ii  \epsilon_a^{\mu_a}\epsilon_b^{\mu_b}p^{\mu_c}_c\me_{\mu_a\mu_b\mu_c}(W_a W_b  W_c)=m_c \epsilon_a^{\mu_a}\epsilon_b^{\mu_b}\me_{\mu_a\mu_b}(W_aW_b \phi_c)
\end{equation}
 we use the vertex in the general form of \fref{eq:3-fr} i.e. without assuming the total antisymmetry of the $f^{abc}$. This gives for the three gauge boson diagram:
\begin{multline}
\epsilon_a^{\mu_a}\epsilon_b^{\mu_b}p^{\mu_c}_c\me_{\mu_a\mu_b\mu_c}(W_a
W_b  W_c)=\\
=(p_a\cdot \epsilon_b)(p_b\cdot\epsilon_a)(f^{abc}+f^{bac}-f^{cab}-f^{cba})-(\epsilon_a\cdot\epsilon_b)\left[f^{acb}(p_a\cdot
p_c)+f^{bca}(p_b\cdot p_c)\right]
\end{multline}
Here all momenta are considered as ingoing, we have used momentum conservation $p_a+p_b+p_c=0$ and the fact that the polarization vectors satisfy $p\cdot\epsilon=0$. 

The diagram with one Goldstone boson gives
\begin{equation}\label{eq:2a-phi-sti}
\epsilon_a^{\mu_a}\epsilon_b^{\mu_b}\me_{\mu_a\mu_b}(W_aW_b \phi_c)=\ii g_{\phi WW}^{cab}(\epsilon_a\cdot\epsilon_b)
\end{equation}
From the WI \eqref{eq:www-wi} we get by matching coefficients
\begin{equation}\label{eq:f_assy_1}
f^{abc}+f^{bac}-f^{cab}-f^{cba}=0
\end{equation}
and
\begin{multline}\label{eq:g2}
m_{W_c} g_{\phi WW}^{cab} =f^{acb}(p_a\cdot
p_c)+f^{bca}(p_b\cdot p_c)\\
=-f^{acb}[m_{W_a}^2+(p_a\cdot p_b)]-f^{bca}[m_{W_b}^2+(p_a\cdot p_b)]
\end{multline}
since the left hand side is independent from the momentum, so must be the right hand side which gives us the condition 
\begin{equation}\label{eq:f_assy_2}
f^{acb}=-f^{bca}
\end{equation}
and finally the Goldstone boson coupling as given in \fref{eq:phi-2w-coupling}
\begin{equation}\label{eq:phi-2w-app}
m_{W_c} g_{\phi WW}^{cab}=f^{abc}(m_{W_a}^2-m_{W_b}^2)
\end{equation}
Since the choice of the unphysical gauge boson is arbitrary, we find from \fref{eq:f_assy_2} that the $f^{abc}$ must be antisymmetric under the exchange of \emph{any} two indices an therefore totally antisymmetric. 

The STI for the triple gauge boson vertex is given by \fref{eq:3w-sti}:
\begin{equation}\label{eq:www-sti}
\onepi{\mathcal{D}_a^\mu(p_a)W_b^\nu(p_b)W_c^\rho(p_c)}=f^{abc}\left[D_{W_b}(p_b)^{-1}-D_{W_c}(p_c)^{-1}\right]
\end{equation}
Here the propagators are in unitarity gauge and we have used the simplification \eqref{eq:unitarity-2point}. Using the results from the Ward Identity, i.e. the total antisymmetry of the $f^{abc}$ and the result for $g_{\phi WW}$ \eqref{eq:phi-2w-app} we find that the STI is satisfied automatically:
\begin{multline}
 \onepi{\mathcal{D}_a(p_a)W_b^\nu(p_b)W_c^\rho(p_c)}\\
=  \ii f^{abc}
\Biggl\{-\left[p_b^\nu p_b^\rho-p_c^\rho p_c^\nu
\right]+\left[(p_b^2-m_{W_b}^2)g_{\nu\rho}-(p_c^2-m_{W_c}^2)g_{\nu\rho}\right]\Biggr\}
\end{multline}
Here we have used the identity
\begin{equation}
  p_a^\rho p_c^\nu-p_b^\rho p_a^\nu = p_b^\rho p_b^\nu -p_c^\rho p_c^\nu
\end{equation}
that follows from momentum conservation.
\subsection{$WHH$ Ward Identity}
The $WHH$ Ward Identity
\begin{equation}
-\ii  p_a^\mu \me_\mu(H_iH_j  W_a)=m_{W_a}\me_\mu(H_iH_j  \phi_a)
\end{equation}
implies after inserting the Feynman rules from \fref{app:feynman} 
\begin{equation}
-\ii T^a_{ij}(p_i-p_j)\cdot p_a =\ii m_{W_a} g_{\phi H^2}^{aij}
\end{equation}
so we find for the Goldstone boson -Higgs coupling, using that the Higgs are on-shell:
\begin{equation}\label{eq:t}
g_{\phi H^2}^{aij}=\frac{1}{m_{W_a}}(m_{H_i}^2-m_{H_j}^2)T^a_{ij}
\end{equation}
This is the same condition as obtained from the gauge invariance of the scalar potential \eqref{eq:phi2h-coupling}.

If the Higgs bosons are off-shell we find, using the result \eqref{eq:t} that
the STI \eqref{eq:hhw-sti} is satisfied automatically: 
\begin{equation}
- \onepi{\mathcal{D}_a(p_a)H_i(p_i)H_j(p_j)}=-\ii T^a_{ij}\left[(p_i^2-m_{H_i}^2)-(p_j^2-m_{H_j}^2)\right]
\end{equation}
\subsection{$\bar f f W$ Ward Identity}
Using the  vector/axial vector notation of \fref{eq:va-coupling} we find for the contraction of the $\bar f f W$ matrix element with the gauge boson momentum
\begin{multline}
-\ii p_a^\mu \me_\mu(\bar\psi_i\psi_j 
 W_a)=-\ii p_a^\mu \bar v_i(p_i)\gamma^\mu(g_{Vij}^a +g_{Aij}^a\gamma^5)u_j(p_j)\\
=-\bar v_i(p_i)[(m_{f_j}-m_{f_i})g_{Vij}^a-(m_{f_j}+m_{f_i})g_{Aij}^a\gamma^5)u_j(p_j)
\end{multline}
Here we have used the Dirac equation for the spinors
\begin{equation}
\fmslash p u_i=m_i u_i\qquad \bar v_i \fmslash p=-m_i\bar v_i
\end{equation}
and $p_a=-(p_i+p_j)$. 
The Goldstone boson matrix element is 
\begin{equation}
\me(\bar\psi_i\psi_j  \phi_a)=\ii v_i(p_2)\left[g_{Sij}^a+ g_{P ij}^a\gamma^5\right]u_j\\ 
\end{equation}
so the WI
\begin{equation}
-\ii p_a^\mu \me_\mu(\bar\psi_i\psi_j 
 W_a)=m_{W_a}\me_\mu(\bar\psi_i\psi_j  \phi_a) 
\end{equation}
gives for the coupling constants
\begin{equation}\label{eq:gphif}
\begin{aligned}
g_{\phi S ij}^a&=\frac{\ii}{m_{W_a}}(m_{f_j}-m_{f_i})g_{Vij}^a\\
g_{\phi P ij}^a&=-\frac{\ii}{m_{W_a}}(m_{f_j}+m_{f_i})g_{Aij}^a
\end{aligned}
\end{equation}
The left and right-handed couplings are therefore
\begin{equation}
\begin{aligned}
  g_{\phi ij}^a&=(g_{\phi S ij}^a-g_{\phi P ij}^a)=-\frac{i}{m_{W_a}}(m_{f_i}\tau_{L ij}^a-m_{f_j}\tau_{R ij}^a)\\
g_{\phi ij}^{a\dagger }&=(g_{\phi S ij}^a+g_{\phi P ij}^a)=-\frac{i}{m_{W_a}}(m_{f_i}\tau_{Rij}^a-m_{f_j}\tau_{L ij}^a)
\end{aligned}
\end{equation}
as derived from the invariance of the Yukawa couplings in \fref{eq:f-phi-coupling}.

Using the result for the fermion-Goldstone boson coupling from \fref{eq:gphif}
we find for off-shell fermions
\begin{multline}
\ii p_a^\mu\onepi{\bar\psi_i(p_i)W_a^\mu(p_a)\psi_j(p_j)}+m_{W_a}\onepi{\bar\psi_i(p_i)\phi_a(p_a)\psi_j(p_j)}\\
=(\fmslash p_i +\fmslash p_j) (g_{Vij}^a+g_{Aij}^a\gamma^5)+\left[(m_{f_i}-m_{f_j})g_{V ij}^a+(m_{f_i}+m_{f_j})g_{A ij}^a\gamma^5\right]\\
=(g_{Vij}^a-g_{Aij}^a\gamma^5)(\fmslash p_j-m_{f_j})+(\fmslash p_i+m_{f_i})(g_{Vij}^a+g_{Aij}^a\gamma^5)
\end{multline}
so the STI \eqref{eq:ffw-sti} is satisfied. 

\subsection{$3W$ WI with two contractions}
The WI for the 3 gauge boson matrix element
\begin{equation*}
\parbox{15mm}{
\begin{fmfchar*}(15,15)
  \fmfleft{A1,A2} \fmfright{A3} \fmf{double}{A1,a} \fmf{photon}{A2,a}
   \fmf{double}{A3,a} \fmfblob{10}{a}
\end{fmfchar*}}=0 
\end{equation*}
involves four matrix elements:
\begin{multline}\label{eq:www-2pow-wi}
\epsilon_a^\mu p_b^\nu p_c^\rho\me_{\mu\nu\rho}(W_a  W_bW_c)=m_{W_b}m_{W_c}\epsilon_a^\mu\me_\mu(W_a 
\phi_b\phi_c)\\
+\ii m_{W_b}p_c^\rho\epsilon_a^\mu\me_{\mu\rho}(W_a 
\phi_b W_c)+\ii m_{W_c}p_b^\nu\epsilon_a^\mu\me_{\mu\nu}(W_a 
 W_b\phi_c )
\end{multline}
We start by calculating the triple gauge boson with two momentum contractions. For later use we record the result before contraction with the polarization vector of the third gauge boson:
\begin{equation}\label{eq:www-2cov}
p_b^\nu p_c^\rho\me_{\mu\nu\rho}(W_a  W_bW_c)
=f^{abc}\left[-p_b^\mu p_a^2+\frac{1}{2}p_a^\mu (p_c^2-p_b^2-p_a^2)\right]
\end{equation}
Multiplying with  $\epsilon_a$ and for  $W_a$ on-shell we get therefore
\begin{equation}\label{eq:www-2pow}
\epsilon_a^\mu p_b^\nu p_c^\rho\me_{\mu\nu\rho}(W_a  W_bW_c)=f^{abc}m_{W_a}^2(\epsilon_a\cdot p_c)
\end{equation}
This gives us for the WI:
\begin{multline}
f^{abc}m_{W_a}^2(\epsilon_a\cdot
 p_c)=m_{W_b}m_{W_c} t_{bc}^a\epsilon_a\cdot(p_b-p_c)\\
-m_{W_b}g_{\phi WW}^{bac}(\epsilon_a\cdot
 p_c)-m_{W_c}g_{\phi WW}^{cab}(p_b\cdot\epsilon_a)
\end{multline}
Using our result for $g_{\phi WW}$ from \fref{eq:phi-2w-coupling} this can be written as
\begin{equation}
f^{abc}m_{W_a}^2=-2m_{W_b}m_{W_c}t^a_{bc}-f^{bac}(m_{W_a}^2-m_{W_c}^2)+f^{cab}(m_{W_a}^2-m_{W_b}^2)
\end{equation}
From this we can determine the Goldstone boson- gauge boson coupling:
\begin{equation}\label{eq:t-ward}
t^a_{bc}=\frac{1}{2m_{W_b}m_{W_c}}f^{abc}(m_{W_a}^2-m_{W_b}^2-m_{W_c}^2)
\end{equation}
This is the result given in \fref{eq:2phi-w-coupling} that we have derived from the Lie algebra and spontanteous symmetry breaking. 

If the third gauge boson is also off-shell we find instead 
\begin{multline}
\onepi{\mathcal{D}_a\mathcal{D}_bW_c^\rho}=-f^{abc}\left[-p_a^\rho
  p_c^2+\frac{1}{2}p_c^\rho (p_b^2-p_a^2-p_c^2)\right]\\
   -m_{W_a}p_b^\rho g^{abc}_{\phi WW}-m_{W_b}p_a^\rho g^{bac}_{\phi
   WW}+m_{W_a}m_{W_b}t_{ab}^c(p_a^\rho-p_b^\rho)\\
=
=f^{abc}\Bigl[p_a^{\rho}(p_c^2-m_c^2)-p_c^\sigma(p_a\cdot p_c)\Bigr] +\frac{1}{2} p_c^\sigma(m_{W_b}^2-m_{W_a}^2-m_{W_c}^2)\\
 =f^{abc}\Bigl[p_a^{\sigma}(p_c^2-m_{W_c}^2)-p_c^\rho(p_a\cdot p_c)\Bigr]
   -p_c^\rho m_{W_a}m_{W_c}t^b_{ac}
\end{multline}
and this is the same result that can be obtained from the STI \eqref{eq:www-sti2}
\subsection{$3W$ WI with 3 contractions}
Here 8 terms have to be taken into account:
\begin{multline}
0=(-\ii)^3 p_a^\mu p_b^\nu p_c^\rho\me_{\mu\nu\rho}(W_a  W_bW_c)-(-\ii)^2m_{W_a}p_b^\nu p_c^\rho\me_{\nu\rho}(\phi_a  W_bW_c)+\dots\\
+(-\ii)m_{W_a}m_{W_b} p_c^\rho\me_{\rho}(\phi_a  \phi_bW_c)+\dots -m_{W_a}m_{W_b}m_{W_c}\me_{\rho}(\phi_a  \phi_b\phi_c)
\end{multline}
The 3 gauge boson vertex contracted with 3 momentum vectors gives, using \fref{eq:www-2cov}
\begin{multline}
p_a^\mu p_b^\nu p_c^\rho\me_{\mu\nu\rho}(W_a  W_bW_c)=f^{abc}\left[-(p_a\cdot p_b) p_a^2+\frac{1}{2}p_a^2 (p_c^2-p_b^2-p_a^2)\right]=0
\end{multline}
where we have used momentum conservation in the form $(p_a\cdot p_b)=\frac{1}{2}(p_c^2-p_a^2-p_b^2)$.

Next we have to add up three diagrams with one Goldstone boson and 2 contractions. This gives
\begin{multline}\label{eq:www-sti3-a}
-(-\ii)^2m_{W_a}p_b^\nu p_c^\rho\me_{\nu\rho}(\phi_a  W_bW_c)+\dots \\
= \ii m_{W_a}g_{\phi WW}^{abc}(p_b\cdot
 p_c)+\ii m_{W_b}g_{\phi WW}^{bac}(p_a\cdot p_c)+\ii m_{W_c}g_{\phi WW}^{cab}(p_a\cdot p_b)
\\
=\ii f^{abc}\Bigl[m_{W_a}^2p_a\cdot(p_b-p_c)+m_{W_b}^2p_b\cdot(p_c-p_a)+m_{W_c}^2p_c\cdot(p_a-p_b)
\end{multline}
The three diagrams with one unphysical gauge boson and 2 Goldstone bosons give
\begin{multline}\label{eq:www-sti3-b}
(-\ii)m_{W_a}m_{W_b} p_c^\rho\me_{\rho}(\phi_a  \phi_bW_c)+\dots\\
=-\ii m_{W_a}m_{W_b}t^c_{ab} p_c\cdot(p_a-p_b)-\ii m_{W_b}m_{W_c} t^a_{bc} p_a\cdot(p_b-p_c)-\ii m_{W_a}m_{W_c}t^b_{ac} p_b\cdot(p_a-p_c)\\
=-\ii\frac{1}{2} f^{abc}\Bigl[(m_{W_c}^2-m_{W_a}^2-m_{W_b}^2) p_c\cdot(p_a-p_b)+(m_{W_a}^2-m_{W_b}^2-m_{W_c}^2) p_a\cdot(p_b-p_c)\\
-(m_{W_b}^2-m_{W_a}^2-m_{W_c}^2) p_b\cdot(p_a-p_c)\Bigr]
\end{multline}
Adding up the 1 and 2 Goldstone boson diagrams \eqref{eq:www-sti3-a} and \eqref{eq:www-sti3-b} we get
\begin{multline}
\ii \frac{1}{2}f^{abc} (m_{W_a}^2+m_{W_b}^2+m_{W_c}^2)\Bigl[p_c\cdot(p_a-p_b)+p_a\cdot(p_b-p_c)\\
+p_b\cdot(p_c-p_a)\Bigr]=0
\end{multline}
Therefore we must have
\begin{equation}
g_{\phi^3}^{abc}=0
\end{equation}
This is also clear on general grounds: since all particles are off-shell we cannot get rid of the momentum dependence in all other diagrams so these \emph{must} add up to zero.
\section{WIs for 4 point function with one contraction}\label{app:4wi}
We now derive the conditions arising from the 4 point Ward Identities with one contraction. Since the 3 point STIs are satisfied, the calculations follow the pattern discussed in \fref{sec:skeleton}: the internal propagators cancel because of the three point STIs and the resulting conditions on the coupling constants are the STIs for the 4 point vertices.
\subsection{$WW\to WW$}\label{app:wwww-wi}
To the amplitude for 4 gauge boson scattering contribute three gauge boson exchange, three Higgs boson exchange diagram and one diagram with a quartic vertex. 
Using the STI for the $WWW$ vertex \eqref{eq:www-sti} we find for the $s$-channel gauge boson exchange diagram:
\begin{equation}\label{4w-s-kanal}
\parbox{15mm}{
\begin{fmfchar*}(15,15)
  \fmftop{A1,A2} \fmfbottom{A3,A4} \fmf{double,label=$B_a$,label.side=right}{A1,a} \fmf{photon,label=$W_c$,label.side=left}{A2,b}
  \fmf{photon}{a,b} \fmf{photon,label=$W_b$,label.side=left}{A3,a} \fmf{photon,label=$W_d$,label.side=right}{A4,b}
 \fmfdot{a}
  \fmfdot{b}  
\end{fmfchar*}}\qquad
\begin{aligned}
=\frac{\ii }{m_{W_a}}f^{abe}f^{cde}[(p_c-p_d)\cdot\epsilon_b
(\epsilon_c\cdot\epsilon_d)+2(\epsilon_{d}\cdot\epsilon_b)(p_d\cdot\epsilon_c)\\
-2(\epsilon_c\cdot\epsilon_b)(p_c\cdot\epsilon_d)]
\end{aligned}
\end{equation}
The diagram with the 4 gauge boson vertex gives
\begin{multline}
\parbox{15mm}{
\begin{fmfchar*}(15,15)
  \fmfleft{A1,A2} \fmfright{A3,A4} \fmf{photon}{A1,a} \fmf{photon}{A2,a}
   \fmf{photon}{A3,a} \fmf{photon}{A4,a}
  \fmfdot{a}
  \fmfv{decor.shape=square,decor.filled=full,decor.size=3}{A1}
\end{fmfchar*}}
=\frac{\ii }{m_{W_a}}(g_{W^4}^{abcd}(p_a\cdot\epsilon_c)(\epsilon_b\cdot\epsilon_d)+g_{W^4}^{abdc}(p_a\cdot\epsilon_d)(\epsilon_b\cdot\epsilon_c)+g_{W^4}^{acbd}(p_a\cdot\epsilon_b)(\epsilon_c\cdot\epsilon_d))
\end{multline}
The Higgs exchange diagrams satisfy the WI for themselves because the 
STI \eqref{eq:hww-sti-contr}) is satisfied even for an off-shell Higgs boson:
\begin{equation}
\parbox{15mm}{
\begin{fmfchar*}(15,15)
  \fmftop{A1,A2} \fmfbottom{A3,A4} \fmf{double,label=$B_a$,label.side=right}{A1,a} \fmf{photon,label=$W_c$,label.side=left}{A2,b}
  \fmf{dashes}{a,b} \fmf{photon,label=$W_b$,label.side=left}{A3,a} \fmf{photon,label=$W_d$,label.side=right}{A4,b}
 \fmfdot{a}
  \fmfdot{b}  
\end{fmfchar*}}\qquad=0
\end{equation}
Adding up $s$, $t$ and $u$-channel diagrams and picking up the  coefficient of $(\epsilon_b\cdot\epsilon_d)$ gives, using momentum conservation
\begin{equation}
2f^{abe}f^{cde}(p_d\cdot\epsilon_c)-f^{ace}f^{bde}(p_a+2p_d)\cdot\epsilon_c+2f^{ade}f^{bce}(p_a+p_d )\cdot\epsilon_c=-g_{W^4}^{abcd}(p_a\cdot\epsilon_c)
\end{equation}
The vanishing of the coefficient of  $(p_d\cdot \epsilon_c)$ now gives the Jacobi identity \eqref{eq:jacobi}:
\begin{equation}
f^{abe}f^{cde}+f^{cae}f^{bde}+f^{ade}f^{bce}=0
\end{equation}
while the coefficient of  $(p_d\cdot \epsilon_c)$ yields the quartic gauge coupling:
\begin{equation}\label{eq:4w-coupling}
g_{W^4}^{abcd}=-f^{ace}f^{bde}-2f^{ade}f^{bce}=f^{abe}f^{cde}-f^{ade}f^{bce}
\end{equation}
The coefficients of the other polarization vectors reproduce these results.

\subsection{$\bar f f \to WW$}\label{app:ffww-wi}
The Ward Identity for gluon boson pair production is the standard example for the application of Ward Identities in unbroken Yang mills theory (see e.g. \cite{Peskin:1995}). The only new features in the spontaneously broken case are the Higgs exchange diagrams and the diagrams with external Goldstone bosons.

Using the STI \eqref{eq:ffw-sti}, the $t$-channel Compton diagram gives
\begin{equation}
\parbox{15mm}{
\begin{fmfchar*}(15,15)
  \fmfleft{f1,f2} \fmfright{A,H} \fmf{fermion,label=$f_j$}{f1,a}
  \fmf{fermion,label=$f_i$, label.side=left}{b,f2}
  \fmf{fermion,label=$f_l$}{a,b} \fmf{double,label=$B_a$, label.side=right}{A,a} \fmf{photon,label=$W_b$,label.side=left}{H,b}
 \fmfdot{a}
  \fmfdot{b}  
 \end{fmfchar*}}
=(\ii )^2\bar v_i(p_i)\fmslash\epsilon_b(\tau_{L il}^b \tau_{Llj}^a(\tfrac{1-\gamma^5}{2})+ \tau_{Ril}^b \tau_{Rlj}^a(\tfrac{1+\gamma^5}{2}))u_j(p_j)
\end{equation}
while the  $u$-channel diagram is obtained by exchanging $a$ and $b$ and applying a sign change since the unphysical gauge boson is inserted into an vertex with an ingoing \emph{anti}-fermion  (see the STI \eqref{eq:ffw-sti}). Therefore the sum of the Compton diagrams is given by
\begin{equation}
\parbox{15mm}{
\begin{fmfchar*}(15,15)
  \fmfleft{f1,f2} \fmfright{A,H} \fmf{fermion,label=$f_j$}{f1,a}
  \fmf{fermion,label=$f_i$, label.side=left}{b,f2}
  \fmf{fermion,label=$f_l$}{a,b} \fmf{double}{A,a} \fmf{photon}{H,b}
 \fmfdot{a}
  \fmfdot{b}  
 \end{fmfchar*}}+
\parbox{15mm}{
\begin{fmfchar*}(15,15)
  \fmfleft{f1,f2} \fmfright{A,H} \fmf{fermion,label=$f_j$}{f1,a}
  \fmf{fermion,label=$f_i$, label.side=left}{b,f2}
  \fmf{fermion,label=$f_l$}{a,b} 
\fmf{phantom}{A,a} \fmf{phantom}{H,b}\fmffreeze 
  \fmf{double}{A,b} \fmf{photon}{H,a}
 \fmfdot{a}
  \fmfdot{b}  
\end{fmfchar*}}
-\bar v_i(p_i)\fmslash\epsilon_b\left([\tau_{L}^b, \tau_{L}^a]_{ij}(\tfrac{1-\gamma^5}{2})+ [\tau_{R}^b, \tau_{R}^a]_{ij}(\tfrac{1+\gamma^5}{2})  \right)u_j(p_j)
\end{equation}
Using the STI \eqref{eq:www-sti}, we find for the $s$-channel gauge boson exchange diagram
\begin{equation}
\parbox{15mm}{
\begin{fmfchar*}(15,15)
  \fmfleft{f1,f2} \fmfright{A,H} \fmf{fermion,label=$f_j$}{f1,a}
  \fmf{fermion,label=$f_i$, label.side=left}{a,f2}
  \fmf{photon,label=$W_c$}{a,b} \fmf{double,label=$B_a$, label.side=right}{A,b} \fmf{photon,label=$W_b$,label.side=left}{H,b}
 \fmfdot{a}
  \fmfdot{b}  
\end{fmfchar*}}
\qquad =-\ii f^{abc}\bar v_i(p_i)\fmslash\epsilon_b(g_{Vij}^c+g_{Aij}^c\gamma^5)u_j(p_j)
\end{equation}
Because of the STI \eqref{eq:ghphiw-wi} the $s$-channel Higgs exchange diagram satisfies the WI by itself:
\begin{equation}
\parbox{15mm}{
\begin{fmfchar*}(15,15)
  \fmfleft{f1,f2} \fmfright{A,H} \fmf{fermion,label=$f_j$,label.side=left}{f1,a}
  \fmf{fermion,label=$f_i$, label.side=left}{a,f2}
  \fmf{dashes,label=$H_k$}{a,b} \fmf{double,label=$B_a$, label.side=right}{A,b} \fmf{photon,label=$W_b$,label.side=left}{H,b}
 \fmfdot{a}
  \fmfdot{b}  
\end{fmfchar*}}\qquad=0
\end{equation}
Adding up the Compton diagrams and the gauge boson exchange diagram we find as the coefficients of $(1-\gamma^5)$ and  $(1+\gamma^5)$ the lie algebra of the representation matrices
\begin{align}
[\tau_{L}^a, \tau_{L}^b]_{ij}-\ii f^{abc}\tau_{Lij}^c&=0\\
[\tau_{R}^a, \tau_{R}^b]_{ij}-\ii f^{abc}\tau_{Rij}^c&=0
\end{align}
\subsection{$\bar f f \to WH$}\label{app:ffwh-wi}
In the Ward Identity for the $\bar f f \to WH$ amplitude, the Yukawa couplings of the Higgs bosons play a essential role, in contrast to the case of gauge boson pair production considered in \fref{app:ffww-wi}. The Yukawa coupling enters in  the Higgs exchange $s$-channel diagram and the Compton-like diagrams. The Higgs exchange diagram can be computed using the STI for the $WHH$ vertex \eqref{eq:hhw-sti}:
\begin{equation}
\parbox{15mm}{
\begin{fmfchar*}(15,15)
  \fmfleft{f1,f2} \fmfright{A,H} \fmf{fermion,label=$f_j$, label.side=left}{f1,a}
  \fmf{fermion,label=$f_i$, label.side=left}{a,f2}
  \fmf{dashes,label=$H_h$}{a,b} \fmf{double,label=$B_b$, label.side=right}{A,b} \fmf{dashes,label=$H_k$,label.side=left}{H,b}
 \fmfdot{a}
  \fmfdot{b}  
\end{fmfchar*}}
\qquad=-\ii \bar v_i(p_i)(g_{H ij}^k(\tfrac{1-\gamma^5}{2})+ g_{Hij}^{k\dagger}(\tfrac{1+\gamma^5}{2}))u_j(p_j)T^b_{hk}
\end{equation}
Using the STI \eqref{eq:ffw-sti} we get for the $t$-channel Compton diagram
\begin{equation}\label{eq:ffwh-t}
\parbox{15mm}{
\begin{fmfchar*}(15,15)
  \fmfleft{f1,f2} \fmfright{A,H} \fmf{fermion,label=$f_j$,label.side=left}{f1,a}
  \fmf{fermion,label=$f_i$, label.side=left}{b,f2}
  \fmf{fermion,label=$f_l$}{a,b} \fmf{double,label=$B_b$, label.side=right}{A,a} \fmf{dashes,label=$H_k$,label.side=left}{H,b}
 \fmfdot{a}
  \fmfdot{b}  
 \end{fmfchar*}}
=(\ii )^2\bar v_i(p_i)(g_{H il}^k \tau_{Llj}^b(\tfrac{1-\gamma^5}{2})+ g_{H
  il}^{k\dagger} \tau_{Rlj}^a(\tfrac{1+\gamma^5}{2}))u_j(p_j)
\end{equation}
The $u$-channel diagram is similarly
\begin{equation}
\parbox{15mm}{
\begin{fmfchar*}(15,15)
  \fmfleft{f1,f2} \fmfright{A,H} \fmf{fermion,label=$f_j$,label.side=left}{f1,a}
  \fmf{fermion,label=$f_i$, label.side=left}{b,f2}
  \fmf{fermion,label=$f_l$}{a,b} 
\fmf{phantom}{A,a} \fmf{phantom}{H,b}\fmffreeze 
  \fmf{double}{A,b} \fmf{dashes}{H,a}
 \fmfdot{a}
  \fmfdot{b}  
\end{fmfchar*}}
=-(\ii )^2\bar v_i(p_i)( \tau_{Ril}^bg_{Hlj}^k(\tfrac{1-\gamma^5}{2})+\tau_{Lil}^b g_{H
   lj}^{k\dagger}(\tfrac{1+\gamma^5}{2}))u_j(p_j)
\end{equation}
Next we turn to the $s$-channel gauge boson exchange diagram. Using the STI for the $HWW$ vertex \eqref{eq:hww-sti}  and using \fref{eq:contr-prop} we can cancel the propagator: 
\begin{multline}\label{eq:ffwh-s}
 \parbox{15mm}{
\begin{fmfchar*}(15,15)
  \fmfleft{f1,f2} \fmfright{A,H} \fmf{fermion,label=$f_j$, label.side=left}{f1,a}
  \fmf{fermion,label=$f_i$, label.side=left}{a,f2}
  \fmf{photon,label=$W_a$}{a,b} \fmf{double,label=$B_b$, label.side=right}{A,b} \fmf{dashes,label=$H_k$,label.side=left}{H,b}
 \fmfdot{a}
  \fmfdot{b}  
\end{fmfchar*}}\qquad=\frac{-(\ii ^2)}{2m_{W_a}^2}\bar v_i(p_i)(\fmslash p_i+\fmslash
p_j)(g_{Vij}^a +g_{Aij}^a\gamma^5)u_j(p_j))g_{HWW}^{kba}\\
=-\ii \frac{1}{2m_{W_a}}\bar v_i(p_i)[g_{\phi ij}^a(\tfrac{1+\gamma^5}{2}) +g_{\phi ij}^{a\dagger}(\tfrac{1-\gamma^5}{2}) ]u_j(p_j))g_{HWW}^{hba}
\end{multline}
In the last step we have used the WI for the $\bar \psi W\psi$ vertex to trade the gauge boson coupling for the Goldstone boson coupling. (We must take care of signs since we consider the momentum of the unphysical gauge boson at the $3W$ vertex as \emph{outgoing}.) 

The sum of the coefficients of  $(1-\gamma^5)$  in all diagrams gives 
\begin{equation}\label{eq:ffwh-wi}
0=-\ii\frac{g_{HWW}^{hba}}{2m_{W_a}}  g_{\phi ij}^a -i(g_{H ij}^h
T^b_{hk})-g_{H il}^k
\tau_{Llj}^b+\tau_{Ril}^bg_{H lj}^k
\end{equation}
That is the result given in \eqref{eq:2fhw-wi}. The coefficient of  $(1+\gamma^5)$ gives the hermitian conjugate equation. 
\subsection{$3WH$}\label{subsec:3wh-wi}
In a theory with a general Higgs sector, there are three diagrams with gauge boson, three with Higgs exchange and one diagram with a quartic vertex contributing to the $3WH$ amplitude. 

The  $t$-channel gauge boson exchange diagram can be computed using the STI for the  HWW vertex \eqref{eq:hww-sti} the relation \fref{eq:contr-prop}  and the Ward Identity for the 3 gauge boson vertex (compare with the comments after \fref{eq:ffwh-s}) leading to:
\begin{multline}\label{eq:3ah-tkanal}
 \parbox{15mm}{
\begin{fmfchar*}(15,15)
  \fmfleft{A1,A2} \fmfright{A3,A4} \fmf{double,label=$B_a$, label.side=left}{A1,a} \fmf{dashes,label=$H_i$, label.side=right}{A2,a}
  \fmf{photon}{a,b} \fmf{photon,label=$W_b$, label.side=right}{A3,b} \fmf{photon,label=$W_c$, label.side=left}{A4,b}
 \fmfdot{a}
  \fmfdot{b}  
\end{fmfchar*}}\qquad = -\ii \frac{g_{HWW}^{iad}}{2 m_{W_d}}g_{\phi WW}^{dbc}=-\ii\frac{g_{HWW}^{iad}}{2 m_{W_d}^2} f^{dbc}(\epsilon_b\cdot\epsilon_c)(m_{W_b}^2-m_{W_c}^2)
\end{multline}

The  $s$- and  $u$  channel gauge boson exchange diagrams give, using the STI for the $3W$ vertex \eqref{eq:www-sti}
\begin{multline}
\parbox{15mm}{\begin{fmfchar*}(15,15)
  \fmfleft{A1,A2} \fmfright{A3,A4} \fmf{double}{A1,a} \fmf{dashes}{A2,b}
  \fmf{photon}{a,b} \fmf{photon}{A3,a} \fmf{photon}{A4,b}
 \fmfdot{a}
  \fmfdot{b} 
\end{fmfchar*}}+
\parbox{15mm}{
\begin{fmfchar*}(15,15)
  \fmfbottom{A1,A2} \fmftop{A3,A4} \fmf{double}{A1,a} \fmf{phantom}{A2,a}
  \fmf{photon}{a,b} \fmf{dashes}{A3,b} \fmf{phantom}{A4,b} \fmffreeze
  \fmf{photon}{A4,a} \fmf{photon}{A2,b}
  \fmfdot{a}
  \fmfdot{b}
\end{fmfchar*}}=
-\ii g_{HWW}^{icd}f^{abd}(\epsilon_{b}\cdot\epsilon_{c})-\ii g_{HWW}^{ibd}f^{acd}(\epsilon_{b}\cdot\epsilon_{c})
\end{multline}
The  $s$- and $u$- channel Higgs exchange diagrams automatically satisfy the WI because of the STI \eqref{eq:hww-sti} for the $HWW$ vertex. 
 \begin{equation}
\parbox{15mm}{
\begin{fmfchar*}(15,15)
  \fmfleft{A1,A2} \fmfright{A3,A4} \fmf{double}{A1,a} \fmf{dashes}{A2,b}
  \fmf{dashes,label=$H$}{b,a} \fmf{photon}{A3,a} \fmf{photon}{A4,b}
  \fmfdot{a}
  \fmfdot{b}
\end{fmfchar*}}+
\parbox{15mm}{
\begin{fmfchar*}(15,15)
  \fmfbottom{A1,A2} \fmftop{A3,A4} \fmf{double}{A1,a} \fmf{phantom}{A2,a}
  \fmf{dashes}{b,a} \fmf{dashes}{A3,b} \fmf{phantom}{A4,b} \fmffreeze
  \fmf{photon}{A4,a} \fmf{photon}{A2,b}
  \fmfdot{a}
  \fmfdot{b}
\end{fmfchar*}}=0
 \end{equation}
This is not the case for the $t$-channel diagram that gives, using the STI \eqref{eq:hhw-sti}
\begin{equation}
\parbox{15mm}{
\begin{fmfchar*}(15,15)
  \fmfleft{A1,A2} \fmfright{A3,A4} \fmf{double}{A1,a} \fmf{dashes}{A2,a}
  \fmf{dashes,label=$H$}{a,b} \fmf{photon}{A3,b} \fmf{photon}{A4,b}
 \fmfdot{a}
  \fmfdot{b}  
\end{fmfchar*}}=
\ii T_{ij}^a g_{HWW}^{jbc}(\epsilon_{b}\cdot\epsilon_{c})
\end{equation} 
The only diagram with a quartic vertex contributing to the WI is the diagram with an external Goldstone boson:
\begin{equation}
\parbox{15mm}{
\begin{fmfchar*}(15,15)
  \fmfleft{A1,A2} \fmfright{A3,A4} \fmf{dashes,label=$\phi$}{A1,a} \fmf{dashes,label=$H$}{A2,a}
   \fmf{photon}{A3,a} \fmf{photon}{A4,a}
  \fmfdot{a} \fmfv{decor.shape=cross,decor.size=5}{A1}
\end{fmfchar*}}= -\ii m_{W_a}(\epsilon_{b}\cdot\epsilon_{c}) g_{H\phi W^2}^{abci}
\end{equation}
Adding up the diagrams results in the condition
\begin{equation}\label{eq:s3-s4} 
m_{W_a}g_{H\phi W^2}^{abci}=-g_{HWW}^{icd}f^{abd}-g_{HWW}^{ibd}f^{acd}-\frac{g_{HWW}^{iad}}{2m_{W_d}}g^{dbc}_{\phi WW}+T^a_{ij}g_{HWW}^{jbc}
\end{equation}
\subsection{$3HW$}
The $3HW$ amplitude is the last Ward Identity with one contraction we have to consider. 
The $s$-channel Higgs exchange diagrams are, using the STI \eqref{eq:hhw-sti}
\begin{equation}\label{3wh-s-channel}
\parbox{15mm}{
\begin{fmfchar*}(15,15)
  \fmftop{A1,A2} \fmfbottom{A3,A4} \fmf{double,label=$B_a$,label.side=right}{A1,a} \fmf{dashes,label=$H_j$,label.side=left}{A2,b}
  \fmf{dashes}{a,b} \fmf{dashes,label=$H_i$,label.side=left}{A3,a} \fmf{dashes,label=$H_k$,label.side=right}{A4,b}
 \fmfdot{a}
  \fmfdot{b}  
\end{fmfchar*}}\qquad+\quad
\parbox{15mm}{
\begin{fmfchar*}(15,15)
  \fmftop{A1,A2} \fmfbottom{A3,A4} \fmf{double}{A1,a} \fmf{dashes}{A2,a}
  \fmf{dashes}{a,b} \fmf{dashes,label}{A3,b} \fmf{dashes}{A4,b}
 \fmfdot{a}
  \fmfdot{b}  
\end{fmfchar*}}\,+\,
\parbox{15mm}{
\begin{fmfchar*}(15,15)
  \fmftop{A1,A2} \fmfbottom{A3,A4} \fmf{double}{A1,a} \fmf{phantom}{A2,a}
  \fmf{dashes}{a,b} \fmf{dashes,label}{A3,b} \fmf{phantom}{A4,b}
\fmffreeze
 \fmf{dashes}{A2,b}\fmf{dashes}{A4,a}
 \fmfdot{a}
  \fmfdot{b}  
\end{fmfchar*}} = \ii g_{H^3}^{jkl}T^a_{il}+\ii g_{H^3}^{ikl}T^a_{jl}+\ii g_{H^3}^{ijl}T^a_{kl}
\end{equation}
The gauge boson exchange diagrams are, using the STI \eqref{eq:hww-sti} and \fref{eq:contr-prop}:
\begin{multline}
\parbox{15mm}{
\begin{fmfchar*}(15,15)
  \fmftop{A1,A2} \fmfbottom{A3,A4} \fmf{double,label=$B_a$,label.side=right}{A1,a} \fmf{dashes,label=$H_j$,label.side=left}{A2,b}
  \fmf{photon}{a,b} \fmf{dashes,label=$H_i$,label.side=left}{A3,a} \fmf{dashes,label=$H_k$,label.side=right}{A4,b}
 \fmfdot{a}
  \fmfdot{b}  
\end{fmfchar*}}\qquad+\,
\parbox{15mm}{
\begin{fmfchar*}(15,15)
  \fmftop{A1,A2} \fmfbottom{A3,A4} \fmf{double}{A1,a} \fmf{dashes}{A2,a}
  \fmf{photon}{a,b} \fmf{dashes,label}{A3,b} \fmf{dashes}{A4,b}
 \fmfdot{a}
  \fmfdot{b}  
\end{fmfchar*}}\,+\,
\parbox{15mm}{
\begin{fmfchar*}(15,15)
  \fmftop{A1,A2} \fmfbottom{A3,A4} \fmf{double}{A1,a} \fmf{phantom}{A2,a}
  \fmf{photon}{a,b} \fmf{dashes,label}{A3,b} \fmf{phantom}{A4,b}
\fmffreeze
 \fmf{dashes}{A2,b}\fmf{dashes}{A4,a}
 \fmfdot{a}
  \fmfdot{b}  
\end{fmfchar*}}\\
= -\ii g_{\phi H^2}^{bjk}\frac{g_{HWW}^{iab}}{2m_{W_b}}-\ii g_{\phi H^2}^{bik}\frac{g_{HWW}^{jab}}{2m_{W_b}}-\ii g_{\phi H^2}^{bij}\frac{g_{HWW}^{kab}}{2m_{W_b}}
\end{multline}
To arrive at this form, we have also used the Ward Identity for the $HHW$ vertex (compare to the remarks after \fref{eq:ffwh-s}).
There is only one contribution to the diagrams with a quartic vertex:
\begin{equation}
\parbox{15mm}{
\begin{fmfchar*}(15,15)
  \fmftop{A1,A2} \fmfbottom{A3,A4} \fmf{dashes,label=$\phi_a$,label.side=left}{A1,a} \fmf{dashes,label=$H_j$,label.side=left}{A2,a}
   \fmf{dashes}{a,b} \fmf{dashes,label=$H_i$,label.side=left}{A3,a} \fmf{dashes,label=$H_k$,label.side=left}{A4,a}
  \fmfdot{a}
  \fmfv{decor.shape=cross,decor.size=5}{A1}
 \end{fmfchar*}}\quad= -\ii m_{W_b}g_{\phi H^3}^{aijk}
 \end{equation}
 so altogether we have derived the condition
 \begin{equation}\label{eq:3hw-wi}
 m_{W_b}g_{\phi H^3}^{aijk}=g_{H^3}^{jkl}T^a_{il}- g_{\phi H^2}^{bjk}\frac{g_{HWW}^{iab}}{2m_{W_b}}+ g_{H^3}^{ikl}T^a_{jl}- g_{\phi H^2}^{bik}\frac{g_{HWW}^{jab}}{2m_{W_b}}+ g_{H^3}^{ijl}T^a_{kl}- g_{\phi H^2}^{bij}\frac{g_{HWW}^{kab}}{2m_{W_b}}
 \end{equation}
\section{4 point WIs with several contractions}\label{app:os_4wi}
In this appendix we turn to the evaluations of the Ward Identities with more than one contraction. The calculations are simplified by the use of the STIs for cubic vertices with 2 contractions \eqref{eq:4point-2cov}. 
\subsection{4$W$ Ward identity with 2 unphysical $W$s}
The $s$-channel gauge boson exchange diagram can again be calculated using the STI \eqref{eq:www-sti2} and the contracted version \fref{eq:amp_www_sti}. We get
\begin{multline}
\parbox{15mm}{
  \begin{fmfchar*}(15,15)
  \fmftop{A1,A2} \fmfbottom{A3,A4} \fmf{double,label=$B_a$,label.side=right}{A1,a} \fmf{photon,label=$W_c$,label.side=left}{A2,b}
  \fmf{photon}{a,b} \fmf{double,label=$B_b$,label.side=left}{A3,a} \fmf{photon,label=$W_d$,label.side=right}{A4,b}
  \fmfdot{a}
  \fmfdot{b}  
\end{fmfchar*}}\qquad
\begin{aligned}
=&\ii f^{abe}f^{cde}\left(
 p_{b\mu_e}+\frac{m_{W_b}}{m_{W_e}}p_{e\mu_e}t^a_{eb}\right)\\
&\times \left[(\epsilon_c\cdot\epsilon_d)(p_c^{\mu_e}-p_d^{\mu_e})
 +2\epsilon_d^{\mu_e}(\epsilon_c\cdot p_d)-2\epsilon_c^{\mu_e}(\epsilon_d\cdot
 p_c)\right]
\end{aligned}\\
=\ii\left(m_{W_b}g_{\phi WW}^{ecd}t^a_{eb}(\epsilon_c\cdot\epsilon_d) \right)\\
+\ii f^{abe}f^{cde}\Bigl[(\epsilon_c\cdot\epsilon_d)p_b\cdot(p_c-p_d)
 +2(\epsilon_d\cdot p_b)(\epsilon_c\cdot p_d)-2(\epsilon_c\cdot p_b)(\epsilon_d\cdot
 p_c)\Bigr]
\end{multline}
For the $t$ and $u$ channel diagrams we have to use the STI for the three gauge boson vertex \eqref{eq:www-sti} for both vertices. The $s$-channel diagram is 
\begin{equation}
 \parbox{15mm}{
  \begin{fmfchar*}(15,15)
    \fmftop{A1,A2} \fmfbottom{A3,A4} \fmf{double}{A1,a} \fmf{photon}{A2,a}
    \fmf{photon}{a,b} \fmf{double}{A3,b} \fmf{photon}{A4,b}
    \fmfdot{a}
    \fmfdot{b}  
\end{fmfchar*}}
\begin{aligned}
&=(-\ii)^3 f^{ace}f^{bde}\epsilon_c^{\mu_e}\left[(p_e^2-m_{W_e}^2)\epsilon_{d\mu_e}-p_{e\mu_e}(p_b\cdot
  \epsilon_d)\right]\\
&=\ii f^{ace}f^{bde}\left[(t-m_{W_e}^2)(\epsilon_c\cdot\epsilon_d)+(p_a\cdot\epsilon_c)(p_b\cdot
  \epsilon_d)\right]
\end{aligned}
\end{equation}
while the $u$ channel diagram is given by
\begin{equation}
\parbox{15mm}{ \begin{fmfchar*}(15,15)
  \fmftop{A1,A2} \fmfbottom{A3,A4} \fmf{double}{A1,a} 
  \fmf{photon}{a,b} \fmf{double}{A3,b} 
\fmf{phantom}{A2,a}
\fmf{phantom}{A4,b}
\fmffreeze
\fmf{photon}{A2,b}
\fmf{photon}{A4,a}
 \fmfdot{a}
  \fmfdot{b}  
\end{fmfchar*}}=
\ii  f^{ade}f^{bce}\left[(u-m_{W_e}^2)(\epsilon_c\cdot\epsilon_d)+(p_a\cdot\epsilon_d)(p_b\cdot
  \epsilon_c)\right]
\end{equation}
The  $s$ channel Higgs exchange diagram gives, using the STI for the  $HWW$-vertex with two unphysical gauge bosons \eqref{eq:hww-sti-2}
\begin{equation}
\parbox{15mm}{
  \begin{fmfchar*}(15,15)
  \fmftop{A1,A2} \fmfbottom{A3,A4} \fmf{double,label=$B_a$,label.side=right}{A1,a} \fmf{photon,label=$W_c$,label.side=left}{A2,b}
  \fmf{dashes}{a,b} \fmf{double,label=$B_b$,label.side=left}{A3,a} \fmf{photon,label=$W_d$,label.side=right}{A4,b}
  \fmfdot{a}
  \fmfdot{b}  
\end{fmfchar*}}\qquad=(\ii)^3 \frac{1}{2}g_{HWW}^{iab}g_{HWW}^{icd}(\epsilon_c\cdot\epsilon_d)
\end{equation}
The $t$- and $u$ channel Higgs exchange  diagrams again satisfy the WI directly. 
\begin{equation}
\parbox{15mm}{
\begin{fmfchar*}(15,15)
  \fmfleft{A1,A2} \fmfright{A3,A4} \fmf{double}{A1,a} \fmf{double}{A2,b}
  \fmf{dashes,label=$H$}{a,b} \fmf{photon}{A3,a} \fmf{photon}{A4,b}
  \fmfdot{a}
  \fmfdot{b}
\end{fmfchar*}}+
\parbox{15mm}{
\begin{fmfchar*}(15,15)
  \fmfbottom{A1,A2} \fmftop{A3,A4} \fmf{double}{A1,a} \fmf{phantom}{A2,a}
  \fmf{dashes}{a,b} \fmf{double}{A3,b} \fmf{phantom}{A4,b} \fmffreeze
  \fmf{photon}{A4,a} \fmf{photon}{A2,b}
  \fmfdot{a}
  \fmfdot{b}
\end{fmfchar*}}=0
 \end{equation}
Only two diagrams with 4 point vertices contribute. The 4 gauge boson diagram is given by
\begin{equation}
 \parbox{15mm}{
  \begin{fmfchar*}(15,15)
    \fmftop{A1,A2} \fmfbottom{A3,A4} \fmf{photon}{A1,a} \fmf{photon}{A2,a}
 \fmf{photon}{A3,a} \fmf{photon}{A4,a}
  \fmfdot{a} 
\fmfv{decor.shape=square,decor.filled=full,decor.size=3}{A1}
\fmfv{decor.shape=square,decor.filled=full,decor.size=3}{A3}
\end{fmfchar*}}\quad
\begin{aligned}
=&(-\ii)^3\Bigl[(f^{abe}f^{cde}-f^{ade}f^{bce})
(p_a\cdot\epsilon_c)(p_b\cdot\epsilon_d)\\
&+(-f^{abe}f^{cde}-f^{ace}f^{bde})(p_a\cdot\epsilon_d)(p_b\cdot\epsilon_c)\\
&+(f^{ace}f^{bde}+f^{ade}f^{bce})(p_a\cdot
 p_b)(\epsilon_c\dot\epsilon_d)\Bigr]
\end{aligned}
\end{equation}
while the $2W2\phi$ diagram is
\begin{equation}
 \parbox{15mm}{
  \begin{fmfchar*}(15,15)
    \fmftop{A1,A2} \fmfbottom{A3,A4} \fmf{dashes}{A1,a} \fmf{photon}{A2,a}
 \fmf{dashes}{A3,a} \fmf{photon}{A4,a}
  \fmfdot{a} \fmfv{decor.shape=cross,decor.size=5}{A1,A3} 
\end{fmfchar*}}=\ii m_{W_a}m_{W_b}g_{\phi^2 W^2}^{abcd}(\epsilon_c\cdot\epsilon_d)
\end{equation}
In the sum of all diagrams we first consider the terms $\propto (\epsilon_c\cdot \epsilon_d)$:
\begin{multline}\label{eq:2cov-4w-coeff}
\ii m_{W_b}g_{\phi WW}^{ecd}t^a_{eb} +\ii f^{abe}f^{cde}p_b\cdot(p_c-p_d)\\
+\ii f^{ace}f^{bde}(t-m_{W_e}^2)+\ii f^{ade}f^{bce}(u-m_{W_e}^2))\\
+\ii (f^{ace}f^{bde}+f^{ade}f^{bce})(p_a\cdot
 p_b)=\ii \frac{1}{2}g_{HWW}^{iab}g_{HWW}^{icd}-\ii m_{W_a}m_{W_b}g_{\phi^2 W^2}^{abcd}
\end{multline}
 Using the Jacobi identity, the momentum dependence drops out and one  can derive the condition
\begin{multline}\label{eq:2pow4w}
m_{W_a}m_{W_b}g_{\phi^2 W^2}^{abcd}-\frac{1}{2}g_{HWW}^{iab}g_{HWW}^{icd}
=  m_{W_b}g_{\phi WW}^{ecd}t^a_{be}\\
+ f^{ace}f^{dbe}(m_{W_d}^2-m_{W_e}^2)- f^{cbe}f^{dae}(m_{W_c}^2-m_{W_e}^2))
\end{multline}
The coefficients of the other terms can be shown to cancel because of the Jacobi identity and momentum conservation.
\subsection{$3WH$ Ward identity with 2 unphysical $W$s}\label{app:app:3wh-2cov}
For the $s$-channel gauge boson exchange diagram we  need the STI for the $WWW$ vertex \eqref{eq:www-sti2} contracted with a propagator. Using \fref{eq:contr-prop} we find
\begin{equation}\label{eq:amp_www_sti}
 \onepi{ \mathcal{D}_a(p_a)\mathcal{D}_b(p_b)W_{c\rho}(p_c)}{G_W^{\rho\mu}}^{-1}(p_c) =\ii
 [ f^{abc}p_b^\mu  +\frac{m_{W_b}}{m_{W_c}}p_c^\mu t^a_{cb}]
\end{equation}
so we get for the diagram:
\begin{multline}
 \parbox{15mm}{
  \begin{fmfchar*}(15,15)
  \fmftop{A1,A2} \fmfbottom{A3,A4} \fmf{double,label=$B_a$,label.side=right}{A1,a} \fmf{dashes,label=$H_i$,label.side=left}{A2,b}
  \fmf{photon}{a,b} \fmf{double,label=$B_b$,label.side=left}{A3,a} \fmf{photon,label=$W_c$,label.side=right}{A4,b}
  \fmfdot{a}
  \fmfdot{b}  
\end{fmfchar*}}\qquad=(\ii)^2 g_{HWW}^{ice} \left[f^{abe}(p_b\cdot
  \epsilon_c)+\frac{m_{W_e}m_{W_b}}{m_{W_e}^2}
  t^a_{eb}(p_i\cdot\epsilon_c)\right]\\
=(\ii)^2\frac{1}{2} g_{HWW}^{ice} f^{abe}\left[-(p_a-p_b)\cdot\epsilon_c+\frac{m_{W_a}^2-m_{W_b}^2}{m_{W_e}^2} (p_a+p_b)\cdot\epsilon_c \right]
\end{multline}
Using the STI for the 3 gauge boson vertex \eqref{eq:www-sti} and the STI for the $HWW$ vertex  \eqref{eq:hww-sti} we find for the $t$-and $u$-channel diagrams
\begin{multline}
 \parbox{15mm}{
  \begin{fmfchar*}(15,15)
    \fmftop{A1,A2} \fmfbottom{A3,A4} \fmf{double}{A1,a} \fmf{dashes}{A2,a}
    \fmf{photon}{a,b} \fmf{double}{A3,b} \fmf{photon}{A4,b}
    \fmfdot{a}
    \fmfdot{b}  
\end{fmfchar*}}+
\parbox{15mm}{ \begin{fmfchar*}(15,15)
  \fmftop{A1,A2} \fmfbottom{A3,A4} \fmf{double}{A1,a} 
  \fmf{photon}{a,b} \fmf{double}{A3,b} 
\fmf{phantom}{A2,a}
\fmf{phantom}{A4,b}
\fmffreeze
\fmf{dashes}{A2,b}
\fmf{photon}{A4,a}
 \fmfdot{a}
  \fmfdot{b}  
\end{fmfchar*}}
=-(-\ii)^2\frac{1}{2}\left [ f^{ace}g_{HWW}^{ibe}(\epsilon_c\cdot p_a)+f^{bce}g_{HWW}^{iae}(\epsilon_c\cdot p_b)\right]
\end{multline}

The $s$-channel Higgs exchange diagram can be calculated using the STI for the $HWW$ vertex with two unphysical gauge bosons \eqref{eq:hww-sti-2}:
\begin{equation}
 \parbox{15mm}{
  \begin{fmfchar*}(15,15)
  \fmftop{A1,A2} \fmfbottom{A3,A4} \fmf{double,label=$B_a$,label.side=right}{A1,a} \fmf{dashes,label=$H_i$,label.side=left}{A2,b}
  \fmf{dashes}{a,b} \fmf{double,label=$B_b$,label.side=left}{A3,a} \fmf{photon,label=$W_c$,label.side=right}{A4,b}
  \fmfdot{a}
  \fmfdot{b}  
\end{fmfchar*}}\qquad=g_{HWW}^{jab} T^c_{ji} (\epsilon_c\cdot p_i)
\end{equation}
Because of the STI for the $HWW$ vertex \eqref{eq:hww-sti}, the  $t$ and $u$ channel Higgs exchange diagrams satisfy the WI by themselves:
 \begin{equation}
 \parbox{15mm}{
  \begin{fmfchar*}(15,15)
    \fmftop{A1,A2} \fmfbottom{A3,A4} \fmf{double}{A1,a} \fmf{dashes}{A2,a}
    \fmf{dashes}{a,b} \fmf{double}{A3,b} \fmf{photon}{A4,b}
    \fmfdot{a}
    \fmfdot{b}  
\end{fmfchar*}}+
\parbox{15mm}{ \begin{fmfchar*}(15,15)
  \fmftop{A1,A2} \fmfbottom{A3,A4} \fmf{double}{A1,a} 
  \fmf{dashes}{a,b} \fmf{double}{A3,b} 
\fmf{phantom}{A2,a}
\fmf{phantom}{A4,b}
\fmffreeze
\fmf{dashes}{A2,b}
\fmf{photon}{A4,a}
 \fmfdot{a}
  \fmfdot{b}  
\end{fmfchar*}}=0
 \end{equation}
The only diagrams with quartic vertices are two $WWH\phi$ diagrams that give
\begin{equation}
  \parbox{15mm}{
  \begin{fmfchar*}(15,15)
    \fmftop{A1,A2} \fmfbottom{A3,A4} \fmf{photon,label=$W_a$,label.side=left}{A1,a} \fmf{dashes,label=$H_i$,label.side=left}{A2,a}
 \fmf{dashes,label=$\phi_b$,label.side=left}{A3,a} \fmf{photon,label=$W_c$,label.side=left}{A4,a}
  \fmfdot{a}  \fmfv{decor.shape=cross,decor.size=5}{A3}
\fmfv{decor.shape=square,decor.filled=full,decor.size=3}{A1}
\end{fmfchar*}}
\quad+\quad \parbox{15mm}{ \begin{fmfchar*}(15,15)
    \fmftop{A1,A2} \fmfbottom{A3,A4} \fmf{dashes,label=$\phi_a$,label.side=left}{A1,a} \fmf{dashes,label=$H_i$,label.side=left}{A2,a}
 \fmf{photon,label=$W_b$,label.side=left}{A3,a} \fmf{photon,label=$W_c$,label.side=left}{A4,a}
  \fmfdot{a}
 \fmfv{decor.shape=square,decor.filled=full,decor.size=3}{A3}
\end{fmfchar*}}\quad =(\ii)^2 \left [m_{W_a}g_{H\phi
    W^2}^{iabc}(p_b\cdot\epsilon_c)+m_{W_b} g_{H\phi W^2}^{ibac}(p_a\cdot\epsilon_c)\right]
\end{equation} 
The coefficients $\propto (\epsilon_c\cdot p_a)$ give:
\begin{multline}
  m_{W_b} g_{H\phi W^2}^{ibac}=\frac{1}{2} g_{HWW}^{ice}
   f^{abe}\left(1-\frac{m_{W_a}^2-m_{W_b}^2}{m_{W_e}^2}\right) +\frac{1}{2}
   f^{ace}g_{HWW}^{ibe}-g_{HWW}^{jab} T^c_{ji}
 \end{multline}
 The terms $\propto (\epsilon_c\cdot p_b)$ give a similar condition.
\subsection{$4W$ Ward identity with 3 unphysical  $W$s}\label{app:4w-3co}
Using the STI for the triple gauge boson vertex \eqref{eq:www-sti} and the contracted STI for the triple gauge boson vertex with two unphysical gauge bosons \eqref{eq:amp_www_sti} the  $s$-channel diagram is 
\begin{multline}
 \parbox{15mm}{
  \begin{fmfchar*}(15,15)
  \fmftop{A1,A2} \fmfbottom{A3,A4} \fmf{double,label=$B_a$,label.side=right}{A1,a} \fmf{photon,label=$W_d$,la.si=left}{A2,b}
  \fmf{photon}{a,b} \fmf{double,label=$B_b$,la.si=left}{A3,a} \fmf{double,label=$B_c$,la.si=right}{A4,b}
  \fmfdot{a}
  \fmfdot{b}  
\end{fmfchar*}}\\
=-(\ii)^2\frac{1}{2}
  f^{abe}f^{cde}\left[-(p_{a\mu}-p_{b\mu})+\frac{m_{W_a}^2-m_{W_b}^2}{m_{W_e}^2}(p_{a\mu}+p_{b\mu})\right]\times\\
\left[(s-m_{W_e}^2)\epsilon_d^\mu+(p_a^\mu
  +p_b^\mu)(p_c\cdot\epsilon_d)\right]\\
=\frac{1}{2}
  f^{abe}f^{cde}\Bigl\{(s-m_{W_e}^2)[-(p_a\cdot\epsilon_d)+(p_b\cdot\epsilon_d)]-(p_c\cdot\epsilon_d)[(p_a^2-m_{W_a}^2)-(p_b^2-m_{W_b}^2)]\Bigr\}
\end{multline}
The  $t$-  and $u$-channel  diagrams are obtained by index exchange.
The Higgs exchange diagrams all satisfy the WI by themselves:
\begin{equation}
 \parbox{15mm}{
  \begin{fmfchar*}(15,15)
  \fmftop{A1,A2} \fmfbottom{A3,A4} \fmf{double,label=$B_a$,label.side=right}{A1,a} \fmf{photon,label=$W_d$,la.si=left}{A2,b}
  \fmf{dashes}{a,b} \fmf{double,label=$B_b$,la.si=left}{A3,a} \fmf{double,label=$B_c$,la.si=right}{A4,b}
  \fmfdot{a}
  \fmfdot{b}  
\end{fmfchar*}}\qquad=0
\end{equation}
The diagrams with quartic vertices contain one diagram with 4 gauge bosons
 \begin{multline}
  \parbox{15mm}{
  \begin{fmfchar*}(15,15)
    \fmftop{A1,A2} \fmfbottom{A3,A4} \fmf{photon}{A1,a} \fmf{photon}{A2,a}
 \fmf{photon}{A3,a} \fmf{photon}{A4,a}
  \fmfdot{a} 
\fmfv{decor.shape=square,decor.filled=full,decor.size=3}{A1}
\fmfv{decor.shape=square,decor.filled=full,decor.size=3}{A3}
\fmfv{decor.shape=square,decor.filled=full,decor.size=3}{A4}
\end{fmfchar*}}
=\begin{aligned}
&(-\ii)^4 \Bigl[ (f^{abe} f^{cde}-f^{ade}f^{bce})(p_a\cdot
  p_c)(p_b\cdot\epsilon_d)\\
&+(f^{ace} f^{bde}+f^{ade}f^{bce})(p_a\cdot p_b)(p_c\cdot\epsilon_d) \\
&-(f^{abe} f^{cde}+f^{ace}f^{bde})(p_b\cdot p_c)(p_a\cdot\epsilon_d) \Bigr]
 \end{aligned}
 \end{multline}
and 3 $WW\phi\phi$ diagrams :
\begin{multline}
  \parbox{15mm}{
  \begin{fmfchar*}(15,15)
    \fmftop{A1,A2} \fmfbottom{A3,A4} \fmf{dashes}{A1,a} \fmf{photon}{A2,a}
 \fmf{dashes}{A3,a} \fmf{photon}{A4,a}
  \fmfdot{a}  \fmfv{decor.shape=cross,decor.size=5}{A1,A3}
\fmfv{decor.shape=square,decor.filled=full,decor.size=3}{A4}
\end{fmfchar*}}
+ \parbox{15mm}{ \begin{fmfchar*}(15,15)
    \fmftop{A1,A2} \fmfbottom{A3,A4} \fmf{dashes}{A1,a} \fmf{photon}{A2,a}
 \fmf{photon}{A3,a} \fmf{dashes}{A4,a}
  \fmfdot{a} \fmfv{decor.shape=cross,decor.size=5}{A1,A4}
 \fmfv{decor.shape=square,decor.filled=full,decor.size=3}{A3}
\end{fmfchar*}}
+ \parbox{15mm}{ \begin{fmfchar*}(15,15)
    \fmftop{A1,A2} \fmfbottom{A3,A4} \fmf{photon}{A1,a} \fmf{photon}{A2,a}
 \fmf{dashes}{A3,a} \fmf{dashes}{A4,a}
  \fmfdot{a} \fmfv{decor.shape=cross,decor.size=5}{A4,A3}
\fmfv{decor.shape=square,decor.filled=full,decor.size=3}{A1}
\end{fmfchar*}}
=\begin{aligned}
&-(\ii)^2  \Bigl [m_{W_a}m_{W_b}g_{\phi^2 W^2}^{abcd}(p_c\cdot \epsilon_d)\\
&+m_{W_b}m_{W_c} g_{\phi^2 W^2}^{cbad}(p_a\cdot \epsilon_d)\\
&+_{W_a}m_{W_c} g_{\phi^2 W^2}^{acbd}(p_b\cdot \epsilon_d)\Bigr]]
 \end{aligned}
\end{multline} 
Eliminating $p_c$ and collecting all terms $\propto (p_a\cdot \epsilon_d)$ we get 
\begin{multline}\label{eq:4w-3cov-lhs}
-\frac{1}{2}
  f^{abe}f^{cde}\Bigl\{(s-m_{W_e}^2)-[(p_a^2-m_{W_a}^2)-(p_b^2-m_{W_b}^2)]+2(p_b\cdot p_c)\Bigr\}\\
-\frac{1}{2}
  f^{ade}f^{bce}\Bigl\{(t-m_{W_e}^2)+[(p_b^2-m_{W_b}^2)-(p_c^2-m_{W_c}^2)]+2(p_a\cdot p_b)\Bigr\}\\
-f^{ace}f^{bde}\Bigl\{(u-m_{W_e}^2)+[(p_a\cdot p_b)+(p_b\cdot p_c)\Bigr\}-m_{W_a}m_{W_b}g_{\phi^2 W^2}^{abcd}+m_{W_c}m_{W_b}g_{\phi^2 W^2}^{cbad}
\end{multline}
A tedious calculation using the Jacobi Identity, the result for the quartic coupling \eqref{eq:2phi-jac} and the identity $s+t+u=p_a^2+p_b^2+p_c^2+m_d^2$ results in the condition
\begin{multline}\label{eq:3pow4w}
f^{ace}f^{ebd}(2m_{W_e}^2-m_{W_a}^2-m_{W_b}^2-m_{W_c}^2-m_{W_d}^2)=\\
-f^{aed}f^{cbe}\left[m_{W_e}^2+\frac{(m_{W_a}^2-m_{W_d}^2)(m_{W_c}^2-m_{W_b}^2)}{m_{W_e}^2}\right]+g_{HWW}^{ibc} g_{HWW}^{iad}- \qquad
    a\leftrightarrow c
\end{multline}
This is the form given in \fref{eq:4w-wi3} and in \cite{Gunion:1991}. To show that this is indeed the $ab$ component of the commutator relation 
\eqref{eq:ssb-kommutator} we work backward:
\begin{multline}
  [t^a,t^c]-[g^a,g^c]=f^{ace}t^e\Rightarrow\quad 
  t^a_{be}t^c_{ed}-t^c_{be}t^a_{ed}-g_{bi}^ag^c_{id}+g^c_{bi}g^a_{id}=f^{ace}t^e_{bd}\\
\Rightarrow \quad \frac{1}{m_{W_e}^2}
  \left[f^{abe}f^{ced}(m_{W_a}^2-m_{W_b}^2-m_{W_e}^2)(m_{W_c}^2-m_{W_e}^2-m_{W_d}^2)\right]-g_{HWW}^{iab}g_{HWW}^{icd} \\
-\quad a\leftrightarrow c  
=2f^{ace}f^{ebd}(m_{W_e}^2-m_{W_b}^2-m_{W_d}^2)
\end{multline}
From this we can derive
\begin{multline}
  2f^{ace}f^{ebd}(m_{W_e}^2-m_{W_b}^2-m_{W_d}^2) +(f^{abe}f^{ced}- f^{cbe}f^{aed})(m_{W_a}^2-m_{W_b}^2+m_{W_c}^2-m_{W_d}^2) \\
 = f^{abe}f^{ced}\left[\frac{(m_{W_a}^2-m_{W_b}^2)(m_{W_c}^2-m_{W_d}^2)}{m_{W_e}^2} +m_{W_e}^2  \right]-g_{HWW}^{iab}g_{HWW}^{icd} -\quad a\leftrightarrow c 
\end{multline}
and then using the Jacobi identity we find \fref{eq:3pow4w} 

\subsection{$\bar f f \to WW$ with 2 unphysical $W$ s}\label{subsec:2os-ffww-wi}
This time we have to use the STI \eqref{eq:ffw-sti} for both vertices so the $t$-channel Compton diagram gives
\begin{equation}
\parbox{15mm}{
\begin{fmfchar*}(15,15)
  \fmfleft{f1,f2} \fmfright{A,H} \fmf{fermion,label=$f_j$}{f1,a}
  \fmf{fermion,label=$f_i$, label.side=left}{b,f2}
  \fmf{fermion,label=$f_l$}{a,b} \fmf{double,label=$B_a$, label.side=right}{A,a} \fmf{double,label=$B_b$,label.side=left}{H,b}
 \fmfdot{a}
  \fmfdot{b}  
 \end{fmfchar*}}\quad
\begin{aligned}
=&\ii\bar v_i(p_i)(g_{Vil}^b -g_{Ail}^b\gamma^5)(\fmslash p_i+\fmslash p_b+m_{f_l})(g_{Vlj}^a+ g_{Alj}^a\gamma^5)u_j(p_j)\\
=&\ii\bar v_i(p_i)\fmslash p_b \left[\tau_{Lil}^b\tau_{Llj}^a(\tfrac{1-\gamma^5}{2})+\tau_{Ril}^b \tau_{Rlj}^a\tfrac{1+\gamma^5}{2})  \right]u_j(p_j)\\
&+m_{W_b}\bar v_i(p_i)\left[g_{\phi il}^{b\dagger}\tau_{Rlj}^a(\tfrac{1+\gamma^5}{2})+g_{\phi il}^b\tau_{Llj}^a(\tfrac{1-\gamma^5}{2})\right]u_j(p_j)
\end{aligned}
\end{equation}
We could obtain the $u$-channel diagram simply by exchanging $a$ and $b$ but it will be more useful below to make the first term in both diagrams proportional to $\fmslash p_b$. Then we obtain for this diagram
\begin{equation}
\parbox{15mm}{
\begin{fmfchar*}(15,15)
  \fmfleft{f1,f2} \fmfright{A,H} \fmf{fermion,label=$f_j$}{f1,a}
  \fmf{fermion,label=$f_i$, label.side=left}{b,f2}
  \fmf{fermion,label=$f_l$}{a,b} 
\fmf{phantom}{A,a} \fmf{phantom}{H,b}\fmffreeze 
  \fmf{double}{A,b} \fmf{double}{H,a}
 \fmfdot{a}
  \fmfdot{b}  
\end{fmfchar*}}\quad
\begin{aligned}
=&\ii\bar v_i(p_i)(g_{Vil}^a -g_{Ail}^a\gamma^5)(-\fmslash p_j-\fmslash p_b+m_{f_l})(g_{Vlj}^b+ g_{Alj}^b\gamma^5)u_j(p_j)\\
=&-\ii\bar v_i(p_i)\fmslash p_b \left[\tau_{Lil}^a\tau_{Llj}^b(\tfrac{1-\gamma^5}{2})+\tau_{Ril}^a \tau_{Rlj}^b\tfrac{1+\gamma^5}{2})  \right]u_j(p_j)\\
&-m_{W_b}\bar v_i(p_i)\left[\tau_{Lil}^ag_{\phi lj}^{b\dagger}(\tfrac{1+\gamma^5}{2})+\tau_{Ril}^ag_{\phi lj}^b(\tfrac{1-\gamma^5}{2})\right]u_j(p_j)
\end{aligned}
\end{equation}
The sum of $t$- and $u$ channel diagrams gives therefore
\begin{multline}\label{eq:compton-2cov}
\parbox{15mm}{
\begin{fmfchar*}(15,15)
  \fmfleft{f1,f2} \fmfright{A,H} \fmf{fermion}{f1,a}
  \fmf{fermion, label.side=left}{b,f2}
  \fmf{fermion}{a,b} \fmf{double}{A,a} \fmf{double,label.side=left}{H,b}
 \fmfdot{a}
  \fmfdot{b}  
 \end{fmfchar*}}+\parbox{15mm}{
\begin{fmfchar*}(15,15)
  \fmfleft{f1,f2} \fmfright{A,H} \fmf{fermion}{f1,a}
  \fmf{fermion}{b,f2}
  \fmf{fermion}{a,b} 
\fmf{phantom}{A,a} \fmf{phantom}{H,b}\fmffreeze 
  \fmf{double}{A,b} \fmf{double}{H,a}
 \fmfdot{a}
  \fmfdot{b}  
\end{fmfchar*}}=
\ii\bar v_i(p_i)\fmslash p_b \left[[\tau_{L}^b,\tau_{L}^a]_{ij}(\tfrac{1-\gamma^5}{2})+[\tau_{R}^b, \tau_{R}^a]_{ij}\tfrac{1+\gamma^5}{2})  \right] u_j(p_j)\\
+m_{W_b} \bar v_i(p_i)\left[(g_{\phi il}^{b\dagger}\tau_{Rlj}^a  -\tau_{Lil}^ag_{\phi lj}^{b\dagger}) (\tfrac{1+\gamma^5}{2})+(g_{\phi il}^b\tau_{Llj}^a-\tau_{Ril}^ag_{\phi lj}^b )(\tfrac{1-\gamma^5}{2})\right]u_j(p_j)
\end{multline}
To compute the $s$-channel gauge boson exchange diagram we need the STI for the $WWW$ vertex  contracted with a propagator \eqref{eq:amp_www_sti}. Therefore the $s$-channel gauge boson exchange diagram is
\begin{equation}\label{eq:2pow-ffww-4}
\parbox{15mm}{
\begin{fmfchar*}(15,15)
  \fmfleft{f1,f2} \fmfright{A,H} \fmf{fermion,label=$f_j$, label.side=left}{f1,a}
  \fmf{fermion,label=$f_i$, label.side=left}{a,f2}
  \fmf{photon,label=$W_c$}{a,b} \fmf{double,label=$B_a$, label.side=right}{A,b} \fmf{double,label=$B_b$,label.side=left}{H,b}
 \fmfdot{a}
  \fmfdot{b}  
\end{fmfchar*}}\qquad
\begin{aligned}
=&(\ii^2)\bar v_i(p_i)\left[f^{abc}\fmslash p_b-\frac{m_{W_b}}{m_{W_c}}(\fmslash p_a+\fmslash p_b) t^a_{cb}\right](g_{Vij}^c+g_{Aij}^c\gamma^5)u_j(p_j)\\
=&-\bar v_i(p_i)\Bigl[(f^{abc}\tau_{Lij}^c\fmslash p_b-\ii m_{W_b}g_{\phi ij}^c t^a_{cb})(\tfrac{1-\gamma^5}{2})\\
&+(f^{abc}\tau_{Rij}^c\fmslash p_b -\ii m_{W_b}g_{\phi ij}^{c\dagger} t^a_{cb})(\tfrac{1+\gamma^5}{2})\Bigr]u_j(p_j)
\end{aligned}
\end{equation}
Using the STI \eqref{eq:hww-sti-2} for the $HWW$ vertex with two unphysical gauge bosons we get for the $s$-channel Higgs-exchange diagram
\begin{equation}\label{eq:2pow-ffww-5}
\parbox{15mm}{
\begin{fmfchar*}(15,15)
  \fmfleft{f1,f2} \fmfright{A,H} \fmf{fermion,label=$f_j$, label.side=left}{f1,a}
  \fmf{fermion,label=$f_i$, label.side=left}{a,f2}
  \fmf{dashes,label=$H_k$}{a,b} \fmf{double,label=$B_a$, label.side=right}{A,b} \fmf{double,label=$B_b$,label.side=left}{H,b}
 \fmfdot{a}
  \fmfdot{b}  
\end{fmfchar*}}\qquad=-\ii g_{HWW}^{kab}\bar v_i(p_i)(g_{H ij}^k(\tfrac{1-\gamma^5}{2})+ g_{H ij}^{k\dagger}(\tfrac{1+\gamma^5}{2}))u_j(p_j)
\end{equation}
Adding up the different contributions we see after using the lie algebra of the fermion generators \eqref{eq:fermion-lie} that the first line of \fref{eq:compton-2cov} cancels the terms with $\fmslash p_b$ in the gauge boson exchange diagram \eqref{eq:2pow-ffww-4}.

The remaining terms $\propto (1-\gamma^5)$  give the condition
\begin{equation}\label{eq:gphif-cond}
\ii g_{\phi il}^b\tau_{Llj}^a  -\ii \tau_{Ril}^ag_{\phi lj}^b= g_{\phi ij}^c t^a_{cb}-g_{H ij}^k \frac{g_{HWW}^{kab}}{2m_{W_b}}
\end{equation}
and the terms $\propto (1+\gamma^5)$ similarly the hermitian conjugate equation. 
 \subsection{$2W2H$ Ward identity with 2 unphysical $W$s}
\subsubsection{Higgs-Exchange}
 Using the STI for the $WWH$ vertex with two contractions \eqref{eq:hww-sti-2} for the $WWH$ vertex, the   $s$-channel Higgs-exchange diagram is given by
 \begin{equation}
 \parbox{15mm}{
  \begin{fmfchar*}(15,15)
  \fmftop{A1,A2} \fmfbottom{A3,A4} \fmf{double,label=$B_a$,label.side=right}{A1,a} \fmf{dashes,label=$H_i$,la.si=left}{A2,b}
  \fmf{dashes}{a,b} \fmf{double,label=$B_b$,la.si=left}{A3,a} \fmf{dashes,label=$H_j$,la.si=right}{A4,b}
  \fmfdot{a}
  \fmfdot{b}  
\end{fmfchar*}}\quad =\quad -ig_{H^3}^{ijk}\frac{g_{HWW}^{kab}}{2}
 \end{equation}

 From the STI \eqref{eq:hhw-sti} the  $t$- and $u$-channel diagrams are found to be
 \begin{multline}
\parbox{15mm}{
  \begin{fmfchar*}(15,15)
    \fmftop{A1,A2} \fmfbottom{A3,A4} \fmf{double}{A1,a} \fmf{dashes}{A2,a}
    \fmf{dashes}{a,b} \fmf{double}{A3,b} \fmf{dashes}{A4,b}
    \fmfdot{a}
    \fmfdot{b}
\end{fmfchar*}}
 + \parbox{15mm}{ \begin{fmfchar*}(15,15)
  \fmftop{A1,A2} \fmfbottom{A3,A4} \fmf{double}{A1,a} 
  \fmf{dashes}{a,b} \fmf{double}{A3,b} 
\fmf{phantom}{A2,a}
\fmf{phantom}{A4,b}
\fmffreeze
\fmf{dashes}{A2,b}
\fmf{dashes}{A4,a}
 \fmfdot{a}
  \fmfdot{b}  
\end{fmfchar*}}\\
=  \ii\left[T^a_{ik}T^b_{kj}((p_a+k_i)^2-m_{H_k}^2)+T^b_{ik}T^a_{kj}((p_a+k_j)^2-m_{H_k}^2)\right]
 \end{multline}
 \subsubsection{Gauge boson exchange}
 The $s$-channel gauge boson exchange diagram can be computed using the STI for the triple gauge boson vertex with 2 contractions in the form of \fref{eq:amp_www_sti}
 \begin{equation}
\parbox{15mm}{
  \begin{fmfchar*}(15,15)
  \fmftop{A1,A2} \fmfbottom{A3,A4} \fmf{double}{A1,a} \fmf{dashes}{A2,b}
  \fmf{photon}{a,b} \fmf{double}{A3,a} \fmf{dashes}{A4,b}
  \fmfdot{a}
  \fmfdot{b}  
\end{fmfchar*}}\quad 
\begin{aligned}
&=\ii T^c_{ij}(k_i-k_j)^\mu
 [ f^{abc}p_b^\mu  +\frac{m_{W_b}}{m_{W_c}}p_c^\mu t^a_{cb}]\\
&=\ii T^c_{ij}
  f^{abc}(k_i-k_j)\cdot p_b +\ii m_{W_b}t^a_{cb}g_{\phi HH}^{cij}
\end{aligned}
\end{equation}
The $u$- and $t$-channel gauge boson exchange diagrams are obtained using the STI for the $HWW$ vertex \eqref{eq:hww-sti}: 
 \begin{multline}
\parbox{15mm}{
  \begin{fmfchar*}(15,15)
    \fmftop{A1,A2} \fmfbottom{A3,A4} \fmf{double}{A1,a} \fmf{dashes}{A2,a}
    \fmf{photon}{a,b} \fmf{double}{A3,b} \fmf{dashes}{A4,b}
    \fmfdot{a}
    \fmfdot{b}
\end{fmfchar*}}
 + \parbox{15mm}{ \begin{fmfchar*}(15,15)
  \fmftop{A1,A2} \fmfbottom{A3,A4} \fmf{double}{A1,a} 
  \fmf{photon}{a,b} \fmf{double}{A3,b} 
\fmf{phantom}{A2,a}
\fmf{phantom}{A4,b}
\fmffreeze
\fmf{dashes}{A2,b}
\fmf{dashes}{A4,a}
 \fmfdot{a}
  \fmfdot{b}  
\end{fmfchar*}}
= \begin{aligned}\ii (\frac{1}{4m_{W_d}^2})g_{HWW}^{iac} g_{HWW}^{jcb}(k_j+p_b)\cdot(k_i+p_a)\\
+\ii(\frac{1}{4m_{W_d}^2})g_{HWW}^{jac}
 g_{HWW}^{icb}(k_i+p_b)\cdot(k_j+p_a)
  \end{aligned}
 \end{multline}
 \subsubsection{Quartic vertices}
The terms with the quartic vertices give
 \begin{equation}
 \parbox{15mm}{ \begin{fmfchar*}(15,15)
  \fmftop{A1,A2} \fmfbottom{A3,A4} \fmf{double}{A1,a} 
  \fmf{double}{A3,a} 
\fmf{dashes}{A2,a}
\fmf{dashes}{A4,a}
 \fmfdot{a}
\end{fmfchar*}}= -\ii g_{H^2 W^2}^{abij}(p_a\cdot p_b)+\ii m_{W_a}m_{W_b} g_{\phi^2 H^2}^{abij}
 \end{equation}
\subsubsection{Condition on the coupling constants}
Adding up the contributions from the diagrams, we obtain the condition
\begin{multline}\label{eq:w2h2-2cov-res}
m_{W_a}m_{W_b} g_{\phi^2 H^2}^{abij}-g_{H^2 W^2}^{abij}(p_a\cdot p_b)\\
=-T^a_{ik}T^b_{kj}((p_a+k_i)^2-m_{H_k}^2)-T^b_{ik}T^a_{kj}((p_a+k_j)^2-m_{H_k}^2)\\
- T^c_{ij}
  f^{abc}(k_i-k_j)\cdot p_b - m_{W_b}t^a_{cb}g_{\phi HH}^{cij} +\frac{1}{4m_{W_d}^2}g_{HWW}^{iac} g_{HWW}^{jcb}(k_j+p_b)^2\\
+\frac{1}{4m_{W_d}^2}g_{HWW}^{jac} g_{HWW}^{icb}(k_i+p_b)^2+g_{H^3}^{ijk}\frac{g_{HWW}^{kab}}{2}
\end{multline}
After another tedious calculation, using  the expression for  $g_{H^2W^2}$ from \fref{eq:2h2w-wi}, we find that the coefficients of the terms without momenta give the condition
\begin{multline}\label{eq:h2phi2-app}
m_{W_a}m_{W_b}g_{\phi^2H^2}^{ijab}= m_{W_a}t^a_{cb}g_{\phi H^2}^{cij}+\frac{1}{2}g_{HWW}^{kab}g_{H^3}^{ijk}
+T^a_{ki}T^b_{kj}\left(m_{H_j}^2-m_{H_k}^2\right)\\
+\frac{m_{H_j}^2}{4m_{W_c}^2} g_{HWW}^{iac}g_{HWW}^{jbc}
+ T^a_{kj}T^b_{ki}\left(m_{H_i}^2-m_{H_k}^2\right) +\ii \frac{m_{H_i}^2}{4 m_{W_c}^2}g_{HWW}^{jac}g_{HWW}^{ibc}
\end{multline}
Inserting the relations for the triple scalar couplings from \eqref{subeq:triple-phi-wi1} and \eqref{subeq:triple-phi-wi2} , we see that this is the same as the invariance condition of the scalar potential \eqref{eq:h2phi2-inv}. 

The remaining terms are all proportional to $p_b\cdot (k_i-k_j)$ and their coefficients are
\begin{multline}
0=-f^{abc}T^c_{ij}-T^a_{ki}T^b_{kj}+T^a_{kj}T^b_{ki} -\frac{1}{4m_{W_c}^2} g_{HWW}^{iac}g_{HWW}^{jbc}+\frac{1}{4 m_{W_c}^2}g_{HWW}^{jac}g_{HWW}^{ibc}
\end{multline}
This is just the commutation relation from \fref{eq:2h2w-wi}
\subsection{$3WH$ Ward identity with 3 unphysical $W$s}
Using the STIs \eqref{eq:hww-sti} and \eqref{eq:amp_www_sti} the $s$-channel gauge boson exchange diagrams is given by
\begin{equation}
 \parbox{15mm}{
  \begin{fmfchar*}(15,15)
  \fmftop{A1,A2} \fmfbottom{A3,A4} \fmf{double,label=$B_a$,label.side=right}{A1,a} \fmf{dashes,label=$H_i$,la.si=left}{A2,b}
  \fmf{photon}{a,b} \fmf{double,label=$B_b$,la.si=left}{A3,a} \fmf{double,label=$B_c$,la.si=right}{A4,b}
  \fmfdot{a}
  \fmfdot{b}  
\end{fmfchar*}}\qquad=-\ii \frac{1}{2}g_{HWW}^{ice}\left[f^{bae}p_a\cdot (p_a+p_b)-\frac{m_a}{m_e}t^b_{ea} s\right]
\end{equation}
The $t$- and $u$-channel diagrams are obtained by permutations of the indices.

Using the STIs for the $HWW$ vertex with one  \eqref{eq:hhw-sti} and two contractions \eqref{eq:hww-sti-2} the $s$-channel Higgs exchange diagrams 
\begin{equation}
 \parbox{15mm}{
  \begin{fmfchar*}(15,15)
  \fmftop{A1,A2} \fmfbottom{A3,A4} \fmf{double,label=$B_a$,label.side=right}{A1,a} \fmf{dashes,label=$H_i$,la.si=left}{A2,b}
  \fmf{dashes}{a,b} \fmf{double,label=$B_b$,la.si=left}{A3,a} \fmf{double,label=$B_c$,la.si=right}{A4,b}
  \fmfdot{a}
  \fmfdot{b}  
\end{fmfchar*}}\qquad=+\ii\frac{1}{2}g_{HWW}^{jab}
  T^c_{ij}(s-m_j^2 ) 
\end{equation}
The diagrams with quartic vertices consist of three
$WWH\phi$ diagrams:
\begin{multline}
  \parbox{15mm}{
  \begin{fmfchar*}(15,15)
    \fmftop{A1,A2} \fmfbottom{A3,A4} \fmf{photon}{A1,a} \fmf{dashes}{A2,a}
 \fmf{dashes}{A3,a} \fmf{photon}{A4,a}
  \fmfdot{a} 
\fmfv{decor.shape=square,decor.filled=full,decor.size=3}{A1,A4}
 \fmfv{decor.shape=cross,decor.size=5}{A3}
\end{fmfchar*}}
+ \parbox{15mm}{ \begin{fmfchar*}(15,15)
    \fmftop{A1,A2} \fmfbottom{A3,A4} \fmf{dashes}{A1,a} \fmf{dashes}{A2,a}
 \fmf{photon}{A3,a} \fmf{photon}{A4,a}
  \fmfdot{a}
 \fmfv{decor.shape=square,decor.filled=full,decor.size=3}{A3,A4}
\fmfv{decor.shape=cross,decor.size=5}{A1}
\end{fmfchar*}}
+ \parbox{15mm}{ \begin{fmfchar*}(15,15)
    \fmftop{A1,A2} \fmfbottom{A3,A4} \fmf{photon}{A1,a} \fmf{dashes}{A2,a}
 \fmf{photon}{A3,a} \fmf{dashes}{A4,a}
  \fmfdot{a}
 \fmfv{decor.shape=square,decor.filled=full,decor.size=3}{A1,A3}
\fmfv{decor.shape=cross,decor.size=5}{A4}
\end{fmfchar*}}\\
=-(\ii)^3 \left [m_{W_a}g_{H\phi
    W^2}^{iabc}(p_b\cdot p_c)+m_{W_b} g_{H\phi W^2}^{ibac}(p_a\cdot p_c)+m_{W_c} g_{H\phi W^2}^{icab}(p_a\cdot p_b)\right]
\end{multline} 
and one $\phi^3 H$ diagram:
\begin{equation}
  \parbox{15mm}{
  \begin{fmfchar*}(15,15)
    \fmftop{A1,A2} \fmfbottom{A3,A4} 
    \fmf{dashes,label=$\phi_a$}{A1,a} 
    \fmf{dashes,label=$H_i$}{A2,a}
    \fmf{dashes,label=$\phi_b$}{A3,a} 
    \fmf{dashes,label=$\phi_c$}{A4,a}
  \fmfdot{a}\fmfv{decor.shape=cross,decor.size=5}{A1,A3,A4} 
\end{fmfchar*}}
=-\ii m_{W_a}m_{W_b}m_{W_c}g_{H\phi^3}^{iabc}
\end{equation} 
To obtain the conditions on the coupling constants, we eliminate $p_i$, $s$,$t$,$u$ and the momentum bilinear $(p_a\cdot p_c)$ by momentum conservation. 
The terms without momentum dependence give the condition
\begin{multline}
m_{W_a}m_{W_b}m_{W_c}g_{H\phi^3}^{iabc}-\frac{1}{2}m_{W_b}m_{H_i^2} g_{H\phi W^2}^{ibac}=-\frac{1}{4}m_{H_i}^2 g_{HWW}^{ibe}\left[f^{cae}-2\frac{m_a}{m_e}t^c_{ea} \right]\\
-\frac{1}{2}m_j^2T^c_{ij} g_{HWW}^{jab}-\frac{1}{2}m_j^2T^a_{ij} g_{HWW}^{jbc}+\frac{1}{2}g_{HWW}^{jac}T^b_{ij}(m_{H_i}^2-m_{H_j}^2 ) 
\end{multline}
Inserting $g_{H\phi W^2}$ from \fref{eq:3wh-wi} results in 
\begin{multline}\label{eq:3wh-3cov}
m_{W_a}m_{W_b}m_{W_c}g_{H\phi^3}^{iabc}=
-\frac{m_{H_i}^2m_b}{2m_e} g_{HWW}^{ice}t^a_{be} +\frac{1}{2}(m_{H_i}^2-m_{H_j}^2) g_{HWW}^{jcb} T^c_{ij}\\
- \frac{1}{2}m_{H_i}^2 g_{HWW}^{ibe}\frac{m_c}{m_e}t^a_{ce}-\frac{1}{2}m_j^2T^a_{ij} g_{HWW}^{jcb}+\frac{1}{2}g_{HWW}^{jac}T^b_{ij}(m_{H_i}^2-m_{H_j}^2 ) 
\end{multline}
This reproduces the invariance condition \eqref{eq:gphi3h}. Using the relations \eqref{eq:3wh-wi} and \eqref{eq:hphi-jac-exp} for $g_{H\phi W^2}$ and the commutation relation \eqref{eq:3wh-wi2} one can show that the momentum dependent terms vanish.
\subsection{$WW\to WW$ with 4 contractions}
\subsubsection{Gauge boson exchange}
Using the STI for the triple gauge boson vertex with two contractions in the forms \eqref{eq:www-sti2a} and \eqref{eq:amp_www_sti}, the $s$-channel diagram is 
\begin{multline}\label{eq:4cov-s}
 \parbox{15mm}{
  \begin{fmfchar*}(15,15)
  \fmftop{A1,A2} \fmfbottom{A3,A4} \fmf{double,label=$B_a$,label.side=right}{A1,a} \fmf{double,label=$B_d$,la.si=left}{A2,b}
  \fmf{photon}{a,b} \fmf{double,label=$B_b$,la.si=left}{A3,a} \fmf{double,label=$B_c$,la.si=right}{A4,b}
  \fmfdot{a}
  \fmfdot{b}  
\end{fmfchar*}}\\
=(\ii)\frac{1}{4}
  f^{abe}f^{cde}\Biggl[(s-m_{W_e}^2)\left((p_a-p_b)\cdot(p_c-p_d)-\frac{m_{W_a}^2-m_{W_b}^2}{m_{W_e}^2}(p_c^2-p_d^2)\right)\\
-\left((p_a^2-p_b^2)-s\frac{m_{W_a}^2-m_{W_b}^2}{m_{W_e}^2}\right)[(p_c^2-m_c^2)-(p_d^2-m_d^2)] \Biggr]
\end{multline}
The  $t$-  and $u$-channel  diagrams are obtained by index exchange.

The $s$ channel Higgs exchange diagram is given by 
\begin{equation}
 \parbox{15mm}{
  \begin{fmfchar*}(15,15)
  \fmftop{A1,A2} \fmfbottom{A3,A4} \fmf{double,label=$B_a$,label.side=right}{A1,a} \fmf{double,label=$B_d$,la.si=left}{A2,b}
  \fmf{dashes}{a,b} \fmf{double,label=$B_b$,la.si=left}{A3,a} \fmf{double,label=$B_c$,la.si=right}{A4,b}
  \fmfdot{a}
  \fmfdot{b}  
\end{fmfchar*}}\qquad =
-\ii\frac{1}{4}g_{HWW}^{iab}g_{HWW}^{icd}(s-m_{H_i}^2)
\end{equation}
The diagrams with quartic vertices contain one diagram with 4 gauge bosons
 \begin{multline}
  \parbox{15mm}{
  \begin{fmfchar*}(15,15)
    \fmftop{A1,A2} \fmfbottom{A3,A4} \fmf{photon}{A1,a} \fmf{photon}{A2,a}
 \fmf{photon}{A3,a} \fmf{photon}{A4,a}
  \fmfdot{a} 
\fmfv{decor.shape=square,decor.filled=full,decor.size=3}{A1,A2,A3,A4}
\end{fmfchar*}}
=\begin{aligned}
&(-\ii)^4 \Bigl[ (f^{abe} f^{cde}-f^{ade}f^{bce})(p_a\cdot
  p_c)(p_b\cdot p_d)\\
&+(f^{ace} f^{bde}+f^{ade}f^{bce})(p_a\cdot p_b)(p_c\cdot p_d) \\
&-(f^{abe} f^{cde}+f^{ace}f^{bde})(p_b\cdot p_c)(p_a\cdotp_d) \Bigr]
 \end{aligned}
 \end{multline}
one quartic Goldstone boson diagram
\begin{equation}
 \parbox{15mm}{ \begin{fmfchar*}(15,15)
  \fmftop{A1,A2} \fmfbottom{A3,A4} \fmf{dashes}{A1,a} 
  \fmf{dashes}{A3,a} 
\fmf{dashes}{A2,a}
\fmf{dashes}{A4,a}
 \fmfdot{a}\fmfv{decor.shape=cross,decor.size=5}{A1,A2,A3,A4}
\end{fmfchar*}}=\ii g_{\phi^4}m_{W_a}m_{W_b}m_{W_c}m_{W_d}
\end{equation}
and 6 $WW\phi\phi$ diagrams :
\begin{multline}
  \parbox{15mm}{
  \begin{fmfchar*}(15,15)
    \fmftop{A1,A2} \fmfbottom{A3,A4} \fmf{dashes}{A1,a} \fmf{photon}{A2,a}
 \fmf{dashes}{A3,a} \fmf{photon}{A4,a}
  \fmfdot{a} 
\fmfv{decor.shape=square,decor.filled=full,decor.size=3}{A2,A4}
\fmfv{decor.shape=cross,decor.size=5}{A1,A3}
\end{fmfchar*}}
+ \dots
=\begin{aligned}
&(\ii)^3  \Bigl [m_{W_a}m_{W_b}g_{\phi^2 W^2}^{abcd}(p_c\cdot p_d)+m_{W_a}m_{W_c} g_{\phi^2 W^2}^{acbd}(p_b\cdot p_d)\\
&+m_{W_a}m_{W_d} g_{\phi^2 W^2}^{adbc}(p_b\cdot p_c)+m_{W_b}m_{W_c} g_{\phi^2 W^2}^{bcad}(p_a\cdot p_d)\\
&+m_{W_b}m_{W_d} g_{\phi^2 W^2}^{bdac}(p_a\cdot p_c)+m_{W_c}m_{W_d} g_{\phi^2 W^2}^{cdab}(p_a\cdot p_b)\Bigr]
 \end{aligned}
\end{multline} 
The terms without momentum dependence result in the condition
\begin{multline}
g_{\phi^4}m_{W_a}m_{W_b}m_{W_c}m_{W_d}=-\frac{m_{H_i}^2}{4}g_{HWW}^{iab}g_{HWW}^{icd}\\
-\frac{m_{H_i}^2}{4}g_{HWW}^{iac}g_{HWW}^{ibd}-\frac{m_{H_i}^2}{4}g_{HWW}^{iad}g_{HWW}^{ibc}
\end{multline}
This is the condition \eqref{eq:4w-4cov} for the quartic Goldstone boson coupling.


\cleardoublepage

\end{fmffile}

\end{document}